%% file: main.tex
\documentclass[showpacs,amsmath,amssymb,aps,twocolumn,superscriptaddress,rmp]{revtex4-2} 
\usepackage{mathtools}
\usepackage{algorithm}
\usepackage{algpseudocode}
\usepackage[pdftex]{graphicx} 
\graphicspath{{./visual_elements/.}}
\usepackage{grffile} 

\usepackage{comment}

\usepackage{physics}

\usepackage{xspace}

\usepackage{float}
\usepackage{dcolumn}
\usepackage{bm}
\usepackage[pdftex,colorlinks=true]{hyperref}
\hypersetup{
	colorlinks=true,
	linkcolor=blue,
	urlcolor=cyan,
}

\usepackage{tikz}
\usetikzlibrary{shapes,arrows,positioning, fit}

\usepackage{array}
\usepackage{amssymb} 
\usepackage{xcolor} 
\usepackage{booktabs} 
\usepackage{colortbl}
\usepackage{adjustbox}
\usepackage{pifont}

\newcommand{\cmarkb}{\textcolor{blue}{\checkmark}}
\newcommand{\cmarkr}{\textcolor{red}{\checkmark}}
\newcommand{\cmark}{\textcolor{black}{\checkmark}}
\newcommand{\xmark}{\textcolor{black}{\texttimes}}

\begin{document}



\title{Reinforcement Learning for Quantum Technology}

\author{Marin Bukov}
\affiliation{Max Planck Institute for the Physics of Complex Systems, N\"{o}thnitzer Str.~38, 01187 Dresden, Germany}
\email{mgbukov@pks.mpg.de}

\author{Florian Marquardt}
\affiliation{Max Planck Institute for the Science of Light, Staudtstrasse 2, 91058 Erlangen, Germany}
\affiliation{Physics Department, Friedrich-Alexander-Universit\"{a}t Erlangen-N\"{u}rnberg, 91058 Erlangen, Germany}
\email{florian.marquardt@mpl.mpg.de}
	

\begin{abstract}
Many challenges arising in Quantum Technology can be successfully addressed using a set of machine learning algorithms collectively known as reinforcement learning (RL), based on adaptive decision-making through interaction with the quantum device.
After a concise and intuitive introduction to RL aimed at a broad physics readership, we discuss the key ideas and core concepts in reinforcement learning with a particular focus on quantum systems. 
We then survey recent progress in RL in all relevant areas. 
We discuss state preparation in few- and many-body quantum systems, the design and optimization of high-fidelity quantum gates, and the automated construction of quantum circuits, including applications to variational quantum eigensolvers and architecture search. 
We further highlight the interactive capabilities of RL agents, emphasizing recent progress in quantum feedback control and quantum error correction, and briefly discuss quantum reinforcement learning as well as applications to quantum metrology. 
The review concludes with a discussion of open challenges -- such as scalability, interpretability, and integration with experimental platforms -- and outlines promising directions for future research.
Throughout, we highlight experimental implementations that exemplify the increasing role of reinforcement learning in shaping the development of quantum technologies.
\end{abstract}

\maketitle

\tableofcontents

\clearpage

\input{./intro}

\input{./rl_nutshell}

\input{./rl4qs}

\input{./rl_vs_qoc}

\input{./qstate_prep}

\input{./gate_engineering}

\input{./circuit_design}

\input{./feedback_ctrl}

\input{./qec}

\input{./qrl}

\input{./sensing}

\input{./outlook}

\acknowledgments{

M.~B.~is grateful to Pankaj Mehta, Markus Schmitt, Dries Sels, Lin Lin, 
Friederike Metz, Jiahao Yao, Ho Nam Nguyen,  Alexandre G.~R.~Day, Pavel Tashev, Stefan Petrov, and Giovanni Cemin  
for various stimulating and enlightening discussions on applying RL to physics problems. Likewise, F.M. thanks Remmy Zen, Jan Olle, Matteo Puviani, Petru Tighineanu, Talitha Lange (Weiß), Riccardo Porotti, and Vittorio Peano  for in-depth discussions in this realm, and above all Thomas Fösel for kickstarting RL for physics in our group.

M.B.~was funded by the European Union (ERC, QuSimCtrl, 101113633) and Deutsche Forschungsgemeinschaft (DFG, German Research Foundation) grant agreement No 544919793. For F.M., this research is part of the Munich Quantum Valley, which is supported by the Bavarian state government with funds from the Hightech Agenda Bayern Plus.

Views and opinions expressed are however those of the authors only and do not necessarily reflect those of the European Union or the European Research Council Executive Agency. Neither the European Union nor the granting authority can be held responsible for them. 
}


\bibliography{
./bib_files/sensing, 
./bib_files/circuit_design,
./bib_files/qec,
./bib_files/feedback,
./bib_files/gate_impl,
./bib_files/general_RL,
./bib_files/qrl,
./bib_files/state_prep
}

\end{document}

%% file: intro.tex
\section{\label{sec:intro}Introduction}

A major effort at the forefront of quantum science is currently devoted to developing quantum technologies, building upon four foundational pillars: 
(i) \textit{Quantum simulation} offers a new paradigm for investigating fundamental questions at the core of modern physics and promises to advance drug and material design on the practical side;
(ii) \textit{Quantum computing} predicts significant speedup for certain quantum-enhanced algorithms, such as the Quantum Fourier transform, Shor's factoring algorithm, and Grover's search algorithm, that could revolutionize how we process and manipulate digital data;
(iii) \textit{Quantum communication} provides novel protocols for cryptography and certification that ensure encrypted data is transmitted securely by exploiting the unique advantages of quantum mechanics; and
(iv) \textit{Quantum sensing \& metrology} have already found applications in microscopy, novel communication technologies, and positioning systems, and are currently considered of primary importance for military security applications.

However, before the full potential of quantum technologies can be realized, significant challenges have to be addressed at nearly every stage of the development pipeline. These include designing noise-resilient state preparation protocols and engineering electromagnetic pulses for quantum gates that surpass fault-tolerance thresholds. They also involve compiling circuits for variational quantum algorithms and transpiling them into native gates that match the constraints of specific quantum hardware platforms. In addition, advances are needed in interactive quantum feedback control, which integrates efficient error correction codes and stabilizes quantum states against decoherence and dissipation.

Current progress relies on confronting these challenges by identifying innovative theoretical concepts and developing more accurate models to advance the state of the art in experimental implementations. Being a data- and computation-intensive field, quantum technologies find a natural partner in the rapidly developing field of modern machine learning (ML)~\cite{mehta2019high,carleo2019machine,carrasquilla2020machine,dunjko2018machine,krenn2023artificial,dawid2022modern}. For instance, the key challenge of developing sophisticated control strategies for quantum devices, often requires feedback loops that apply operations based on measurement outcomes. However, optimal strategies are typically unknown, making this a particularly challenging yet important application area for machine learning approaches~\cite{ma2025machine}.

Modern machine learning is about making predictions from, and learning about, data, and comprises three main branches:
(i) \textit{supervised learning} deals with labeled data; perhaps the most well-known application is image recognition, which can be used in physics to classify phases of matter~\cite{carrasquilla2017machine,van2017learning}.
(ii) \textit{unsupervised learning} is about approximating the distribution that produced a given dataset, from which new data points can be generated. Prominent examples include generative AI, and include composing music, drawing paintings, and writing text; in quantum physics, a widespread application is neural quantum states~\cite{carleo2017solving,schmitt2020quantum}.
Finally, (iii) \textit{reinforcement learning} (RL) stands out as the subfield of ML where an agent learns as a result of repeated deliberate interactions with the outside world~\cite{sutton_barto_book}, instead of passive training on static data. 
Distinctive features of RL are the discovery of a strategy to interactively control the behavior of physical systems, and the ability to adapt the latter to identified changes in the dynamics on the fly.
As we will discuss in detail, reinforcement learning holds the promise to elevate quantum technologies to new heights by harnessing the capabilities of deep learning to enable progress in situations where no simple yet sufficiently accurate models for the physical system are known.

\begin{figure}[t!]
\centering
\includegraphics[width=0.95\columnwidth]{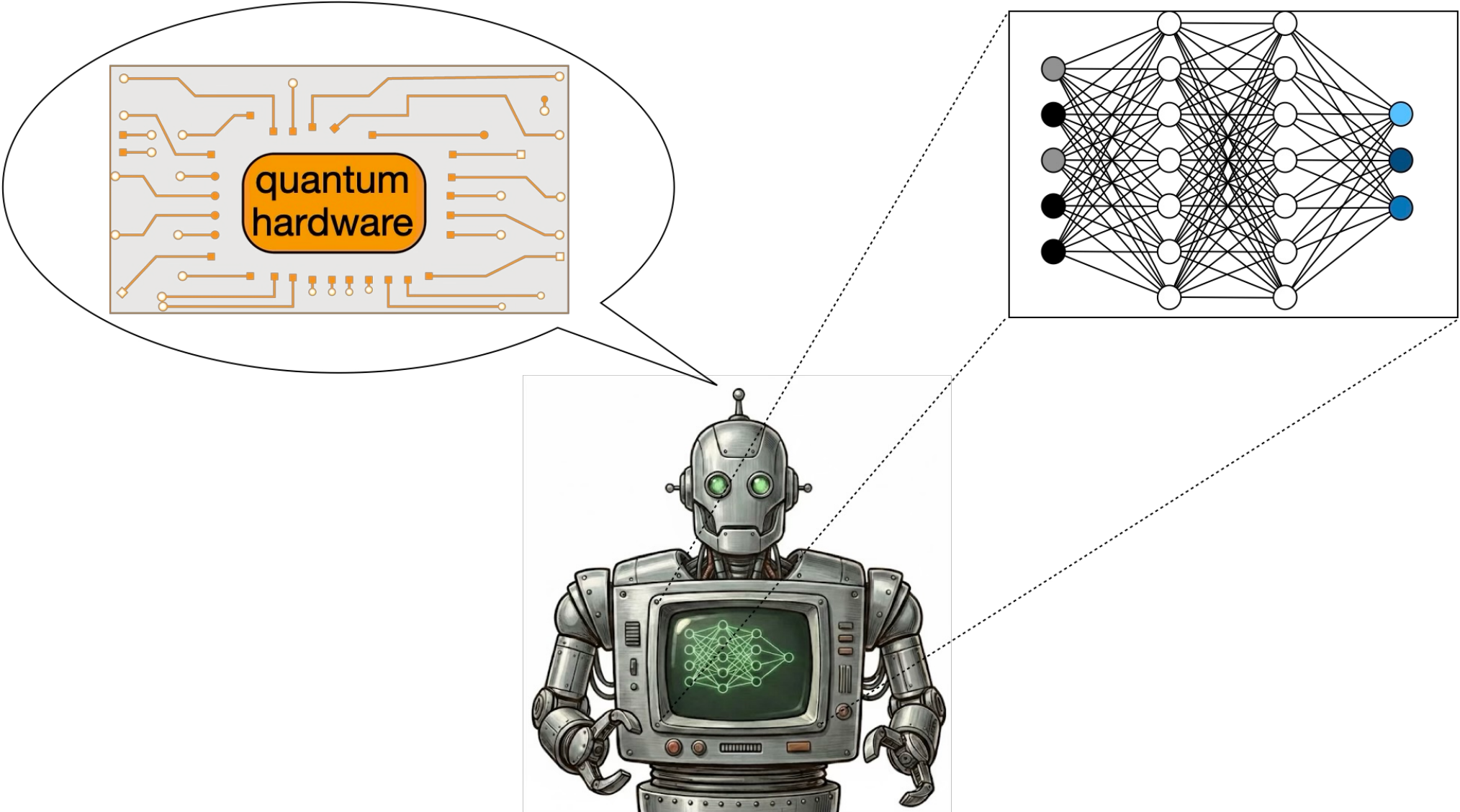}
\caption{
    \textbf{Reinforcement Learning} (RL) is a framework of (deep) learning algorithms that can be used for interactive feedback control of quantum devices. Illustration created in part using Google Gemini.
}
\label{fig:RL_schematic}
\end{figure}

\textbf{What is RL?}
The reinforcement learning framework consists of an agent that learns how to solve a task by interacting with a physical system, called the environment\footnote{\textit{Environment} in RL refers to the entire physical system, which is different from the notion of \textit{environment} in open systems.}. The agent can select actions that modify the state of its environment; as a consequence, a reward is given to the agent, and the process repeats in a positive feedback loop. The objective is to maximize the sum of expected rewards, and thereby accomplish the task. In doing so, the agent learns a strategy that can be employed under similar yet different circumstances.  
Besides its machine learning features, RL is intimately related to, and shares a common framework with, the modern theory of Optimal Control~\cite{todorov2006optimal}. Thus, many of the concepts inherent to RL are borrowed from control theory\footnote{RL has also borrowed elements from behavioral psychology~\cite {sutton_barto_book}.}, and hence the two fields can be considered the two sides of the same coin. That said, RL algorithms come with their own distinctive features, as we discuss in this review.  

\textbf{When is RL useful?}
A primary advantage of reinforcement learning algorithms is that they are model-free, i.e., they do not require a model for the physical system the agent interacts with. This allows agents to learn the relevant features of the system directly `from the source', thus circumventing unwanted effects caused by using inaccurate or approximate models. In this sense, we can say that model-free RL agents require no pre-knowledge of the controlled physical system and, therefore, produce unbiased strategies. 
A second noteworthy feature of deep RL agents is their adaptivity: once optimization is complete, the trained agent is ready to be deployed in new setups without further optimization, e.g., starting from different initial states; this can point to unexpected connections between seemingly unrelated systems. Moreover, the RL framework is particularly suitable for solving optimization problems in the presence of stochasticity.   
Third, RL is interactive, and as such, it is designed for feedback control. The ability to make decisions online that can optimize the outcome in the longer run, can be used to mitigate and counteract the effects of noise on the fly. 
And last but not least, RL agents are autonomous, which makes them good candidates for automating complex protocols, along the lines of some of the most prominent applications of modern ML; this is expected to have an impact on the future design of experimental apparatuses. 

These features are reflected in perhaps the most famous applications of RL outside physics nowadays -- finding hitherto unseen strategies to play board and video games. In 1995, Tesauro demonstrated the first application of an RL agent learning to play Backgammon without using a model~\cite{tesauro1995temporal}. With the advent of deep learning, this line of research was elevated to train the first RL agents to play Atari~\cite{mnih2015human} and other more complex video games. Soon after, the AlphaZero and AlphaGo algorithms demonstrated that deep RL agents can surpass professional human performance in playing board games as complex as Go and Chess~\cite{silver2017mastering}, which was recently contested~\cite{wang2023adversarial}. This clearly showcased RL's ability to discover highly efficient strategies for complex tasks, and attracted attention in other natural sciences. 

As of today, RL has found direct applications in designing molecules with desired properties for new drugs~\cite{popova2018deep}, floorplanning in chip development~\cite{mirhoseini2021graph}, high-performance flight controllers~\cite{bellemare2020autonomous}, as well as robotic engineering~\cite{singh2019end} and locomotion~\cite{margolis2024rapid}. RL applications in physics beyond quantum mechanics are widespread and include:
(i) navigating thermal currents~\cite{reddy2016learning} and turbulent flows~\cite{colabrese2017flow} in hydrodynamics; 
(ii) controlling plasma dynamics~\cite{degrave2022magnetic};
(iii) sampling rare trajectories~\cite{rose2021reinforcement,gillman2022reinforcement}, searching for ground states in spin glasses~\cite{fan2023searching}, and free energy learning~\cite{wu2019solving} in statistical mechanics;
(iv) optimal control of reaction-diffusion problems~\cite{schenk2024model} in biophysics;
and (v) exploring vacua in string theory~\cite{halverson2019branes}, just to name a few. 
Such interesting topics are unfortunately beyond the scope of this review; below, we will rather focus on applications to quantum technology.

\textbf{Why RL for quantum technology?}\
The probabilistic nature of measurements in quantum mechanics makes quantum systems a natural match for RL's ability to explore and control stochastic environments. Since stochasticity is intrinsic to quantum mechanics via the uncertainty principle, RL offers a natural framework for addressing outstanding challenges in quantum technologies.

In particular, RL is especially useful when the controlled quantum system has unknown noise sources (e.g., decoherence or dissipation), or when the exact Hamiltonian/Lindbladian is unknown and high-precision control is still required (e.g., in controlling complex quantum materials or quantum computing platforms).

Moreover, when designing strategies to manipulate quantum systems using weak and quantum non-demolition measurements, feedback control becomes essential, as strong projective measurements destroy the quantum state and are thus unsuitable for extracting information about the system. RL agents, as we will see, can learn from partial observations.

Last but not least, deep learning captures abstract features without a physical model; combined with RL's model-free decision-making, this can reveal effective degrees of freedom for quantum control. For instance, we learn to pour water into a glass without solving Newton's equations for the dynamics of $10^{23}$ molecules. However, in quantum many-body dynamics -- where macroscopic behavior is often poorly understood -- we often lack such intuition. Thus, RL can help identify effective macroscopic degrees of freedom in interacting quantum matter.

\textbf{Why now?} 
Recent progress in the development of various platforms for quantum technology has brought us to the doorstep of an era where quantum data is ubiquitous and abundant. Quantum simulators and quantum computers are in a constant struggle to produce measurement datasets faster and with higher quality than ever before. Moreover, quantum interactive dynamics, which relies on feedback and feedforward control based on intermittent midcircuit measurements, has emerged as a promising tool to manipulate correlated quantum states. It is, therefore, the right time to develop data-efficient RL algorithms tailored to the needs of quantum systems, and consider the potential of this framework to address outstanding challenges in quantum technology.

Inspired by the success of reinforcement learning algorithms in outperforming the best human players in complex games, such as Chess and Go, RL techniques have, in recent years, generated a large body of original theoretical work across the entire spectrum of quantum technologies. Diverse applications now span a wide range of modern problems, from preparing quantum states and optimizing expectation values of observables [Sec.~\ref{sec:state_prep}], to pulse engineering for implementing quantum gates in quantum computing devices [Sec.~\ref{sec:gates}], quantum circuit synthesis and compilation [Sec.~\ref{sec:circuit_synth}], as well as interactive quantum dynamics involving feedback control [Sec.~\ref{sec:feedback_ctrl}] and quantum error correction [Sec.~\ref{sec:QEC}].
In addition, reinforcement learning algorithms have been shown to benefit from quantum computing architectures [Sec.~\ref{sec:QRL}], and have emerged as a powerful tool in quantum metrology and sensing [Sec.~\ref{sec:metrology-comm}], where agents optimize measurement strategies and adaptively enhance sensitivity under resource constraints or experimental noise.

In the past few years, researchers have also started to apply reinforcement learning in real-world experiments on quantum devices. Among the most remarkable applications of RL in quantum technology are:  
(1) the experimental preparation of spin-squeezed states in spin-1 Bose-Einstein condensates across a quantum critical point~\cite{guo2021faster};  
(2) the optimization of electromagnetic pulses that implement entangling gates with fidelities an order of magnitude higher than state-of-the-art pulses for superconducting platforms~\cite{baum2021experimental};  
(3) training RL agents that construct arbitrary disentangling quantum circuits on trapped-ion quantum computers~\cite{tashev2024reinforcement};  
(4) the implementation of real-time quantum feedback protocols for a superconducting qubit, enabled by a sub-microsecond-latency neural network on a field-programmable gate array~\cite{reuer2023realizing};  
(5) the discovery and experimental implementation of previously unknown quantum error-correcting codes that stabilize a logical qubit with enhanced quantum coherence~\cite{sivak2023real}; and  
(6) the use of photonic circuits to obtain a quantum speedup of reinforcement learning algorithms~\cite{saggio2021experimental}.
We will examine these and related studies in greater detail.

Evidently, the RL toolbox has found numerous direct applications in quantum technologies, largely inspired by success stories outside physics. However, research over the last decade has shown that advancing solutions to outstanding challenges in quantum technologies requires RL frameworks that respect the laws and intrinsic properties of quantum mechanics. Examples include the inability to directly observe quantum states and the irreversible wavefunction collapse caused by projective measurements.
This necessity to design algorithms that comply with quantum constraints makes the straightforward transfer of existing ML ideas into quantum physics inappropriate or insufficient, driving the field along its own unique course. 
The purpose of this review is to survey the rapid development of reinforcement learning applications in quantum physics, focusing on the current state of the art, distilling the most successful approaches, emphasizing promising ideas, and identifying key outstanding challenges for the community to address in the near future.

This review is organized as follows. We start with a brief overview of reinforcement learning, introducing the core concepts in an intuitive manner~[Sec.~\ref{sec:RL_nutshell}]. This is followed by a more detailed exposition of the RL formalism and a discussion of the most widely used algorithms, with particular emphasis on their application to quantum systems~[Sec.~\ref{sec:RL_basics}].
Then comes a short discussion comparing RL to traditional quantum optimal control in Sec.~\ref{subsec:RL_vs_OC}.
We then explore the use of RL for quantum state preparation in few- and many-body systems~[Sec.~\ref{sec:state_prep}], before turning to quantum gate engineering and implementation on quantum computing platforms~[Sec.~\ref{sec:gates}]. This leads naturally to applications in quantum circuit synthesis and compilation~[Sec.~\ref{sec:circuit_synth}]. Subsequently, we examine RL-based approaches to interactive quantum dynamics involving feedback control~[Sec.~\ref{sec:feedback_ctrl}], including the implementation and discovery of quantum error-correcting codes~[Sec.~\ref{sec:QEC}].
In addition, we consider how RL algorithms can benefit from quantum computing architectures~[Sec.~\ref{sec:QRL}], and highlight their emerging role in quantum metrology and sensing, where they are used to optimize measurement strategies and enhance sensitivity under experimental constraints~[Sec.~\ref{sec:metrology-comm}]. We conclude with an outlook and a discussion of open challenges for future research~[Sec.~\ref{sec:outro}].

%% file: rl_nutshell.tex
\section{\label{sec:RL_nutshell}Reinforcement Learning in a Nutshell}

We open up the discussion with a simple, intuitive analogy of what reinforcement learning (RL) is, and how it works. In doing so, we introduce the major concepts and ideas behind RL. To keep the discussion light on the math side, we postpone any formal definitions to Sec.~\ref{sec:RL_basics}.

\begin{figure}[t!]
\centering
\includegraphics[width=1.0\columnwidth]{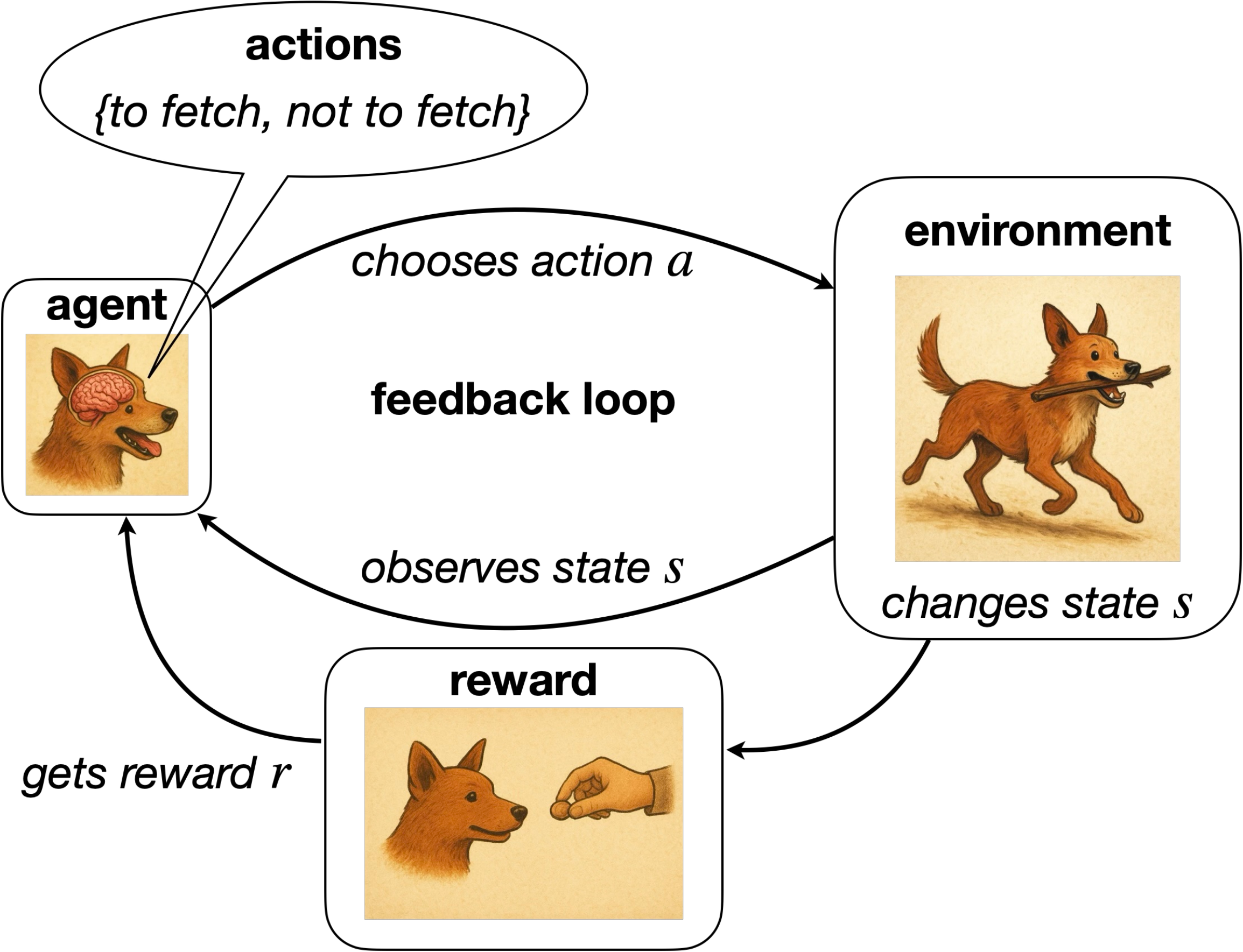}
\caption{
    \textbf{Illustration of the basic principle of reinforcement learning.}
    A puppy is trained to fetch a stick. At each new attempt, the puppy decides whether to fetch or not. As a consequence of the chosen action, its environment responds and changes its state. A treat is given to the puppy in the form of a reward to incentivize the desired behavior. 
    Illustrations created using OpenAI’s DALL·E model.
}
\label{fig:puppy_tamer}
\end{figure}

Suppose a tamer wants to teach a puppy to behave in a certain way upon a command: for instance, they may want to teach it the paw request, or something as simple as sitting down or fetching a stick; let us focus on the latter case [see Fig.~\ref{fig:puppy_tamer}], and see how this works in practice. The tamer says \textit{``fetch''} a large number of times, and the puppy may or may not do what the tamer expects it to (especially in the beginning); but when it does so, the tamer gives the puppy a treat to reward it for doing what it was asked to do and, at the same, time to incentivize it to do this again in the future. 

In the abstract context of RL~\cite{sutton_barto_book}, the puppy's mind (or brain) is an agent that interacts with its environment comprising the puppy's body, the stick, the tamer, and the rest of the outside world. The training process can be subdivided into discrete subsequent steps, each representing a new attempt to make the puppy fetch the stick. In an idealized scenario, at each step the puppy is presented with only two actions: \textit{to go and fetch the stick} or \textit{not to fetch the stick}, from which it selects one. Based on this choice, the body of the puppy reacts, and the puppy either goes after the stick or not: as a consequence, the environment changes its state. The tamer plays the role of an arbiter, which, based on the new state of the environment, produces a reward (here, \textit{treat} or \textit{no treat}).
As this process repeats over and over again, the puppy eventually learns that going to fetch the stick upon hearing the command is the right thing to do if it wants to get the treat. This feedback loop ultimately establishes a positive correlation between the action the agent takes and the rewards it receives.  

\begin{figure}[t!]
\centering
\includegraphics[width=1.0\columnwidth]{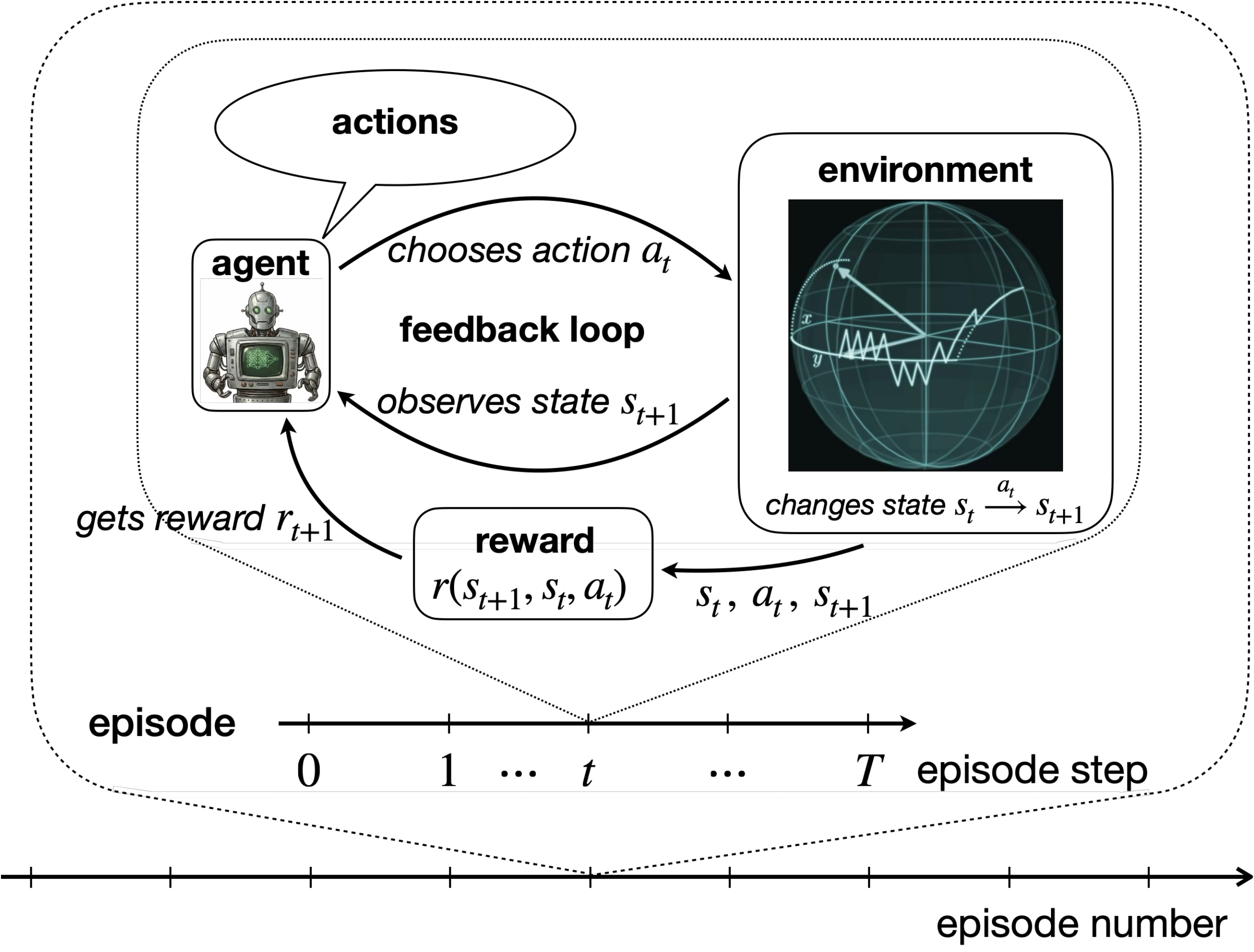}
\caption{
    \textbf{Training process in RL.}
    RL agents are usually trained in episodic tasks, where each training iteration comprises an episode, which consists of discrete steps. At each step, the agent selects an action that changes the state of the environment. The goal for the agent is, based on the observation of the current environment state, to select that action which maximizes the sum of the rewards within the episode. 
}
\label{fig:agent-env-interface}
\end{figure}

More formally, in RL an agent has to solve a task by interacting with its environment [Fig.~\ref{fig:agent-env-interface}]. Unlike the simple example with the puppy, complex tasks often require an entire sequence of steps to be taken before they can be solved. These steps form a so-called learning episode.
RL episodes start with the environment in an initial state, and end in a terminal state. 
At each episode step, the agent observes the state of the environment $s$; it is then presented with a number of actions $a$ to choose from, each of which causes a change in the environment state. The environment responds according to the laws governing its dynamics, which may be deterministic or stochastic. 
Depending on the new environment state, a reward $r$ is given back to the agent. 
This process repeats until the environment enters the terminal state, for which the task is considered complete, or the sum of rewards (so-called return) exceeds a fixed threshold; the episode comes to an end, after which the environment is reset to the initial state and a new episode begins. 
The goal for the agent is, at each step $t$, by observing the state $s_t$ of the environment, to select that action $a_t$ which maximizes the expected cumulative reward over the entire episode $\mathbb{E}\left[\sum_t r_t\right]$; the expectation here is over any uncertainty caused by the environment dynamics and the action selection. 

The RL agent selects actions based on a strategy, called a policy and denoted $\pi(a|s)$; it encodes the probability of selecting a given action $a$, upon observing the environment in the state\footnote{In many real-world problems, the agent only has access to partial observations of the state; in such cases, the policy is conditioned, and depends on, the observation.} $s$. 
The agent updates the policy based on the rewards it receives, in an attempt to maximize the expected return; the precise update rules constitute the underlying RL algorithm. The optimal policy results in the maximum achievable expected return. 

The environment of the RL agent encodes the physical system and the complete set of laws that govern it. It differs from, and should not be confused with, the environment of a physical system used, e.g., to describe open systems. In particular, the state of the RL environment defines a complete description of the physical state underlying the system of interest. The RL agent can observe the state of its environment, e.g., by doing measurements on the system. Actions, on the other hand, cause the environment to change state, according to the laws that govern the dynamics of the physical state.  
RL environments can either be simulated or represent real physical systems in the lab.

An important problem that arises when defining the RL framework for a given task is credit assignment, or how to choose rewards. The reward signal encodes a figure of merit that determines the degree to which the task has been solved successfully. 
Notice that, while the environment obeys physical laws that govern the transition dynamics of its state, it does not define the reward function; the environment is what the agent observes and acts upon\footnote{In practice, whether the assignment of rewards is included as part of the environment is largely a matter of taste; this is frequently done, e.g., when implementing the environment as a computer program.}.  
It is rather the arbiter (the tamer, in the example with the puppy) who examines the task and defines a reward function that assigns rewards based on the instantaneous environment dynamics. 
For the same environment, another arbiter with a different task in mind would assign different rewards.
Importantly, the reward function is fixed before training begins and cannot be changed as the agent proceeds from one episode to the next.

The reward function $r(s_{t+1},s_t,a_t)$ may depend on any combination of the current state $s_t$, the action chosen $a_t$, and the next state $s_{t+1}$ of the environment (i.e., the state to which the environment has transitioned after being exposed to the effect of the action). The reward has to be carefully designed to reflect the task the agent is supposed to solve; in particular, if two policies produce the same reward, they should be equally good at solving the task. Moreover, the reward function should prevent the agent from behaving in unwanted ways: e.g., inappropriate combinations of rewards and penalties may lead to cancellation in the cumulative expected reward $\mathbb{E}\left[\sum_t r_t\right]$, which the RL algorithm maximizes.
Unfortunately, there is no one-size-fits-them-all recipe for designing reward functions. That said, we recommend identifying the \textit{dimensionless} physical quantities in the problem first (e.g., typical energy, length, and time scales) and using (some of) them to construct the reward.

\begin{figure}[t!]
\centering
\includegraphics[width=1.0\columnwidth]{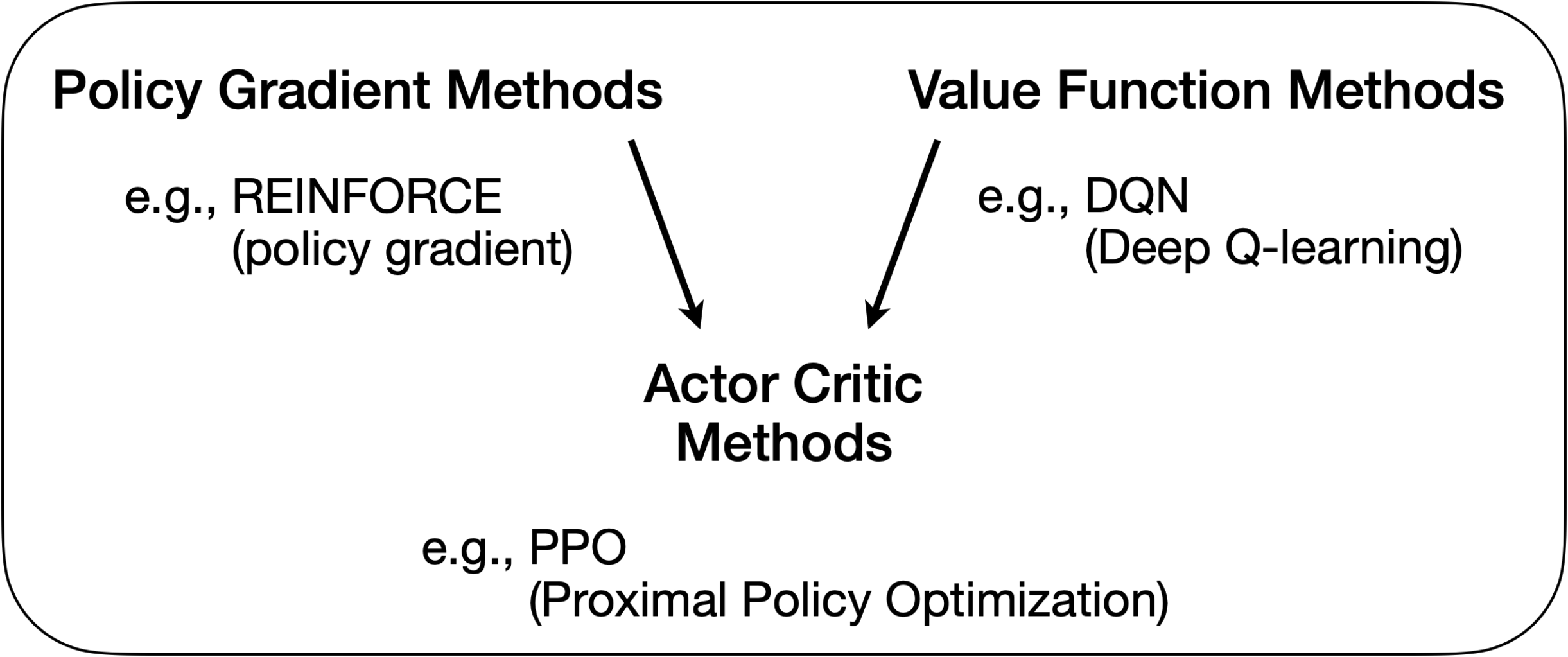}
\caption{
    \textbf{Overview of RL algorithms.}
    RL algorithms fall into two overlapping categories. Policy gradient methods have the primary objective to improve the policy $\pi$. Value-function methods, on the other hand, learn a value function (e.g., the value function $v_\pi$, the Q-function $Q_\pi$, or the advantage function $A_\pi$). 
    Actor-critic methods are a hybrid class of algorithms that learn both a policy and a value function. 
}
\label{fig:algo_overview}
\end{figure}

Given the agent-environment interface described above [Fig.~\ref{fig:agent-env-interface}], reinforcement learning comprises a set of algorithms to learn a policy that maximizes the expected return. While all these algorithms strive to find an optimal policy, in practice, one rarely has the guarantee of having done so; the most frequently encountered scenario is to converge to some local optimum. Nevertheless, RL agents can find interesting solutions that are beyond the reach of other commonly used optimization techniques.  
RL algorithms can be roughly divided into two overlapping categories: policy gradient and value function methods [see Fig.~\ref{fig:algo_overview}].

\textbf{Policy gradient methods} are easier to comprehend by the physicist's mind: here one parametrizes the policy $\pi\approx\pi_\theta$ using some variational parameters $\theta$; one then runs gradient ascent (the return is being maximized) in this parameter space. As the name suggests, policy gradient methods rely on (1) the ability to efficiently compute (or estimate) the gradient of the return w.r.t.~$\theta$, and (2) the expressivity of the variational ansatz. A common choice is to represent the policy as a deep neural network that accepts RL states as inputs, and outputs the probability for each available action. In this case, the variational parameters $\theta$ are given by the weights and biases. Although deep neural networks are known to be universal approximators, it is conceivable that other variational ans\"atze may be more suitable, depending on the problem at hand.
Examples of policy gradient algorithms include the vanilla Policy Gradient (PG) algorithm, also known as REINFORCE~\cite{williams1992simple}.

\textbf{Value function methods} follow a radically different approach. To gain an intuition about them, we will consider video games as another example. Video games can be placed within the RL framework as follows: the agent can select actions, according to those available in the controller; the environment is the game itself, with dynamical laws being the rules, and the state is the current configuration as shown on the screen; finally, the reward is the score.  

Now, suppose we took the game and paused it in some fixed configuration. If we asked an expert player to pick the controller and play the game, what they would (maybe unconsciously) do before they start, is look at the state of the game and try to estimate the largest possible score that they can obtain from the given initial configuration, when applying their expert policy $\pi$ for selecting all subsequent actions. This procedure assigns to every state $s$ of the game a value $v_\pi(s)$, corresponding to the expected return that can be obtained starting from $s$ and following the policy $\pi$ afterwards. 

However, knowing the value of every state $v_\pi(s)$ is still insufficient to formulate the best policy $\pi$. To do that, we define the $Q_\pi(s,a)$ (or quality) function: it is similar to the value function, except this time we inquire about the expected return of a state $s$ given that we select the specific action $a$, and then follow the policy $\pi$ afterwards. Compared to the value function, the resolution over actions allows us to consider the $Q$-function of a fixed state $s$ over all possible actions, and select the action that maximizes it: $a=\text{argmax}_{b} Q_\pi(s,b)$. This introduces a self-consistency condition: the best policy must be such that it always selects the optimal action based on the $Q$-function that itself refers to this policy. This allows us to construct a (deterministic) policy, as good as the original policy $\pi$. The self-consistency relation presents the basic idea behind the Q-learning algorithm~\cite{watkins1992q-learning}.
Most deep RL algorithms based on the value-function approach use function approximations, such as neural networks, to find approximations to the optimal $Q_\pi$ function. The most famous example is Deep Q-Learning~\cite{mnih2013playing}. 

\textbf{Hybrid methods.} 
Nowadays, the most prominent and widely used RL algorithms combine the best of both worlds, value functions and policy gradients. These are known as Actor-Critic methods~\cite{konda1999actor,haarnoja2018soft}, and modern state-of-the-art examples include Trust-Region Policy Optimization (TRPO)~\cite{schulman2015trust}, Proximal Policy Optimization (PPO)~\cite{schulman2017proximal}, and Deep Deterministic Policy Gradient (DDPG)~\cite{silver2014deterministic,lillicrap2015continuous}.  
When in doubt about which algorithm to choose, PPO is a robust, generic modern RL algorithm that covers many use cases and, therefore, presents an excellent starting point.
We also refer the interested readers to excellent tutorials on reinforcement learning for quantum technology~\cite{giannelli2022tutorial, duncan2025taming}.

%% file: rl4qs.tex
\section{\label{sec:RL_basics}Reinforcement Learning for Quantum Systems}

This section introduces the mathematical framework behind reinforcement learning (RL). We first give a formal definition of the major concepts introduced in Sec.~\ref{sec:RL_nutshell} above. We also emphasize specific restrictions related to applications in real-world quantum systems. This is followed by a discussion on deep reinforcement learning and some commonly used state-of-the-art algorithms. Finally, we contrast model-free with model-based RL, and compare RL methods with optimal control techniques.
Readers primarily interested in the applications of RL to quantum technology may prefer to skip this section on a first reading and refer back to it when needed.

\subsection{Reinforcement Learning Framework}
\label{subsec:RL_framework}

In Sec.~\ref{sec:RL_nutshell}, we outlined the notions of the agent and the RL environment, which form the RL framework, and provided an intuitive way to think about them. Here, we present the formal definitions. In doing so, we focus the discussion on quantum systems and discuss the associated challenges. 

Compared to classical systems, quantum systems present a number of restrictions:
(i) The state of a quantum system is described by an abstract wavefunction; importantly, the wavefunction itself cannot be measured (full-state tomography is exponentially costly in the number of particles);
(ii) Information about the state is revealed by measuring local observables; however, quantum mechanics obeys the Heisenberg uncertainty principle, which introduces a fundamental intrinsic uncertainty that cannot be overcome by the Central Limit Theorem;
(iii) Individual projective measurements collapse the state of the system; hence, any attempt to extract information from the system invariably affects its state;
(iv) The measurement outputs are discrete numbers (corresponding to the eigenvalues of the measured observable): in the simplest case of a qubit, individual measurements provide binary information (rather than real numbers);
(v) Individual measurement outputs are fundamentally probabilistic, reflecting the uncertainty principle mentioned above. 
These aspects set quantum systems apart from their classical counterparts. As a result, additional care needs to be taken when designing RL frameworks for quantum problems.

Let us now give a formal definition of the quantities appearing in the agent-environment interface, cf.~Fig.~\ref{fig:agent-env-interface}. We first introduce the general framework, and then place the emphasis on the specifics of quantum systems. 

\subsubsection{\label{subsubsec:SOAR}Environment, States, Observations, Actions, and Rewards}

As explained in Sec.~\ref{sec:RL_nutshell}, an RL agent learns to solve a given task by interacting with its environment (not to be confused with the environment in the context of open systems). Learning iterations typically proceed in episodes, and each episode is divided into time steps. At each episode step, the agent takes actions that alter the state of its environment. Depending on a combination of the old state, the action, and the new state, a reward is fed back to the agent. The agent strives to maximize the expected return over the entire episode rather than optimizing the individual rewards at each timestep. In doing so, it finds (close to) optimal strategies to solve the task. 

The agent-environment interface consists of the agent, the actions, the states of the environment (and potentially their observations), and the reward.     

The \textbf{environment} can be thought of as the physical system the agent interacts with: this can either be a simulator or a real experimental system, see Sec.~\ref{subsubsec:sim_vs_exp} for a detailed discussion. In particular, the RL environment contains 
(1) a \textit{state} $s$ that provides a complete description of the physical system, and 
(2) a set of rules that prescribe how the state changes (so-called dynamical laws), as a result of applying actions to the physical system\footnote{In some cases, even if the agent applies no action, the system may evolve and hence its state will undergo a change.}. The dynamical laws define the so-called \textit{transition probabilities} $p(s'|s,a)$ -- the probability for the state $s$ to transition to the $s'$ after the environment is acted upon with the action $a$. Whenever the transition probability is a delta function, the environment is called deterministic (e.g., evolution of entire wavefunctions or density matrices); otherwise, it is called stochastic (e.g., the probabilistic dynamics of individual quantum trajectories). 
Note that in many cases, we are not interested in dealing with a complete description of the physical world (as in the wavefunction of the entire universe) but rather in a property of a system that can be completely described by a set of effective degrees of freedom, which can also be used to define the state of the environment (e.g., the reduced density matrix of this system). 
The set of all admissible states defines the \textit{state space} $\mathcal{S}$, and can in general be continuous, discrete, or mixed. For most quantum systems, the state space is the space of density matrices (or wave functions), and is hence continuous.
Importantly, we emphasize that RL states are properties of the environment, not the agent. 

\textbf{Observations} $o$, as the name suggests, represent the information about the environment state that is accessible or visible to the agent. The agent receives the observation based on which it selects the new action.
In the context of quantum control, e.g., the state can be the wave function while observations can be the expectation values of observables; likewise, in quantum error correction, the RL state can be the wave function of the physical qubits, while the observations would be the syndrome measurements. In the case of a fully observable environment, there is no distinction between observations and states, and the agent has access to the state of the environment. Whenever this is not the case, the environment is called \textit{partially observable}. Similar to the state space, we can also define an observation space $\mathcal{O}(s)$, which may depend on the state via some function $O(o|s)$. Note that the same state can give rise to different observations; and conversely, different states can give rise to the same observation (this is similar to different errors giving rise to the same syndrome in quantum error correction). Therefore, learning from observations is more difficult than learning from states, yet it is more common. 

Incidentally, we note that it can be useful to train one agent to solve a whole family of tasks simultaneously. This is easily implemented by adding to the observation a vector describing the task, e.g., containing a few parameters characterizing the target state in a state preparation protocol. The advantage of this approach is that the agent can learn more rapidly by generalizing from what it has learned in related tasks, and the fully trained agent can then be applied across the whole range of tasks without retraining.

\textbf{Actions}, $a$, can be thought of as external controls to be applied to the physical system. In the context of quantum systems, these are quantum channels arising from, e.g., applied fields, gates and other unitary operations, measurement operations, etc. In contrast to classical systems, the decision by the agent to perform a measurement would also be an action, because it changes the state of the RL environment instead of being merely a passive observation. Often, physical controls are parametrized by a continuous parameter (e.g., the angle of rotation, the intensity or frequency of the laser field, or the strength of a weak measurement): in such a case, actions can be indicated via the specific parameter values. 
The \textit{action space} $\mathcal{A}$ defines the set of available actions; it can be a discrete set (e.g., in the case of a few fixed gates), a continuous space (e.g., the amplitude of an electromagnetic pulse), or any mixed combination; it can also be higher dimensional, e.g., if the agent can apply multiple fields simultaneously. Moreover, the action space may be state- (observation-) dependent, $\mathcal{A}(s)$: this corresponds to a setup where not all actions are always allowed, e.g., because of additional constraints.  

The \textbf{reward function} $r_t = r(s_{t+1},s_t,a_t)$ determines the rewards and is used to train the agent. 
In general, it is a function of the current environment state $s_t$, the selected action $a_t$, and the state $s_{t+1}$ that the environment transitioned into; it may be deterministic or stochastic. 
If the reward is extracted from an experiment, it can often only be assigned in the final episode step, e.g., when a measurement on the resulting quantum state is carried out; in such cases, the state preparation fidelity, the entanglement, or the expectation value of an observable can be estimated (averaging over many repeated episodes). In other cases, it may be possible to assign immediate rewards at each time step (e.g., in quantum error correction, stabilizer measurements can tell if the state left the code manifold); in these cases, the goal of the agent is to maximize the cumulative sum of rewards. 

In the simplest setup, the \textbf{return} is merely $\sum_{k=0}^T r_k$, where $T$ is the total number of episode steps. However, this prescription can be refined. At time step $t$, the optimal policy selects an action that helps maximize this total return; but equivalently, it maximizes the return calculated as a sum from time step $t$ onwards, since future actions obviously cannot influence past rewards. Adopting such a modified definition reduces fluctuations and aids training. Taking this idea one step further, but now introducing an approximation, one may decide to suppress the influence of rewards that are produced much later in the episode and thus favor more 'greedy' strategies that aim for short-term rewards. This can be encoded in the \textit{discounted return} (elsewhere sometimes also denoted '$R$'):
\begin{equation}
    \label{eq:return}
    G_t = \sum_{k=t+1}^T \gamma^{k-t-1} r_k.
\end{equation}
Here, $\gamma\in[0,1]$ is a discount factor used to determine the relative weight of future rewards.  

The primary purpose of the return $G_t$ is to serve as a figure of merit for the degree to which the agent has completed the task. In particular, it is essential that the return is maximized only by the strategy that solves the task; using ambiguous reward functions can result in the agent learning the wrong behavior. However, note that the concept of a reward does not require knowledge of the optimal strategy, i.e., one can define a reward function for a given task without knowing the optimal policy itself.  
Furthermore, it is important to emphasize that the immediate reward $r_t$ is not used by the agent to determine the next action, see Sec.~\ref{subsubsec:policy-value_fns}; rather, the agent learns using the return $G_t$. 
We also emphasize that the return is integrated/summed over time, and hence represents a time-nonlocal quantity. As a result, the agent does not maximize the immediate reward, but the long-term return; this feature allows it to take seemingly suboptimal actions at a given time step $t$ that, however, unlock higher-reward actions in the longer run (similar to sacrificing a queen to win a chess game). 
Finally, note that the concept of rewards and returns is opposite in spirit to that of a cost function, and negative rewards are often used as penalties.

\subsubsection{\label{subsubsec:policy-value_fns}Policy, Value, and Action-Value ($Q$-) Functions}

While the above definitions set up the agent-environment interface, they do not explain how the agent takes actions. Actions are taken according to a \textbf{policy} $\pi(a|s)$, which describes the probability of taking action $a$ if the environment is observed in the state $s$. While there can be a number of reasons why such a stochastic approach to action selection can be beneficial, the most prominent are: 
(i) Action probabilities are continuous even when the actions themselves are discrete, and hence they are central for enabling gradient-ascent-based RL algorithms.
(ii) Stochastic environments need not have deterministic optimal policies; this is because the underlying uncertainty in the environment dynamics can be used to obtain a higher expected return, as compared to deterministic policies;
(iii) Stochastic policies provide a natural way to explore new actions, which is vital to finding optimal policies. 

A central element of RL is the so-called \textit{Exploration-Exploitation dilemma}, which states that an agent should strike a balance between exploiting its knowledge about the environment and exploring new actions that can potentially bring higher returns. 
RL algorithms are designed to optimize the policy for a given task to maximize the expected return.
Since stochasticity is native to quantum mechanics via the uncertainty principle,
the concept of a policy provides a natural framework for manipulating quantum systems.

As we discussed at the end of Sec.~\ref{sec:RL_nutshell}, learning an optimal policy directly can be difficult. In such cases, RL offers an alternative through the so-called value functions. The \textbf{value function} of a state $s$ corresponding to the policy $\pi$ is defined as~\cite{sutton_barto_book}
\begin{eqnarray}
    \label{eq:value_fn}
    v_\pi(s) &=& \mathbb{E}_\pi\left[G_t\big| s_t=s \right] \\
             &=&\sum_{a\in\mathcal{A}}\pi(a|s)\sum_{s'\in\mathcal{S}}  p(s'|s,a) [r(s,s',a) + \gamma v_\pi(s')]. \nonumber
\end{eqnarray}
It quantifies the expected return $G_t$ if the environment is initialized in the state $s$ at timestep $t$, following the fixed policy $\pi$ afterwards. The expectation value here is taken both under the transition dynamics $p$ of the environment and the agent's policy $\pi$. The second line in Eq.~\eqref{eq:value_fn} is known as the \textit{Bellman equation}; it defines an implicit equation for $v_\pi$, which occurs on both sides.

While value functions provide a measure of how much a state is worth under a given policy, it is not possible to infer the policy from them. To this end, we define the action-value, or $Q$-function:
\begin{eqnarray}
    \label{eq:Q_fn}
    Q_\pi(s,a) &=& \mathbb{E}_\pi\left[G_t\big| s_t=s,a_t=a \right] \\
             &=&\sum_{s'\in\mathcal{S}}  p(s'|s,a) [r(s,s',a) + \gamma v_\pi(s')]. \nonumber
\end{eqnarray}
which gives the expected return starting from state $s$ and taking action $a$ at time step $t$, and then following the policy $\pi$ afterwards. 
An important difference is the resolution over actions; one can use it to extract a deterministic policy $\pi'(s) = \text{argmax}_a Q_\pi(s,a)$ from the $Q$-function that is at least as good as the original policy $\pi$~\cite{sutton_barto_book}. Under certain conditions, this can be employed to design algorithms that can reconstruct the optimal policy from the optimal $Q$-function, see Sec.~\ref{subsec:RL_algos}.  Equation~\ref{eq:Q_fn} is the Bellman equation for the action-value function $Q_\pi$; again, it represents an implicit equation whose solution yields the $Q$-function for the optimal policy.

\subsubsection{\label{subsubsec:MDPs}Markov Decision Processes and the Reinforcement Learning Objective}

As we discussed in Sec.~\ref{subsubsec:SOAR}, the environment dynamics is encoded in the transition probabilities $p(s'|s,a)$. In particular, this rule assumes that the new state of the environment $s'$ is completely determined by the previous state $s$ and the selected action $a$, and does not depend on any previous states -- a property is known as \textit{markovianity}. More precisely, reinforcement learning is a \textbf{Markov Decision Process} (MDP), i.e., a Markov process defined on the state-action space $\mathcal{S}\times\mathcal{A}$. 
Here, `decision' refers to the policy $\pi(a|s)$, from which the actions are sampled. While the goal for the agent is to improve its policy, it has no control over the transition probabilities $p(s'|s,a)$ since the laws of physics and the properties of the specific environment are fixed; instead, the agent learns a policy to maximize the expected return, consistent with the transition probabilities of the environment. 

Any MDP naturally gives rise to trajectories of state-action-reward tuples. A \textbf{trajectory} 
\begin{equation}
    \label{eq:traj}
    \tau = [(s_0,a_0,r_1), (s_1,a_1,r_2), \cdots, (s_{T-1},a_{T-1},r_T), s_T]
\end{equation}
is a finite sequence of such triples, starting from an initial state $s_0$ at the initial environment step $t=0$, and ending in a terminal state $s_T$ at the final time step $T$. The initial and final steps delineate the boundaries of an episode. Once an episode comes to an end, the environment is reset to its initial state, and the agent starts over again. Such tasks are called \textit{episodic} (or finite-horizon), in contrast to \textit{non-episodic} (infinite-horizon) tasks where $T\to\infty$. 

We can now formulate the \textbf{reinforcement learning objective} as follows:
\begin{equation}
    \label{eq:RL_obj}
    J = \mathbb{E}_{\tau\sim P_\pi}\left[ G_0(\tau) \big| s=s_0 \right],
\end{equation}
where the expectation value is taken over the probability $P_\pi(\tau) = p_0(s_0)\prod_{t=0}^{T-1} \pi(a_t|s_t)p(s_{t+1}|s_{t},a_t)$ for trajectory $\tau$ to occur under the policy and the environment dynamics; here $p_0$ is the initial state distribution. 
In other words, the agent seeks to maximize the total \textbf{expected return} $J$, starting from an initial state $s_0$, and afterwards following the policy $\pi$ in an environment determined by the transition probabilities $p$. 

We emphasize that RL algorithms optimize in expectation, which offers certain benefits:
(i) They are naturally designed for stochastic optimization problems which, as we shall see, are abundant in quantum physics.
(ii) The presence of the expectation value in Eq.~\eqref{eq:RL_obj} relaxes the conditions on the continuity and differentiability of the reward function $r$; in particular, RL is applicable with discrete reward functions that otherwise cannot be dealt with using gradient methods. 
(iii) In addition, the RL objective $J$ contains the expected return, which is a time-nonlocal quantity; hence, it can, at least in principle, capture nonlocal correlations between state-action pairs occurring at different times within an episode. Physically, this means that RL agents are `capable of planning' ahead of time. Note that this does not compromise markovianity in much the same way as one can sample long-range correlated ferromagnetic states of the 2D Ising model using Markov-Chain Monte Carlo at low enough temperatures.

We note that the foregoing discussion is immediately applicable to the important situation of a partially observable Markov decision process, where the agent observes only parts of the environment state. This merely means that the policy $\pi(a|o)$ is conditioned only on the observation, not on the complete state. This is the standard situation in applications of RL to quantum technologies. We refer the interested reader to the relevant literature~\cite{hausknecht2015deep}.

\subsubsection{\label{subsubsec:sim_vs_exp}Simulated versus Real Environments}

Training RL agents usually proceeds either (i) in a simulated environment, (ii) on a real system, or (iii) in some mixed form. Below, we discuss in detail the first two approaches. 

In the first case, a model for the environment dynamics is used to define the transition probabilities $p(s'|s,a)$; using this model, one simulates the dynamics of the physical system. In the simplest case, for a quantum system, the transition probabilities are deterministic and determined by the Schr\"odinger equation. Importantly, despite utilizing this simulation, in the context of the model-free RL algorithms we are focusing on here, the simulated environment is treated as a black box, and the agent typically does not know anything about the underlying model. Rather, it infers the information relevant for maximizing the return from the interaction with the environment [see Sec.~\ref{subsec:model_based_RL} for details on the alternative of model-based RL]. 
Since complete knowledge of the state is available in a simulation, one can construct very sophisticated rewards, much better than what one would have available in the real system. In addition, one can easily train the agent on observations that are much more detailed than would be the case in experiment\footnote{That said, if one is trying to learn a feedback strategy in simulation to apply it in the lab later on, one will have to modify the framework (e.g., using two-stage training or learning from demonstration~\cite{li2025robust}) to obtain an agent that can be deployed in the real world with less information.}. Another important advantage of training in a simulated environment is that one may be able to generate training data at a much higher rate, which is essential to the speed of learning\footnote{However, in some quantum systems (e.g, quantum many-body systems), this classical simulation may actually turn out slower than using a quantum simulator.}. Moreover, the timescale at which the agent interacts with the environment and the timescale at which the environment responds to actions selected by the agent, are typically limited by computational power. 


Whereas training in a simulation is usually preferred in many theoretical studies, it has the major disadvantage of assuming a model for the environment. If the trained agent is to be deployed on a real physical system after training is complete, the agent will be faced with dealing with a somewhat different environment. For instance, quantum devices are not precisely calibrated, and their Hamiltonians or dissipative dynamics are governed by slightly different parameters than those assumed in our model. 
A major advantage of RL is the ability to learn from real data; as such, the agent figures out policies that naturally comply with the behavior of the system (known or unknown). 

When it comes to real-world quantum systems, a number of additional considerations have to be made. On the level of the agent-environment interface, the timescales of the agent to select an action and the timescale of the quantum system to respond may be very different: for instance, currently gates for superconducting (SC) qubit platforms are executed on timescales of $10 ns$ and even measurements require only a few hundred nanoseconds (while atomic platforms are typically much slower, with gate times on the order of $100 \mu s$ for trapped ions, with recent advances pushing this down to $\mu {\rm s}$ in some cases, and $\mu s$ for Rydberg atoms); at the same time the time for the agent to compute the action may be significantly longer, depending on the number of parameters in the neural network encoding the policy or the value function. The latter process is additionally slowed down if data (i.e., measured observations, rewards, etc.) has to be transferred physically. Recent work managed to overcome this difficulty by solving the challenges of implementing an RL agent for a single superconducting qubit as a sub-microsecond-latency neural network on a field-programmable gate array (FPGA)~\cite{reuer2023realizing}.

While most of the above considerations hold true also for classical systems, real-world quantum systems come with their own restrictions, which are inherited by the corresponding RL environment. First, recall that quantum states are unmeasurable theoretical constructs: this automatically renders the environment partially observable. Second, (strongly) measuring any observable collapses the quantum state, which introduces a strong backaction mechanism. From the quantum Zeno effect, we know that measurements repeated at a fast rate arrest the state and do not let it evolve, which makes it difficult for the agent to control the system. Third, because of the wave function collapse, a reward cannot be given at each time step of an episode; typically, rewards have to be restricted to the final step (so-called sparse rewards), and the reward coincides with the return. Sparse rewards contain less information, and hence constitute more challenging learning problems per se.        

In fact, as a consequence of the uncertainty principle and the wavefunction collapse, already the simplest quantum system -- the qubit -- has a minimally observable environment, since any measurement output is necessarily binary and contains a single bit of information (the Pauli matrices used to define any observable have eigenvalues $\pm 1$). In general, there are two approaches to dealing with these quantum features: 
(1) if the timescale for the agent to select an action is much longer than the response time of the quantum system, one can perform repeated measurements and collect statistics; this allows one to define non-binary observations and rewards from the measured expectation values of observables. However, one has to keep in mind that this approach compromises the use of pure states, since the slightest amount of noise will give rise to ensemble averages. 
(2) if the two timescales are comparable, one can proceed and learn from single-shot measurement data. In practice, this comes at the cost of sufficiently large batch sizes during training to reduce the variance in stochastic gradient descent. One also has to explicitly make sure that the return, averaged over the batch, will be maximal when the task is considered solved (e.g., in the case of state preparation, the batch-averaged return has to be maximized by the target state).

A common approach to mitigate the wave function collapse problem is the use of ancilla(s)~\cite{foesel2018reinforcement,sivak2022model-free,sivak2023real}. Ancillary qubits are widely used in quantum computing, both for gate implementation and to do measurements in an arbitrary basis. 
For example, in the case of a single qubit, one can add an additional ancilla qubit to the environment. The RL state is then the joint quantum state of the qubit and the ancilla, while actions can be defined on either or both of them.
If the qubit and the ancilla are weakly entangled, projective measurements performed on the ancilla reveal partial information about the state of the qubit (so-called weak qubit measurement). Note that this comes at the cost of perturbing the state of the qubit; the more information the ancilla measurement contains, the stronger the collapse of the qubit state, and the more strongly the observation affects the environment state. 
In this setup, the observations fed into the agent correspond to the binary measurement outcomes of the projective ancilla measurements at each episode step. Once the final episode step has been reached, a binary projective measurement of the qubit is used as a return. 
The above procedure can be formalized into a quantum partially observable Markov decision process~\cite{barry2014quantum}.

\begin{figure*}[t!]
\centering
\includegraphics[width=.65\textwidth]{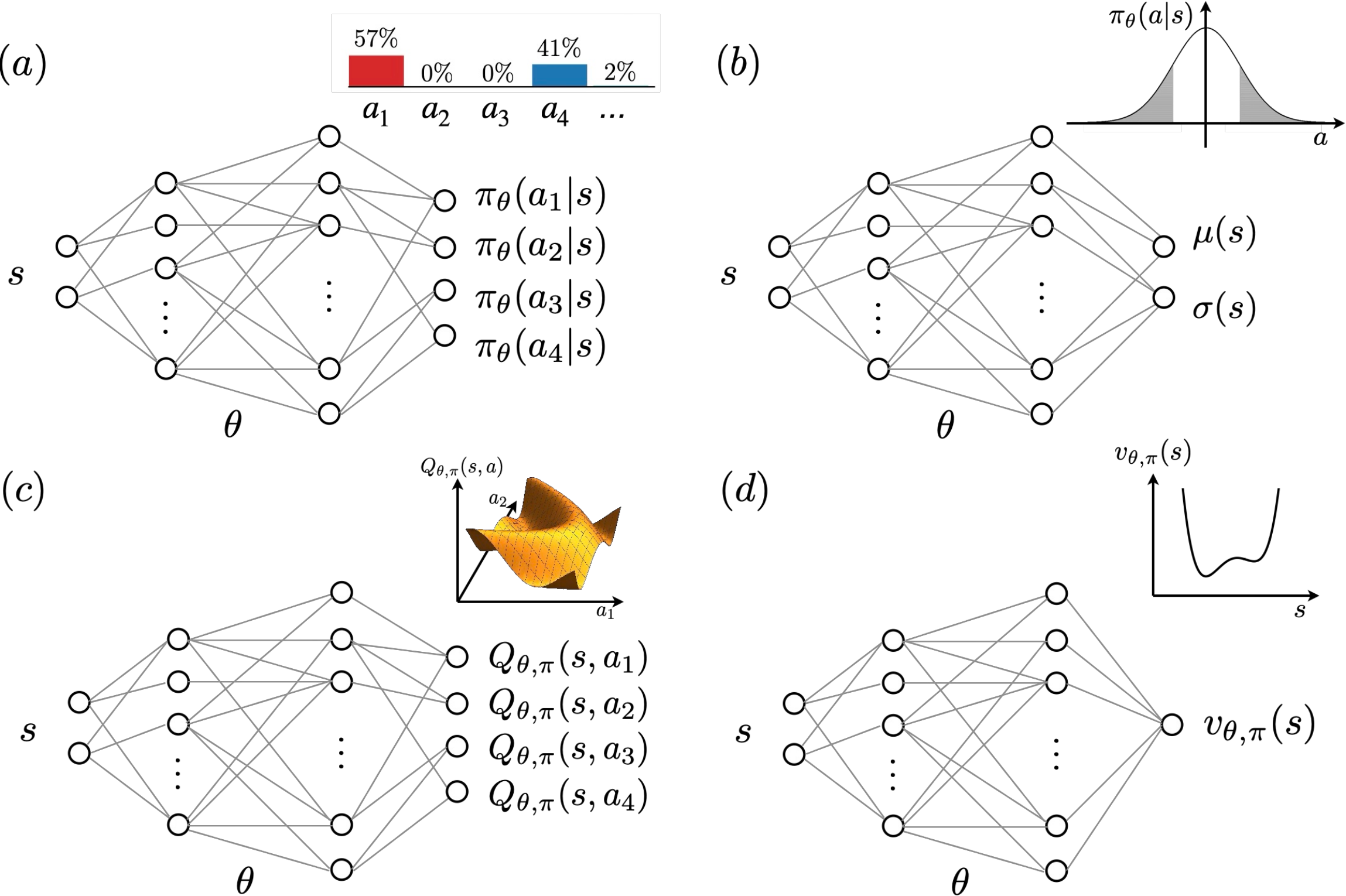}
\caption{
    \textbf{Common deep learning architectures used in RL.}
    (a) The RL policy $\pi_\theta(a|s)$ corresponds to a histogram over the actions for a discrete action space, for each input state $s$.
    (b) For continuous actions, the policy ansatz learns the state-dependent parameters of a continuous probability density (e.g., the mean $\mu(s)$ and variance $\sigma^2(s)$ of a Gaussian). 
    (c) A typical architecture for the $Q$-function has RL states as input, and outputs the $Q_{\theta,\pi}(s,a)$ value for each of the actions. 
    (d) Value functions $v_{\theta,\pi}(s)$ take in states $s$ and output a single scalar. 
}
\label{fig:RL_architectures}
\end{figure*}

\subsection{\label{subsec:DRL}Deep Reinforcement Learning}

Reinforcement learning agents learn by exploring the state space $\mathcal{S}$ of the environment. If the state space is finite and small, a sufficiently long training procedure guarantees that the agent will eventually learn how to act optimally from every state, thus mastering the optimal policy. 

A common issue with applying RL to real-life problems is caused by the size $|\mathcal{S}|$ of the state space; it can be either exponentially large in some system size parameter (e.g., in board games such as Chess or Go, as well as for Ising spin-like configurations), or even continuous (e.g., phase space, or Hilbert space). In such a scenario, it is infeasible or even impossible for an agent to visit all states during training. This raises the question as to how an RL agent is supposed to `know' which action to select when the environment enters into a hitherto unencountered state.  

A way out of this problem is to use a variational approximation for the policy or the $Q$-function. The basic idea is to approximate the policy $\pi(a|s)\approx \pi_\theta(a|s)$, the value $v_\pi(s) \approx v_{\theta,\pi}(s)$, or the $Q$-function $Q_\pi(s,a)\approx Q_{\theta,\pi}(s,a)$ using some variational parameters $\theta$. The variational ans\"atze
\begin{equation}
    \label{eq:varl_ansatz}
    \theta \mapsto \pi_\theta(a|s),\qquad \theta\mapsto Q_{\theta,\pi}(s,a),\qquad \theta\mapsto v_{\theta,\pi}(s)
\end{equation}
are functions of our choice; in principle, they can be selected freely, so long as they satisfy the defining properties of the approximated target function (e.g., a policy is a proper probability distribution, etc.). 
Needless to say, the choice of ansatz will affect the ability of the agent to learn. 
Choosing an ansatz that can accept continuous states $s$ as input allows the RL agent to interpolate and extrapolate the policy or value functions of new, unseen states. 

Choosing an appropriate variational ansatz often reflects any additional information one might have about the approximated quantity. For instance, if we expect it to be periodic, we may expand in the Fourier basis and view the expansion coefficients as the variational parameters. Selecting a good variational ansatz is similar to choosing the ``right basis'' before truncating the approximation to finitely many parameters.
A frequently used family of variational ans\"atze is given by various deep neural networks, shown to be universal approximators. While this ad-hoc solution works universally, it may well be that a physics-inspired ansatz performs better for one problem or the other: e.g., recently tensor networks have been suggested as alternatives to deep neural networks in approximating the $Q$-function in controlling quantum spin chains~\cite{metz2023self}.   

When it comes to using deep neural networks to approximate the policy, a common choice is to use a logsoftmax layer to enforce the probability constraint. Learning directly the logarithm of the policy can be advantageous not only to get a better resolution between low-probability actions, but also in training, as it appears when estimating the gradient of the RL objective [cf.~Sec.~\ref{subsec:RL_algos}]. In problems with discrete action spaces, the policy network accepts a state (an observation), and outputs a vector of (log-) probabilities corresponding to each discrete action [Fig.~\ref{fig:RL_architectures}(a)]. When the action space is continuous, the model typically learns the state-dependent mean $\mu(s)$ and variance $\sigma(s)$ of a normal distribution $\mathcal{N}(\mu(s),\sigma^2(s))$ from which a continuous action is sampled [Fig.~\ref{fig:RL_architectures}(b)]; it is also possible to consider multivariate and even correlated Gaussian distributions in the case of multiple continuous actions (e.g., when the agent controls a few fields simultaneously); for continuous actions of a finite range, one can use the Beta distribution or sample from the sigmoid-Gaussian distribution~\cite{yao2022noise}.

Approximating value functions with neural networks is straightforward; for $Q$-functions, one typically chooses to either define the input as the states (observations) with the network outputs being a vector approximating the value of the $Q$-function for each discrete action $Q(s,\cdot)$, or one considers the tuple $(s,a)$ as input, and learns the value $Q(s,a)$ [Fig.~\ref{fig:RL_architectures}(b)]. Combining $Q$-functions with continuous actions is also possible using the DDPG algorithm, see Sec.~\ref{subsubsec:value_algos}. 
Similarly, one can also train neural networks to approximate the value function, cf.~Fig.~\ref{fig:RL_architectures}(d).

Irrespective of which RL quantity one decides to work with, a few general guidelines have emerged. 
The first few input layers should reflect the structure of the input data (states/observations, etc.). In particular, if these share any local structure (e.g., syndrome measurements in the 2D toric code are local), it may be advantageous to use convolutional layers whose filters can learn local patterns. More generally, if the input data has a known symmetry (e.g., translation or permutation invariance), learning may be facilitated by building it into the variational ansatz. While the value function is a scalar and the corresponding network architecture should be invariant under the symmetry operation, the policy $s\mapsto\pi(a|s)$ and the $Q$-function $s\mapsto Q(s,a)$ architectures are vector-valued over the action space and should thus be designed equivariant (i.e., a permutation within the input state $s$ should also permute the output action dimensions accordingly to enforce the symmetry property). For instance, a permutation equivariant policy was constructed to disentangle multiqubit states in Ref.~\cite{tashev2024reinforcement} using a transformer architecture. For a handy introduction to equivariant neural networks, we refer the reader to Refs.~\cite{white2021deep,bronstein2021geometric}.    

An interesting open question is how to construct deep reinforcement learning architectures that can learn efficient representations of the quantum state of the environment, from binary observations (i.e., quantum data). In particular, designing embedding layers that are capable of reconstructing the state from the observation to the extent possible within the restrictions of information theory, can prove handy in applying RL agents to experimental setups.   

\subsection{\label{subsec:RL_algos}Overview of Reinforcement Learning Algorithms}

We are now in a position to discuss the training of the RL agent. As mentioned briefly in Sec.~\ref{subsubsec:policy-value_fns}, the historical development of RL algorithms places them in two categories: Policy Gradient and Value Function methods, cf.~Fig.~\ref{fig:algo_overview} for an overview. That said, most modern state-of-the-art approaches combine and mix these concepts, so a clear-cut separation can no longer be established. Below, we briefly introduce the most commonly used algorithms in the field. Because the field is rapidly developing, we place the focus on the major concepts and ideas, rather than presenting specific pseudocodes; for the latter, we refer the interested reader to the abundant literature. 

Most RL algorithms proceed in two stages. 
In the so-called \textit{policy evaluation stage}, the current policy is executed to generate states (observations) and rewards; this involves routines to sample actions and apply them to the environment state. Sometimes, this is also referred to as the \textit{roll-out stage}. 
Next comes the \textit{policy improvement stage}, where the RL algorithm uses the previously collected data to update the policy; the latter can happen directly (by updating the parameters of the policy model) or indirectly (e.g., by updating the $Q$-function). It is this stage that brings in the `learning' component. 
The frequency of alternating policy evaluation with policy improvement determines the behavior of the RL algorithm. 

Before we dive into the description of various deep RL algorithms, a word of caution is in place. As we explained in Sec.~\ref{subsec:DRL} above, challenging problems have exponentially large or infinite state and observation spaces, which requires us to approximate the main quantities of interest: the policy $\pi$, the value function $v$, the action-value function $Q$, or any combination thereof. While necessary, this may compromise the convergence properties of the respective RL algorithm: typically, mathematical proofs of convergence are available for tabular methods or linear function approximation which do not offer the desired expressive power~\cite{sutton_barto_book}.   

\subsubsection{\label{subsubsec:PG_algos}Policy Gradient Methods}

Policy gradient methods comprise a set of algorithms where, as the name suggests, the central object is a policy $\pi$ that is being optimized. The simplest and perhaps one of the more intuitive RL algorithms for physicists is REINFORCE~\cite{williams1992simple}. The basic idea is to parameterize the policy $\pi_\theta$ using a set of variational parameters $\theta$, compute the gradient $\nabla_\theta J(\theta)$ of the RL objective 
\begin{equation}
\label{eq:J_theta}
    J(\theta) = \mathbb{E}_{\tau\sim P_{\pi_\theta}} \left[ \sum_{t=0}^{T-1} r(s_t,a_t,s_{t+1}) \bigg| s_{t=0}=s_0 \right],
\end{equation}
and use it to do gradient ascent in $\theta$-space. Note that (unlike energy landscapes) in RL the return is being maximized, hence ascent. 

To make a useful algorithm out of this natural idea, we need a way of computing the gradients $\nabla_\theta J(\theta)$. The difficulty here lies with the expectation value in Eq.~\eqref{eq:J_theta} which cannot be evaluated exactly, due to the infeasibility to exhaust all possible trajectories; indeed, for an RL problem with $|\mathcal{A}|$ actions and $T$ time steps, there are a total of $|\mathcal{A}|^T$, i.e., exponentially many trajectories. 

As we know from statistical physics, a way out is to construct a Monte Carlo estimator for $\nabla_\theta J(\theta)$. To this end, consider the probability for a trajectory $\tau$ to occur, cf.~Sec.~\ref{subsubsec:policy-value_fns},
\begin{equation}
\label{eq:prob_trajs}
    P_{\pi_\theta}(\tau) = p(s_0)\prod_{t=0}^{T-1} \pi_\theta(a_t|s_t) p(s_{t+1}|s_t,a_t) , 
\end{equation}
where $p(s_0)$ is the initial state distribution. One can now formally write the expected return as
\begin{eqnarray}
    J(\theta) &=& \mathbb{E}_{\tau\sim P_\pi} \left[ \sum_{t=0}^{T-1} r(s_t,a_t,s_{t+1}) \bigg| s_{t=0}=s_0 \right] \nonumber\\
    &=& \int\mathcal{D}\tau\;  P_{\pi_\theta}(\tau)\; G(\tau),
\end{eqnarray}
where the (path) integral is over all trajectories, and the return of a trajectory $G(\tau)$ is defined in Eq.~\eqref{eq:return} (we consider an undiscounted task, $\gamma=0$, for simplicity). 
Taking the derivative $\nabla_\theta J(\theta)$ results in
\begin{eqnarray}
\label{eq:nabla_J}
\nabla_\theta J(\theta) &=& \int\mathcal{D}\tau\; \nabla_\theta P_{\pi_\theta}(\tau) G(\tau) \nonumber\\
&=& \int\mathcal{D}\tau\; P_{\pi_\theta}(\tau)\; \nabla_\theta \log P_{\pi_\theta}(\tau)\; G(\tau)  \nonumber\\
&=& \mathbb{E}_{\tau\sim P_\pi} \left[\nabla_\theta \log P_{\pi_\theta}(\tau)\; G(\tau)\right],
\end{eqnarray}
where we used the logarithmic derivative to cast the final expression as an expectation value. 

The key insight is that Eq.~\eqref{eq:nabla_J} allows us to directly estimate the gradient $\nabla_\theta J(\theta)$ using Monte-Carlo (MC) samples: 
\begin{eqnarray}
\label{eq:RL_grad_sampled}
    \nabla_\theta J &{=}& \mathbb{E}_{\tau\sim P_\pi} \left[\nabla_\theta \log \pi_\theta(\tau)\; G(\tau)\right]  \nonumber \\
    &{\approx}& \frac{1}{N_\text{MC}}\sum_{j=1}^{N_\text{MC}} \nabla_\theta \log \pi_\theta(\tau^{(j)})\; G(\tau^{(j)}) \\
    &{=}& \frac{1}{N_\text{MC}}\!\!\sum_{j=1}^{N_\text{MC}} \left( \sum_{t=0}^{T-1} \nabla_\theta \log\pi_\theta(a^{(j)}_t|s^{(j)}_t) \!\!\sum_{t'=0}^{T-1} r(a^{(j)}_{t'},s^{(j)}_{t'}) \right),  \nonumber
\end{eqnarray}
where the superscript $j$ denotes index of the trajectory sampled using the policy $\pi_\theta$, and $t$ is the episode step; the total number of sampled trajectories is $N_\text{MC}$. In the last line of Eq.~\eqref{eq:RL_grad_sampled} we used the definition of the return and the identity $\nabla_\theta \log P_\pi(\tau^{(j)}) = \sum_{t=0}^{T-1} \nabla_\theta\log \pi_\theta(a^{(j)}_t|s^{(j)}_t)$ which follows directly from Eq.~\eqref{eq:prob_trajs} since neither the transition probabilities $p(s_{t+1}|s_t,a_t)$ nor the initial state distribution $p(s_0)$ depends explicitly on the parameters $\theta$.

Finally, the policy parameters are updated using gradient ascent via
\begin{equation}
\label{eq:grad_ascent}
    \theta \leftarrow \theta + \alpha \nabla_\theta J(\theta),
\end{equation}
where in practice we use the MC estimate of the gradients. Another practical trick when computing gradients is to use the pseudoloss function
\begin{equation*}
    J_\text{pseudo} {=}  \frac{1}{N_\text{MC}}\!\!\sum_{j=1}^{N_\text{MC}}\!\! \left( \sum_{t=0}^{T-1} \log \pi_\theta(a^{(j)}_t|s^{(j)}_t)\! \sum_{t'=0}^{T-1} r(a^{(j)}_{t'},s^{(j)}_{t'}) \right),
\end{equation*}
which satisfies the relation $\nabla_\theta J_\text{pseudo} \approx \nabla_\theta J$ within the MC approximation.

Before we move on, let us briefly discuss the special case of a correlated multivariate Gaussian policy over the $d$-dimensional continuous action space $\mathcal{A}=\mathbb{R}^d$ (e.g., think of the actions $\vec a$ as being the values of $d$ different control fields):
\begin{equation}
    \pi_\theta(\vec a|s) = \frac{1}{\sqrt{(2\pi)^d\det\Sigma_\theta(s)}}\; \mathrm e^{-\frac{1}{2}(\vec a-\vec\mu_\theta(s))^T\Sigma_\theta^{-1}(s)(\vec a-\vec\mu_\theta(s))}.
\end{equation}
Here $\vec\mu_\theta(s)\in\mathbb{R}^d$ is the vector of state-dependent means, and $\Sigma_\theta\in\mathbb{R}^{d\times d}$ is the symmetric positive-definite state-dependent covariance matrix; both are parametrized by the variational parameters $\theta$, e.g., through a suitable deep neural network architecture. 
A straightforward calculation using the relation $2\log\pi_\theta = -\log \det\Sigma_\theta -(\vec a-\vec\mu_\theta)^T\Sigma_\theta^{-1}(\vec a-\vec\mu_\theta)+const.,$ gives the policy gradients as~\cite{yao2020policy}
\begin{eqnarray}
    \frac{\partial \log \pi_\theta(\vec a|s)}{\partial \mu_\theta(s)} &=& \Sigma_\theta^{-1}(s)(\vec a-\vec\mu_\theta(s)), \\
    \frac{\partial \log \pi_\theta(\vec a|s)}{\partial \Sigma_\theta(s)} &=&
    \frac{1}{2}\bigg(-\Sigma^{-1}_\theta(s) \nonumber\\
    && + \Sigma^{-1}_\theta(s)(\vec a-\vec\mu_\theta(s))(\vec a-\vec\mu_\theta(s))^T\Sigma^{-1}_\theta(s)\bigg). \nonumber
\end{eqnarray}
Using Eq.~\eqref{eq:RL_grad_sampled}, it is straightforward to apply the gradient ascent rule from Eq.~\eqref{eq:grad_ascent}. Other continuous probability density distributions can also be considered, in particular, when the actions have to be bounded within some compact interval.

Back to REINFORCE, a common problem is the variance of the gradient estimator. If gradients are not estimated correctly, finding an optimal policy is likely to fail. There are two common tricks to reduce variance:
(i) the sum over $t'$ in Eq.~\eqref{eq:RL_grad_sampled} contains a priori rewards with $t'<t$ which clearly cannot influence selecting the action at step $t$; in other words, the agent only reinforces actions based
on future and not the past returns. It is therefore common to constrain $t'\in\{t,t+1,\dots, T-1\}$.
This is a special case of a more general observation: (ii) shifting the return by an arbitrary constant (commonly referred to as a baseline, $b$) does not change the value of the gradient of the expected return: 
\begin{eqnarray}
    \nabla_\theta J(\theta)
    \approx \frac{1}{N_\text{MC}}\sum_{j=1}^{N_\text{MC}} \sum_{t=0}^{T-1} && \nabla_\theta \log \pi_\theta(a^{(j)}_t|s^{(j)}_t)\times \\
    &&\times \left[\sum_{t'=t}^{T-1} r(a^{(j)}_{t'}|s^{(j)}_{t'}) - b(s)\right]. \nonumber
\end{eqnarray}
One can show that the most general baseline can depend on the state (but not the action). A common choice for a state-independent baseline is the batch-averaged return $b_t = N_\text{MC}^{-1}\sum_j G_t(\tau^{(j)})$.

An interesting candidate for a state-dependent baseline is the value function $b(s)=v_{\pi}(s)$. Since the value function corresponding to the policy $\pi_\theta$ is not known, we have to extend the above algorithm. We can approximate $v_\pi(s)\approx v_{\eta,\pi} (s)$ by a different variational model with parameters $\eta$, and use a linear regression on the return data with the cost function:
\begin{equation}
    J_\text{critic}(\eta) = \frac{1}{N_\text{MC}}\sum_{j=1}^{N_\text{MC}} \sum_{t=0}^{T-1}  \| v_{\eta,\pi}(s^{(j)}_t) - G_t(s^{(j)}_t)  \|^2, 
\end{equation}
where $G_t(s^{(j)}_t)$ is the return at time step $t$ when trajectory $\tau^{(j)}$ is in state $s^{(j)}_t$ [following a MC estimate of the first line in Eq.~\eqref{eq:value_fn}].
This fruitful idea has led to an entire class of hybrid RL algorithms, known as Actor-Critic~\cite{konda1999actor,haarnoja2018soft}. The actor is the agent policy, and the critic is the value function. 
The optimization then proceeds in alternating updates of the policy and the value function. 

A particularly widely used generalization of REINFORCE is Proximal Policy Optimization (PPO)~\cite{schulman2017proximal}. The advantage of the family of PPO actor-critic algorithms is that they allow us to recycle part of the data collected during the agent-environment interaction before we dispose of it. Through a number of sophisticated but numerically light-to-implement checks, PPO ensures that the current policy does not deviate excessively from the old policy used to collect data. This makes PPO one of the preferred state-of-the-art deep RL algorithms to date. 

Another popular extension of policy gradient is Trust Region Policy Optimization (TRPO)~\cite{schulman2015trust}. The idea behind TRPO is to apply natural policy gradient~\cite{kakade2001natural}, which uses estimates of the Fisher information matrix of the policy $\pi_\theta$ to improve the gradient update. However, this requires estimating the inverse Hessian of the KL divergence between the policies at the current and previous iterations, which can become prohibitively expensive for policy models with a large number of variational parameters. A similar problem occurs in Stochastic Reconfiguration used to train neural quantum states; recently, a nice workaround was proposed, known as MinSR~\cite{chen2024empowering}, which essentially exchanges the roles of the number of samples and the number of parameters, and works in the large-parameter regime. It is an interesting open problem whether MinSR can also give rise to an efficiency boost in TRPO.

Two additional widely used actor-critic algorithms worth mentioning are Asynchronous Advantage Actor-Critic (A3C)~\cite{mnih2016asynchronous} and Soft Actor Critic (SAC)~\cite{haarnoja2018soft}. We refer the interested reader to the original literature. 

Let us conclude this discussion with two practical remarks. 
In tasks where an episode is terminated when a quantity falls below a fixed threshold or when the agent is incentivized to look for the shortest protocol, it often happens that different trajectories in the batch terminate after a different number of steps. In order to speed up the training process, one can run multiple trajectories in parallel in a vectorized environment: if a trajectory comes to an end, its environment is reset to the initial state individually and continues without waiting for the other trajectories to terminate, assuring data contiguity. To train the agent in such a parallel pipeline, one can define an update stage after a fixed number of segments; once a training stage is over, each trajectory picks up from where it paused before the update. Such an implementation of PPO was used in~\cite{tashev2024reinforcement}. 

Last but not least, it often happens that a policy converges to a delta function in the limit of large training iterations. This may or may not be advantageous, depending on the problem at hand. In particular, if one is after learning an optimal policy where multiple actions are equally likely from a given state, this ``symmetry'' will likely be broken by statistical fluctuations during training. A common way to prevent this behavior from happening is to add the entropy of the policy distribution to the reward function. This closely resembles the free energy in statistical mechanics, where there is a competition between energy (here, the return) and entropy (here, the policy entropy), and physicists can transfer the intuition they have gained. This powerful technique is known as entropy regularization~\cite{levine2018reinforcement,tashev2022developing}.

\subsubsection{\label{subsubsec:value_algos}Value Function Methods}

For some problems, learning the optimal value function can be easier than learning an optimal policy. Moreover, different degenerate optimal policies can have the same value/$Q$-function. It is therefore useful to devise algorithms that maximize the expected return by searching for an optimal $Q$-function. 

Recall that one can extract a deterministic policy from any $Q$-function using the assignment $\pi(s) = a_\ast = \text{argmax}_{a'}Q(s,a')$. To enable exploration, one can soften this greedy policy, by adding a small probability $\varepsilon$ to take a suboptimal action:
\begin{equation}
    \pi_\varepsilon(a|s) =
    \begin{cases}
        1 - \varepsilon - \frac{\varepsilon}{|\mathcal{A}(s)|}, \qquad &\text{if \ } a=a_\ast  \\
        \frac{\varepsilon}{|\mathcal{A}(s)|} , \qquad &\text{if \ } a\neq a_\ast.
    \end{cases}
\end{equation}
There are also alternatives: (i) one can use a Boltzmann policy, i.e., $\pi_\beta(a|s) \propto \exp(+\beta Q(s,a))$ for some inverse temperature $\beta$ (in RL we maximize the expected return, hence the $+$ sign); (ii), sometimes it is advantageous to use noise that is correlated in time, e.g., generated by an Ornstein-Uhlenbeck process~\cite{lillicrap2015continuous} which describes the stochastic motion of an overdamped oscillator in the presence of thermal fluctuations; or, (iii) one can add parametric noise to the neural network parameters~\cite{fortunato2017noisy}. 

In Sec.~\ref{subsubsec:policy-value_fns} we introduced the concepts of value and $Q$-functions, associated with a fixed policy $\pi$. The optimal value functions $v_\ast(s), Q_\ast(s,a)$ are formally defined by taking the maximum over all possible policies:
\begin{equation}
    v_\ast(s) = \max_{\pi} v_\pi(s),\qquad Q_\ast(s,a) = \max_\pi Q_\pi(s,a).
\end{equation}
In particular, taking the maximum over policies on both sides of Eq.~\eqref{eq:Q_fn}, we arrive at the recursive relation~\cite{sutton_barto_book}
\begin{equation}
\label{eq:Bellman_eq}
    Q_\ast(s,a) = \sum_{s'\in\mathcal{S}} p(s'|s,a)\left(r(s,a,s')+\gamma \max_{a'}Q_\ast(s',a')\right),
\end{equation}
which defines an implicit equation for the optimal $Q$-function, and is known as the Bellman optimality equation. 

We can turn Eq.~\eqref{eq:Bellman_eq} into an iterative algorithm as follows: 
(i) we initialize a $Q$-function $Q(a,s)$ arbitrarily; 
(ii) we then sample a transition $(s,a,s',r)$ using some soft policy, e.g., $\varepsilon$-greedy w.r.t.~$Q(a,s)$;
(iii) finally, we update the $Q$-function according to the rule $Q^{(n+1)}(s,a) = Q^{(n)}(s,a)-\alpha \delta^{(n)}$, where $\alpha\in[0,1]$ is some small step size, and the Bellman error is given by
\begin{equation}
\label{eq:Bellman_error}
    \delta^{(n)}(s,a) = Q^{(n)}(s,a) - \left(r(s,a,s')+\gamma \max_{a'}Q^{(n)}(s',a')\right).
\end{equation}
One can show that, as $n\to\infty$, this procedure converges to the optimal value function $Q^{(n)}(s,a)\to Q_\ast(s,a)$, provided each state can be visited arbitrarily often. 
This algorithm was first proposed by Watkins, and is known as (tabular) Q-learning~\cite{watkins1992q-learning}. 

Observe that the concept of episodes is not strictly needed in $Q$-learning; in particular, learning occurs from time-local transitions $(s,a,s',r)$, rather than complete trajectories. This property is useful in tasks with excessively long episodes. It also allows us to construct a deep-learning version of Q-learning, where the $Q$-function is approximated by a neural network $Q_\theta$. To this end, we first use any policy to sample a dataset of transitions (called a replay buffer), $\mathcal{B} = \{(s_j,a_j,r_j,s'_j) \}_j$. We can then formulate a regression problem using the cost function
\begin{equation}
    \mathcal{L}(\theta) = \frac{1}{2}\sum_{j\in \mathcal{B}'} \|Q_\theta(s_j,a_j) - y_j \|^2, 
\end{equation}
where $y_j = r(s'_j,s_j,a_j)+\gamma \max_{a'}Q_\theta(s'_j,a')$, and $\mathcal{B}'\subset\mathcal{B}$ is a subset of all buffer transitions. Notice that, within this regression problem, we treat $y_j$ as $\theta$-independent (otherwise our labels will not be well-defined). This gives rise to the following update rule for the variational parameters
\begin{eqnarray}
\label{eq:Q_grad_descent}
    \theta\leftarrow\theta - \alpha \sum_{j\in \mathcal{B}'} \nabla_\theta Q_\theta(s_j,a_j) \delta(s_j,a_j),
\end{eqnarray}
where $\nabla_\theta$ acts only on the $Q$-function and the Bellman error is assumed $\theta$-independent. 
In practice, we use stochastic gradient descent by independently sampling a batch of transitions from the buffer. This is the core idea behind the DQN algorithm~\cite{mnih2013playing,mnih2015human}.

Although this algorithm may look deceptively similar to solving a supervised learning problem, the update rule in Eq.~\eqref{eq:Q_grad_descent} may not converge due to the so-called running target problem: indeed, the target $y_j$ changes every time we update the $Q_\theta$ network. To tackle this issue, one typically introduces a second auxiliary $Q$-network $Q_{\varphi}$ which is initialized as $\varphi=\theta$. Then, we always make sure to compute the Bellman error in Eq.~\eqref{eq:Q_grad_descent} using the parameters $\varphi$, which are held fixed while updating the target network parameters $\theta$. In every iteration, the auxiliary parameters are updated towards the target network parameters by a small amount: $\varphi\leftarrow \zeta \varphi + (1-\zeta)\theta$ with $\zeta\lesssim 1$. This `adiabatic' learning has a stabilizing effect on the regression problem for the $Q_\theta$-function. Another useful trick to stabilize training is to clip the gradients on the right-hand side of Eq.~\eqref{eq:Q_grad_descent}.  

A common issue with $Q$-learning is that the learned $Q$-values may not be accurate. This problem can be traced back to the $\max$ operation in Eq.~\eqref{eq:Bellman_error} which can lead to a biased estimate of the expected return for the new state $s'$. To mitigate it, a common approach is Double Q-learning~\cite{sutton_barto_book}. The key idea is based on the identity $\max_{a'} Q (s',a') = Q(s', \text{argmax}_{a'}Q(s',a'))$. We can now introduce a second $Q$-network, which can provide an unbiased estimate of the optimal $Q$-function in the state $s'$. Suppose the two networks have parameters $\theta_1, \theta_2$; then the Double learning update rule reads as
\begin{eqnarray}
    \delta_1(s,a) &=& Q_{\theta_1}(s,a) \nonumber\\
    && - \left[r(s,a,s')+\gamma Q_{\theta_2}\left(s',\text{argmax}_{a'}Q_{\theta_1}(s',a')\right)\right] \nonumber\\
    \delta_2(s,a) &=& Q_{\theta_2}(s,a) \nonumber\\
    && - \left[r(s,a,s')+\gamma Q_{\theta_1}\left(s',\text{argmax}_{a'}Q_{\theta_2}(s',a')\right)\right] \nonumber\\
     \theta_k &\leftarrow& \theta_k - \alpha \sum_j \nabla_{\theta_k} Q_{\theta_k}(s_j,a_j) \delta_k(s_j,a_j),\quad k=1,2. \nonumber
\end{eqnarray}
Similar considerations are behind a variety of state-of-the-art deep $Q$-learning algorithms.
Particularly noteworthy are Deep Deterministic Policy Gradient (DDPG)\footnote{Despite its name, DDPG uses a $Q$-function to generate actions, and hence belongs to the value-based algorithms; it also features an auxiliary deterministic policy to aid convergence.}~\cite{lillicrap2015continuous} and Twin-Delayed DDPG (TD3)~\cite{fujimoto2018addressing}, which can handle continuous actions.

Finally, we remark that oftentimes in applications to quantum technologies, one is interested in solving a task while, in addition, keeping the number of episode steps minimal. A useful feature to keep in mind is that using $Q$-learning with strictly negative rewards produces short protocols as a byproduct since the agent is trying to minimize the penalty\footnote{Note that this is not necessarily the case for policy gradient methods, where the baseline often centers the return.}.

\subsubsection{\label{subsubsec:other_algos}Other Related Algorithms}

Three other related classes of algorithms deserve brief mention. 

The first one is \textbf{Evolutionary Methods}~\cite{eiben2015introduction}, with Genetic Algorithms (GA) forming an important subclass. The core idea here is to run an entire population of \textit{individuals} that undergo evolution-inspired processes such as mutation, crossover, and selection as they search for the solution. Although GAs can be generally formulated within the policy improvement framework, a typical distinction to RL is that the RL agent observes its environment and bases the action selection process on this observation. Moreover, Genetic Algorithms typically run faster and are more data-consuming.   

The second class is that of \textbf{Search Algorithms}~\cite{russell-norvig_book}. Examples include Monte Carlo Tree Search (MCTS), Beam Search, or $A^\ast$-Search. Search algorithms can be used to produce data to pretrain policies and $Q$-functions using supervised learning -- a process known as \textit{imitation learning}. 
Perhaps the most prominent example is the AlphaZero algorithm, which uses an MCTS to train the policy~\cite{silver2017mastering}. 
In general, search algorithms help explore vast state spaces and are often used to run so-called planning routines for RL agents. 

Finally, \textbf{Projective Simulation} (PS)~\cite{briegel2012projective} is a physics-motivated framework for artificial intelligence that models agent cognition through a stochastic network of episodic memory units called clips, comprising percept clips that encode environmental state representations, and action clips that represent the agent's behavior. During the agent-environment interaction, sensory input activates percept clips, which propagate activation through weighted edges to action clips via a random walk process governed by transition probabilities. Similar to RL, the agent's deliberation is modeled as a Markov chain with action selection determined by the stationary distribution. Learning occurs through reinforcement-driven plasticity that modulates edge weights based on reward feedback -- successful sequences strengthen transition probabilities while unsuccessful paths undergo decay -- incorporating physics-inspired principles such as detailed balance conditions or thermodynamic learning dynamics. The PS framework exhibits emergent properties, including adaptive forgetting through weight damping, compositional learning through clip combination, and generalization via shared network pathways, providing a unified approach that bridges episodic memory models, network theory, and reinforcement learning.

\subsection{\label{subsec:model_based_RL}Model-Free versus Model-Based Reinforcement Learning}

We have focused our description on so-called model-free reinforcement learning, where the RL approach does not need access to an underlying model of the system. This is obviously an advantage whenever dealing with an experimental system that is more or less a black box. Even if training is performed on a computer and a simulation is available, maybe that simulation code is a black box itself and cannot easily be adapted to work together with modern automatic differentiation techniques or whatever would be required for the model-based approach.

However, if a system is well characterized and a faithful simulation is available that has been written in a modern framework (like jax, tensorflow, or pytorch), it is possible to automatically take gradients through the simulation representing the dynamics of the physical system. In that case, one can directly take gradients of the overall return with respect to the trainable parameters characterizing the RL agent. This drastically reduces the number of samples required in training, in comparison to model-free techniques that essentially have to estimate those gradients by comparing many runs with slightly different actions. In the domain of quantum physics, early articles advocating this approach include \cite{schafer2020differentiable} and \cite{abdelhafez2019gradient}, with extensions to feedback in \cite{schafer2021control}, improving sample efficiency~\cite {khalid2023sample}, and eventually for the most general case in \cite{porotti2023gradient}.

We remark that in the general literature (e.g., in computer science and robotics) the term "model-based RL" is applied to many different situations, not only when gradients are taken directly through a simulation. It may also encompass scenarios where a differentiable model of the environment is first learned and then gradients are applied to that model, or where a model-free approach is applied to a simulation that has previously been learned.

\begin{table}[t!]
\centering
\begin{tabular}{l*{8}{c}}
& \rotatebox{70}{\textbf{REINFORCE}}
& \rotatebox{70}{\textbf{PPO}}
& \rotatebox{70}{\textbf{TRPO}}
& \rotatebox{70}{\textbf{DQN}}
& \rotatebox{70}{\textbf{DDPG}}
& \rotatebox{70}{\textbf{A3C}}
& \rotatebox{70}{\textbf{SAC}}
& \rotatebox{70}{\textbf{TD3}} \\
\toprule
\rowcolor{gray!10} {\color{blue}on}/{\color{red}off}-policy         & \cmarkb & \cmarkb & \cmarkb & \cmarkr & \cmarkr & \cmarkb & \cmarkr & \cmarkr \\
{\color{blue}on}/{\color{red}off}-line              & \cmarkr & {\color{blue}(\cmarkb)} & {\color{blue}(\cmarkb)} & \cmarkb & \cmarkb & \cmarkr & \cmarkb & \cmarkb \\
\rowcolor{gray!10} policy-based     & \cmark & \cmark & \cmark & \xmark & \xmark & \cmark & \cmark & \xmark \\
value-based                            & \xmark & \xmark & \xmark & \cmark & \cmark & \xmark & \xmark & \cmark \\
\rowcolor{gray!10} actor-critic        & \xmark & \cmark & \cmark & \xmark & \cmark & \cmark & \cmark & \cmark \\
continuous actions                     & \cmark & \cmark & \cmark & \xmark & \cmark & \cmark & \cmark & \cmark \\
\rowcolor{gray!10} discrete actions    & \cmark & \cmark & \cmark & \cmark & \xmark & \cmark & (\cmark) & \xmark \\
experience replay                      & \xmark & \xmark & \xmark & \cmark & \cmark & \xmark & \cmark & \cmark \\
\rowcolor{gray!10} distributed         & \xmark & (\cmark) & \xmark & \xmark & \xmark & \cmark & \xmark & \xmark \\
\bottomrule
\end{tabular}
\caption{
\textbf{Properties of commonly used RL algorithms.} 
Checkmarks \cmark, \cmarkb, \cmarkr\ indicate the algorithm supports a property, while crosses \xmark\ indicate it does not; parenthesized checkmarks (\cmark) indicate the existence of non-standard or less common variants.
\textit{Policy-based} algorithms use a policy to generate actions, whereas \textit{value-based} algorithms -- a $Q$-function; \textit{actor-critic} algorithms involve both (action-)value functions and policies to aid convergence during training. 
\textit{Experience replay} refers to using a data buffer for training.
The \textit{distributed} property refers to whether an algorithm is designed to run across multiple processes, threads, or machines simultaneously during training.
}
\label{tab:rl_algorithms}
\end{table}

\subsection{\label{subsubsec:algos_comparison}Choosing the Appropriate Algorithm}

A particularly important concept regarding RL algorithms is the distinction between \textit{on-policy} and \textit{off-policy} algorithms~\cite{sutton_barto_book}. On-policy algorithms require fresh data from the current policy to improve it. A primary example is REINFORCE, where the current policy is updated once complete trajectories are drawn from it; the next update uses new trajectories, etc. This means that, once used, training data is discarded; hence, every update iteration requires drawing new data from the current policy. Intuitively, this makes sense: old data comes from an outdated policy and is not necessarily representative of the current policy distribution. 

By contrast, off-policy algorithms can use any data for training, old and new. A primary example of an off-policy algorithm is Q-learning, where the $Q$-function is updated from transitions between the states that are local in time. As such, one can re-use any old data to improve the $Q$-function (and with it the policy). 
This makes off-policy algorithms much more data efficient compared to on-policy algorithms. In particular, the off-policy feature also allows one to decouple the data collection process from the policy improvement step, which results in so-called asynchronous updates; this means one can add new transitions to the buffer $\mathcal{B}$ and update the $Q$-function at different rates.
Such features can become very handy if data points are scarce or take a long time to collect.

Another direction of comparison is the online-offline axis~\cite{sutton_barto_book}. \textit{Offline algorithms} require the episode to come to an end before they can improve the policy. Examples include REINFORCE or DQN. By contrast, \textit{online algorithms} such as Q-learning improve the policy at each episode step. Clearly, making use of already available information to improve the policy may be beneficial for learning. 
Table~\ref{tab:rl_algorithms} lists the most commonly used RL algorithms and some of their most relevant properties.

Last but not least, we draw the reader's attention to the trade-off between sample efficiency and runtime. 
Sample efficiency is a primary criterion to determine the suitability of RL algorithms for use in quantum experiments. Depending on the setup, collecting data from the environment may become a bottleneck; in such cases, off-policy algorithms such as Deep Q-Learning may be a good choice. On the other hand, when training using a simulator, especially if it is more efficient or faster to generate data than training the agent, Policy Gradient and Actor-Critic algorithms are typically considered a good starting point. 

\begin{figure}[t!]
\centering
\includegraphics[width=.3\textwidth]{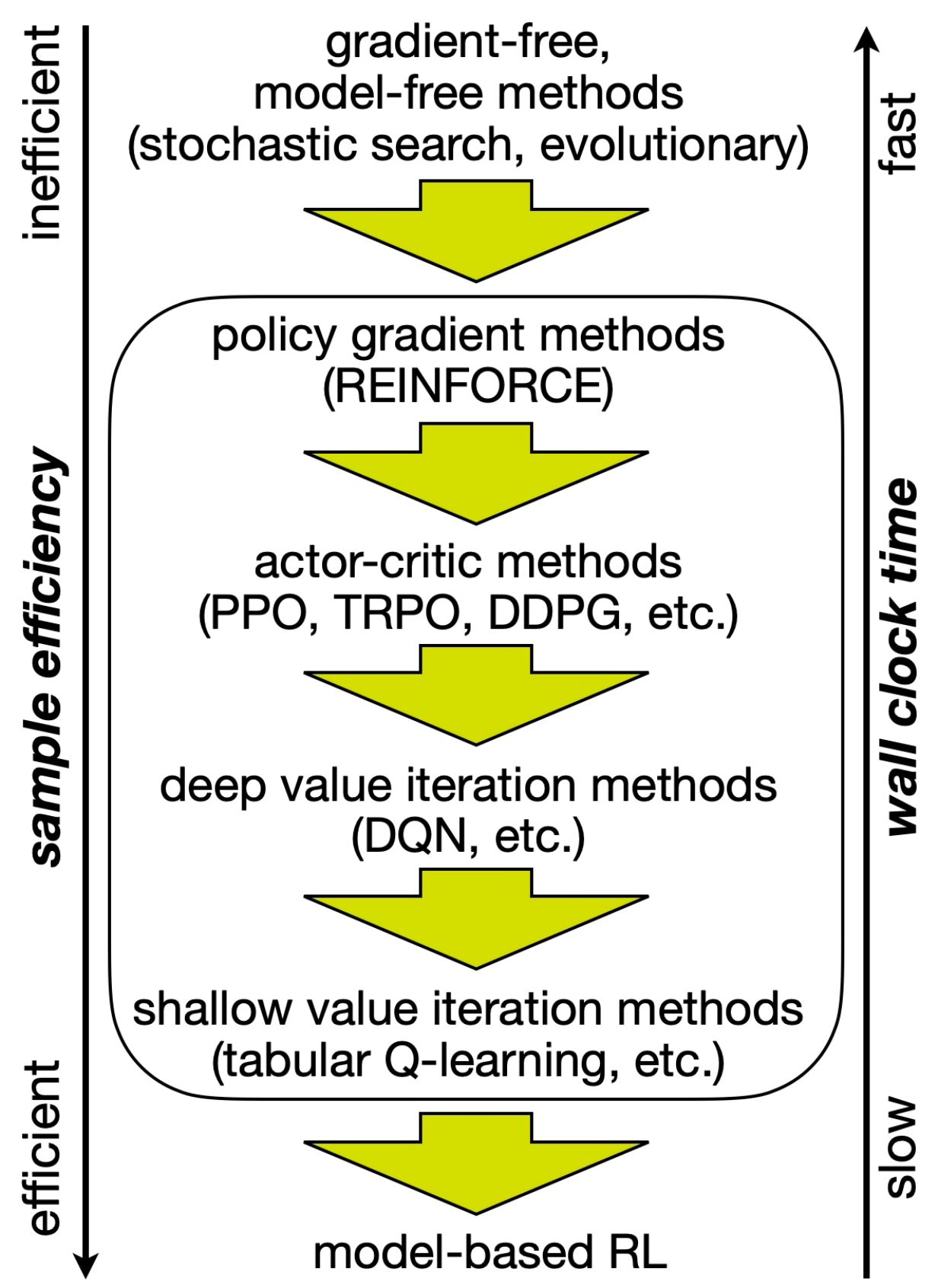}
\caption{
\textbf{Conceptual placement of RL algorithms by sample efficiency and wall clock runtime.} 
Sophisticated RL algorithms are typically more sample-efficient but also slower [see text].  
The oval contains traditional model-free deep RL algorithms. 
}
\label{fig:sample-efficiency}
\end{figure}

Figure~\ref{fig:sample-efficiency} shows a heuristic qualitative estimate of the relative sample efficiency of the algorithms discussed in the previous sections. 
Learning a model for the transition dynamics and using planning often reduces interactions with the environment; however, model-based methods [Sec.~\ref{subsec:model_based_RL}] can be computationally expensive (think of planning, model learning, etc.), so they commonly reside in the sample-efficient but slower wall-clock region.
Moving up, shallow value iteration methods (tabular Q-learning or policy iteration, etc.) are extremely sample-efficient and fast for small state-space/tabular problems, but they do not scale to high-dimensional observation spaces without function approximation.
Deep model-free methods (e.g., DQN) are relatively sample-inefficient in the sense that they require many environment steps to generate the data stored in the buffer; that said, training decouples from data generation since they are off-policy. Depending on the implementation, they can be computationally expensive (e.g., require GPUs for training) but are easily parallelizable. Therefore, they are placed towards sample-inefficient, and somewhere on the moderate to fast wall-clock side. 
Actor-critic methods (e.g., PPO, A2C, SAC, etc.) are online, which generally makes them more sample efficient compared to vanilla policy-gradient (REINFORCE), which notably also has higher variance.
At the opposite end of the diagram, we find gradient-free/evolutionary methods, cf.~Sec.~\ref{subsubsec:other_algos}. They are often quite sample inefficient since they evaluate many complete episodes, but can be wall-clock competitive if massively parallelized.
To sum it up, as a rule of thumb, the runtime typically increases with improved sample efficiency, with gradient-free methods, such as stochastic searches and evolutionary methods, running much faster than model-based RL algorithms. 

%% file: rl_vs_qoc.tex
\section{\label{subsec:RL_vs_OC}Reinforcement Learning \& Quantum Optimal Control}

As we explained already in Sec.~\ref{sec:intro}, reinforcement learning and optimal control can be formulated within the same framework and viewed as the two sides of the same coin~\cite{todorov2006optimal}. 
Nevertheless, a natural question that often arises is when one should use RL as opposed to conventional quantum control techniques~\cite{koch2022quantum}. We now turn our attention to discussing some of the advantages and disadvantages of RL algorithms. 

An important feature of the RL algorithms discussed in Sec.~\ref{subsec:RL_algos} is that they belong to the so-called gradient-free methods. However, since we use gradient descent/ascent to train the policy and the $Q$-function, this statement requires some clarification: by `gradient-free' methods, we mean algorithms that do not require taking the derivative of the cost w.r.t.~the controls (i.e., actions). As a consequence, (model-free) RL algorithms work even when no detailed model of the quantum dynamics is available, and they apply to non-differentiable and even discontinuous reward functions; this is particularly handy for applications in quantum technologies, where single-qubit measurements produce binary data. 

Moreover, RL agents learn in expectation (over the policy dynamics and the transition probabilities of the environment). This allows them to naturally handle various types of uncertainty, but also discontinuous reward functions, including stochasticity in the environment dynamics as well as the reward signal. As a result, RL-optimized policies are natively resilient to different sources of noise, and can readily accommodate the probabilistic nature of quantum measurements. Interestingly, the stochastic nature of exploration can even be beneficial for deterministic tasks, as it leads to policies that are naturally robust to perturbations.

Perhaps the most prominent difference between RL and conventional optimal control algorithms is the leftover learned object, i.e., the trained policy or $Q$-function. While optimal control requires re-optimization once the environment changes, RL agents are capable of exploiting their knowledge after training is complete without further optimization: for instance, they are capable of generalizing to new, so far unseen, initial states and can self-correct errors inflicted by the effect of noise on the fly~\cite{metz2023self}. On a fundamental level, an RL agent can learn an optimal policy that entails multiple (degenerate) optimal protocols at once, e.g., by assigning equal probabilities to multiple actions.   

That said, the computational efficiency of RL algorithms is not as good as that of conventional optimal control: RL requires more iterations since it starts without prior knowledge about the environment (except for model-based RL, discussed above in Sec.~\ref{subsec:model_based_RL}). RL relies on a steady supply of data to train the underlying deep-learning architectures. As a result, especially model-free RL agents may not be as efficient in minimizing the cost function as some of their gradient-based optimal control counterparts.

In a nutshell, RL brings in four major advantages:
(i) model-free RL requires no previous knowledge of the controlled physical system: in quantum technologies, this can be particularly useful when learning how to mitigate the effects of unknown sources of noise (e.g., to improve the fidelity of quantum operations), or when the Hamiltonian is not fully known (e.g., when using approximate effective descriptions such as Schrieffer-Wolff transformations, but also in quantum solids where it is impossible to know all terms in the Hamiltonian with the precision required for high-fidelity operations).
(ii) RL agents are adaptive, as they allow transferring the acquired knowledge during training between similar systems. This is expected to prove useful in identifying connections between unrelated phenomena. 
(iii) RL agents are interactive and naturally designed for feedback control -- a feature particularly useful for quantum control, which requires learning to deal with backaction effects arising from measurement collapse in strong or weak measurements. 
And finally, (iv) RL algorithms are autonomous; thus, they can provide novel insights into automating complex manipulation protocols in experiments without the need for human intervention.

In the following, we will exemplify diverse applications of RL to quantum technology. Much of the referred literature contains studies that compare RL to optimal control or provide other justification for the use of RL in the given study. We therefore invite the interested readers to consult the corresponding research literature.

%% file: qstate_prep.tex
\section{\label{sec:state_prep}Quantum State Preparation}

By far and large, the most natural application of RL in quantum technology is in quantum control~\cite{koch2022quantum,duncan2025taming, giannelli2022tutorial}. This is motivated by the structure of the RL framework, where the agent takes actions to manipulate its environment to achieve a given task.

Perhaps the most common application of RL in quantum control is quantum state preparation, where the goal is to transfer the population from an initial to a target state in some fixed duration $T$. This is usually done by applying external magnetic and electric fields, whose time dependence defines the so-called control protocol. Depending on the experimental system of interest, additional constraints may exist regarding the magnitude or frequency of the applied field and the kind of terms it couples to. The figure of merit that evaluates the success of the state preparation process is usually the fidelity $F$ between the evolved state $\ket{\psi(T)}$ and the target state $\ket{\psi_\ast}$:
\begin{equation}
    F=|\bra{\psi_\ast}\ket{\psi(T)}|^2,\qquad F\in[0,1],
\end{equation}
where $\ket{\psi(T)} = \mathcal{T}\exp\left(-i\int^T_0\mathrm dt H(t)\right)$ is the time-evolved state using a Hamiltonian $H(t)$ that contains the control protocol. 
The closer the fidelity is to unity, the closer the final state is to the target; alternatively, many studies prefer to minimize the infidelity $1-F$.

\begin{figure*}[t!]
\centering
\includegraphics[width=1.0\textwidth]{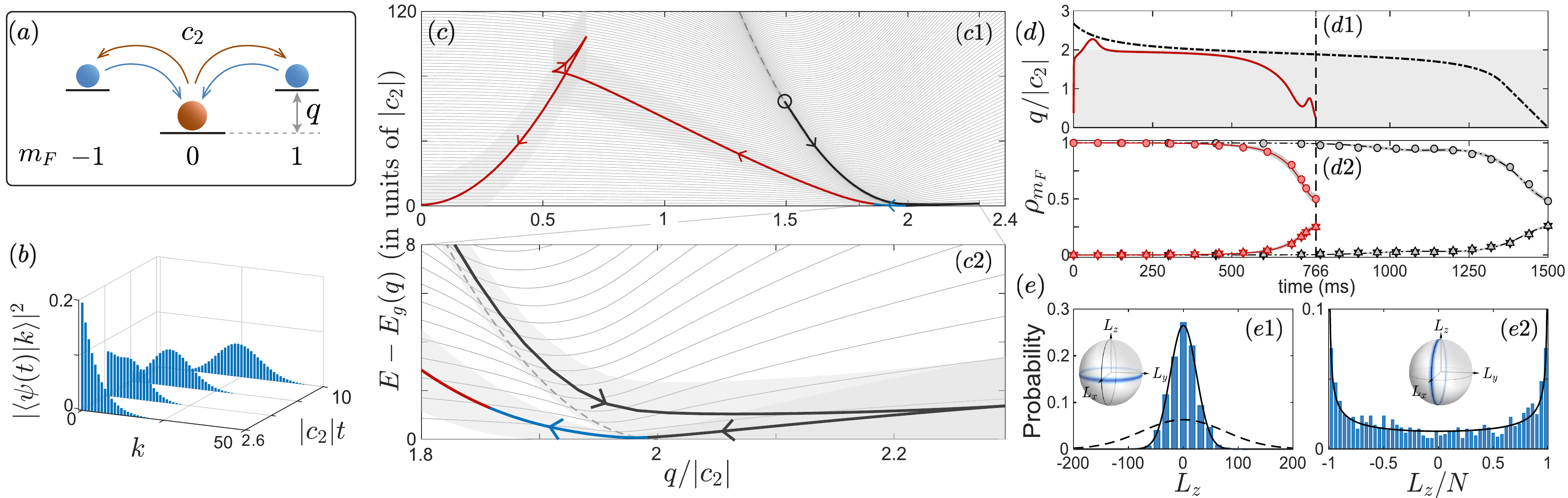}
\caption{
\textbf{RL agent prepares a Dicke state in a spin-$1$ Bose-Einstein condensate (BEC).}
(a) Schematic representation of the three hyperfine states $m_F=-1,0,+1$ for each atom; the BEC state (red) corresponds to $m_F=0$. 
(b) Time evolution of the instantaneous probability $|\braket{\psi(t)}{k}|^2$ density over Fock basis $\ket{k}$, during Stage 2 -- blue line in (c2) [see text]: the state has
been transformed to a Gaussian-like wavepacket. 
(c) Pre-training solution: 
excess energy above the instantaneous ground state during RL protocol (c1) with a zoom into the critical region (c2): the three stages are marked in grey, blue, red; arrows show the direction of time; $E=\langle\psi(t)|H(q)|\psi(t)\rangle$, and $E_g(q)$ is the instantaneous ground state energy at $q$.
(d) Experiment: control protocol (d1) and hyperfine states occupation (d2) as a function of time. The red line in (d1) corresponds to the RL protocol; the dashed-dotted black line to an empirical optimal control solution. Gray-shaded region corresponds to the Dicke phase. 
The experimental measurements of hyperfine occupation in (d2) are as follows: circles ($m_F=0$) and up/down-triangles ($m_F=\pm 1$, respectively); color code corresponds to the protocols in (d1). 
(e) Probability distributions in total spin space against $L_z$ (e1) and against $L_z/N$ after a $\pi/2$ rotation about the $x$-axis (e2), reveal the prepared Dicke (spin-squeezed) state. 
Solid lines correspond to theoretical expectation, and blue bars -- to 1057 continuous experimental runs. 
Figure adapted with permission from Ref.~\cite{guo2021faster}. Copyrighted by the American Physical Society.
}
\label{fig:state_prep_BEC}
\end{figure*}

To cast quantum state preparation in the RL framework, one typically maps the control parameters to actions, while the fidelity (or some function thereof) is used to define the rewards for the agent. However, depending on the particular application setup, the fidelity may not be directly accessible on the quantum device. In such cases, one either works with an approximation to the fidelity (e.g., when multiple copies of the system can be run in parallel or the protocol and subsequent measurement are repeated), or estimates it using the outcome of projective quantum measurements. Alternatively, if a parent Hamiltonian for the target state is known, one can indirectly prepare the state by minimizing its energy.

The definition of the RL state requires particular care since, as we discussed in Sec.~\ref{subsec:RL_framework}, quantum states are not directly measurable. In this case, one either uses a tomographic approximation when the system size is not too large, or formulates a partially observable MDP using measurements of observables or their expectation values. This fundamental limitation of partial observability represents one of the key challenges in applying RL to quantum systems, and necessitates the development of sophisticated approaches to extract meaningful state information from limited measurement data.

\textbf{Featured experimental application.} Recently, \cite{guo2021faster} deployed RL for fast preparation of an entangled Dicke state across a quantum phase transition in an experiment with a spinor Bose-Einstein condensate (BEC). The difficulty in this setup arises from a spectral gap closing at the critical point, which leads to the uncontrolled creation of excitation that renders adiabatic state preparation ineffective. Moreover, for their experimental system of a few to several thousand particles, the adiabatic timescales are beyond the coherence times of the experiment, and three-body losses start playing an insurmountable role.
By contrast, as we now discuss, the RL agent identifies new protocols that create just the right kind of excitations when passing through the gap; this allows it to reach the target state with excellent fidelity in the presence of experimental imperfections such as noise. 

To appreciate this, let us now introduce the experimental setup in more detail. Consider a spinor BEC made out of $^{87}$Rb atoms, with the ground state populating the $F=1$ hyperfine manifold ($m_F=0,\pm 1$), see Fig.~\ref{fig:state_prep_BEC}(a).
If we denote the creation operator of a boson in a given hyperfine state by $\hat a^\dagger_{m_F}$, the total particle number operator is $\hat N=\sum_{m_F}\hat N_{m_F}$, where $\hat N_{m_F}=\hat a^\dagger_{m_F}\hat a_{m_F}$. 
The Hamiltonian of the system is given by
\begin{equation}
    \hat H=\frac{c_2}{2N}\mathbf{\hat L}^2 - q(t)\hat N_0,
\end{equation}
with spin exchange interaction $\mathbf{\hat L}^2$ and a quadratic Zeeman shift term $\hat N_0=\hat N_{m_F=0}$ with controllable time-dependent coupling $q(t)$; $c_2$ is a parameter that sets the spin exchange energy scale per particle. Here the collective spin operator is $\mathbf{\hat L}=\sum_{\mu,\nu}\hat a^\dagger_\mu \mathbf{F}_{\mu\nu} \hat a_\nu$, with the vector of spin-1 matrices $\mathbf{F}$. Note that the magnitude of the spin operator is $||\mathbf{\hat L}||=N$.
We also define the magnetization of the atoms as measured by $\hat L_z=\hat N_{+1}-\hat N_{-1}$. 

The spin-exchange interaction $\mathbf{\hat L}^2$ contains processes that depopulate the BEC and magnetize the system, of the form $\hat a^\dagger_{+1}\hat a^\dagger_{-1}\hat a^2_0$ (plus hermitian conjugates). This is an effective all-to-all interaction for the atoms occupying a single BEC mode. Therefore, setting the Zeeman shift to zero, $q=0$, the system maximizes the exchange interaction $\mathbf{\hat L}^2$, and hence the ground state spans a $(2N+1)$-fold degenerate subspace $|N,L_z\rangle$ (the maximal angular momentum value is $||\mathbf{\hat L}||=N$). 
One of these degenerate ground states is the highly balanced Dicke state $|\psi_\text{Dicke}\rangle=|N,L_z=0\rangle$, which is highly entangled in terms of the original atomic degrees of freedom; it enables measurement precision close to the Heisenberg limit and is therefore particularly useful for quantum sensing. 
On the other hand, in the opposite limit $q/|c^2|\gg 1$, the ground state is the polar BEC state $|N_{-1},N_0,N_{+1}\rangle=|0,N,0\rangle$ with all atoms populating the $m_F=0$ state, which is easy to prepare in experiments.    
The two states are separated by a critical point at $q_c/|c^2|\approx 2$ where the spectral gap closes as $N^{-1/3}$.  

The goal of \cite{guo2021faster} is to use deep RL to prepare the Dicke state $|\psi_\text{Dicke}\rangle$ in a finite time, starting from the polar BEC state $|0,N,0\rangle$. 
To this end, they define the RL observations $o_t$ to consist of a tuple containing the BEC density $\rho_0{=}\langle N_0\rangle/\langle N\rangle$, and its fluctuations $\langle\delta N_0^2\rangle/N^2$ as mean-field parameters, as well as the absolute value $|\langle \hat a^\dagger_{+1}\hat a^\dagger_{-1}\hat a^2_0\rangle|/N^2$, and the phase $\theta{=}\arg(\langle \hat a^\dagger_{+1}\hat a^\dagger_{-1}\hat a^2_0\rangle)$ of the first-order quantum correction term. Crucially, observations have a well-defined large-$N$ limit which allows for pre-training agents on smaller system sizes and deploying them on larger ones. 
The RL actions determine the coupling strength of the (normalized) quadratic Zeeman shift $q(t)/|c_2|\in[-3,3]$, and are bounded and continuous. They allow the agent to construct piece-wise constant protocols $q(t)$. 

The authors use an actor-critic version of PPO, using the following pipeline: 
(i) first, the agent is trained in a simulator on exact dissipationless dynamics using as a reward the fidelity differences between consecutive timesteps resolved on a logarithmic scale: $r_t=F_t-F_{t-1}$, where $F_t=|\langle\psi_\text{Dicke}|\psi(t)\rangle|^2$.
Next, (ii), the agent is trained in the presence of dissipation by simulating quantum trajectories to reach larger system sizes; here, a (in principle) experimentally accessible reward function is used, namely the entanglement-enhanced three-mode SU(2) interferometric sensitivity: $r_t=f_t-f_{t-1}$, where $f_t=\langle\delta\theta^2\rangle_\text{rel}=1/(6\langle\delta (L_z)^2_{\theta=0}\rangle+1)$. The ideal target Dicke state has $f=1$, while the initial polar BEC state has $f=0$.
Finally, (iii), the authors test the protocol found by the RL agent in an experiment with $\sim 10^{4}$ $^{87}$Rb atoms.
The target Dicke state is reached with fidelity $F\approx 0.99$ over a time of $|c_2|\tau = 15.5$, see Fig.~\ref{fig:state_prep_BEC}(d, e). This is significantly shorter than the linear adiabatic sweep, which requires $|c_2|\tau \approx 600$, or a nonlinear sweep optimized for local adiabaticity that demands $|c_2|\tau \approx 350$ for the same fidelity level.

Interestingly, analyzing the nonadiabatic protocol found by the agent, see Fig.~\ref{fig:state_prep_BEC}(c), it appears to take advantage of excited-level dynamics. In particular, three interpretable stages are discernible in the protocol, depending on the features of the excitation spectrum of the Hamiltonian.
While the initial BEC state is given, the agent chooses to start the protocol on the Dicke side of the transition $q(0)<q_c$: this enables it to cross the transition twice, creating just the necessary amount of excitations that can be attenuated to reach the target Dicke state at the end of the protocol, as we now explain. 
(1) The first stage of the RL protocol crosses the critical point and terminates at $q_\mathrm{max}/|c_2|\approx 2.3$. The agent successively depletes the excitations via controlled destructive interference, but does not completely reach the instantaneous ground state at $q_\mathrm{max}$: the state at the end of the stage is a superposition of a few low-lying excitations. This stage serves to prepare a `better' initial state for the subsequent dynamics. 
(2) During the second stage, the agent sweeps through the critical point $q_c$ again, this time adiabatically. This increases the population of the low-lying excitations at a rate proportional to the gap, $\propto N^{-1/3}$. The end of the second stage is set by $q/|c_2|\approx 1.86$. The state has been transformed to a Gaussian-like wavepacket over the Fock basis $|k,N-2k,k\rangle$, cf.~Fig.~\ref{fig:state_prep_BEC}(b).
(3) The last, third, stage corresponds to a rapid translation of the Gaussian wavepacket in Fock space. Whereas in the beginning the wavepacket of excitation is localized at small $k$, the protocol ends at $q/|c_2|=0$ very close to the Dicke state, where the Gaussian wavepacket of excitations is located at $k\sim N/4$ ($\approx 500$, in the simulation).

To summarize, the RL agent reveals the existence of a fast-sweeping protocol that takes advantage of the dynamics of excited levels. This is enabled by the choice of a sophisticated RL framework that uses experimentally accessible observations consisting of the values of observables with well-defined values in the thermodynamic limit. It allows pre-training the RL agent on smaller system sizes before extrapolating to the experimentally relevant system sizes. 
RL has by now turned into a common tool to prepare states in the Dicke model~\cite{o2025bounding}.

\textbf{Breadth and scope of this section:} 
Beyond ultracold atoms, RL has found diverse applications for quantum state preparation in general.

Perhaps one of the first applications of learning algorithms to quantum state preparation is reported in \cite{rabitz1992teaching}, which uses a genetic algorithm to learn to excite specified rotational states in a diatomic molecule in one of the first examples of learning in an experimental quantum system. More recently, RL has been used to find optimal tweezer steering~\cite{praeger2021playing}, shape laser pulses~\cite{capuano2025shaping}, design a self-tuning laser source of dissipative solitons~\cite{kuprikov2022deep}, enable hardware-designed optimal control~\cite{ding2025hardware}, and perform incoherent state control of quantum systems with wavefunction-controllable subspaces using projective measurements~\cite{dong2008incoherent}.

Moreover, deep RL concepts have proven useful for characterizing the noise impacting a quantum chip and emulating it during simulations~\cite{bordoni2024quantum}, as well as for precise atom manipulation in nanofabrication experiments at the atomic scale via scanning tunneling microscopy~\cite{chen2022precise}. On the many-body side, notable applications include designing quantum chain fillings that achieve efficient transfer of excitations~\cite{sgroi2024reinforcement} and preparing prescribed current states~\cite{haug2021machine}. Additionally, it has been shown how to devise quantum imaginary time evolution for ground state preparation using RL agents~\cite{cao2022quantum}.

Broadly speaking, we can categorize the types of state preparation techniques developed using RL by the size of the controlled quantum system. Below, we survey the literature starting with few-body states of qubits and $\Lambda$-systems, and other systems containing up to three degrees of freedom or particles. We then move to many-body state preparation, similar to the BEC example above, where we discuss the advances brought by RL algorithms for controlling spin-squeezed states, spin chains, and Hubbard-type models.

\subsection{\label{subsec:few-body-QSP}Few-Body Systems}

\textbf{Control of $\Lambda$-systems.}
In an early application of RL before the onset of the deep learning revolution, \cite{chen2013fidelity} used tabular (probabilistic) Q-learning for state preparation in a spin-1/2 system and a $\Lambda$-type atomic system. This work compares the discretized Bloch sphere and state vector representations for RL states. The actions are chosen as predefined unitary gates, while the reward is the fidelity.

Building on these foundations, several later works have focused on control in $\Lambda$-type systems and multi-level atomic systems. \cite{paparelle2020digitally} studied the control of stimulated Raman adiabatic passage (STIRAP) in a $\Lambda$ system in the presence of dissipation modeled by Lindblad dynamics, using PPO. They define the RL state to be the density matrix of the system and use the values of the Rabi frequencies for the actions. The reward is designed as a combination of the state populations to be maximized and additional penalty terms. This work shows that RL agents can find fast and flexible nonadiabatic protocols for integer and fractional population transfer.

A physics-informed RL algorithm incorporating priors about the desired time scales of the quantum state dynamics, in addition to realistic control signal limitations, was proposed by \cite{ernst2025reinforcement}. Their PPO agent is trained to prepare states and gates in a multi-level $\Lambda$ system, a Rydberg atom, and a superconducting transmon, subject to noise, and outperforms the state-of-the-art in fidelity and robustness. The authors use the fully observable density matrix as an RL state; the actions are the control fields and the reward is based on the Uhlmann fidelity.

Similarly, \cite{guatto2024improving} used RL to improve the robustness of quantum feedback control in a three-level system. They showed that the controls learned by their agent are more resilient to exposure to unmodeled perturbations compared to a simple feedback strategy. To this end, they use estimates of the density matrix as observations, whereas the actions are the external control fields. They considered two reward functions: the fidelity (when the environment is fully observable), and expectation values of quantum observables (for partially-observable environments).

\cite{brown2021reinforcement} applied REINFORCE with a baseline to search for protocols implementing coherent population transfer in three-level quantum systems; they consider the case of fixed coupling rates but time-varying detunings in the presence of nonunitary (Lindblad) dynamics. Their results reveal the existence of efficient protocols that are competitive with standard Raman, stimulated Raman adiabatic passage, and other adiabatic schemes. The RL states in this study consist of the time steps parsed to an LSTM; the RL agent has access to actions that model the continuous values of the two detuning parameters. They used the target state fidelity as a reward.

The challenge of dissipation has been a recurring theme in quantum control applications. Control in the presence of dissipative Lindblad dynamics was also considered by \cite{an2021quantum}. They used an actor-critic PPO algorithm to design a multi-level quantum control framework capable of identifying optimal strategies with restricted control parameters. In particular, starting from the ground state of the Hamiltonian of an $N$-level system and targeting the most excited state, they show that RL agents prove resilient to perturbations in the presence of dissipation. The RL state is defined as the complete density matrix, the actions represent a binary control field, while the reward is the target state fidelity.

\textbf{Quantum dots control.}
Moving to solid-state quantum systems, several studies have explored RL applications in semiconductor quantum dots. A DQN agent for the preparation of fixed single- and two-qubit target states from arbitrary initial states was presented by \cite{he2021deep}. They considered semiconductor double quantum dots with only a few discrete control pulses. The DQN agent outperforms traditional optimization approaches using gradients and finds protocols robust to stochastic perturbations such as charge and nuclear noise. A complete quantum state vector parametrization in hyperspherical coordinates is used for the RL states; the actions are the control fields, and the reward is the fidelity.

For efficient measurement in quantum dot devices, \cite{nguyen2021deep} employed a dueling DQN agent. This work proposes an approach to the efficient measurement of quantum devices based on deep reinforcement learning, focusing on double quantum dot devices and demonstrating a fully automatic identification of specific transport features called bias triangles. The agent learns to navigate large parameter spaces to automatically locate these characteristic transport features, offering advantages in automating decision processes for quantum device characterization.

\cite{porotti2019coherent} also examined adiabatic passage through an array of semiconductor quantum dots. They use an actor-critic version of TRPO to design an RL agent and discover a control sequence that outperforms the state of the art (the so-called "counterintuitive" control sequence). The RL state is chosen as the reduced density matrix; bounded Rabi frequencies are used as actions, and the reward is a combination of state populations and penalizing terms.

Another approach to quantum dot systems involves spin control. \cite{mackeprang2020reinforcement} studied a central spin coupled to two nuclear spins. The goal for their agent is to prepare the nuclear spins in a Bell state by performing projective measurements in the presence of drift terms. Using DQN, their analysis displays learning features in preparing arbitrary two-qubit entangled states and delivers successful action sequences that generalize previously found human solutions. They use the full wave function amplitudes of the three-spin system as the RL state. The actions are projective measurements along each axis of the central spin and the identity operation, and the reward is the target state fidelity.


\textbf{Learning from projective measurements.} The studies discussed so far focused on the fidelity as a reward and assumed the RL agent has access to a sufficiently accurate estimate of it. However, in realistic experimental scenarios, direct access to quantum state fidelity is often impractical or impossible, leading researchers to explore alternative reward formulations based on measurable quantities.

Recognizing the projective nature of quantum measurements, \cite{bukov2018kapitza} considered the use of binary rewards. The author used tabular Q-learning to prepare the inverted position state of the quantum Kapitza oscillator in the presence of various sources of noise. The RL states consist of the history of taken actions: whereas this avoids the issue of using unmeasurable quantum states, this framework makes it difficult for the agent to generalize to different initial states. The actions are chosen from a discrete set of kick strengths. The reward is sparse, i.e., given at the end of the protocol, and consists of a binary number drawn according to the underlying target state fidelity.

Similarly, building on the concept of using realistic measurement data, \cite{sivak2022model-free} employed binary measurements to qubit control using tomographically complete POVMs. The key idea is to use quantum certification protocols, which are implemented in practice based on probabilistic reward measurements, to estimate the target state fidelity: $\text{argmax}_{|\psi\rangle} \mathbb{E}[R]=|\psi_\text{target}\rangle$. 

Despite these advances in making rewards experimentally accessible, both approaches face significant challenges. Such sparse rewards are given at the end of the RL episode, and pose a challenge when the episode length becomes significant, since the RL algorithm does not pick up sufficient signal to learn from.


\textbf{Arbitrary target state control.} Another interesting prospect offered by RL algorithms is the ability to generalize. While this capability is naturally built into the RL framework when it comes to finding protocols for different initial states, it is significantly more challenging to design agents that learn to prepare arbitrary target states -- a problem that requires the agent to understand the underlying control landscape rather than simply optimize for a single objective.

Addressing this challenge, \cite{haug2020classifying} used PPO to teach an agent to construct protocols that can reach different target states. To this end, they employed a neural network that automatically groups the protocols into similar types, enabling the identification of general control strategies. In particular, they showcased the ability of their agent to generate arbitrary superposition states for the electron spin in a multi-level nitrogen-vacancy (NV) center; the NV is modeled by a multilevel system that undergoes Lindblad dynamics. They used the Bloch sphere coordinates of the current and the target state as an RL state. The actions are the control parameters, and the reward is the fidelity.

Taking a complementary approach to arbitrary state preparation, \cite{wang2024arbitrary} developed a framework that explicitly integrates information about both the initial and the target state. They use a DQN agent that feeds the neural network with the target state information to find control trajectories between two arbitrary quantum states. This direct encoding of the target enables the agent to learn a more general mapping from state pairs to control sequences. The framework is applied to a semiconductor double quantum dots system in both single- and two-qubit setups. The RL states here are the POVM representation of the density matrix. The actions are the control fields, whereas the reward is the target state fidelity.

In \cite{hutin2025preparing}, a model-based reinforcement learning approach was used to learn state preparation simultaneously for a whole family of target states, in a qubit-microwave-cavity system. In contrast to many other works, the neural network was asked to output a number of parameters characterizing the entire pulse shape, after having been given the parameters characterizing the target state. After training on a simulation, these pulses were then tested on an experimental superconducting circuit setup.


\textbf{Quantum thermodynamics.} When it comes to applications of RL in quantum systems, a large body of work has been devoted to quantum thermodynamics. Since these studies, at their core, concern few-body state preparation in one way or another, we now provide a brief survey.

Entropy and energy control have been at the forefront of applications of RL in quantum thermodynamics. \cite{sgroi2021reinforcement} used REINFORCE to design an RL agent that reduces entropy production and energy dissipation in a closed quantum system brought out of equilibrium. They applied the framework to single- and two-particle systems. The RL states combine measurements at the input and the end of the episode with the time step using an LSTM architecture. The actions are the external fields. Finally, the reward is the negative quantum relative entropy of the final state and instantaneous Gibbs state.

Moving to practical thermodynamic engines, several studies have explored quantum heat engines. In a related study, \cite{sordal2019deep} investigated applications of RL to quantum Szilard engines that can convert one bit of Shannon information obtained by a binary measurement into useful work. To model the quantum Szilard engine, they consider a quantum particle in a box with a controllable strength of a fixed off-center delta function potential. The goal for their DQN agent is to find a protocol that manipulates the barrier strength, which brings the particle into an equal superposition of the lowest two eigenstates at the end of the ramp. Using RL, their study shows that asymmetric nonadiabatic Szilard engines can operate with the same efficiency as the traditional Szilard engine with adiabatic insertion of a central barrier. The RL state consists of the barrier strength and the time step. The actions are the rate of change of the delta barrier, while the reward function is based on the desired final state populations.

A key challenge in quantum heat engine optimization is that the full optimization requires operating at high power, high efficiency, and low power fluctuations, yet these objectives cannot be optimized simultaneously. Addressing this fundamental trade-off, \cite{erdman2023pareto} proposed an RL framework based on the Soft Actor-Critic algorithm to identify Pareto-optimal cycles for driven quantum heat engines that achieve a trade-off between power, efficiency, and fluctuations. The specific platform they target is quantum dots. They use a combination of the density matrix, traceless Hermitian operators, and controls to define the state space. The actions consist of the control field and the temperature. The reward function is a combination of the relative power of the heat engine and the relative cooling power of the refrigerator.

Building on this approach, \cite{erdman2023model} introduced a general model-free RL framework to identify out-of-equilibrium thermodynamic Pareto-optimal cycles. The authors discuss applications to a refrigerator based on a superconducting qubit modeled by a 2LS, and to a heat engine based on a quantum harmonic oscillator. The RL states consist of the control history; the actions are the control fields ($x$-field and oscillator frequency) and can be either discrete or continuous. The reward is a combination of the power of the heat engine and the cooling power of the refrigerator.

Finally, the concept of Pareto optimality has been extended beyond thermodynamic performance metrics. Pareto optimality in the trade-off between process fidelity and energetic cost was demonstrated by \cite{fauquenot2024eo}. The authors proposed to use techniques from quantum optimal control to simultaneously optimize the energetic cost and the process fidelity of a quantum unitary gate. They proposed a method for pulse engineering based on energy-optimized deep RL, and showed that, by itself, it performs worse than gradient-based methods, highlighting the continued importance of comparing RL approaches with traditional optimization techniques.

Extending RL applications to quantum energy storage devices, \cite{erdman2024reinforcement} applied reinforcement learning to optimize the charging protocols of Dicke quantum batteries. Their approach addresses a fundamental challenge in quantum battery design: while the Dicke model promises collective charging advantages through N two-level systems coupled to a common cavity mode, its inherently chaotic dynamics significantly limit the extractable work (ergotropy). The authors employed RL agents to discover optimal time-dependent control strategies for two key parameters: the atom-cavity coupling strength and the system-cavity detuning. Through adaptive control protocols learned by the RL agent, they demonstrated substantial improvements in both the stored energy and charging precision compared to conventional constant-parameter charging schemes. The framework successfully preserves the quantum collective speedup characteristic of Dicke batteries while mitigating the detrimental effects of quantum chaos, offering a pathway toward practical quantum energy storage applications.


\textbf{Miscellaneous.}
Finally, RL has been applied to more complex quantum systems, including cavity optomechanical systems and photonic models. To deterministically generate non-classical states in cavity optomechanical systems, \cite{liu2025deterministic} identified the carrier-wave resonance conditions and used DDPG to find the optimal driving pulses for state preparation. In particular, their agent was tasked with optimizing pulsed driving fields in a way that suppresses the induced undesired transitions facilitated by dissipation. The authors demonstrate that their RL agent succeeds in high-fidelity preparation of both (superpositions of) phononic Fock states and two-mode entangled states, in single- and two-resonator optomechanical systems, respectively. The RL states are the fully observable reduced density matrices, and the actions are the continuous control fields. The reward function is based on the target state fidelity.

\subsection{\label{subsec:many-body-QSP}Many-Body Systems}

State preparation in quantum many-body systems is significantly more complicated, due to the typically exponentially large state and configuration spaces. As a result, in general, the fidelity also decreases exponentially with increasing the number of degrees of freedom. Nevertheless, studies based on RL have demonstrated the advantages of the framework for cooling, preparation of spin-squeezed states, and ground states of many-body spin chains, as we now discuss.

\textbf{Cooling.} 
The application of RL to cooling protocols represents one of the most direct implementations of quantum state preparation. Using an RL agent to control the dynamics of cold atomic gases in magneto-optical traps was proposed by \cite{reinschmidt2023reinforcement}. Using DDPG, the RL agent enables the preparation of predefined numbers of atoms in the atom cloud. The actions control (i) the detuning of the laser frequency with respect to atomic resonance, and (ii) the magnetic field gradient at the center of the trap along the field's symmetry axis. The RL observations represent fluorescence images, fed into a neural network. The reward function is computed based on the absorption image of the atom.

In a different approach to cooling, \cite{kalita2024domino} used actor-critic algorithms to learn individual control of the evolution of coupled harmonic oscillators in a linearly-coupled oscillator network. The agent is tasked to perform indirect active cooling of the oscillators to their thermal ground states. To this end, the actions implement modulated feedback on the external oscillators to cool them down through a unique series of phase kicks. The observations consist of Monte-Carlo-estimated expectation values of the single-oscillator observables $p^2_j, q^4_j, p_jq_j$. The reward function is based on the expectation values of the particle number operator.


In one of the first RL applications to optomechanics, \cite{sommer2020prospects} employed a policy-gradient reinforcement learning algorithm to simultaneously damp multiple vibrational modes of a mechanical resonator. The agent's objective is to actively reduce the total energy of the system using only partial observations of collective mode quadratures. The actions implement discrete force pulses on the resonator, chosen from a finite set of amplitudes, to optimally lower the system’s energy. The observations are given by continuous-valued projections of the positions and momenta of the modes. The reward function is defined as the negative change in total energy, encouraging the agent to discover control sequences that effectively cool all modes concurrently.

\textbf{Spin-squeezed states.}
Beyond cooling applications, as we have already seen, RL has proven particularly effective for preparing spin-squeezed states, which contain entanglement useful for quantum sensing close to the Heisenberg limit. \cite{zhao2024prepare} used a DQN agent to prepare quantum spin squeezed states of a large collective spin. The study considers a spin-coherent initial state in a stochastic environment modeled by a Lindblad equation with dissipation and dephasing. The physical system of interest is the Lipkin-Meshkov-Glick (LMG) model. The actions correspond to the $x$-field control strength, and the RL states contain the reduced density matrix. Finally, the reward function is implemented as a penalty related to the inverse square $z$-squeezing parameter.

Extending this idea to related quantum states, \cite{chen2019manipulation} use a PPO agent to prepare a twin-Fock state, a variant of a spin-squeezed state, from a spin-$1$ atomic system. Applied to a mean-field system and a two-body quantum system, the resulting framework is shown to outperform the adiabatic dynamics in the many-body spin model. The RL state comprises a selection of local observables. The actions are the amplitude values of the nonlinear $zz$-interaction. The reward is the fidelity difference at consecutive time steps, given at each time step.

\textbf{Quantum spin systems.} 
A set of systems widely used to investigate many-body quantum state preparation are quantum spin models. These systems provide an ideal testbed for RL algorithms due to their rich dynamics and well-understood theoretical foundations. 

Among the first investigations of many-body state preparation in spin chains using RL was \cite{bukov2018reinforcement}. The authors use tabular Q-learning to prepare the ground state of a ferromagnetic Ising model. They reveal that RL agents typically get stuck in local optima of the control landscape. The RL state is defined as a combination of the control field value at a given time step, and time. The actions are binary values of the control field (bang-bang control), while the reward is the target state fidelity.

In a related work, \cite{august2018taking} applied PPO to train agents parameterized by long short-term memory (LSTM) networks, trained using stochastic policy gradients. These structures are reported to be particularly useful in controlling quantum spin chains. The RL states consist of the control protocol sequence; the actions are comprised of the control field, and the reward is given by the many-body fidelity.

Extending beyond ground state preparation to dynamic processes, \cite{zhang2018automatic} deployed DQN to study quantum state transfer in a one-dimensional XY spin chain exposed to a transverse $z$-field. The agent is tasked with finding the shortest high-fidelity protocols, and is shown to find faster protocols than previously known ones. The RL state is chosen as the real and imaginary part of the wave function components in the $z$-basis. The actions are the strength of the external $z$-field, and are taken binary (bang-bang protocols). Finally, the reward is a combination of the fidelity and protocol duration, to optimize for speed.

To address the challenge of scaling to larger systems, \cite{metz2023self} used DQN using a tensor-network learning architecture (matrix product states and operators) to train an RL agent to perform self-correcting many-body state preparation. The RL states are defined as a compressed MPS representation of the quantum state. The discrete actions are given by the control fields, whereas the reward is the instantaneous logarithmic many-body fidelity.

Moving beyond discrete control schemes, \cite{zhou2023auxiliary} investigate the effectiveness of continuous DDPG control for controlling quantum systems. The authors propose an auxiliary task-based deep RL framework to tackle the sparse reward problem of a quantum environment. They provide applications to single and multi-spin systems, including Bell-state preparation and computational state initialization (i.e., moving a set-to-unity bit in the computational basis states of a many-body chain). The RL states are defined by the real and imaginary parts of the quantum wave function amplitude. The actions are given by the continuous control field. The reward is based on the fidelity.

To systematically evaluate different approaches, a comparative study in the problem of preparing a desired quantum state on the efficacy of various control algorithms is provided by \cite{zhang2019does}. These include tabular Q-learning, deep Q-learning, and policy gradient, as well as the non-ML algorithms stochastic gradient descent and Krotov algorithm. DRL algorithms are reported to adaptively reduce the number of control steps to find optimal solutions efficiently, whereas SGD and Krotov require a fixed number of steps. The RL states are given by discretized values of the wave function amplitudes. The actions are piecewise constant control pulses, whereas the reward is based on a logarithmic resolution of the instantaneous state fidelity.

Beyond direct state preparation, RL algorithms have also made their way into the study of simulated annealing. \cite{mills2020finding} used PPO to search for the ground state of spin Hamiltonians defined on the interaction graph of variable connectivity, for both classical and quantum problems. The RL observations are either the spin states (classical setup) or based on the outcome of destructive measurements (quantum setup). The actions are the values for the total inverse temperature change of the annealing schedule. The reward is the minimum negative energy; it is sparse, i.e., given at the end of the episode.

Exploring more complex quantum phenomena, \cite{ye2024controlling} investigate the controllability of nonergodicity in quantum many-body systems, using the example of a one-dimensional Fermi-Hubbard model. To this end, they construct a deep RL framework, and demonstrate that PPO protocols offer enhanced control robustness. The RL observations are given by the tuple of spin-resolved atom imbalance, subsystem fidelity, and the full-state fidelity. The continuous actions are the values of the tilt field and the interaction strength. The reward function is based on the fidelity and the imbalance.

Taking a different approach to quantum state preparation, recent work has focused on optimizing quantum circuit architectures themselves for preparing thermal states in complex many-body systems~\cite{kundu2025improving}. In particular, the Sachdev-Ye-Kitaev (SYK) model presents a challenging testbed where RL has been employed to optimize both the quantum circuit and its parameters for thermal state preparation. This approach demonstrates the effectiveness of the RL framework in both noiseless and noisy quantum hardware environments, with remarkable efficiency gains that reduce the number of CNOT gates by two orders of magnitude for systems $N \geq 12$, compared to traditional methods. The RL framework uses the current circuit architecture as the state representation, while actions comprise CNOT and parametrized single-qubit rotations ($X$, $Y$, $Z$). The reward function is based on the free energy, and the optimization is carried out using DDQN with convolutional neural networks to handle the complex circuit topology optimization problem.

\textbf{Learning representations from small amounts of data} is essential for the control of quantum many-body systems using RL, since quantum states are only partially observable. Addressing this fundamental challenge, \cite{zhu2024controlling} propose a representation-guided reinforcement learning framework that tackles the state preparation problem when only limited measurement information is available. The authors develop an ML algorithm that uses a small amount of measurement data to construct an internal representation of the quantum many-body state. The algorithm compares the data-driven representations with the representation of the target state, and uses a PPO agent to find the appropriate control operations. 

This innovative approach demonstrates the efficient control of uncharacterized, intermediate-scale quantum systems using a limited set of quantum measurements, making it particularly valuable for realistic experimental scenarios where complete state information is unavailable. The framework is applied to two distinct quantum systems: (1) ground state preparation across different phases of the anisotropic XXZ chain with dimerized interactions, where the RL agent was tasked with preparing the ground state starting from various phases of the model: symmetry-broken, trivial, and symmetry-protected topological (SPT); (2) preparation of continuous-variable cat states, demonstrating the versatility of the approach across different quantum state classes.

The technical implementation involves RL observations comprised of (i) an estimate of a POVM which is randomly chosen and independent of the target, and (ii) a vector of measurements. The actions control the strength of the various interactions, while the reward is based on the norm difference of the representations of the current and the target state.

\textbf{Meta-learning.}
A distinct approach within the RL framework for quantum state preparation involves meta-learning, where the RL agent optimizes the learning process itself rather than directly controlling physical parameters. \cite{jae2024reinforcement} use an actor-critic algorithm to meta-learn the hyperparameters for a variational quantum circuit, which aims to transform a target few-qubit state to the product state $\ket{0}^{\otimes}$. The learning task (different from the RL agent's task) here amounts to finding the optimal gate angles sampled from a normal distribution. The role of the RL agent is to find the optimal training schedules for the variance of this distribution and the step size for gradient descent. To do this, the authors use curriculum learning and demonstrate that the RL agent manages to reduce the number of data points needed in training to learn the quantum circuit, outperforming quantum state tomography. Interestingly, an agent trained on the 3-qubit problem generalizes up to 5 qubits.
The meta-RL agent receives as an observation a one-hot vector of the measurement outcome, and learns the discrete probability to take a point from a discretized grid of hyperparameter space. The reward signal is $0$, if the loss for the learning task falls below a preset threshold, and the agent incurs a penalty of $-1$ otherwise.

%% file: gate_engineering.tex
\section{\label{sec:gates}Pulse-Shape Engineering for Quantum Gate Implementation}

The quest for reliable quantum computing is inextricably linked to the accuracy of logical quantum operations implemented by quantum gates. While modern software for NISQ devices allows users to directly select desired unitary operations, gate operations are implemented at the hardware level via Hamiltonian evolution following optimized pulse shapes: externally driven microwave-mediated operations for superconducting qubits, and external laser fields for trapped-ion and Rydberg-atom platforms. 
A key obstacle arises from the sensitivity of these protocols to noise and fabrication variability, often resulting in imperfect quantum gates. This is exacerbated by the difficulty of microscopically characterizing system properties and stringent engineering constraints. 

This section reviews recent progress in RL for quantum gate synthesis and gate set design. Preparing quantum gates is more subtle than the joint state preparation of an independent basis set, as gates encode information about relative phases between these orthogonal states. A commonly used figure of merit is the average gate fidelity
\begin{equation}
\label{eq:gate_fidelity}
    \mathcal{F} = \int\mathrm d\psi_0\; |\bra{\psi_0}U_\ast^\dagger U(T)\ket{\psi_0}|^2\; ,
\end{equation}
where the integral is taken over initial states $\ket{\psi_0}$ distributed according to the Haar measure on the relevant Hilbert space. In practice, the integral is difficult to evaluate exactly; one estimates the gate fidelity by sampling initial states from the Haar measure or some unitary $k$-design, or restricts to using an orthonormal basis of initial states.  

\begin{figure*}[t!]
\centering
\includegraphics[width=1.0\textwidth]{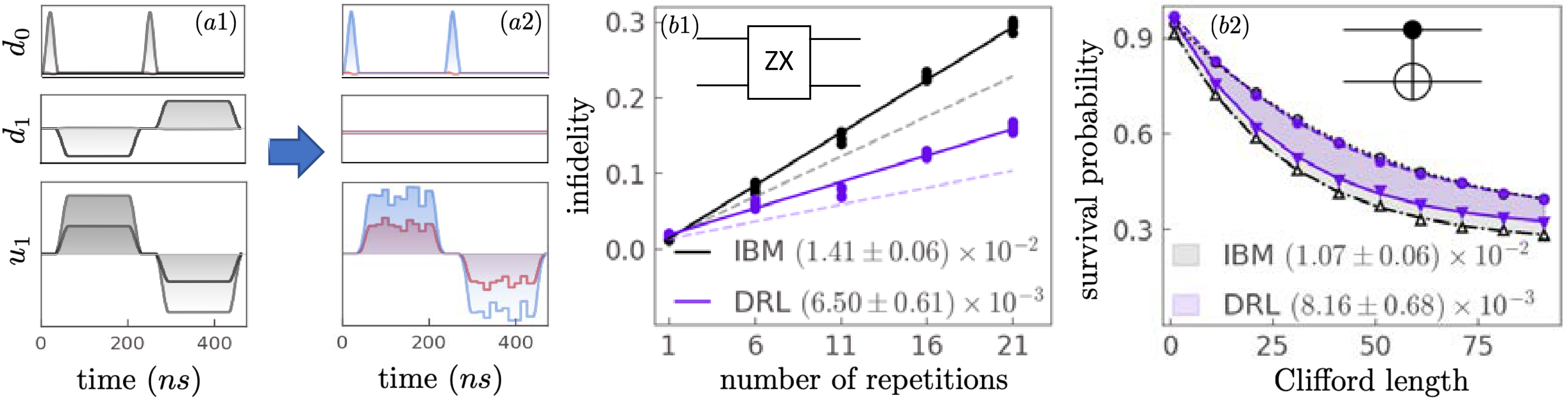}
\caption{
\textbf{RL optimized cross-resonance (ZX) and CNOT two-qubit gates on a superconducting qubit quantum computer.}
(a) Protocols for the three control channels ($d_0,d_1,u_1$) -- electromagnetic pulse shapes used to implement a ZX($-\pi/2$) gate:
state-of-the-art theory protocol (a1) achieves infidelity $(1.05\pm 0.07)\times 10^{-2}$;
RL optimized protocol (a2) achieves infidelity $(4.40\pm 0.58)\times 10^{-3}$. 
(b) Robustness of RL-optimized two-qubit gates after training was complete, with gate infidelities shown in the legend.
(b1) Infidelity of ZX($-\pi/2$) gate against number of repetitions immediately after optimization (dashed lines) compared to 25 days after optimization (solid lines).  
(b2) Survival probability of the initial state against the length of the Clifford circuits in interleaved randomized benchmarking 25 days after RL optimization. Circles represent standard randomized benchmarking sequences; triangle sequences interleave the standard sequence with CNOTs. 
Figure adapted with permission from Ref.~\cite{baum2021experimental}. 
Copyrighted by the American Physical Society.
}
\label{fig:gates}
\end{figure*}

Reinforcement learning offers several advantages for quantum gate synthesis. First, designing low-level controls can compensate for hardware errors due to miscalibration or malfunction. Using a model-free framework avoids relying on accurate noise models, which are often unavailable. RL algorithms do not require knowledge of specific Hamiltonian models, controls, or underlying error processes; successfully trained RL agents can learn platform-specific imperfections from measurement data. Second, RL optimization can discover new gate protocol sequences accounting for leakage to other levels, parameter drifts over time, or terms beyond theoretical approximations. Finally, RL agents can provide pulse protocols to realize compound unitary gates that are shorter than transpilation using a universal gate set.
Let us now discuss a concrete example.

\textbf{Featured experimental application.} An early application of RL to quantum gate design was presented in \cite{baum2021experimental}. The authors design a universal set of error-robust quantum logic gates on a superconducting quantum computer. Their RL agent synthesizes RX$(\pi/2)$ single-qubit rotations up to 3 times faster than the state-of-the-art DRAG protocol; it also implements ZX$(-\pi/2)$ entangling gates using a cross-resonance pulse applied to the control qubit, up to 2 times faster than the default implementation reaching a fidelity $F_\text{ZX}> 99.5\%$. Both exhibit robust calibration-free performance over a few weeks. The work also discusses RL strategies to directly optimize SWAP gates in situ. 

To appreciate the advantages offered by RL, recall the conventional paradigm where one starts with a Hamiltonian and a noise model that are assumed to describe the system's behavior; one then uses analytical and numerical methods to optimize the waveforms of the underlying pulse shapes. However, this requires a detailed and accurate model; in particular, making small approximations can often lead to significant degradation of gate fidelity. 

By contrast, the RL approach is model-free and unbiased; it takes into account additional degrees of freedom (e.g., higher levels) and drift terms. However, a key difficulty is the design of an RL framework compatible with the limited observability of the quantum state and its unitary evolution on the quantum device. This raises the question of how to supply the RL algorithm with data during episodic training.  \cite{baum2021experimental} overcome this issue by performing simplified state tomography to estimate the density matrix of the state by repeating each intermediate measurement $256$ times (and $1024$ times for the final episode state). At each step $t$ during the RL episode, they first reinitialize the qubits and repeat the exact sequence of actions through step $t-1$, and then apply the new action at step $t$ before moving on to the next step. 

To estimate the gate fidelity at the end of the episode, the authors use a reward built out of a weighted sum of different repetitions to amplify the gate error and overcome state preparation and measurement errors. They evaluate the gate performance using so-called Clifford randomized benchmarking to estimate average error per gate~\cite{baum2021experimental}.  

To learn the pulse sequence that implements a gate, the total fixed gate duration is discretized into small time steps, see Fig.~\ref{fig:gates}(a2) for an example. The RL agent then constructs piece-wise constant waveforms by selecting the strength of the control fields according to its policy without the application of additional cross-talk cancellation tones. In the case of transmon qubits, gate synthesis requires simultaneous access to multiple complex-valued control channels ($u_1, d_0, d_1)$, decomposed into amplitude and phase optimization.

The RL agent is trained using a variant of REINFORCE. The algorithm is executed in runtime on experimental superconducting qubit hardware. The agent itself is hosted on a cloud server, and is given access to the quantum computer via the cloud for fixed windows of time. To account for the mismatch of timescales on the superconducting quantum platform and the classical computer where the training is performed, the authors batch the data from several episodes executed on the hardware during training. The algorithm is reported to converge within as few as $500$ episodes.   

Figure~\ref{fig:gates}(a) shows the waveforms of the three control channels used to implement an entangling cross resonance gate ZX$(-\pi/2)$ according to an optimized theory model (a1) and after model-free RL optimization (a2). The authors report an infidelity decrease from $(1.05\pm 0.07)\times 10^{-2}$ for the theory protocol, to $(4.40\pm 0.58)\times 10^{-3}$ for the RL pulse shape. Interestingly, the RL agent introduces additional features in the protocol that likely help mitigate noise. 
Figure~\ref{fig:gates}(b) demonstrates the resilience of the RL optimized ZX$(-\pi/2)$ gate after training was complete. The gate infidelity is measured after a fixed number of gate repetitions immediately after training (dashed lines) and 25 days later (b1). 
In Fig.~\ref{fig:gates}(b2), the authors show randomized benchmarking results for the CNOT gate implementations. Here, a standard sequence of Clifford gates is compared to a sequence with interleaved CNOT gates, and the survival probability of the initial state is plotted against the circuit depth.  
Although the fidelity is expected to degrade, the RL gate clearly outperforms the state-of-the-art theory gate pulse. 

In summary, \cite{baum2021experimental} report one of the first experimental applications of model-free deep RL for the efficient and autonomous design of error-robust quantum logic gates on a superconducting quantum computer. They show that RL presents an efficient toolbox for finding error-robust gate sets. Moreover, their results demonstrate that RL agents can outperform state-of-the-art human-designed gate implementations by incorporating the effects of unknown terms via direct interactions with experimental hardware.

\textbf{Breadth and scope of this section:} The remainder of this section is devoted to a survey of recent applications of RL to pulse shape engineering for quantum gates. We review the simpler case of single-qubit gates in Sec.~\ref{subsec:1q_gates} before moving on to the more challenging two-qubit entangling gates in Sec.~\ref{subsec:2q_gates}.

\subsection{\label{subsec:1q_gates}Single-Qubit Gates}

Single-qubit gate synthesis represents a fundamental application of RL in quantum control, with approaches ranging from general unitary approximation to hardware-specific optimization and noise-robust protocols.

\textbf{General Unitary Synthesis.}
\cite{moro2021quantum} developed an RL strategy for approximating arbitrary single-qubit unitaries using single precompilation, training a PPO agent on Haar-random single-particle gates. The approach uses observations $O_n = U_\ast U_n$, where $U_n$ is the gate operation up to time step $n$ and $U_\ast$ is the inverse target gate, which effectively measure distance to the target. Actions consist of $\pi/128$-rotations along $\pm x,\pm y,\pm z$-axes, with rewards based on the number of remaining steps plus negative gate fidelity (averaged over the Haar ensemble and normalized by the total number of steps) at episode termination. The RL-synthesized gates demonstrate reduced execution time and improved tradeoffs between protocol length and execution time, enabling potential real-time operations.

\textbf{Hardware-Optimized Control.}
\cite{wright2023fast} address the gate speed and leakage challenges simultaneously through a custom deep RL algorithm employing two specialized agents: one constructing in-phase control pulses and another determining out-of-phase pulses for leakage reduction. Their proof-of-concept experiments on superconducting qubit hardware demonstrated X and $\sqrt{\text{X}}$ gate synthesis with various durations. Within approximately 200 training iterations, the agents can produce novel control pulses up to twice as fast as state-of-the-art gates while maintaining equivalent fidelity and leakage rates. The approach uses experimental observation signals (both in-phase and quadrature components) directly as RL observations, bypassing quantum state classification. Actions control the $x$ and $y$-quadrature components of control pulses, with rewards constructed from estimates of measured signal components and provided at intermediate times to circumvent experimental inaccessibility of gate fidelity.

\textbf{Noise-Robust Digital Control.}
Fast and robust digital pulse synthesis under Lindblad noise was investigated by \cite{ding2021breaking}, who combined shortcuts to adiabaticity (STA) with RL. A PPO agent learned bounded $z$-field protocols achieving robust digital quantum control with operation times constrained by STA-dictated quantum speed limits, while exhibiting robustness against systematic errors. The RL state comprises excited state population, renormalized previous actions, and a normalized timestep, with continuous actions constructing piecewise constant control fields. Different instantaneous rewards are employed during pretraining (action-related) and training (fidelity-related).

The approach was experimentally validated by \cite{ai2022experimentally} on a trapped $^{171}$Yb$^+$ ion using PPO. The piecewise constant RL protocol was transformed into frequency-modulated microwave drives, with protocol strength determining the value of frequencies at each timestep. State-dependent fluorescence measurements determined excited state probabilities, confirming that the agent discovers fast and robust digital quantum controls with operation times consistent with STA estimates. Notably, the study demonstrated protocol robustness against over-rotations and off-resonance errors without prior knowledge input.

\textbf{Measurement-Based Learning.}
\cite{ding2023closed} also employed PPO to learn single-qubit gates from observations under Lindbladian noise modeling detuning, dephasing, and relaxation. The setup preserves the qubit wavefunction while introducing slight perturbations through weak ancilla measurements for information extraction. The agent learns to flip qubit states using finite-time X-pulse sequences followed by weak continuous $z$-basis measurements of the ancilla. Training demonstrates successful transfer learning adaptation to new noise patterns without retraining from scratch. The RL state is complex and incorporates the action, timestep, ancilla projective measurement output, and absolute values of qubit density matrix entries. Actions control the normalized Rabi frequency, while rewards derive from deviations of the excited state population from unity.

\subsection{\label{subsec:2q_gates}Two-Qubit Gates}

While RL frameworks enhance single-qubit gate fidelity in noisy environments, their primary advantage emerges in synthesizing entangling two-qubit gates where exact analytical solutions are typically unavailable. The complexity of two-qubit systems necessitates sophisticated optimization approaches that can navigate high-dimensional control landscapes while simultaneously addressing fidelity, speed, and robustness constraints.

\textbf{Superconducting Transmon Systems.}
Superconducting transmon architectures have emerged as a primary testbed for RL-based two-qubit gate synthesis. \cite{niu2019universal} developed an early TRPO framework simultaneously optimizing speed and fidelity against leakage and stochastic control errors in two-state and three-state transmon models. Their agent achieved a two orders of magnitude reduction in average gate error over stochastic gradient descent solutions, and up to an order of magnitude reduction in gate time compared to optimal synthesis baselines. The framework uses gate unitaries as states, piecewise constant pulse shapes from learnable Gaussian distributions as actions, and composite rewards incorporating Haar-averaged gate fidelity, transmon leakage, control field strength, and gate duration.

\cite{nguyen2024reinforcement} employed DDPG to design cross-resonance ZX and CNOT gates for superconducting qubits without prior physical system knowledge. Their approach discovered novel control solutions fundamentally different from conventional gate-error suppression techniques while surpassing direct and echoed control schemes in both fidelity and execution time. The method uses quantum states as RL observations, continuous actions controlling real and imaginary components of transmon control channels, and log-resolved Haar-averaged gate fidelity as reward. 

\cite{dalgaard2020global} applied AlphaZero to construct cross-resonance gates in transmon systems, investigating digital, constrained, and unconstrained optimization of controlled quantum dynamics. Their approach represents states as current unitaries, uses actions controlling complex-valued pulse strengths (smoothed via Gaussian filtering), and employs gate fidelity as reward. A noteworthy feature of their action implementation is that, while the protocol is discretized in time steps, the physical system receives smooth pulse shapes, mimicking real experimental hardware. 

\textbf{General Gate Synthesis Frameworks.}
Several studies have developed platform-agnostic approaches to two-qubit gate synthesis. 

\cite{an2019deep} applied gradient-free dueling double DQN to synthesize CNOT gates alongside single-qubit Hadamard gates, using RL states containing real and imaginary parts of the gate at each timestep, discrete actions modifying control fields, and logarithmic gate fidelity as reward. \cite{shindi2023model} extended this approach using the same algorithm for Hadamard, CNOT, and T gates, relying exclusively on end-of-episode measurements to enable optimal control policy discovery without quantum state access during learning. Their framework uses control field values and remaining timesteps as observations, with actions controlling Hamiltonian couplings and rewards computed from log-resolved minimum gate fidelity over test state ensembles.

\cite{hu2022deep} enhanced the efficiency of gate synthesis by incorporating step penalties in the reward function, employing TD3 to automatically discover shorter control sequences than traditional gradient-based and evolutionary algorithms while maintaining high accuracy for single-qubit and CNOT gates. Their approach uses real and imaginary gate components as states, continuous actions controlling piecewise-constant control field strengths, and affine functions of log-resolved gate fidelity as rewards.

\cite{xu2024robust} investigated PPO for computer-assisted optimal experiment design in implementing $\text{R}_\text{YY}(\pi/4)$ entangling gates; they also analyzed the necessity of human expert guidance for the effective RL implementation. They concluded that their framework requires no physics insights, such as Hamiltonian details. RL states comprise current unitary, previous action, and timestep tuples; actions control single-qubit X/Y and Z-field coupling strengths plus ZZ-interaction strength; rewards use logarithmic gate infidelity at episode termination.

\textbf{Miscellaneous.}
\cite{ivanova2024discovery} optimized exchange-gate sequences for CNOT and CZ gates in exchange-only quantum computation using DDPG agents manipulating six spins representing two logical qubits. Their approach achieved significant total gate time improvements over previously known results. The state space consists of 35 values parametrizing the $\text{SWAP}^\alpha$ gate parameters ($\alpha \in [0,2]$), from which exchange pulse sequences are constructed. Actions either increment $\alpha$ values or apply optimization routines to existing pulses. Rewards combine total timesteps and Fong-Wandzura gate distance. 

Last but not least, \cite{zhang2025meta} developed meta-reinforcement learning quantum control algorithms addressing decoherence and control pulse imperfections in high-fidelity quantum gate realization. Their hierarchical approach employs an inner RL network for concrete optimization decision-making and an outer meta-learning network adapting to varying environments while providing inner network feedback. The framework successfully realizes Hadamard, $\pi/8$-phase, phase, and CNOT gates across diverse environmental conditions.

Now that we have discussed how to implement quantum gates using RL, we move on to the construction of entire circuits. 

%% file: circuit_design.tex
\begin{figure*}[t!]
\centering
\includegraphics[width=1.0\textwidth]{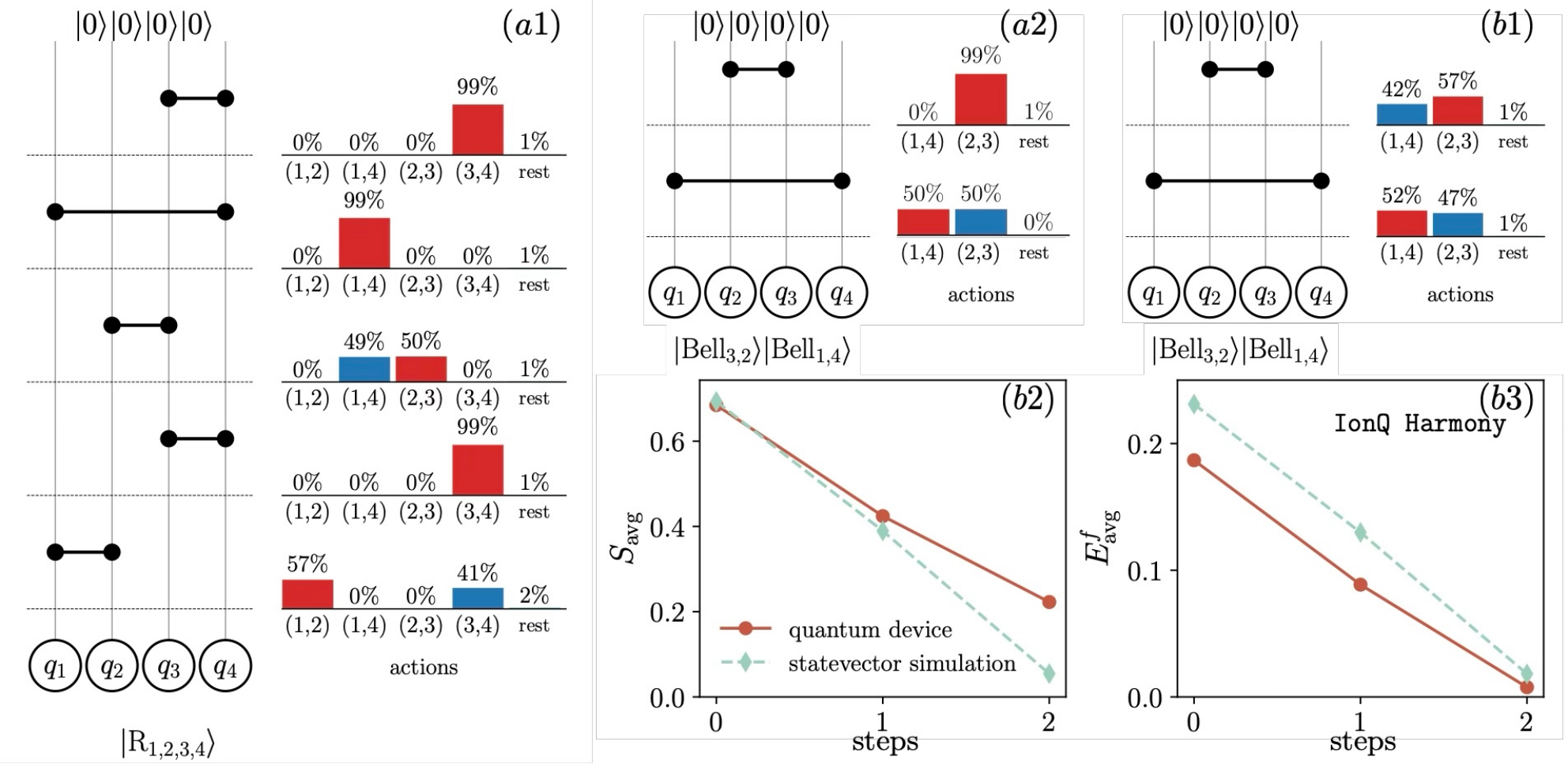}
\caption{
\textbf{RL-designed disentangling circuits for 4-qubit states, implemented on a trapped ion quantum computer.}
A trained RL agent can disentangle arbitrary 4-qubit states using at most 5 state-dependent two-qubit gates (black dots connected by solid line). At each step (marked by a horizontal dotted line), the agent receives as an observation all two-qubit reduced density matrices, and selects the pair of qubits to place an optimally disentangling gate according to its policy, shown as a histogram over qubit pairs (selected action marked in red). 
A Haar-random state $\ket{\text{R}_{1,2,3,4}}$ (a1) and a product of Bell states $\ket{\text{Bell}_{3,2}}\ket{\text{Bell}_{1,4}}$ (a2). 
The agent is then tested on a trapped-ion quantum computer (b1), where noise degrades the quality of the observations and the policy becomes less certain. Whereas the average single-qubit entanglement remains finite at the end of the circuit (red line, b2) [corresponding to setup in (b1)], the agent brings the entanglement of formation $E^f_\text{avg}$ to zero (red line, b3), indicating that the qubit state becomes entangled with its physical environment. 
Figure adapted with permission from Ref.~\cite{tashev2024reinforcement}. 
}
\label{fig:disentangling_circuit_4q}
\end{figure*}

\section{\label{sec:circuit_synth}Quantum Circuit Design}

The design and optimization of quantum circuits present fundamental challenges in modern quantum computing, with direct implications for the scalability and practical implementation of quantum algorithms on near-term devices. As quantum systems continue to grow in complexity and size, the traditional approaches to circuit design -- often relying on analytical methods or heuristic optimization -- become increasingly inadequate for navigating the exponentially large space of possible quantum gate sequences and architectures. This challenge is exacerbated in the development of noisy intermediate-scale quantum (NISQ) devices, where circuit depth has to be minimized to reduce the accumulation of errors, while simultaneously achieving optimal performance for specific computational tasks.

Reinforcement learning has emerged as a particularly promising paradigm for addressing quantum circuit design challenges, offering a systematic framework for exploring vast combinatorial spaces and learning optimal control policies either in simulations or even directly through interaction with quantum systems. The appeal of RL in this context arises from its ability to handle sequential decision-making problems with delayed rewards, cf.~Sec.~\ref{subsec:RL_framework}, making it naturally suited for the step-by-step construction of quantum circuits where the final performance depends on the entire sequence of gate operations. Moreover, as we shall see, RL algorithms can adapt to different hardware constraints, noise models, and optimization objectives, providing a flexible approach that can be tailored to specific quantum computing platforms and applications.

The intersection of reinforcement learning and quantum circuit design has given rise to a rich landscape of research directions, encompassing problems ranging from fundamental quantum state preparation to sophisticated applications in quantum chemistry and optimization. These efforts can be broadly categorized into three key areas: the engineering of quantum entanglement through optimal control strategies [Sec.~\ref{subsec:ent_ctrl}], the enhancement of variational quantum algorithms through learned circuit architectures [Sec.~\ref{subsec:VQE}], and the development of intelligent compilation methods that can automatically synthesize efficient quantum circuits for specific hardware platforms [Sec.~\ref{subsec:arch_search}]. Each of these areas presents challenges and opportunities, requiring careful consideration of the trade-offs between circuit complexity, quantum resource requirements, and algorithmic performance.

Another significant area where RL is being used to optimize quantum computing is in quantum error correction strategies, which will be discussed in Sec.~\ref{sec:QEC}.


\textbf{Featured experimental application.} 
Before delving into these details, we highlight a recent application of RL to construct disentangling quantum circuits for multi-qubit pure states. The circuits, built from two-qubit gates, unitarily map arbitrary initial quantum states to the $\ket{0}^{\otimes L}$ state [Fig.~\ref{fig:disentangling_circuit_4q}(a1)]. Reversing disentangling circuits provides a method for preparing arbitrary states on quantum computers, making them highly relevant for future digital quantum simulation applications. The RL agent we are about to discuss, was trained in simulation and subsequently tested experimentally on a trapped-ion quantum computer in a proof-of-concept demonstration~\cite{tashev2024reinforcement}.

The state disentangling problem can be formulated as entanglement minimization over circuit parameters. Rather than pre-defining the circuit architecture, the RL agent constructs it dynamically, one gate at a time, using reduced two-qubit density matrices as partial observations. In the generic problem, at each timestep, the optimization is two-fold: (i) selecting the best from the discrete set of all qubit pairs on which to apply the two-qubit gate, and (ii) optimizing over the continuous set of gate parameters that define the gate. \cite{tashev2024reinforcement} solve the discrete optimization using RL while employing an analytical solution for the continuous optimization. Crucially, both optimizations depend on the current circuit state, which determines the actual disentangling gate applied.

The reward function is designed to minimize the average single-qubit entanglement entropy -- the entanglement between each qubit and the rest of the system, averaged over all qubits -- which vanishes if and only if the pure state is a product state. For practical purposes, a state is considered disentangled when the average single-qubit entanglement drops below a threshold (here $10^{-3}$). Additionally, a penalty term incentivizes the agent to maintain circuit brevity.

As mentioned in Sec.~\ref{subsec:RL_framework}, measuring exact quantum states is infeasible due to the exponential growth of Hilbert space size with qubit number $L$. In contrast, estimating two-qubit reduced density matrices scales as $\propto L^2$. The authors therefore use all two-qubit reduced density matrices for a given state as RL observations. The action space encompasses all qubit pairs for gate placement, with no restrictions on using neighboring qubits. Hence, qubits can be thought of as residing on the vertices of an $L$-vertex graph.

The agent is trained in simulation using an actor-critic PPO algorithm, starting from Haar-random states with fixed qubit numbers. Notably, the policy network utilizes attention layers, rendering it permutation equivariant: if two qubits are permuted in the input state, the corresponding output actions adjust automatically. This feature significantly facilitates training and enhances agent generalizability.

Figure~\ref{fig:disentangling_circuit_4q}(a1) shows a disentangling circuit constructed by the trained RL agent for a Haar-random 4-qubit initial state. Analysis of the emergent structure reveals that while the specific gates applied are state-dependent, the five-gate circuit architecture is universal~\cite{tashev2024reinforcement}. Specifically, any 4-qubit state can be exactly disentangled to $\ket{0}^{\otimes 4}$ using at most 10 CNOT gates. An application to a product of two Bell pairs is shown (a2), demonstrating the RL agent's ability to identify entanglement distribution among qubits and leverage this information to solve the disentangling problem. 
One can then see that disentangling circuits remove correlations between all qubits as the agent involves the same qubit in consecutive disentangling operations. Moreover, different topological classes can be assigned to circuit segments where the circuit is incompressible. 

For practical applications, the RL agent outperforms deterministic disentangling algorithms -- which require full state knowledge -- for initial states where entanglement distribution exhibits structure (e.g., some qubits are more weakly entangled than others). To scale to larger systems, the authors relax the constraint that initial states are Haar-distributed~\cite{tashev2024reinforcement}.

The authors also explore the resilience of the trained agent to noise, highlighting their potential for real-world quantum computing applications. Whereas training on the quantum device would require multiple runs to collect quantum statistics, the use of only local observations makes the RL agent directly applicable to modern NISQ devices. Among these, trapped-ion computers can implement high-fidelity two-qubit gates on arbitrary qubit pairs. In a proof-of-concept implementation [Fig~\ref{fig:disentangling_circuit_4q}(b1-b3)], the RL agent demonstrates robustness to experimental noise and successfully disentangles 4-qubit entangled initial states~\cite{tashev2024reinforcement}. To detect true entanglement levels in the presence of decoherence and dissipation in the experiment, the authors use the entanglement of formation metric.

In summary, \cite{tashev2024reinforcement} demonstrate that partial quantum state knowledge can effectively control multiqubit entanglement. More broadly, this work showcases reinforcement learning agents as a suitable framework for investigating interactive quantum dynamics, with potential to address outstanding challenges in quantum state preparation and compression, quantum control, and quantum complexity.

\textbf{Breadth and scope of this section:}
In the remainder of this section, we provide a comprehensive survey of recent advances in applying reinforcement learning to quantum circuit design, examining both the theoretical foundations and practical implementations across different quantum computing paradigms. We explore how RL agents are trained to discover novel circuit architectures, optimize existing quantum algorithms, and navigate the complex landscape of quantum compilation while respecting hardware constraints and noise limitations. Throughout this section, we aim to highlight the transformative potential of reinforcement learning approaches in quantum circuit design and identify promising directions for future research in this rapidly evolving field.

\subsection{Entanglement Control}
\label{subsec:ent_ctrl}

The generation and stabilization of quantum entanglement lies at the heart of many quantum technologies, from quantum communication protocols to quantum sensing applications. However, creating and maintaining entangled states in realistic quantum systems presents significant challenges due to decoherence, measurement backaction, and the inherent complexity of many-body quantum dynamics. Reinforcement learning offers a powerful framework for addressing these challenges by learning optimal control strategies that can adapt to system-specific constraints and noise characteristics.

\textbf{Continuous Control Approaches.}
\cite{li2024reinforcement} proposed an RL-based control framework to enhance entanglement by modulating the two-photon-driven amplitude in the Rabi model. They argue that entanglement can reveal properties of the underlying phase transition and use a DQN agent to find temporal sequences of control pulses to increase entanglement in the presence of dissipation. Their framework works across different parameter regimes and is robust against dissipation. The RL observations consist of expectation values of spin and photon observables and their variances. The actions control the strength with which two-photon transitions are driven, and take three discrete values. The reward is based on the partial transpose criterion for entanglement; it is designed to incentivize the agent to entangle the state.

A particularly sophisticated approach to entanglement engineering combines optimal feedback control with weak continuous measurements under partial state observation. \cite{ye2024entanglement} developed an RNN-based agent trained using the actor-critic PPO algorithm for entanglement generation in quantum optomechanical systems. The framework addresses both linear and nonlinear photon-phonon interactions, with the PPO agent sequentially interacting with one or multiple parallel quantum optomechanical environments to maximize accumulated rewards for creating and stabilizing quantum entanglement over arbitrary time periods.

The RL framework in this context is particularly noteworthy: observations comprise expectation values of photon number occupation and photocurrent measurements, while actions modulate the laser amplitude within physical bounds. The reward function quantifies deviations from target state values, creating a direct connection between the RL objective and the desired quantum state properties. This approach demonstrates how RL can effectively handle the partial observability inherent in realistic quantum systems while maintaining robust performance.

Moving beyond single quantum systems, \cite{almanakly2024deterministic} explored RL applications in distributed quantum architectures, focusing on maximizing absorption efficiency in chiral quantum interconnects. Their work addresses the critical challenge of distributing entanglement between non-local computational modules, with a PPO agent trained to generate remote four-qubit $\ket{\text{W}}$-states between separated modules.

The RL implementation incorporates simultaneous measurements of data qubit populations using single-shot readout, while continuous actions modify the amplitude, phase, and frequency of six control pulses. The reward function targets specific two-qubit states ($\ket{01}$ and $\ket{10}$), directly optimizing for the desired entanglement distribution. This work illustrates how RL can address the complex dynamics of multi-node quantum networks, opening pathways toward scalable quantum communication architectures.

\textbf{Discrete Gate-Based Entanglement Engineering.}
In contrast to continuous-control approaches, discrete gate-based methods focus on constructing specific entangled states through optimal sequences of quantum gates.
\cite{giordano2022reinforcement} employed tabular Q-learning to generate representative 4-qubit entangled states, revealing connections between entanglement features and the quantum gates required to realize them. Their analysis shows that the resulting quantum circuits are optimal with respect to the chosen gate set, providing both implementable circuits and theoretical insight into the structure of multipartite entanglement.

In their RL framework, the environment corresponds to the space of 4-qubit quantum states, represented in a discrete form by the list of computational basis terms with nonzero amplitudes, rather than by full complex wavefunctions. The agent’s actions consist of applying quantum gates from a predefined set, which is progressively expanded, from single-qubit and two-qubit gates such as Hadamard, X, and CNOT, to multi-qubit gates including Toffoli and Controlled-Hadamard, as required to reach more complex entanglement classes. The reward function employs a sparse structure: nonzero rewards are assigned only when the application of a gate transforms a state directly into the desired target state. This design guides the agent to discover minimal, optimal gate sequences that prepare the target entangled states within the selected gate set.

Addressing the scalability limitations imposed by classical simulation requirements, \cite{bao2024reinforced} developed an innovative approach using efficient disentanglers on random Clifford circuits. Their PPO agent characterizes patterns of optimal disentanglers, defined as sets of projective measurements applied to two-qubit gates arranged in brick-wall patterns. Remarkably, their results suggest that fewer measurements are required to disentangle random quantum circuits than predicted by theoretical studies of measurement-induced phase transitions.

The RL state representation employs a binary matrix encoding projective measurements along the circuit, while actions correspond to bit-flipping operations that add or remove measurements at specified locations. The reward function combines the averaged von Neumann entropy of the final state with measurement configuration costs, creating a multi-objective optimization that balances disentanglement efficiency with resource utilization.

An interesting new approach to entanglement control is Entangling Games~\cite{morral-yepes2024entanglement}, where two agents (entangler and disentangler) compete to prevent the build-up of entanglement, starting from a unitary state. Each turn, a biased coin toss determines which of the two agents plays. The agents then place the gates randomly on the qubits. The bias probability controls the entanglement content of the long-time steady-state dynamics. 

\cite{cemin2025learning} propose to use RL to make the disentangling agent `smart' by letting it select the bond on which to place the disentangling gate, while leaving the entangler to act randomly. 
Using Clifford circuits, they showed that active feedback control can change the nature of the long-time steady state from equilibrium to non-equilibrium; this alters the critical properties of the state, reducing its entanglement content.   
The study also investigates the role of information contained in the RL observations, used to control entanglement. The authors mask the Clifford tableau encoding the state of the circuit to construct RL observations; the actions are the bonds defining where to place the disentangling gate on the qubit chain, while the reward is the negative entanglement, integrated over each bond of the chain. The trained agent is demonstrated to successfully master the individual control of up to 128 qubits -- a prime example of genuine many-body control.  
This work highlights the active feedback control aspect of RL, and its suitability to study how partial information enhances control strategies.

Finally, let us mention that projective simulation was also employed to design complex photonic quantum experiments that produce high-dimensional entangled multiphoton states. \cite{melnikov2018active} demonstrate that it can autonomously design novel photonic quantum experiments for generating high-dimensional entangled multiphoton states. The agent's episodic and compositional memory represents experimental building blocks (beam splitters, phase shifters, nonlinear crystals, detectors) as clips in a network, with actions corresponding to selecting and arranging these optical components. The agent receives rewards based on whether the designed setup produces the target entanglement class efficiently. Remarkably, the system not only optimized known configurations but autonomously rediscovered advanced experimental techniques that have only recently become standard in quantum optics—techniques that were never explicitly programmed, and even proposed setups that human researchers initially deemed impossible, demonstrating potential for creative contributions to experimental physics methodology.

These diverse approaches to entanglement control demonstrate the versatility of reinforcement learning in addressing different aspects of quantum entanglement engineering, from fundamental state preparation to complex distributed quantum systems. The success of these methods across various platforms and objectives suggests that RL-based entanglement control will play an increasingly important role in the development of practical quantum technologies. 
A major challenge is that RL requires many repeated measurements at each step when building circuits experimentally, particularly for systems too large or complex to simulate

\subsection{\label{subsec:VQE}Variational Quantum Eigensolvers}

Variational quantum eigensolvers (VQE) represent a class of promising near-term quantum algorithms for studying ground-state properties of quantum many-body systems on NISQ devices. The fundamental challenge is to find the shortest-depth unitary circuit of native gates that, when applied to an easily prepared initial state $\ket{\psi_i}$, produces the target ground state of a given Hamiltonian $H$. The circuits typically consist of two-qubit entangling gates and single-qubit rotations native to the quantum platform of interest. 

By parametrizing the individual gates with a set of parameters $\alpha$, one can formulate the optimization problem through the cost function
\begin{equation}
    C(\alpha) = \bra{\psi_i}U^\dagger(\alpha)HU(\alpha)\ket{\psi_i},
\end{equation}
where minimizing $C(\alpha)$ yields the desired circuit $U(\alpha)$ since the energy expectation is bounded from below by the ground-state energy.

Reinforcement learning offers three distinct approaches to tackle the VQE paradigm: (i) training agents to minimize the cost $C(\alpha)$ directly for fixed circuit architectures with unknown gate parameters, constituting a continuous optimization problem; (ii) solving the discrete combinatorial optimization of finding the optimal circuit structure $U(\alpha)$ (gate sequence and qubit assignments), where RL agents select gate types while auxiliary continuous optimizers handle gate parameter optimization; and (iii) hybrid approaches combining both strategies. In practice, most studies focus on either approach (i) or (ii) for simplicity.

\textbf{RL-enhanced Quantum Approximate Optimization Algorithm (QAOA).}
The Quantum Approximate Optimization Algorithm has emerged as a particularly successful VQE algorithm, with several RL-enhanced implementations demonstrating significant improvements over conventional optimization methods.
 
\cite{chen2022robust} employed the TD3 algorithm to improve QAOA performance under realistic experimental constraints. Recognizing that full quantum state observation is experimentally infeasible due to exponential measurement scaling, they utilized noisy estimates of state coordinates as RL observations. The framework treats gate angles as continuous actions and constructs rewards from the norm difference between measured and target state observations, requiring only partial state information rather than complete quantum state tomography.

Complementing this direct optimization approach, \cite{yao2021reinforcement} developed a PPO framework specifically designed for noise-resilient variational parameter optimization. Their key insight was using an ansatz based on counter-diabatic driving to determine the gate set defining the action space. The agent learns to compile state preparation circuits using energy-based rewards, with observations consisting of action histories. This approach proved particularly effective for preparing ground states of Heisenberg and Ising-like spin Hamiltonians.

The same research group extended this work through several follow-ups: first, developing analytical expressions for multivariate correlated Gaussian policy gradients~\cite{yao2020policy}, and then creating a hybrid policy gradient algorithm capable of simultaneously optimizing continuous and discrete variables using autoregressive policy architectures~\cite{yao2022noise}. Then, to address the formidable challenge of hybrid optimization in QAOA, they combined Monte-Carlo Tree Search with natural policy gradient solvers~\cite{yao2022monte}. These frameworks demonstrated superior performance compared to existing optimization algorithms for quantum state preparation under Hamiltonian noise or measurement uncertainty, using binary rewards determined by target state fidelity.

\cite{herreramarti2022policy} further advanced policy gradient methods by parametrizing policies with correlated multivariate Gaussians for variational quantum circuit compilation. The approach proves more competitive than gradient-free methods for both noiseless and noisy circuits, using fully observable wave function amplitudes as states and continuous gate angles as actions, with rewards based on circuit fidelity averaged over random initial states.

Several studies have explored more specialized QAOA applications. 
\cite{an2024learning} deployed DQN to classify quantum phase transitions using variational disentangling circuits, achieving accurate critical point identification in the transverse-field Ising and XXZ models. Their approach demonstrates remarkable scalability by learning to recognize Kramers-Wannier duality in entanglement structures. The agent constructs incrementally the circuit by selecting actions from a predefined universal gate set containing single-qubit rotations, Hadamard, and CNOT gates. The RL observations contain a representation of the quantum circuit (initialized as the identity operation) with rewards based on entanglement entropy differences.

\cite{wauters2020reinforcement} developed a feedback-controlled PPO framework for QAOA that converged to protocols corresponding to optimal adiabatic solutions of transverse field Ising chains. The method proved effective across different system sizes and in the presence of disorder, using ZZ and X observable measurements as observations and sparse energy-based rewards.

\cite{patel2024curriculum} developed a curriculum-based Double DQN for quantum architecture search in noisy NISQ devices, outperforming state-of-the-art on quantum chemistry VQA tasks. RL states consist of tensor-based binary encodings of parametrized quantum circuit gate structures. Actions comprise CNOT and single-qubit rotations about all $x,y,z$-axes. Reward functions combine final state energies with penalties and bonuses to incentivize learning, with continuous circuit parameters optimized separately for reward computation.

Comprehensive algorithmic comparisons have also been conducted. \cite{altmann2023challenges} tested various deep RL algorithms for quantum gate sequence discovery, developing frameworks capable of controlling universal sets of continuously parameterized gates for both state preparation and unitary compilation. Their fully observable approach utilizes complete state vectors or unitary matrices as states, with actions selecting an axis ($x-$ or $z-$) to perform controlled and uncontrolled operations, or terminating the episode. The reward consists of a penalty when exceeding certain threshold values for the available actions, and the target fidelity at the final episode step.

Recent advances have looked into incorporating modern neural network architectures. \cite{kundu2024kanqas} employed curriculum DQN with Kolmogorov-Arnold networks for quantum architecture search in quantum chemistry applications, reporting superior performance compared to multilayer perceptron architectures. The framework uses tensorial one-hot encodings for gate-qubit relationships as observations and predefined gate sets as actions, with intermediate fidelity-based rewards and threshold bonuses.

\textbf{Combinatorial Optimization Applications: Max-Cut.} 
The Max-Cut problem has served as a canonical testbed for QAOA-RL integration. \cite{garcia2019quantum} adapted DQN for continuous actions to solve Max-Cut instances, achieving optimal results for problems up to $N=21$ qubits. Their framework uses local operator expectation values ($X_i, Z_i$) as observations, QAOA angles as actions, and the max-cut objective as terminal rewards.

\cite{khairy2019reinforcement} developed a sophisticated PPO approach focused on generalization to unseen problem instances. Training on small Max-Cut instances in classical simulators, their learned policies remarkably generalize to larger instances and outperform other optimization algorithms on superconducting NISQ devices. The framework incorporates energy expectation histories and QAOA angle sequences as observations, relative angle changes as actions, and instantaneous energy differences as rewards.

Extending beyond Max-Cut, \cite{fodera2024reinforcement} created a PPO framework for autonomously generating quantum circuits for various combinatorial problems, including Maximum Clique and Minimum Vertex Cover. Their agent discovers new solution families, achieving higher approximation ratios than state-of-the-art methods, using fully observable quantum wavefunctions as states, gate selection as actions, and combined Hamiltonian expectation and depth penalty rewards.

\textbf{Circuit Architecture Discovery.} 
Beyond parameter optimization, RL has shown remarkable success in discovering optimal circuit architectures. \cite{kundu2024easy} developed Gadget RL (GRL), a Double Deep Q-learning agent that first learns composite gates and subsequently uses these as enhanced actions to navigate the exponentially large gate space more efficiently. Applied to ground state preparation in the transverse field Ising model, GRL discovered compact circuits suitable for real hardware implementation, reducing errors by a few orders of magnitude compared to pure RL approaches. The framework uses tensor encodings of parameterized quantum circuits as states, with actions chosen from native elementary gate sets and penalty-based rewards.

Building on this success, \cite{sadhu2024quantum} applied similar double deep Q-learning techniques to variational quantum state diagonalization, training agents to devise entanglement-guided ansätze for quantum architecture search. Given a quantum state, their RL agent discovers circuit representations of optimal diagonalizing unitaries using the same framework as the GRL approach.

For larger quantum many-body systems, \cite{bolens2021reinforcement} used DQN to construct optimized quantum circuits for digital quantum simulation with strict gate count constraints. Their agent successfully builds circuits reproducing physical observables with fewer than three entangling gates for systems up to 16 qubits. Applications include long-range Ising chains and lattice Schwinger models. Using fixed alternating entangling and rotation layer architectures from trapped ion gate sets, the framework employed full quantum wavefunctions as observable states, continuous gate durations as actions, and sparse rewards computed from relative entropy between final and target reduced density matrices.


The diverse applications and methodologies we reviewed so far demonstrate RL's versatility in addressing the multifaceted challenges of variational quantum algorithms, from parameter optimization and architecture discovery to specialized problem-solving and circuit compilation, establishing it as a powerful tool for advancing quantum algorithm development in the NISQ era.

\subsection{Quantum Circuit Compilation and Architecture Search}
\label{subsec:arch_search}

While variational quantum eigensolvers focus on finding circuits that prepare desired states, quantum circuit transpilation or synthesis tackles the complementary problem of decomposing a given unitary operation into elementary gates from another gateset. The success of RL in classical analog circuit design~\cite{settaluri2020autockt} naturally raises questions about its application to quantum circuit architecture problems.

The theoretical foundation for transpilation lies in the Solovay-Kitaev theorem, which states that any set of single-qubit quantum gates generating a dense subgroup of $SU(2)$ can approximate any desired single-qubit quantum gate through a short gate sequence that can be found efficiently. Consequently, a quantum circuit of $m$ constant-qubit gates\footnote{Constant-qubit gates are quantum gates that operate on a fixed, small number of qubits, most commonly one or two qubits, which does not change as the total size of the quantum system increases.} can be approximated (in operator norm) to error $\varepsilon$ by a quantum circuit of $O(m\log^{c}(m/\varepsilon))$ gates from a desired finite universal gate set, where $c<1$ is a constant~\cite{nielsen2010quantum}. However, this remarkable theorem does not specify how to find the shortest approximating gate sequence -- a crucial consideration since quantum circuit architecture greatly affects algorithm performance on NISQ devices.

Unitary transpilation aims to identify quantum circuits representing a given unitary while minimizing circuit depth, total gate count, or specific gate counts (e.g., T gates). Often, the goal is shortening the circuit depth of known decompositions -- a process called unitary synthesis. Both quantum transpilation and unitary synthesis involve sequential processes with multiple compilation steps and numerous optimization passes, making them naturally suited yet challenging problems for RL approaches.

For instance, \cite{sadhu2024quantum} investigate reinforcement learning-assisted quantum architecture search applied to the variational quantum state diagonalization problem. Their Double DQN agent with an $\varepsilon$-greedy policy searches for ansatz circuits that diagonalize random 2-qubit density matrices by maximizing fidelity via conditional entropy and concurrence–based reward signals. The RL observations encode the current ansatz configuration through tensor-based representations of the circuit structure. Actions correspond to appending quantum gates to the ansatz, while rewards incorporate entanglement-guided incentives, including concurrence bounds and performance bonuses, often weighed relative to circuit depth and gate count. This work shows that RL agents can efficiently discover variational quantum circuits that diagonalize 2-qubit density matrices.

\textbf{Clifford+T Gate Set Synthesis.}
Several studies have focused on circuit synthesis using the Clifford+T gate set, a universal gate set particularly relevant for fault-tolerant quantum computing. 

\cite{kolle2024reinforcement} introduced a comprehensive RL environment for quantum circuit synthesis where circuits are constructed from Clifford+T gates to prepare specific target states. Their PPO agent successfully designs minimal quantum circuits for 2-qubit Bell states, with RL states defined from real and imaginary wavefunction components of both current and target quantum states. Actions correspond to quantum gates, while rewards impose fixed penalties at each episode step unless the current fidelity falls below a preset threshold.

Building on this foundation, \cite{weiden2024learning} developed the Diagonalizing Value Network for Unitaries -- a model-based RL approach that reduces general unitary synthesis to diagonal unitary synthesis problems. Their method achieves up to 16.8\% reduction in T-gate counts compared to competitive approaches by casting circuit synthesis as a diagonalization problem. The MDP formulation uses states $s_t = L_t U_\ast^\dagger R_t$ where $L,R$ are learnable unitary circuits and $U_\ast$ is the target unitary, with the objective of transforming the final state $s_T$ into a diagonal unitary. Actions comprise discrete Clifford+T gates, with rewards providing $-1$ penalties per step except when $s_T$ becomes diagonal within a threshold.

\cite{rietsch2024unitary} investigated a related problem using Gumbel AlphaZero tree search for exact unitary synthesis into Clifford+T circuits. The Gumbel extension proves particularly suitable when the action space size exceeds the available tree search budget. Their agent synthesizes circuits on up to five qubits from randomized circuits with up to 60 gates, outperforming state-of-the-art methods. The RL observation is the product $U_\ast U_t^\dagger$, and $U_t$ the current time step unitary. Actions are Clifford+T alphabet gates, with rewards providing $-1$ penalties for each gate placement and $0$ upon exact synthesis.

Another approach in T-count minimization was reported by Google DeepMind~\cite{ruiz2024quantum}. Their AlphaTensor-Quantum algorithm works similar to AlphaZero and builds on a Monte-Carlo tree search (MCTS) guided by a deep policy network. The authors demonstrate that it can outperform the state of the art for T-count optimization (i.e., the optimal placement of T gates among Clifford gates in a given quantum circuit) on a set of arithmetic benchmarks; in particular, the agent finds the best human-designed solutions for relevant arithmetic computations used in Shor's algorithm and for quantum chemistry simulations. The agent was tasked with minimizing the number of T gates needed to implement a given circuit by encoding information about the non-Clifford components into a binary signature tensor, with the goal of reaching the all-zero tensor in as few moves as possible. The RL observations consist of the signature tensor and all past played actions, while actions define vectors used to modify binary components of the signature tensor corresponding to quantum gate applications. The reward is a simple penalty of $-1$ at each step.

\textbf{Hardware-Aware Circuit Optimization.}
A few approaches also address hardware-specific constraints and optimization objectives. 

\cite{wang2024quantum} developed an RL-based quantum compiler for superconducting processors using DQN-guided search algorithms capable of discovering short-depth circuits. On a 9-qubit superconducting processor, their agent finds unit-fidelity circuits with seven CZ gates for three-qubit quantum Fourier transform implementation, outperforming state-of-the-art methods while considering device hardware topological constraints. The framework uses supervised training with RL observations tracking current unitaries and actions selected from native gate sets based on circuit topology.

\cite{foesel2021quantum} demonstrate that deep convolutional PPO agents can autonomously learn optimization strategies for arbitrary circuits on specific hardware architectures. Their agent begins with complete but inefficient circuits and progressively minimizes their depth through circuit transformation sequences. RL states represent quantum circuits while actions are circuit transformations, with rewards combining circuit depth and gate count minimization objectives. In contrast to other approaches, where RL training needs to be re-run for each optimization task, the agent (once trained) can be applied to an arbitrary circuit in a single shot. For 12-qubit random circuits, they achieved 27\% depth reduction and 15\% gate count reduction on average, and due to the convolutional nature of the agent, the agent can be immediately applied to larger circuits (50 qubits were demonstrated). 

The nearest neighbor architecture constraint problem was tackled by \cite{li2023quantum} using model-based RL for policy and value improvement in reconfigurable architecture design. Their method achieves up to 62.5\% reduction in SWAP gate usage compared to state-of-the-art algorithms. Similar to previous studies, RL states represent quantum circuits, with actions including both qubits (control/target assignment) and gates from the set (X, CNOT, $\sqrt{\text{CNOT}}$, $\sqrt{\text{CNOT}^\dagger}$). Rewards are based on appropriately chosen nearest-neighbor cost functions.

\textbf{Graph-Theoretic and Tree-Search Approaches.}
ZX-diagrams represent a graphical language for quantum computation reasoning, and every quantum circuit can be translated into such a diagram.  \cite{nagele2024optimizing} introduced RL for ZX diagram simplification, employing PPO with graph neural network architectures to optimize ZX diagrams using graph-theoretic simplification rules. RL states are ZX-diagrams, with action spaces comprising diagram modification rules plus stop actions. Rewards are based on ZX diagram node count reduction. In \cite{riu2023reinforcement}, the same strategy was followed, but now with translation from and to quantum circuits, with rewards based on gate count differences before and after action application, enabling both total and two-qubit gate count reductions. Their agent, trained on small system sizes, consistently improves state-of-the-art performance while maintaining computational competitiveness.

\cite{mattick2025optimizing} also proposed an RL-based framework that operates on ZX-graph representations of quantum circuits using graph neural networks and tree search. With actions defined as ZX-calculus rewrite rules, the agent learns to minimize two-qubit (CNOT) gate count by exploring arbitrary optimization sequences. The method is competitive with state-of-the-art optimizers and generalizes effectively across diverse circuits, offering a flexible, learned alternative to fixed heuristic rules.

Single- and two-qubit compilation was studied by \cite{chen2022efficient} using deep Q-learning combined with A$^*$ search and data generation. The agent outperforms existing algorithms in both circuit depth and inference time. RL states are built from quantum circuit gate representations, with action spaces containing hardware-specific universal gate sets (both inverse-closed and inverse-free). Rewards are set by norm differences between compiled and target unitaries.

Beyond traditional approaches, \cite{zhang2020topological} developed efficient algorithms for transpiling arbitrary single-qubit gates into elementary gate sequences using weighted $A^*$ search (cf.~Sec.~\ref{subsubsec:other_algos}) guided by policy and value networks. Applied to topological Fibonacci anyon compilation, this hardware-agnostic, ancilla-free approach produces near-optimal braiding sequences for arbitrary single-qubit unitaries using fully observable wavefunction amplitudes and backward operation strategies with specialized search heuristics rather than traditional reward structures.

\textbf{Advanced Circuit Synthesis and Compilation.}
A couple of studies have explored more sophisticated synthesis approaches. 

\cite{he2021variational} employed Double Q-learning to automatically design variational quantum circuits without human intervention. Their agent selects quantum gates from native alphabets to explore different circuit structures, discovering exact compilations with fewer gates than state-of-the-art methods while reducing decoherence and gate noise errors on NISQ devices. RL states are defined as tuples of gate type, gate qubits, and total gate numbers, with action sets comprising predefined alphabet gates and target qubits. Rewards are based on Hilbert-Schmidt norms between final and target unitaries.

\cite{kremer2024practical} integrated RL into quantum transpilation workflows, enhancing synthesis and routing of quantum circuits. This PPO agent provides near-optimal synthesis of Linear Function, Clifford, and Permutation circuits, achieving significant reductions in two-qubit gate depth and count. Starting with desired unitary implementations, their RL framework constructs unitaries $U$ from predefined native gate sets such that $UU_\ast=1$ for a target $U_\ast$; inverting $U$ produces the desired transpiled circuit. Curriculum learning was used to dynamically adjust input gate difficulty as agents improve. RL observations use quantum gate representations with actions from predefined gate sets, while reward functions contain small per-step penalties (depending on single vs.~two-qubit gates) and large positive rewards upon target achievement.

\textbf{Fault-Tolerant and Multi-Core Compilation.}
Advanced compilation strategies have been developed for fault-tolerant and distributed quantum computing architectures. 

\cite{zen2024quantum} used PPO combined with transfer learning to discover compact, hardware-adapted fault-tolerant quantum circuits.Exploring different qubit connectivities, their agent discovers circuits with fewer gates and ancillary qubits than state-of-the-art methods, with and without hardware constraints for up to 15 physical qubits. For fixed target logical states of specified quantum error correction codes, gate sets, and qubit connectivity, agents find circuits preparing logical states fault-tolerantly using so-called flag qubits. RL states are given by stabilizer tableaus, with actions comprising discrete Clifford gates (Hadamard, Phase, CNOT) respecting qubit connectivity constraints. Reward functions are based on successive differences in fidelity, energy, or tableau distance.

Multi-core quantum architectures present unique compilation challenges. \cite{russo2024attention} considered quantum processors built from smaller cores, using policy gradient methods with transformer encoders and graph neural networks to provide insights into autoregressive RL agent design for multi-core quantum compilation. RL states combine action sequences with learnable network hidden states, while actions consist of possible cores for logical qubit allocation. Rewards are based on the total inter-core communications required for qubit assignment.

\cite{pastor2024circuit} addressed circuit partitioning and quantum circuit mapping challenges using maskable PPO agents (where masks selectively determine policy network update regions during training). Their maskable models outperform state-of-the-art algorithms by minimizing nonlocal communication. RL observations contain current and previous qubit-to-core assignments, lookahead weights, and binary variables indicating whether all interactions occur within the same core. Actions are swaps shifting qubits between cores, with reward functions based on average nonlocal communication numbers.

\textbf{Specialized Compilation Tasks.}
Several specialized compilation tasks have been addressed using RL approaches. 

In the context of distributed quantum computing, \cite{promponas2024compiler} introduced Double Deep Q-learning compilers prioritizing expected execution time reduction by jointly managing EPR pair generation and routing, scheduling remote operations, and injecting SWAP gates to facilitate local gate execution. RL states are tuples of logic qubit positions within quantum processing units, qubit availability times, and operations required for initial circuit completion. Action sets consist of identity, CNOT, and SWAP gates, with sophisticated action-dependent reward functions scoring agent compilations.

\cite{lanore2024automated} present an automated framework for designing photonic implementations of nonlocal realizations using homodyne detections and quantum state heralding. They combine a PPO agent with efficient circuit simulation and numerical optimization to generate practical photonic circuits that violate the Clauser-Horne-Shimony-Holt (CHSH) inequality. To do this, they formulate the search for photonic Bell states as an RL problem: the RL states are given by the expansion coefficients of the quantum state in a Gaussian state basis; actions are drawn from a predefined set of photonic gates (beam splitter, phase shifter, single-mode squeezer, two-mode squeezer). The reward function is zero except at the last timestep, where it is a function of the so-called CHSH score that is maximized over all continuous gate parameters. The episodic task is set to terminate once the circuit depth reaches a maximum predefined value.

\cite{preti2024hybrid} used projective simulation with LSTM networks to find circuit depth reduction strategies through hybrid discrete-continuous optimization across continuous gate sets tailored to underlying hardware architectures. Their algorithm uses gradient descent for continuous circuit parameters while deep RL agents optimize gate ordering using projective simulation. Agent-compiled circuits reproduce known unitary processes and significantly reduce relevant quantum circuit sizes for trapped-ion devices for unknown processes. Actions consist of parametrized gate sets (M\o lmer-Sorensen, XY, Z) with gate angles optimized after circuit architecture completion. Rewards are based on gate fidelity figures of merit.

Qubit routing -- modifying quantum circuits to satisfy target quantum computer connectivity constraints by inserting SWAP gates -- was addressed by \cite{pozzi2022using} using deep Q-learning. Their agent proposes qubit routing procedures minimizing circuit depth, increased by adding SWAP gates, outperforming state-of-the-art procedures on random and realistic circuits across near-term architecture sizes up to 50 qubits. RL states use tuples containing qubit locations, targets, and current circuit progress, with actions containing individual gate/swap scheduling or layer scheduling. Rewards are based on circuit depth overhead and ratio concepts.

\cite{quetschlich2023compiler} adopted PPO to design frameworks for developing optimized quantum circuit compilation flows using only native gates while considering device topology. Their agent outperforms state-of-the-art compilers in expected fidelity, critical depth, and combinations thereof. RL observations contain quantum circuit features such as qubit numbers, circuit depth, and composite program communication features (critical depth, entanglement ratio, parallelism, liveness). Actions are chosen among platform selection, device selection, synthesis, mapping, and optimization, with reward functions relying on expected fidelity and critical circuit depth combinations.

Finally, \cite{dubal2025pauli} explored Pauli network circuit synthesis, extending the toolkit of RL-based quantum compilation methods to this specialized domain.

%% file: feedback_ctrl.tex
\section{\label{sec:feedback_ctrl}Quantum Feedback Control}

\begin{figure*}[t!]
\centering
\includegraphics[width=1.0\textwidth]{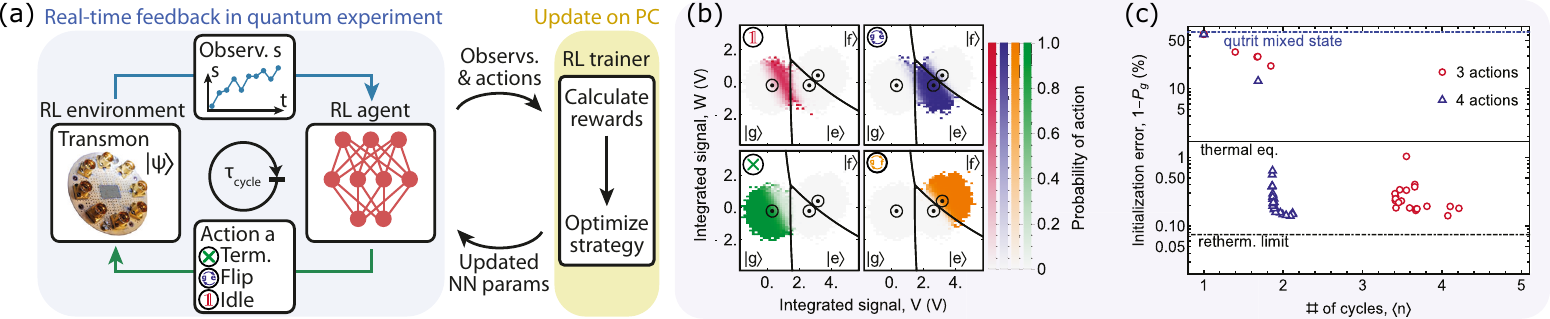}
\caption{
\textbf{Deep neural network agent for real-time feedback control of a superconducting qubit in an experiment.} (a) Outline of the experimental setup and its interaction with the RL agent. In this setup, a response time of less than one microsecond was achieved. (b) Analysis of resulting strategy for ground state preparation in a three-level system (qutrit). The measurement signal is projected onto two variables $V$ and $W$, and the panels illustrate how the probabilities for various actions (microwave pulses) depend upon the observed measurement result, as suggested by the fully trained agent. (c) Initialization error vs. average number of feedback cycles (``3 actions": the agent is not allowed to flip the transmon directly from state f to state g).
Figure adapted with permission from Ref.~\cite{reuer2023realizing}.
}
\label{fig:feedback}
\end{figure*}

The standard description of RL, with an agent selecting actions based on observations, immediately brings to mind a physical feedback scenario. In previous sections we have seen examples of observations including the most recent action taken or the quantum unitary currently being considered in an optimization task. In contrast to that, a true feedback situation in the physical sense arises when the agent's observation corresponds to a real measurement on a quantum system. Such scenarios are particularly in need of the power of RL, since feedback strategies live in an exponentially larger space than mere control strategies: If there are $N$ measurement steps so far and $M$ measurement outcomes in each of those steps, a (non-Markovian) feedback strategy needs to provide a suitable action for each of the $M^N$ possible outcome sequences, while a pure control ("open loop") strategy only needs to provide a single suitable action. This also means that the agent that emerges after training has to be able to map such (possibly high-dimensional) observations to actions during the execution of an experiment, while for the tasks discussed in previous chapters the outcome of RL may be e.g. one fixed optimized quantum circuit which can simply be deployed on a quantum computer.

For such physical feedback scenarios, an important additional challenge arises in experiments: in many cases, the feedback has to happen in real time, i.e. on the time scales of the individual steps of an experimental procedure. Often, these times can be very short, e.g. in a qubit experiment, where the agent has to come up with a response to the most recent observation on a time much shorter than the qubit's coherence time. Depending on this coherence time, it can be hard to implement a neural-network agent that can provide real-time feedback on that time scale. There are also other, much more lenient, physical feedback scenarios, where feedback needs to be given only once a single experimental run has been completed, e.g. in active calibration strategies – in such cases, one can afford longer waiting times and the implementation of the agent becomes less critical.

\textbf{Featured experimental application.}  The challenge of implementing and training a real-time neural-network based RL agent in an experiment has been addressed in ~\cite{reuer2023realizing}. In that work, an agent was trained to initialize the state of a superconducting qubit.

Superconducting qubits represent one of the leading platforms for quantum computation. Individual gate times are on the order of nanoseconds, and even a single measurement can be completed during a few hundred nanoseconds. Therefore, any real-time feedback agent needs to operate ideally with a cycle time less than a microsecond, in order not to appreciably delay the actions. This makes that platform a particularly challenging benchmark case for neural-network based RL.

On these time scales, it would not be possible to transfer data back and forth to a CPU or GPU. Rather, the network of  ~\cite{reuer2023realizing} was implemented on a field-programmable gate array (FPGA). However, this alone would not be sufficient to reach the required speeds. Therefore, an additional general idea was introduced: the evaluation of the network and the extraction of measurement data proceeded simultaneously. While the activations of the next layer were calculated, additional data was streaming in and added to those activations. As a result, the overall evaluation of the network only added a latency of 50 nanoseconds, corresponding to the last layer, beyond what would have been needed anyway for the entire measurement. This enabled the agent to achieve  less than 1 microsecond response time. 

As a test case, that agent was trained to initialize thermally excited qubits and qutrits in their ground state by a sequence of measurements and pulses. After about 1000 real-time actions performed in the experiment, the acquired  data was uploaded to a PC, where a second copy of the network was updated using the PPO algorithm, and the revised network parameters were then send back down to the FPGA. The entire training was completed within three minutes. This constituted the first deep neural network agent for real-time feedback in a quantum computing experiment trained via reinforcement learning. Applications of this setting to several qubits would enable more advanced applications like quantum state preparation in a multi-qubit system or elements of quantum error correction.

\textbf{Breadth and scope of this section:} \\
In the remainder of this section, we recount some more illustrative examples of RL being applied to problem settings that require feedback control.

Feedback control of open quantum systems was investigated by~\cite{erdman2024maxwelldemon} using a variant of the soft actor-critic algorithm, with the objective to automate and capture the role of a quantum Maxwell’s demon. Their RL agent is tasked with discovering optimal feedback control strategies in qubit systems by maximizing the trade-off between measurement-powered cooling and measurement efficiency. The agent finds counterintuitive interpretable strategies in a framework that employs weak or projective quantum measurements to explore different regimes based on the hierarchy of timescales set by thermalization, measurement, and unitary feedback.
The RL states are fully observable and consist of the reduced density matrix describing the quantum system conditioned on the measurement outcomes. Actions comprise continuous control parameters (coupling strength or unitary gate) in addition to a discrete choice from the set \{thermalize, measure, unitary feedback\}. The reward is a function of the cooling power and the dissipation.

\cite{sivak2022model-free} train a PPO agent to learn characteristics of a control circuit through trial-and-error interaction with the quantum system, using binary measurement outcomes as the only source of information about the quantum state. They apply it to control a harmonic oscillator coupled to an ancilla qubit: by applying logical gates on the encoded qubits, the agent is tasked with the preparation of different quantum states using both unitary control and adaptive measurement-based quantum feedback. The authors show that their framework does not make use of state tomography or fidelity estimation, and significantly outperforms state of the art model-free methods in terms of sample efficiency.
It is based on a POMDP where the RL agent measures an ancilla qubit to control the oscillator state; the RL observations are the binary output of the ancilla $Z$ measurements. The set of actions comprises continuously parametrized Krauss maps: e.g., unitary control is accomplished using the selective number-dependent arbitrary phase (SNAP) and displacement gates. The circuit architecture is fixed, so the agent has to learn the continuous gate angles. The reward function is general so long as it satisfies the property $\text{argmax}_{\ket{\psi}} \mathbb{E}[R]=\ket{\psi_\ast}$, e.g., a dichotomic POVM that estimates fidelity.

The PPO algorithm was also employed by \cite{porotti2022deep} to discover feedback control strategies for Lindblad dynamics without prior knowledge. The physical system of interest here is a cavity subject to quantum-non-demolition detection of photon number, and it is controlled by a linear drive. The agent has to prepare and stabilize high-fidelity Fock states. The study shows that it can even reach superposition states if the measurement rates of different Fock states can be individually controlled.
The RL states consist of the reduced density matrix and are thus fully observable. The actions are given by the real and imaginary parts of the displacement operator applied to the cavity model, and are of bang-bang type. Training is facilitated by reward engineering (the reward function is given by the eighth power of the target state fidelity,provided at every episode time step).


\textbf{Model-based RL.}
If the system dynamics is known, one can in principle gain efficiencies in many problems by applying model-based RL, where the idea is to take gradients through a simulation. However, in quantum feedback situations, that simulation will have to contain measurements with their stochastic outcomes. On a technical level, this is particularly challenging if we are talking about discrete outcomes, as found in a projective qubit measurement, since it is impossible to take gradients through those. This was resolved in \cite{porotti2023gradient}, by using a combination of Monte Carlo sampling, mathematical concepts from model-free policy gradient RL, and automatic differentiation. The resulting algorithm, termed "feedback GRAPE", was illustrated on a range of problems including state preparation and stabilization in qubit-cavity systems and the discovery of robust strategies. It has by now been applied successfully to discover a significantly enhanced version of the GKP-state quantum error correction protocol applied in \cite{sivak2023real}, in the form of a non-Markovian feedback strategy \cite{puviani2025non}.



%% file: qec.tex
\section{\label{sec:QEC}Quantum Error Correction}

Whereas to design quantum circuits that perform computations, it is necessary to know how to initialize qubit states, implement high-fidelity quantum gates, and apply them, this is not yet sufficient for reliable quantum computing. Decoherence and depolarization due to the coupling between the qubits and their environment (including during the measurement process) lead to a limited ability to preserve the information stored in the quantum state. 

\begin{figure*}[t!]

\centering
\includegraphics[width=.8\textwidth]{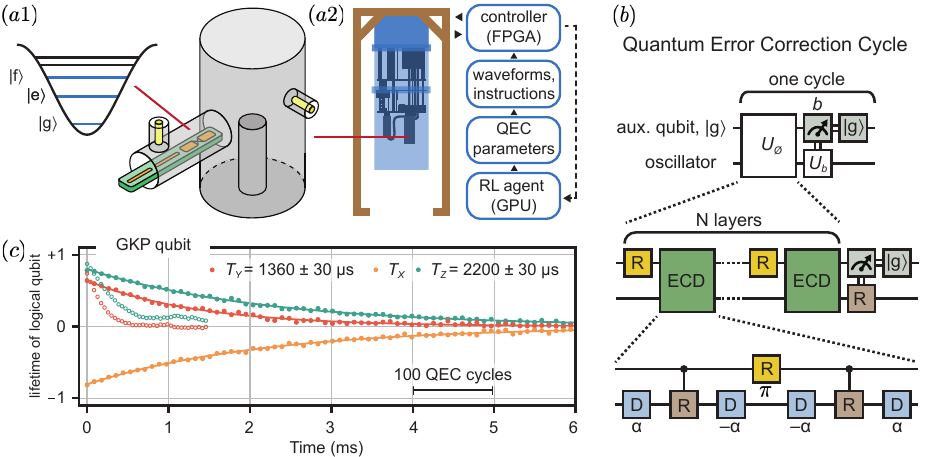}
\caption{
\textbf{RL-enhanced quantum error correction (QEC) for a Gottesman-Kitaev-Preskill (GKP) code.}
(a) The GKP code is realized in a cavity oscillator mode (large gray cylinder); the logical qubit is interacted with using an external auxiliary transmon modeled by a nonlinear oscillator (orange chip), see (a1). The device is placed in a cryostat (a2). The RL agent interacts with the system by optimizing the parameters of a controller placed on an FPGA.
(b) The quantum error correction cycle includes entangling gates between the auxiliary transmon and the cavity oscillator, single-qubit gates, and transmon measurement readout. The entangling operation contains a repeated sequence of $N$ layers consisting of single-qubit rotations $\text{R}(\varphi)$ and entangling echoed conditional displacement (ECD) gates, where $\text{D}(\alpha)$ is the displacement operator for the oscillator mode. The RL agent learns the continuous gate parameters of this circuit. 
(c) Logical Pauli operator expectation values $T_j,\ j\in\{X,Y,Z\}$ as a function of time show an increase of the logical GKP qubit lifetime by a factor of $2.27 \pm 0.07$. Filled circles correspond to RL-enhanced error correction; open circles represent evolution in the absence of QEC.
Figure adapted with permission from Ref.~\cite{sivak2023real}. Reproduced with permission from Springer Nature. 
}
\label{fig:qec}

\end{figure*}

Quantum error correction (QEC) algorithms have been invented to counteract these detrimental effects. Broadly speaking, the idea is to use multiple physical qubits to encode a single logical qubit in a two-dimensional subspace of the total Hilbert space, known as the code space. By measuring multi-qubit observables of the physical qubits (so-called syndromes), it is possible to detect changes in the logical quantum state (decoding) and revert them. The net effect is a quantum error-correcting operation, which increases the lifetime of the logical qubit.  
Reliable decoding of quantum error-correcting codes is essential for protecting quantum information from decoherence and depolarization; this requires a good characterization of the error channels corrupting the logical quantum state and a means to provide this information to the decoder.

A crucial challenge in QEC comes from the inability to accurately quantify and characterize all error channels relevant for a given NISQ device; oftentimes, noise parameters even change with time. Another notable difficulty is the lack of guarantees for a unique decoding operation: indeed, multiple errors can give rise to the same measurable syndrome. Suppose we add to these the unique peculiarities of each experimental platform: it quickly becomes clear that a theory-only solution cannot provide the required accuracy to achieve fault-tolerant quantum computing. 
In this context, RL has emerged as a valuable building block for state-of-the-art QEC codes deployed in real devices~\cite{acharya2024quantum}. 

\textbf{Featured experimental application.} A recent experiment has demonstrated that an RL agent can achieve both full code stabilization and quantum error correction, increasing the lifetime of the logical qubit by a factor of $G= 2.27 \pm 0.07$~\cite{sivak2023real}. The experiment uses the framework of Gottesman–Kitaev–Preskill (GKP) codes --  an efficient way of encoding a logical qubit in an infinite tower of harmonic oscillator states.  
The quantum oscillator describes an electromagnetic mode of a superconducting cavity whose state is controlled using an auxiliary transmon qubit, see Fig.~\ref{fig:qec}(a). We emphasize that the logical qubit is contained in the oscillator Hilbert space, while the auxiliary qubit is merely used to control it and read off the logical state. 
The experimental setup allows for real-time classical processing and measurement-based feedback; this enabled the authors to train an RL agent to learn the optimal QEC circuit parameters in situ, i.e., to ensure their adaptation to the real error channels and control imperfections of the experimental system. 

More precisely, they used a field-programmable gate array (FPGA) as a QEC controller, whose parameters are optimized by an RL agent implemented on a graphics processing unit (GPU). The control of the logical state is indirectly achieved by the classical RL-optimized controller via the auxiliary transmon qubit.
The agent is trained using data from binary measurements of Pauli operators, potentially corrupted by state preparation and measurement errors. Importantly, the QEC protocol uses a theoretical model based on Kraus operators describing the noise on the logical qubit, and the RL agent is tasked with learning the relevant correction to it.

Error correction is achieved by applying a repetitive QEC cycle, see Fig.~\ref{fig:qec}(b).
The QEC cycle is a circuit parametrized by $45$ parameters that include the amplitudes of various primitive pulses in the circuit decomposition, to be learned by the agent. The circuit itself is built out of multiple layers, containing parametrized controlled rotation gates (based on the state of the auxiliary qubit) and displacement gates, which together can entangle the states of the auxiliary qubit and the oscillator state; thus, by measuring the auxiliary, it is possible to infer (partial) information about the logical qubit state. 
At the beginning of each cycle, the joint state of the oscillator and the auxiliary (transmon) qubit is reset to $\ket{g}_\text{aux}\ket{0}_\text{GKP}$. Then, the logical qubit is set to an eigenstate of either the logical Pauli $\text{X}_L$ or $\text{Z}_L$ operator. 
The reward is based on a measurement of either $\text{X}_L$ or $\text{Z}_L$ (depending on the initial state) and is stochastic due to the nature of quantum measurement; it is designed to incentivize the agent to find the QEC protocol with the longest logical qubit lifetime. The entire candidate QEC protocol is repeated $160$ cycles to enhance the signal-to-noise ratio of the reward. 

To train their agent, the authors use the PPO algorithm; they report that it outperforms alternative approaches in simulation when applied to high-dimensional problems with stochastic objective functions. As usual for continuous-policy RL, the policy is modeled by an architecture containing both LSTM and fully connected layers to learn the mean and the variance of independent Gaussian distributions, one for each of the $45$ QEC cycle angles. Actions correspond to the values of the angles drawn from the policy. The RL observations consist only of a one-hot variable that keeps track of the wall clock time during a QEC cycle (no real-time feedback/closed-loop control is involved). As mentioned above, the reward is the binary output of the auxiliary qubit measurement at the end of each episode.
This framework for model-free learning of quantum control policies is explicitly tailored to the stochasticity and minimalistic quantum observability~\cite{sivak2022model-free}.

Training exhibits two stages: in the first hundred training epochs, the agent learns to correct large errors in the initial angle values; in the subsequent few hundred epochs, it carries on to fine-tune the circuit parameters and achieve the highest performance. Upon analyzing the behavior of the trained agent, the authors find that it tends to reduce the intermediate photon number in the cavity and improve the performance of QEC at the cost of using much slower gates. Ultimately, the RL agent increases the logical qubit lifetime by a factor of $2.27 \pm 0.07$, see Fig.~\ref{fig:qec}(c). This work provides clear experimental evidence for the advantages offered by RL in quantum error correction.

\textbf{Breadth and scope of this section:} The remainder of this section discusses recent work on RL for improving and building decoders for stabilizer codes in Sec.~\ref{subsec:RL_decoding}, as well as applications of RL to discover the error correcting codes via combinatorial optimization, cf.~Sec.~\ref{subsec:QEC_discovery}.

\subsection{\label{subsec:RL_decoding}Quantum Decoders}

Several works have focused on optimizing decoder performance through RL-based calibration and parameter learning. 

\cite{sivak2024optimization} develop a method for calibrating error channel characterization and the information provided to the decoder, aiming to minimize the logical error rate. They use a multi-agent pipeline based on a PPO-type gradient estimator to obtain significant improvement of the decoding accuracy in repetition and surface code memory experiments, outperforming the state of the art. The authors model the prior probability distribution of errors as a hypergraph in which nodes correspond to detectors and hyperedges connecting clusters of nodes --- to various error mechanisms that activate detectors. This formulation allows them to use small error-correcting codes as local sensors of the error hypergraph, with the RL agent learning sensor error graph parameters sampled from multiple independent Gaussians. There are no RL states in this framework; the actions are the parameters (means and variances) of the Gaussian distribution, and the reward is based on the logical error rate of an MWPM decoder.

A suitable modification of this framework was recently applied to a superconducting quantum processor realizing the distance $d=5$ surface code~\cite{sivak2025reinforcement}. Key to this success is the definition of a surrogate objective function $C$ which is the number of error detection events, directly available from the data. The authors argue that, since $C\propto\varepsilon$ is proportional to the average physical error rate $\varepsilon$, within this model one can express the gradient of the logical error rate $\varepsilon_L\propto (\varepsilon_\text{th}/\varepsilon)^{-d/2}$ [with $\varepsilon_\text{th}\gg\varepsilon$ the QEC threshold] via the relation $\grad\log\varepsilon_L = (d+1)/2\grad\log C$, thus enabling training from estimates of $C$ without knowing the logical qubit state, and bypassing the exponential suppression of $\varepsilon_L$ with code distance $d$. They apply their idea to a surface code experiment. The RL framework improves the logical error rate by a factor of 2.4 (3.5 with additional decoder steering). Importantly, this work demonstrates that RL agents trained on error detection can be used for simultaneous QEC and real-time recalibration of gate parameters; moreover, the framework is scalable, as was shown for a $d=15$ surface code with $40,000$ gate control parameters.  

\cite{freire2025optimizing} investigate optimizing the performance of hypergraph product codes against the quantum erasure channel using projective simulation to improve decoder performance relative to the state of the art. The RL states comprise the space of Tanner graphs associated with binary parity-check matrices of fixed row and column weights, while actions consist of picking two edges of the graph and swapping their endpoints to preserve weight conditions. The reward is based on a Monte Carlo estimation of the logical error rate due to the quantum erasure channel.

Addressing the challenge of time-varying noise in NISQ devices, \cite{matekole2022decoding} use double deep Q-learning with CNN layers and probabilistic policy reuse. Their agent learns surface code decoding strategies for quantum environments with varying noise sources, providing numerical evidence for the ability to adjust as noise parameters drift while reducing computational complexity. The RL states are built from the history of actions and faulty syndromes, actions correspond to physical bit and phase flip operations, and the reward function is set to unity if the state is restored and zero otherwise, from which the qubit lifetime can be inferred.

\cite{park2024enhancing} train a PPO agent to find circuit-level measurement-free error-correcting protocols based on local error information, considering faulty multi-qubit gates for both syndrome extraction and error removal. Their RL framework takes a fixed set of faulty gates as inputs and returns an optimized QEC circuit, outperforming the state of the art on the 2D classical Ising model and the 4D toric code at improving logical qubit lifetime. The framework uses no observation due to the assumption of absent midcircuit measurement, with the agent taking local error correction gate operations as actions and the reward function constructed using the decoding success rate at the end of the decoding sequence.

Decoding topological quantum codes can be naturally reformulated as repeated interactions between a decoding agent and a code environment. \cite{andreasson2019quantum} used a CNN architecture to train a DQN agent for quantum error correction of bit-flip errors on the toric code. The performance of their agent approaches that achieved by the Minimum Weight Perfect Matching algorithm (MWPM) for code distances up to $d=7$. They used error syndromes as observations and Pauli X and Z gates applied on physical qubits as actions, with a penalty of $-1$ at each episode step. In a follow-up study, \cite{fitzek2020deep} train a decoding DQN agent for depolarizing noise, observing that by using correlations between bit-flip and phase-flip errors, the RL decoder outperforms MWPM, achieving higher success rate and higher error threshold for code distances up to $d=9$. Moreover, an RL decoder trained on depolarizing noise exhibits close to optimal performance for uncorrelated noise but provides suboptimal decoding for biased noise models. While using the same observations and actions as above, the reward is constructed from the difference in the number of defects in a syndrome at consecutive episode steps and includes a large bonus in the final episode step.

A similar approach was followed by \cite{sweke2020reinforcement}, who formulate the problem of decoding within fault-tolerant quantum computation as a reinforcement learning problem, training decoding DQN agents using phenomenological noise models. Their RL framework uses syndrome measurements as partial observations and allows the agent to apply Pauli X and Z operations on any physical qubit, with the reward function based on agent success and input from a referee decoder.

\subsection{\label{subsec:QEC_discovery}Discovery of Error Correcting Codes}

Besides improving QEC decoders, RL has also been used to discover quantum error correction codes and encoding strategies tailored to specific qubit hardware platforms.

In a pioneering work, \cite{foesel2018reinforcement} showed that a natural policy gradient agent can discover complete quantum-error-correction strategies to protect a collection of qubits against noise using feedback adapted to measurement outcomes. To address the combinatorially large search space, the authors propose a two-stage learning procedure with a fully state-aware teacher network and observation-only student network, along with a reward that quantifies the capability to recover quantum information stored in a multiqubit state. The RL observations consist of binary qubit measurement results, while actions include unitary gates (CNOT, X) and measurements on ancilla qubits. The reward function is based on a newly defined quantity, the "recoverable information", which measures the persistence of the initial quantum information at intermediate times -- this addresses the challenge that using fidelity does not work due to the scarcity of trajectories with a large final-time fidelity at early stages of training. Due to the general formulation of the problem, strategies other than stabilizer encoding can also be discovered by the same framework, e.g. in the form of noise mitigation.

To determine the most suitable encoding for unknown error channels or specific laboratory setups, \cite{nautrup2019optimizing} employed a projective simulation framework to optimize and fault-tolerantly adapt quantum error correction codes. Their RL agent modifies a family of surface code memories until a desired logical error rate is achieved, using transfer learning to find near-optimal solutions for up to 70 data qubits. The RL states contain the code structure and the underlying qubit lattice information, with actions corresponding to placing qubits to grow the quantum code lattice according to prescribed rules. A reward of unity is given only if the code error rate falls below a threshold after a fixed number of steps.

\cite{olle2023simultaneous} train an RL agent to discover from first principles both QEC codes and their encoding circuits for a given gate set, qubit connectivity, and error model. Using an actor-critic PPO algorithm, they scale the system to 20 physical qubits and distance $d=5$ codes through a noise-aware meta-agent that learns to produce encoding strategies simultaneously for a range of noise models. The RL observations consist of a representation of the generators of the underlying stabilizer group, while actions are gates from a predefined set restricted to the available gate set and platform connectivity. The reward uses a weighted sum of the Knill-Laflamme conditions given at each step, allowing the agent to favor specific code types during optimization. In a follow-up study, the authors used scalable RL for systematic discovery and introduction of composite Clifford gates~\cite{olle2025scaling}.

Last but not least, addressing the implementation costs and errors introduced by measurement weights in QEC codes, \cite{he2025discovering} use a PPO agent to improve code design by reducing the weight of stabilizer codes. Their agent discovers low-weight codes that outperform the state of the art in practically relevant parameter regimes of moderately large code distance, leading to an order of magnitude reduction in physical qubit overhead for weight-6 codes accessible in experiments. The environment is defined by the Tanner graph of a stabilizer code, with the agent tasked to minimize the maximum degree of variable and check nodes while preserving code distance. The RL state represents the Tanner graph state, actions correspond to adding or removing graph edges, and the reward is based on code distance and weight.

%% file: qrl.tex
\section{\label{sec:QRL}Quantum Reinforcement Learning}

In a seminal paper, \cite{dunko2016quantum} propose a classification of reinforcement learning frameworks based on the properties of the agent and its environment. In particular, both the agent and the environment can be either classical (C) or quantum (Q). The vast majority of studies in the field of reinforcement learning then fall into the CC category, where a classical agent learns in a classical environment, with AlphaGo representing a prominent example [see Sec.~\ref{sec:intro} for a longer list]. 

By contrast, in the previous sections of this review, we explored the CQ category -- the application of classical reinforcement learning algorithms to environments describing the transition dynamics of quantum systems. In particular, even though we have tasked the agent with learning to manipulate quantum behavior, the RL agent has so far been implemented using a classical algorithm. 
A notable shortcoming of classical RL agents is that they are not always efficient in solving quantum tasks. For instance, recall the problem with the full observability of quantum states that cannot be measured in the lab due to the exponential scaling of full-state tomography with the number of quantum degrees of freedom. In such cases, a quantum RL agent can offer potential advantages over its classical counterpart, e.g., a more natural framework for the RL agent to interact with a quantum environment. 

We now turn our attention to setups where the RL agent itself, i.e., the policy or the (action-)value function, can be a quantum mechanical object, and its optimization is subject to the laws of quantum mechanics. This setup is known as quantum reinforcement learning (QRL), and comprises the two remaining categories: 
(QC), where a classical environment can be navigated more efficiently by exploiting quantum resources by a quantum agent (using, e.g., a quantum computer to learn strategies for the game of Go more quickly); and, 
(QQ), where a quantum environment is navigated by a quantum RL agent, an example of which we shall discuss below.

A similarity between RL agents and quantum circuits becomes discernible if one looks at them from the perspective of a variational ansatz: indeed, RL agents stand for $Q$-, value-, and policy functions, parametrized using variational parameters. The same is true about variational quantum circuits (sometimes called quantum neural networks), where one typically optimizes for unknown gate parameters. It is thus natural to ask the question of whether one can implement RL agents using a quantum circuit\footnote{Of course, one still has to overcome the familiar difficulties encountered with optimizing quantum circuits, such as barren plateaus or vanishing gradients.}. 

Quantum devices allow for executing certain operations in parallel, which can result in a quantum speedup during the decision-making process, which is at the core of the RL framework; this immediately raises the possibility for a faster convergence of the RL algorithm. As we will discuss in detail below, under certain conditions, granting agents access to quantum hardware via classical communication allows actions to be executed quadratically faster, although this may not reduce the actual learning time. 
However, due to the early stage of universal quantum computing at present, it is currently unknown whether QRL can offer an advantage over classical RL \textit{in practice}, i.e., beyond artificial problem formulations and proof-of-concept implementations.

\textbf{Featured experimental application.} 
Due to an increased demand for ML algorithms that can learn quickly and efficiently in practical applications, a natural question arises as to whether quantum devices can help reduce the learning time, besides merely speeding up the decision-making process.
Motivated by this, \cite{saggio2021experimental} addressed the question of how fast RL agents learn; they demonstrated that the learning process in RL can be sped up using an agent that interacts quantum mechanically with its environment via a quantum channel. 
To evaluate the improvement, they combine quantum and classical communication to control the agent's learning progress optimally. They implement this quantum-enhanced hybrid agent on a compact and fully tunable integrated nanophotonic processor, allowing for both quantum and classical information transfer.
We will now discuss this experiment in more detail. 

\begin{figure*}[t!]
\centering
\includegraphics[width=1.0\textwidth]{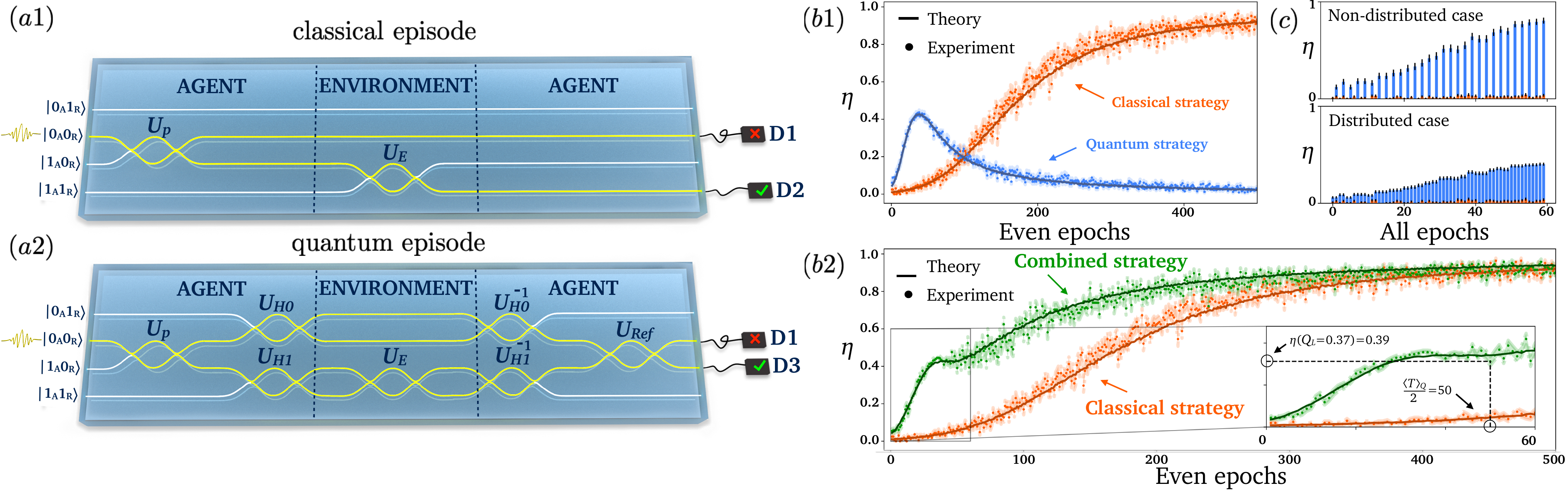}
\caption{
    \textbf{Quantum RL agent learns faster via photon-exchange quantum communication.}
    (a) A QRL agent implemented using a programmable nanophotonic processor can be trained in a classical or quantum way to implement a quantum search for positive-reward states. Waveguide modes are used to implement the action and reward spaces, resulting in a $4$-state Hilbert space over actions $A$ and rewards $R$. The agent can be trained using fully classical episodes (a1), or by alternating quantum (a2) and classical (a1) episodes [see text]. A binary reward is given only at the end of classical episodes if a photon is found in detector D2. Photon detection at the end of quantum episodes implements a Grover-like search maximizing the probability of positive-reward states.
    (b) Average reward $\eta$ against the number of training episodes. The classical strategy (b1, orange data) leads to slower training than the quantum strategy initially (b1, blue); however, the latter exhibits a peak at an optimal episode number $j_\ast$, following which its performance declines [see text]. This motivates the hybrid classical-quantum strategy (b2, green data), where the agent uses a quantum strategy until an optimal reward $\eta\approx 0.39$ is reached (b2, inset), and then switches over to a purely classical strategy (b2). 
    (c) Since the quantum strategy contains alternating quantum-classical episodes, for a fair comparison of the two strategies, the reward in the quantum-classical strategy is evenly distributed between the two kinds of episodes in a post-processing step. 
    Figure adapted with permission from Ref.~\cite{saggio2021experimental}. Reproduced with permission from Springer Nature. 
}
\label{fig:qrl}
\end{figure*}

Consider first a single-step episodic task where an agent receives a reward $r\in\{0,1\}$. The episode consists of an RL state $s$, an action $a\in\{0,1\}$, and a reward $r$ [for a generalization to multi-step episodes, see~\cite{saggio2021experimental}]. If an episode has $r(a){=}1$, we call the final state winning.
Both ${ s}={ s}({ a})$ and $r=r({ a})$ are completely determined by the action ${ a}$.
The agent draws the action from its policy $\pi(a|s)$, according to:
\begin{equation}
    a= 
    \begin{cases} 
        1 &  \text{with probability}\quad \varepsilon \\
        0 &  \text{with probability}\quad 1-\varepsilon ,
    \end{cases}
\end{equation}
where $\varepsilon$ is the probability of having a winning final state, which is parametrized by an angle $\xi$, and is defined by
\begin{equation}
\label{eq:winning_prob}
    \varepsilon=\sin^2\xi=\sum_{\{{ a} \,:\, r({ a})>0\}} \pi({a}|s),\qquad\qquad \xi\in[0,2\pi)\, .
\end{equation} 

Consider now a communication in the form of a single-photon exchange between the RL agent and its environment, based on a fixed discrete alphabet. If the communication occurs via the classical channel of the nanophotonic processor, the information is encoded in classical bits using a fixed preferred basis, e.g., the vertical (or horizontal) photon polarization. By contrast, in the quantum channel, superposition states of arbitrary polarization can be exchanged. 
We can model both the classical and the quantum behavior using a quantum-enhanced hybrid agent that interacts with its environment by exchanging quantum states $\ket{s_i}, \ket{a_i}, \ket{r_i}$ representing the RL states, actions, and rewards, respectively. Such agents can behave both classically and quantum-mechanically, as will become clear below. The policy $\pi(a_i|s_i)$ in \cite{saggio2021experimental} is updated in a classical optimization using projective simulation.

The behavior of the underlying deterministic environment consists in the application of a controlled unitary operation $U_\text{E}$  acting on the action register $A$ and reward register $R$ as follows:
\begin{equation}
	U_\text{E}\ket{{a}}_A\ket{0}_R =  
	\begin{cases} 
		 \ket{{a}}_A\ket{1}_R,  &  \text{if}\quad r({ a})>0 \\
		 \ket{{a}}_A\ket{0}_R,  &  \text{if}\quad r({ a})=0 .
	\end{cases}
\end{equation}
The application of this unitary means that whenever $r({a})>0$, the reward state is flipped, and otherwise it is left intact.
By exploiting the behavior of $U_\text{E}$, the agent has to learn the optimal action by implementing a quantum search for positive-reward states. To this end, it can choose between classical and quantum episodes, as follows.  

In \textit{classical} episodes, the agent prepares the state $\ket{{a}}_A\ket{0}_R$, and sends it to the environment. The environment applies $U_\text{E}$ and returns a new state to the agent to measure. 
Hence, classical epochs end with measurements that test the performance of the agent. 

\textit{Quantum} episodes proceed as follows:
(i) the agent prepares the initial state $\ket{{a}}_A\ket{{-}}_R$ and sends it to the environment. Here, $\ket{{-}}_R=(\ket{0}-\ket{1})/\sqrt{2}$, and 
$$\ket{{a}}_A=\sum_{{a}}\sqrt{p({a})}\ket{{a}}_A=\cos\xi \ket{\ell}_A+\sin\xi\ket{w}_A$$ 
is a superposition of winning (i.e., rewarded) $\ket{w}_A$ and losing (i.e., non-rewarded) $\ket{l}_A$ states; 
(ii) the environment applies $U_\text{E}$ and flips the sign of the winning state:  $$U_\text{E}\ket{\bf{a}}_A\ket{\bf{-}}_R =\left(\cos\xi \ket{\ell}_A-\sin\xi\ket{w}_A\right)\ket{\bf{-}}_R.$$ 
The environment then returns the resulting state to the agent.
Finally, (iii) the agent performs the reflection $U_{\rm Ref}=2\left|\psi\right\rangle\left\langle\psi\right|_A-1_A$ over the initial state $\ket{\psi}_A$, resulting in amplitude amplification as in Grover's search algorithm. This increases the probability of finding winning trajectories by re-injecting the quantum state as an action in the subsequent interaction with the environment.
Unlike their classical counterparts, quantum episodes do not produce rewards. 

The agent alternates between quantum and classical episodes, and the policy is updated once a reward is obtained.
Using this framework, \cite{saggio2021experimental} demonstrated experimentally that quantum agents find a close-to-optimal strategy faster and more efficiently than classical agents; the learning speed is quantified by the average number of episodes required to achieve a certain fixed winning probability.

The experimental implementation uses a fully programmable nanophotonic processor, interfaced with photons at telecommunication wavelengths -- a platform capable of implementing active feedback control. 
Mach-Zehnder interferometers equipped with two tunable phase shifters implement beam splitters that enable the coherent implementation of sequences of quantum gates. 

The authors represent the winning and losing action states as $\ket{1}_A=\ket{w}_A$ and $\ket{0}_A=\ket{\ell}_A$. They use another qubit to encode the reward. This results in a $4$-level system, where each level corresponds to a mode of a different waveguide, see Fig.~\ref{fig:qrl}(a). 
Starting from the state $\ket{0}_A\ket{0}_R$, the agent first applies a unitary $U_\text{P}$ to obtain the superposition $\propto (\cos\xi \ket{0}_A + \sin\xi \ket{1}_A)\ket{0}_R$.

For the classical episodes, the environment applies $U_\text{E}$ to flip the reward qubit only if the action qubit is in the winning state $\ket{w}_A$ [Fig.~\ref{fig:qrl}(a1)]. Next, the photon is read out in either detector D1 or D2 with probability $\cos^2\xi$ and $\sin^2\xi$, respectively. If it is found in D2 (that is, the agent has been rewarded), the policy $\pi$ is improved using feedback by updating the winning probability. 
The classical test episode is implemented only in software.

In quantum episodes, see Fig.~\ref{fig:qrl}(a2), after creating the superposition state by applying $U_\text{P}$, the reward qubit is first rotated to $\ket{-}_R$ using two Hadamard-like operations $U_\text{H0}, U_\text{H1}$ (the subscript $0,1$ denotes the action state). Next, the environment acts as an oracle (cf.~Grover's algorithm) applying $U_\text{E}$. Last, the agent applies the inverse operations $U^\dagger_\text{H0}, U^\dagger_\text{H1}$, and performs the reflection $U_\text{Ref}$. By measuring in the computational basis of the action register, a rewarded action sequence is read off in D3 with increased probability $\sin^2 3\xi$, see~\cite{saggio2021experimental}.

The binary measurement output of each classical episode is recorded after every alternating quantum-classical pair of episodes when training the quantum strategy. By contrast, the classical strategy contains only classical epochs. The two strategies are compared after a fixed number of measurements (i.e., the quantum strategy contains twice as many episodes in total). The rewards are averaged over $10^4$ different agents. 

Figure~\ref{fig:qrl}(b) shows the average reward $\eta$ against the number of training episodes. 
The Grover-like quantum policy found by the RL agent improves up to an optimal episode number, and then degrades [Fig.~\ref{fig:qrl}(b1), blue curve]. This results in a drop in amplitude amplification past this optimal epoch $j_\ast$. In practice, the authors identify the optimal probability $\varepsilon_{j_\ast}$ up to which the quantum strategy offers an advantage over its classical counterpart. Thus, beyond training epoch $j_\ast$, using an entirely classical strategy is more advantageous.

As we saw, the average reward in the quantum strategy starts decreasing at $\eta{=}0.39$ [Fig.~\ref{fig:qrl}(b2), inset]. The hybrid setup allows for switching to the entirely classical strategy past that point, resulting in a so-called combined strategy, which outperforms both the purely quantum and the purely classical strategies [Fig.~\ref{fig:qrl}(b2), green curve]. 
This demonstrates that hybrid agents can experience a quadratic speedup in their learning time~\cite{saggio2021experimental}.

\textbf{Breadth and scope of this section:}
In the remainder of this section, we barely scratch the surface of quantum RL. It is not the purpose of this review to present a complete discussion of that field. Instead, we refer the interested reader to the literature, such as
\cite{chen2024introduction} for an introduction to QRL,
\cite{corli2024quantum} for a review on anomaly detection using quantum RL, 
\cite{dunjko2017advances} for a discussion of using quantum mechanics to enhance classical RL metalearners, and  \cite{dunjko2018machine}  for an early review including QRL. Therefore, we now give a brief survey of a select number of studies using QRL.

\textbf{Quantum Advantage for Classical RL.}
The most common applications of quantum RL use quantum circuits to enhance the efficiency of classical ML algorithms.

Several studies have addressed the challenge of continuous action spaces in QRL. \cite{jerbi2021quantum} demonstrated the potential of quantum computers to enhance deep RL performance in large action spaces by introducing deep energy-based models that outperform standard deep RL, with corresponding quantum algorithms providing speedups on near-term quantum devices. \cite{jin2025ppo} tackled continuous action spaces and high-dimensional state spaces by introducing hybrid quantum-classical networks for actor and critic components in PPO algorithms, achieving parameter reduction and state-of-the-art performance when trained on superconducting quantum devices. Similarly, \cite{wu2025quantum} proposed quantum DDPG algorithms efficiently addressing both classical and quantum sequential decision problems with continuous action spaces, demonstrating learning capabilities in quantum control tasks.

Comparative studies have explored extensions of classical RL algorithms to quantum domains. \cite{lockwood2020reinforcement} investigated pure and hybrid quantum algorithms for DQN and Double DQN, showing that both approaches can solve RL tasks with reduced parameter spaces. \cite{kruse2025benchmarking} provided comprehensive comparisons between three QRL frameworks: parameterized quantum-circuit-based QRL (policy gradient and Q-learning variants), free-energy-based QRL, and amplitude-amplification-based QRL, analyzing their respective strengths and limitations.

\textbf{Data Encoding and NISQ Constraints.}
Data encoding represents a major challenge in QRL implementations alongside quantum computer utilization during inference. 

\cite{liu2024qtrl} addressed these issues by applying the Quantum-Train method to RL, training classical policy networks using quantum machine learning models to achieve polylogarithmic parameter reduction. This framework mitigates data encoding problems while reducing classical policy network training parameters, requiring only classical computers during inference.

The limited qubit availability in NISQ devices poses significant constraints for QRL applications. \cite{chen2022variational} developed a deep quantum RL framework based on gradient-free evolutionary optimization, employing state-amplitude encoding with hybrid variational quantum circuit architectures based on tensor networks. This approach enables handling input dimensions exceeding available qubit counts.

\cite{kolle2024study} developed quantum circuit-based RL techniques using quantum PPO frameworks incorporating data re-uploading, input scaling, output scaling, and exponential learning rate decay. Their findings indicate that data re-uploading and exponential learning rate decay significantly enhance hyperparameter stability and performance, while output scaling effectively manages agent greediness, increasing learning speed and robustness. Building on these results, \cite{kolle2024optimizing} integrated metaheuristic algorithms including Particle Swarm Optimization, Ant Colony Optimization, Tabu Search, Genetic Algorithm, Simulated Annealing, and Harmony Search into QRL, highlighting particular potential for Particle Swarm Optimization and Simulated Annealing in efficient QRL.

\textbf{Circuit Architecture Optimization.}
Quantum circuit architecture search has emerged as a promising QRL application. 

\cite{wang2024rnn} considered solving the classical MNIST classification problem on a quantum computer, where data is encoded into quantum state amplitudes -- an example of quantum ML. The task for their RL agent is to find an optimal architecture for the underlying quantum circuit. They developed a policy-gradient approach with an LSTM architecture for layer-based search to improve computational efficiency. The RL agent observes variational circuit structures at each timestep, with actions corresponding to target qubits and quantum gates (single-qubit rotations at arbitrary angles and CNOT gates), while gate parameters are optimized iteratively. Rewards derive from circuit accuracy on validation sets.

\cite{kolle2024architectural} combined metaheuristic algorithms with QRL for variational quantum circuit optimization that supports multiple agents. \cite{rapp2024reinforcement} addressed heuristic circuit architecture selection by generating problem-specific encoding circuits to improve quantum machine learning model performance, developing model-based RL algorithms to reduce required circuit evaluations during architecture search while demonstrating sample efficiency. Model-based offline quantum RL was further explored by \cite{eisenmann2024modelbased}.

\textbf{Photonic QRL Platforms.}
Quantum optics platforms have attracted increasing interest for QRL agent implementation due to their high controllability. 

\cite{konaka2025scalable} explored photonic systems for RL task solving with emphasis on scalability, utilizing photon orbital angular momentum to solve competitive multi-armed bandit problems. Their findings demonstrate that quantum optics systems can solve problems with scalable bandit arms while exceeding existing technique performance.

\cite{piera2024synthesizing} simulated agent behavior as RL algorithms making `bets' on informationally-complete measurement outputs, adjusting decisions to maximize expected returns. This study investigated how realistic but non-ideal agent decision-making deviates from Born's rule in experimental implementations using heralded single photons.

\textbf{Quantum Communication} has also proven a useful playground for QRL. 
The ability of RL agents to identify key quantum protocols that realize teleportation, entanglement purification, and quantum repeaters was explored in \cite{wallnofer2020machine}. The study finds improved solutions to problems in long-distance communication that go beyond the state of the art. 

\textbf{Quantum Speedups in Active Learning.}
Complex optimization tasks in real-world RL scenarios present significant challenges, particularly when efficient environment navigation requires rapid responses within limited timeframes. \cite{paparo2014quantum} leveraged quantum mechanics to achieve quadratic speedups in active reinforcement learning, with frameworks particularly beneficial for highly complex and dynamic environments.

\cite{sriarunothai2018speeding} provided proof-of-principle experimental demonstrations of quantum speedups for learning agents using small-scale quantum information processors. Their implementation employed two-qubit systems for QRL agent decision-making within the framework of projective simulation, utilizing radiofrequency-driven trapped ions. The experiments demonstrated quadratic improvements in QRL agent deliberation time compared to classical RL agents.

%% file: sensing.tex
\section{\label{sec:metrology-comm}Quantum Metrology: Parameter Estimation \& Sensing}

Measurement and estimation of model parameters are essential for quantum technologies. In the final section of this review, we briefly survey recent progress in applying RL to systematically design high-precision protocols for probing physical systems. 

In a nutshell, quantum metrology exploits quantum protocols and probe states to enhance the precision of estimating unknown parameters. A central task is the preparation of entangled states that achieve accuracies beyond the standard quantum limit (SQL), where the variance of an estimated parameter scales as $N^{-1/2}$ with $N$ the number of quantum degrees of freedom; the ultimate benchmark is the Heisenberg limit with scaling $N^{-1}$. Prominent examples of quantum probe states include squeezed states, GHZ states, and critical ground states of many-body systems, all of which are actively explored for quantum-enhanced sensing due to their increased sensitivity to external perturbations. While ground states can often be prepared via adiabatic evolution, this method becomes inefficient near criticality, making alternative state-preparation strategies (see Sec.~\ref{sec:state_prep} and Sec.~\ref{subsec:ent_ctrl}) essential for realizing practical quantum sensors that rely on critical ground states. 

However, access to a state featuring ``useful'' entanglement alone is not sufficient. The goal is to accurately estimate physical parameters of interest, such as the strength of a magnetic field or an optical phase shift. To assess the achievable precision, one typically needs to compute the quantum Fisher information (QFI) of a parameter-dependent state, which sets a fundamental bound on the precision via the Cram\'er--Rao inequality. Quantum parameter estimation is thus concerned with both the design of quantum probe states and the optimization of measurement strategies to attain the QFI limit. Adaptive protocols, where measurement settings are updated based on prior outcomes, play a particularly important role here, and RL provides a natural framework to automate and optimize such feedback. 

To the best of our knowledge, a compelling experimental demonstration of RL for quantum parameter estimation, metrology, or sensing beyond the current state of the art, is yet to be realized. Nevertheless, a growing body of theoretical work indicates that RL can discover nontrivial adaptive strategies, optimize entangled probe states, and design measurement feedback policies. Although the literature is still smaller compared to, e.g., quantum state preparation, there is significant methodological overlap, and many techniques developed in other contexts can be directly transferred to quantum sensing tasks. 

\textbf{Parameter Estimation.}  
A central challenge in quantum metrology is to identify control protocols for probe states that enable high-precision adaptive estimation.  

\cite{belliardo2024application} develop a model-aware RL framework for Bayesian quantum metrology based on particle filtering, where training is enhanced by automatic differentiation. In a follow-up study, \cite{belliardo2024model} present further applications of the framework, including the estimation of magnetic fields, hyperfine interactions, and phases of coherent photonic states. They also demonstrate estimation of decoherence times for electronic spins in diamond using multiple Ramsey measurements.  

Since present-day NISQ devices undergo parameter drifts, accurate modeling of their environment is computationally costly, while direct measurements may be expensive. \cite{crosta2024automatic} showed that a model-free control loop based on RL can continuously recalibrate quantum device parameters.  

It is well established that optimal controls can significantly improve the precision of parameter estimation. However, \cite{xu2019generalizable} argue that computing such controls is demanding, since they depend on the unknown parameter and must be updated after each estimate. They present an RL agent that efficiently identifies controls to improve measurement precision and investigate its generalization capabilities.  

\cite{fallani2022learning} use RL to find feedback control strategies with enhanced estimation precision. They perform frequency estimation for a single bosonic field evolving under a squeezing Hamiltonian, continuously monitored via homodyne detection. The feedback strategies discovered by their agent outperform both control-free and standard open-loop policies in the long-time regime.  

\cite{qiu2022efficient} present an RL framework to optimize state preparation protocols that accelerate one-axis twisting dynamics for entanglement generation, using the QFI as a figure of merit. The protocols discovered satisfy Heisenberg scaling and are implementable with current experimental capabilities.  

\cite{xiao2022parameter} adopt a geometric perspective for time-dependent parameter estimation. Their RL reward function incorporates noise bounds together with a physics-inspired linear time-correlated control ansatz, which improves training. They validate the performance of both time-dependent and time-independent estimation schemes in noise-free and noisy scenarios.  

\textbf{Quantum Sensing.}  
RL has also been applied to broader sensing contexts, where the goal is to maximize the sensitivity of quantum devices to external perturbations, which may involve time-dependent or static fields that are not part of the device itself.  

\cite{schuff2020improving} employ the cross-entropy method in RL to optimize the strength and position of nonlinear control pulses proposed to enhance quantum-chaotic sensors. Their RL-optimized pulses mitigate decoherence and improve measurement precision, outperforming periodic protocols.  

\cite{chih2022train} use RL to design optimal sequences for an optical-lattice gyroscope. Their agent generates end-to-end lattice shaking protocols that optimize sensitivity to rotational signals. Interestingly, the RL-discovered design differs from the conventional Mach--Zehnder-type interferometer and yields a 20-fold improvement over traditional Bragg interferometry.  

Finally, quantum reinforcement learning (see Sec.~\ref{sec:QRL}) can enhance sensing protocols in critical quantum many-body systems. \cite{xu2025towards} show that their QRL agent identifies protocols that saturate the finite quantum speed limit while generalizing to large systems, achieving Heisenberg and even super-Heisenberg scaling. This work highlights the potential of QRL agents to enable precise state preparation for highly accurate quantum-critical sensing.

Reinforcement learning offers a versatile toolbox for both optimizing quantum probe states and designing adaptive measurement strategies, bridging the gap between theoretical precision limits and experimentally implementable protocols. While experimental demonstrations in metrology and quantum sensing remain limited, the growing theoretical literature highlights RL's potential to discover nontrivial adaptive controls, accelerate entanglement generation, and enhance sensitivity in complex quantum systems. Future work may integrate RL with hybrid classical-quantum optimization, robust feedback under realistic noise, and scalable many-body sensing platforms. Overall, the RL framework is well-suited to become an increasingly valuable paradigm for realizing the full potential of quantum-enhanced parameter estimation and sensing.

%% file: outlook.tex
\section{\label{sec:outro}Open Challenges \& Future Directions}

We close this review by discussing some of the pressing challenges in the field, and identify avenues for future research. The discussion below progresses systematically from algorithmic foundations to implementation challenges and concludes with broader scientific implications.

\textbf{Design of Reward Functions and Learning Objectives.}
The formulation of appropriate reward functions plays a central role in the success of RL, and can determine the difference between breakthrough and failure. In the context of quantum technologies, a critical frontier is \textit{physics-informed reward engineering} that systematically incorporates physical constraints, symmetries and related conservation laws, and the experimental limitations inherent to near-term quantum devices. While, as we have seen, initial studies have started incorporating these elements, developing frameworks for reward design tailored to quantum problems remains largely unexplored. 

Related, and equally important, is the capability of RL frameworks to learn from sparse observations. This is crucial for experimental applications. By contrast, if we train RL agents in simulations, the environment is fully observable; however, the learned policy may not be applicable in a real-world experiment without additional modifications later on, particularly when it implements active feedback. 
Hence, RL frameworks have to address the fundamental challenge of learning from sparse quantum measurements while maintaining sample efficiency -- a prerequisite for bridging the gap between theoretical algorithms and experimental implementation.

Of similar significance is the development of RL agents capable of rapid adaptation to new system Hamiltonians, target states, or experimental conditions without extensive retraining. This adaptability will prove critical for practical deployment across diverse quantum platforms, from trapped ions to neutral atoms and superconducting circuits. A particularly challenging open question in quantum control is \textit{multi-objective optimization}, where agents must simultaneously balance their performance on competing tasks such as maximizing gate fidelity, keeping the protocol resilient to decoherence, and meeting experimental runtime constraints. In this context, meta-learning approaches may offer promising solutions to this inherently multifaceted optimization landscape~\cite{oh2025discovering, beck2025tutorial}.

\textbf{Next-Generation RL Architectures.}
The unique characteristics of quantum systems require specialized RL architectures that extend beyond conventional ML approaches. A fundamental challenge arises from the frequent lack of perfect Markovianity in quantum environments, where current states often depend on measurement history and environmental interactions. Addressing this requires sophisticated memory architectures: transformer-based policies with attention mechanisms for capturing long-range temporal correlations, recurrent networks, or graph neural networks designed to capture the spatial-temporal structure of quantum correlations.

Beyond memory considerations, well-characterized experimental quantum systems like quantum computers enable \textit{model-based approaches} that incorporate predictive modeling for quantum planning and control. The integration of tree search algorithms with quantum-specific heuristics represents a particularly promising direction. Furthermore, the distributed nature of many quantum technologies suggests exploring \textit{coordinated multi-agent architectures}~\cite{chalkiadakis2003coordination}, where multiple RL agents collaborate to control different subsystems or optimize complementary objectives. While some of these architectural innovations have been suggested in other domains, their specific benefits for quantum applications remain largely unexplored.

\textbf{Scaling to Many-Body Quantum Systems.}
The transition from few- to many-body quantum systems presents perhaps one of the most fundamental challenges for RL in quantum simulation. As the number of controlled degrees of freedom increases, the application of RL can be affected in different ways which can originate from both the environment used to generate training data and the agent architecture itself.

The underlying quantum environment has a Hilbert space dimension that grows exponentially with the number of quantum degrees of freedom. This spells trouble for the design of numerical simulators from which RL agents learn (both model-free and model-based). 
Addressing the rapid increase of the state space dimension in classical simulations, requires leveraging the inherent structure of quantum many-body systems. \textit{Equivariant neural network architectures} that respect physical symmetries (permutation invariance, gauge redundancies, or spatiotemporal symmetries) can dramatically reduce the effective dimensionality of both policy and $Q$-function representations. Tensor network parameterizations offer an alternative approach, providing compressed representations that naturally capture entanglement structure.
An elegant resolution to this curse-of-dimensionality problem in the near future may be offered by using quantum simulators, or by training RL agents directly on the real quantum many-body system of interest.

Whereas the underlying state space typically grows exponentially with the number of controlled quantum degrees of freedom, the observation space need not. Physical observables typically exhibit spatial locality (e.g., they are sums of few-body operators), and their number scales only polynomially with the volume of the system. Therefore, the fundamental \textit{learning problem} to be solved is that of learning from sparse, partial, and experimentally accessible observations. Note the additional complexity brought by quantum mechanics via the incompatibility of simultaneous measurements of noncommuting observables, and the invasive character of the observation process which changes the underlying quantum state a posteriori. 

When scaling the control from few to many quantum degrees of freedom, the action space also expands. Here, the natural open research question is whether control of individual particles at a microscopic level is still required in the many-body regime. Open problems include the design of RL action spaces that allow for a parallel application of actions when possible (e.g., we do not flip qubits individually but all at once whenever possible), and the identification of compound actions. For action space complexity, \textit{hierarchical control strategies} show promise through automated discovery of reusable control primitives -- so-called ``quantum gadgets'' that can be composed into more complex protocols.

The emergence of collective phenomena in interacting many-body systems presents both challenges and opportunities. These systems exhibit behaviors fundamentally absent in few-body regimes, e.g., quantum phase transitions and critical phenomena. A compelling open question is whether RL agents can learn to identify and exploit emergent degrees of freedom, such as order parameters, for enhanced control, potentially enabling critical point manipulation and the discovery of novel quantum simulation protocols. This connects to the broader challenge of \textit{transfer learning across system sizes}: developing theoretical frameworks to understand when and how control policies learned on smaller systems can be systematically scaled to larger architectures.

\textbf{Real-Time Implementation on Quantum Hardware.}
Transitioning from theoretical advances to practical deployment on quantum hardware introduces constraints that fundamentally reshape the RL problem. 

\textit{Real-time deployment constraints:} Direct implementation on quantum hardware faces severe limitations from classical-quantum communication latencies and finite feedback bandwidths. Modern quantum control systems can operate on nanosecond to microsecond timescales, while classical computation and data transfer still introduce delays on the order of microseconds to milliseconds. Developing low-latency architectures and edge computing solutions specifically for quantum RL represents a major engineering challenge.

\textbf{Partial observability and measurement limitations:} Quantum states cannot be fully observed, and full tomographic reconstruction scales exponentially with system size. Therefore, RL algorithms must learn effective policies from fundamentally incomplete measurement data while minimizing experimental overhead. Shadow tomography, randomized measurement protocols, and Bayesian inference techniques for uncertainty quantification offer promising directions for efficient state learning and reward estimation.

\textit{Sample-efficient learning:} The high cost of quantum measurements requires algorithms that achieve strong performance with minimal sample complexity. Hybrid model-based/model-free approaches may reduce measurement requirements by leveraging prior physical knowledge while maintaining the flexibility to adapt to unknown system characteristics.

\textbf{Physics Discovery through Reinforcement Learning.}
Beyond robust experimental control, the most transformative potential of RL in quantum technology lies in its capacity for scientific discovery. The development of \textit{interpretable quantum control policies} represents a crucial frontier: extracting human-readable insights from trained agents to potentially reveal new physical principles or identify optimal control strategies that go beyond human intuition.

This interpretability challenge requires specialized tools, including visualization techniques for identifying critical components in neural network policies and symbolic regression methods for distilling learned behaviors into analytical forms. This could transform RL from a mere optimization tool into an \textit{instrument for fundamental scientific discovery}. The extent to which RL can uncover new dynamical principles, identify novel quantum protocols, or reveal hidden conservation laws remains an exciting open question.

\textbf{Integration with the Broader AI Ecosystem.}
Realizing RL's full potential in quantum physics discovery requires strategic integration with advances across the broader AI landscape. 

The emergence of \textit{Large Language Models} as "foundation models" covering a wide range of abilities~\cite{bommasani2021opportunities} presents unprecedented opportunities. These models could provide symbolic reasoning about quantum protocols and theoretical insights, enable natural language interfaces for experimental design and hypothesis formulation, and facilitate automated literature synthesis and protocol suggestion. Such integration might accelerate the translation of theoretical advances into experimental implementations. In addition, the techniques behind LLMs, like the attention mechanism, as well as the concept of training ``foundation models" that cover a wider range of tasks in a particular domain, are starting to be translated back into quantum science \cite{rende2025foundation,zaklama2025attention}. 

Perhaps of greatest practical importance in the present context, however, are the outstanding coding abilities of modern LLMs. These LLMs are able to code an entire small RL project from scratch within a matter of minutes and (with some additional human guidance) can then arrive at a robust and tested solution within some hours, dramatically accelerating the development cycle for RL and, in the future, democratizing access to the approach of RL for quantum science.

RL has also been crucial for performance improvements in LLMs themselves. Indeed, \textit{Human-AI collaboration} through Reinforcement Learning from Human Feedback (RLHF)~\cite{ouyang2022training} has been the decisive step in producing useful LLMs. In principle, this technique would then also offer mechanisms to incorporate further expert physicist intuition, experimental heuristics, and domain-specific constraints directly into agent training (whether the agent is an LLM or merely a smaller neural network). This approach may prove particularly valuable given the nature of quantum phenomena and the importance of physical intuition in experimental design, although it requires considerable human-time investment.

Another, more recent breakthrough where RL has been crucial in boosting the capabilities of LLMs is through RL with verifiable rewards (RLVR), where the final answers can easily be checked (e.g., for math puzzles) and the LLM learns better reasoning strategies on its own in this way. This underlies modern strong reasoning models. In principle, specializing this to the domain of quantum science could produce high-performing LLM reasoning agents for that domain. Downsides of LLMs, however, are that currently the strongest performing models are closed source, and that training an LLM (whether in pretraining or later RL) usually comes with tremendous costs and hardware requirements that are often not affordable in the academic domain.

\textit{Multimodal learning approaches} that combine RL with computer vision offer additional pathways, particularly in automated experimental setup optimization and real-time diagnostics of quantum devices. The integration of visual feedback with quantum measurements could enable more comprehensive and adaptive quantum research platforms, potentially identifying correlations between system performance and environmental factors that escape human observation.

Looking forward, the convergence of quantum technologies with advances in AI suggests possibilities that extend far beyond current applications. As both fields mature, their intersection may yield entirely new paradigms for scientific discovery, experimental design, and alter our fundamental understanding of complex quantum systems.

%% file: main.bbl
\begin{thebibliography}{239}%
\makeatletter
\providecommand \@ifxundefined [1]{%
 \@ifx{#1\undefined}
}%
\providecommand \@ifnum [1]{%
 \ifnum #1\expandafter \@firstoftwo
 \else \expandafter \@secondoftwo
 \fi
}%
\providecommand \@ifx [1]{%
 \ifx #1\expandafter \@firstoftwo
 \else \expandafter \@secondoftwo
 \fi
}%
\providecommand \natexlab [1]{#1}%
\providecommand \enquote  [1]{``#1''}%
\providecommand \bibnamefont  [1]{#1}%
\providecommand \bibfnamefont [1]{#1}%
\providecommand \citenamefont [1]{#1}%
\providecommand \href@noop [0]{\@secondoftwo}%
\providecommand \href [0]{\begingroup \@sanitize@url \@href}%
\providecommand \@href[1]{\@@startlink{#1}\@@href}%
\providecommand \@@href[1]{\endgroup#1\@@endlink}%
\providecommand \@sanitize@url [0]{\catcode `\\12\catcode `\$12\catcode
  `\&12\catcode `\#12\catcode `\^12\catcode `\_12\catcode `\%12\relax}%
\providecommand \@@startlink[1]{}%
\providecommand \@@endlink[0]{}%
\providecommand \url  [0]{\begingroup\@sanitize@url \@url }%
\providecommand \@url [1]{\endgroup\@href {#1}{\urlprefix }}%
\providecommand \urlprefix  [0]{URL }%
\providecommand \Eprint [0]{\href }%
\providecommand \doibase [0]{https://doi.org/}%
\providecommand \selectlanguage [0]{\@gobble}%
\providecommand \bibinfo  [0]{\@secondoftwo}%
\providecommand \bibfield  [0]{\@secondoftwo}%
\providecommand \translation [1]{[#1]}%
\providecommand \BibitemOpen [0]{}%
\providecommand \bibitemStop [0]{}%
\providecommand \bibitemNoStop [0]{.\EOS\space}%
\providecommand \EOS [0]{\spacefactor3000\relax}%
\providecommand \BibitemShut  [1]{\csname bibitem#1\endcsname}%
\let\auto@bib@innerbib\@empty
\bibitem [{\citenamefont {Abdelhafez}\ \emph {et~al.}(2019)\citenamefont
  {Abdelhafez}, \citenamefont {Schuster},\ and\ \citenamefont
  {Koch}}]{abdelhafez2019gradient}%
  \BibitemOpen
  \bibfield  {author} {\bibinfo {author} {\bibnamefont {Abdelhafez},
  \bibfnamefont {Mohamed}}, \bibinfo {author} {\bibfnamefont {David~I}\
  \bibnamefont {Schuster}}, and\ \bibinfo {author} {\bibfnamefont {Jens}\
  \bibnamefont {Koch}}} (\bibinfo {year} {2019}),\ \bibfield  {title} {\enquote
  {\bibinfo {title} {Gradient-based optimal control of open quantum systems
  using quantum trajectories and automatic differentiation},}\ }\href
  {https://doi.org/10.1103/PhysRevA.99.052327} {\bibfield  {journal} {\bibinfo
  {journal} {Physical Review A}\ }\textbf {\bibinfo {volume} {99}}~(\bibinfo
  {number} {5}),\ \bibinfo {pages} {052327}}\BibitemShut {NoStop}%
\bibitem [{\citenamefont {Acharya}\ \emph {et~al.}(2024)\citenamefont
  {Acharya}, \citenamefont {Aghababaie-Beni}, \citenamefont {Aleiner},
  \citenamefont {Andersen}, \citenamefont {Ansmann}, \citenamefont {Arute},
  \citenamefont {Arya}, \citenamefont {Asfaw}, \citenamefont {Astrakhantsev},
  \citenamefont {Atalaya} \emph {et~al.}}]{acharya2024quantum}%
  \BibitemOpen
  \bibfield  {author} {\bibinfo {author} {\bibnamefont {Acharya}, \bibfnamefont
  {Rajeev}}, \bibinfo {author} {\bibfnamefont {Laleh}\ \bibnamefont
  {Aghababaie-Beni}}, \bibinfo {author} {\bibfnamefont {Igor}\ \bibnamefont
  {Aleiner}}, \bibinfo {author} {\bibfnamefont {Trond~I}\ \bibnamefont
  {Andersen}}, \bibinfo {author} {\bibfnamefont {Markus}\ \bibnamefont
  {Ansmann}}, \bibinfo {author} {\bibfnamefont {Frank}\ \bibnamefont {Arute}},
  \bibinfo {author} {\bibfnamefont {Kunal}\ \bibnamefont {Arya}}, \bibinfo
  {author} {\bibfnamefont {Abraham}\ \bibnamefont {Asfaw}}, \bibinfo {author}
  {\bibfnamefont {Nikita}\ \bibnamefont {Astrakhantsev}}, \bibinfo {author}
  {\bibfnamefont {Juan}\ \bibnamefont {Atalaya}},  \emph {et~al.}} (\bibinfo
  {year} {2024}),\ \bibfield  {title} {\enquote {\bibinfo {title} {Quantum
  error correction below the surface code threshold},}\ }\href
  {https://arxiv.org/abs/2408.13687} {\bibinfo  {journal} {arXiv preprint
  arXiv:2408.13687}\ }\BibitemShut {NoStop}%
\bibitem [{\citenamefont {Ai}\ \emph {et~al.}(2022)\citenamefont {Ai},
  \citenamefont {Ding}, \citenamefont {Ban}, \citenamefont
  {Mart{\'\i}n-Guerrero}, \citenamefont {Casanova}, \citenamefont {Cui},
  \citenamefont {Huang}, \citenamefont {Chen}, \citenamefont {Li},\ and\
  \citenamefont {Guo}}]{ai2022experimentally}%
  \BibitemOpen
\bibfield  {journal} {  }\bibfield  {author} {\bibinfo {author} {\bibnamefont
  {Ai}, \bibfnamefont {Ming-Zhong}}, \bibinfo {author} {\bibfnamefont
  {Yongcheng}\ \bibnamefont {Ding}}, \bibinfo {author} {\bibfnamefont {Yue}\
  \bibnamefont {Ban}}, \bibinfo {author} {\bibfnamefont {Jos{\'e}~D}\
  \bibnamefont {Mart{\'\i}n-Guerrero}}, \bibinfo {author} {\bibfnamefont
  {Jorge}\ \bibnamefont {Casanova}}, \bibinfo {author} {\bibfnamefont
  {Jin-Ming}\ \bibnamefont {Cui}}, \bibinfo {author} {\bibfnamefont {Yun-Feng}\
  \bibnamefont {Huang}}, \bibinfo {author} {\bibfnamefont {Xi}~\bibnamefont
  {Chen}}, \bibinfo {author} {\bibfnamefont {Chuan-Feng}\ \bibnamefont {Li}},
  and\ \bibinfo {author} {\bibfnamefont {Guang-Can}\ \bibnamefont {Guo}}}
  (\bibinfo {year} {2022}),\ \bibfield  {title} {\enquote {\bibinfo {title}
  {Experimentally realizing efficient quantum control with reinforcement
  learning},}\ }\href {https://doi.org/10.1007/s11433-021-1841-2} {\bibfield
  {journal} {\bibinfo  {journal} {Science China Physics, Mechanics \&
  Astronomy}\ }\textbf {\bibinfo {volume} {65}}~(\bibinfo {number} {5}),\
  \bibinfo {pages} {250312}}\BibitemShut {NoStop}%
\bibitem [{\citenamefont {Almanakly}\ \emph {et~al.}(2024)\citenamefont
  {Almanakly}, \citenamefont {Yankelevich}, \citenamefont {Hays}, \citenamefont
  {Kannan}, \citenamefont {Assouly}, \citenamefont {Greene}, \citenamefont
  {Gingras}, \citenamefont {Niedzielski}, \citenamefont {Stickler},
  \citenamefont {Schwartz} \emph {et~al.}}]{almanakly2024deterministic}%
  \BibitemOpen
  \bibfield  {author} {\bibinfo {author} {\bibnamefont {Almanakly},
  \bibfnamefont {Aziza}}, \bibinfo {author} {\bibfnamefont {Beatriz}\
  \bibnamefont {Yankelevich}}, \bibinfo {author} {\bibfnamefont {Max}\
  \bibnamefont {Hays}}, \bibinfo {author} {\bibfnamefont {Bharath}\
  \bibnamefont {Kannan}}, \bibinfo {author} {\bibfnamefont {Reouven}\
  \bibnamefont {Assouly}}, \bibinfo {author} {\bibfnamefont {Alex}\
  \bibnamefont {Greene}}, \bibinfo {author} {\bibfnamefont {Michael}\
  \bibnamefont {Gingras}}, \bibinfo {author} {\bibfnamefont {Bethany~M}\
  \bibnamefont {Niedzielski}}, \bibinfo {author} {\bibfnamefont {Hannah}\
  \bibnamefont {Stickler}}, \bibinfo {author} {\bibfnamefont {Mollie~E}\
  \bibnamefont {Schwartz}},  \emph {et~al.}} (\bibinfo {year} {2024}),\
  \bibfield  {title} {\enquote {\bibinfo {title} {Deterministic remote
  entanglement using a chiral quantum interconnect},}\ }\href
  {https://arxiv.org/abs/2408.05164} {\bibinfo  {journal} {arXiv preprint
  arXiv:2408.05164}\ }\BibitemShut {NoStop}%
\bibitem [{\citenamefont {Altmann}\ \emph {et~al.}(2023)\citenamefont
  {Altmann}, \citenamefont {B{\"a}rligea}, \citenamefont {Stein}, \citenamefont
  {K{\"o}lle}, \citenamefont {Gabor}, \citenamefont {Phan},\ and\ \citenamefont
  {Linnhoff-Popien}}]{altmann2023challenges}%
  \BibitemOpen
\bibfield  {journal} {  }\bibfield  {author} {\bibinfo {author} {\bibnamefont
  {Altmann}, \bibfnamefont {Philipp}}, \bibinfo {author} {\bibfnamefont
  {Adelina}\ \bibnamefont {B{\"a}rligea}}, \bibinfo {author} {\bibfnamefont
  {Jonas}\ \bibnamefont {Stein}}, \bibinfo {author} {\bibfnamefont {Michael}\
  \bibnamefont {K{\"o}lle}}, \bibinfo {author} {\bibfnamefont {Thomas}\
  \bibnamefont {Gabor}}, \bibinfo {author} {\bibfnamefont {Thomy}\ \bibnamefont
  {Phan}}, and\ \bibinfo {author} {\bibfnamefont {Claudia}\ \bibnamefont
  {Linnhoff-Popien}}} (\bibinfo {year} {2023}),\ \bibfield  {title} {\enquote
  {\bibinfo {title} {Challenges for reinforcement learning in quantum
  computing},}\ }\href {https://arxiv.org/abs/2312.11337} {\bibinfo  {journal}
  {arXiv preprint arXiv:2312.11337}\ }\BibitemShut {NoStop}%
\bibitem [{\citenamefont {An}\ \emph {et~al.}(2024)\citenamefont {An},
  \citenamefont {Cao}, \citenamefont {Xu},\ and\ \citenamefont
  {Zhou}}]{an2024learning}%
  \BibitemOpen
\bibfield  {journal} {  }\bibfield  {author} {\bibinfo {author} {\bibnamefont
  {An}, \bibfnamefont {Zheng}}, \bibinfo {author} {\bibfnamefont {Chenfeng}\
  \bibnamefont {Cao}}, \bibinfo {author} {\bibfnamefont {Cheng-Qian}\
  \bibnamefont {Xu}}, and\ \bibinfo {author} {\bibfnamefont {DL}~\bibnamefont
  {Zhou}}} (\bibinfo {year} {2024}),\ \bibfield  {title} {\enquote {\bibinfo
  {title} {Learning quantum phases via single-qubit disentanglement},}\ }\href
  {https://doi.org/10.22331/q-2024-07-22-1421} {\bibfield  {journal} {\bibinfo
  {journal} {Quantum}\ }\textbf {\bibinfo {volume} {8}},\ \bibinfo {pages}
  {1421}}\BibitemShut {NoStop}%
\bibitem [{\citenamefont {An}\ \emph {et~al.}(2021)\citenamefont {An},
  \citenamefont {Song}, \citenamefont {He},\ and\ \citenamefont
  {Zhou}}]{an2021quantum}%
  \BibitemOpen
  \bibfield  {author} {\bibinfo {author} {\bibnamefont {An}, \bibfnamefont
  {Zheng}}, \bibinfo {author} {\bibfnamefont {Hai-Jing}\ \bibnamefont {Song}},
  \bibinfo {author} {\bibfnamefont {Qi-Kai}\ \bibnamefont {He}}, and\ \bibinfo
  {author} {\bibfnamefont {D.~L.}\ \bibnamefont {Zhou}}} (\bibinfo {year}
  {2021}),\ \bibfield  {title} {\enquote {\bibinfo {title} {Quantum optimal
  control of multilevel dissipative quantum systems with reinforcement
  learning},}\ }\href {https://doi.org/10.1103/PhysRevA.103.012404} {\bibfield
  {journal} {\bibinfo  {journal} {Phys. Rev. A}\ }\textbf {\bibinfo {volume}
  {103}},\ \bibinfo {pages} {012404}}\BibitemShut {NoStop}%
\bibitem [{\citenamefont {An}\ and\ \citenamefont {Zhou}(2019)}]{an2019deep}%
  \BibitemOpen
  \bibfield  {author} {\bibinfo {author} {\bibnamefont {An}, \bibfnamefont
  {Zheng}}, and\ \bibinfo {author} {\bibfnamefont {DL}~\bibnamefont {Zhou}}}
  (\bibinfo {year} {2019}),\ \bibfield  {title} {\enquote {\bibinfo {title}
  {Deep reinforcement learning for quantum gate control},}\ }\href
  {https://doi.org/10.1209/0295-5075/126/60002} {\bibfield  {journal} {\bibinfo
   {journal} {Europhysics Letters}\ }\textbf {\bibinfo {volume}
  {126}}~(\bibinfo {number} {6}),\ \bibinfo {pages} {60002}}\BibitemShut
  {NoStop}%
\bibitem [{\citenamefont {Andreasson}\ \emph {et~al.}(2019)\citenamefont
  {Andreasson}, \citenamefont {Johansson}, \citenamefont {Liljestrand},\ and\
  \citenamefont {Granath}}]{andreasson2019quantum}%
  \BibitemOpen
  \bibfield  {author} {\bibinfo {author} {\bibnamefont {Andreasson},
  \bibfnamefont {Philip}}, \bibinfo {author} {\bibfnamefont {Joel}\
  \bibnamefont {Johansson}}, \bibinfo {author} {\bibfnamefont {Simon}\
  \bibnamefont {Liljestrand}}, and\ \bibinfo {author} {\bibfnamefont {Mats}\
  \bibnamefont {Granath}}} (\bibinfo {year} {2019}),\ \bibfield  {title}
  {\enquote {\bibinfo {title} {Quantum error correction for the toric code
  using deep reinforcement learning},}\ }\href
  {https://doi.org/10.22331/q-2019-09-02-183} {\bibfield  {journal} {\bibinfo
  {journal} {Quantum}\ }\textbf {\bibinfo {volume} {3}},\ \bibinfo {pages}
  {183}}\BibitemShut {NoStop}%
\bibitem [{\citenamefont {August}\ and\ \citenamefont
  {Hern{\'a}ndez-Lobato}(2018)}]{august2018taking}%
  \BibitemOpen
  \bibfield  {author} {\bibinfo {author} {\bibnamefont {August}, \bibfnamefont
  {Moritz}}, and\ \bibinfo {author} {\bibfnamefont {Jos{\'e}~Miguel}\
  \bibnamefont {Hern{\'a}ndez-Lobato}}} (\bibinfo {year} {2018}),\ \bibfield
  {title} {\enquote {\bibinfo {title} {Taking gradients through experiments:
  Lstms and memory proximal policy optimization for black-box quantum
  control},}\ }\bibfield  {booktitle} {\emph {\bibinfo {booktitle} {High
  Performance Computing: ISC High Performance 2018 International Workshops,
  Frankfurt/Main, Germany, June 28, 2018, Revised Selected Papers 33}},\ }\href
  {https://doi.org/10.1007/978-3-030-02465-9_43} {\bibinfo  {journal}
  {Springer}\ ,\ \bibinfo {pages} {591--613}}\BibitemShut {NoStop}%
\bibitem [{\citenamefont {Bao}\ \emph {et~al.}(2024)\citenamefont {Bao},
  \citenamefont {Furuya},\ and\ \citenamefont {Suer}}]{bao2024reinforced}%
  \BibitemOpen
\bibfield  {journal} {  }\bibfield  {author} {\bibinfo {author} {\bibnamefont
  {Bao}, \bibfnamefont {Ning}}, \bibinfo {author} {\bibfnamefont {Keiichiro}\
  \bibnamefont {Furuya}}, and\ \bibinfo {author} {\bibfnamefont {Gun}\
  \bibnamefont {Suer}}} (\bibinfo {year} {2024}),\ \bibfield  {title} {\enquote
  {\bibinfo {title} {Reinforced disentanglers on random unitary circuits},}\
  }\href {https://arxiv.org/abs/2411.09784} {\bibinfo  {journal} {arXiv
  preprint arXiv:2411.09784}\ }\BibitemShut {NoStop}%
\bibitem [{\citenamefont {Barry}\ \emph {et~al.}(2014)\citenamefont {Barry},
  \citenamefont {Barry},\ and\ \citenamefont {Aaronson}}]{barry2014quantum}%
  \BibitemOpen
\bibfield  {journal} {  }\bibfield  {author} {\bibinfo {author} {\bibnamefont
  {Barry}, \bibfnamefont {Jennifer}}, \bibinfo {author} {\bibfnamefont
  {Daniel~T.}\ \bibnamefont {Barry}}, and\ \bibinfo {author} {\bibfnamefont
  {Scott}\ \bibnamefont {Aaronson}}} (\bibinfo {year} {2014}),\ \bibfield
  {title} {\enquote {\bibinfo {title} {Quantum partially observable markov
  decision processes},}\ }\href {https://doi.org/10.1103/PhysRevA.90.032311}
  {\bibfield  {journal} {\bibinfo  {journal} {Phys. Rev. A}\ }\textbf {\bibinfo
  {volume} {90}},\ \bibinfo {pages} {032311}}\BibitemShut {NoStop}%
\bibitem [{\citenamefont {Baum}\ \emph {et~al.}(2021)\citenamefont {Baum},
  \citenamefont {Amico}, \citenamefont {Howell}, \citenamefont {Hush},
  \citenamefont {Liuzzi}, \citenamefont {Mundada}, \citenamefont {Merkh},
  \citenamefont {Carvalho},\ and\ \citenamefont
  {Biercuk}}]{baum2021experimental}%
  \BibitemOpen
  \bibfield  {author} {\bibinfo {author} {\bibnamefont {Baum}, \bibfnamefont
  {Yuval}}, \bibinfo {author} {\bibfnamefont {Mirko}\ \bibnamefont {Amico}},
  \bibinfo {author} {\bibfnamefont {Sean}\ \bibnamefont {Howell}}, \bibinfo
  {author} {\bibfnamefont {Michael}\ \bibnamefont {Hush}}, \bibinfo {author}
  {\bibfnamefont {Maggie}\ \bibnamefont {Liuzzi}}, \bibinfo {author}
  {\bibfnamefont {Pranav}\ \bibnamefont {Mundada}}, \bibinfo {author}
  {\bibfnamefont {Thomas}\ \bibnamefont {Merkh}}, \bibinfo {author}
  {\bibfnamefont {Andre~R.R.}\ \bibnamefont {Carvalho}}, and\ \bibinfo {author}
  {\bibfnamefont {Michael~J.}\ \bibnamefont {Biercuk}}} (\bibinfo {year}
  {2021}),\ \bibfield  {title} {\enquote {\bibinfo {title} {Experimental deep
  reinforcement learning for error-robust gate-set design on a superconducting
  quantum computer},}\ }\href {https://doi.org/10.1103/PRXQuantum.2.040324}
  {\bibfield  {journal} {\bibinfo  {journal} {PRX Quantum}\ }\textbf {\bibinfo
  {volume} {2}},\ \bibinfo {pages} {040324}}\BibitemShut {NoStop}%
\bibitem [{\citenamefont {Beck}\ \emph {et~al.}(2025)\citenamefont {Beck},
  \citenamefont {Vuorio}, \citenamefont {Zheran~Liu}, \citenamefont {Xiong},
  \citenamefont {Zintgraf}, \citenamefont {Finn},\ and\ \citenamefont
  {Whiteson}}]{beck2025tutorial}%
  \BibitemOpen
  \bibfield  {author} {\bibinfo {author} {\bibnamefont {Beck}, \bibfnamefont
  {Jacob}}, \bibinfo {author} {\bibfnamefont {Risto}\ \bibnamefont {Vuorio}},
  \bibinfo {author} {\bibfnamefont {Evan}\ \bibnamefont {Zheran~Liu}}, \bibinfo
  {author} {\bibfnamefont {Zheng}\ \bibnamefont {Xiong}}, \bibinfo {author}
  {\bibfnamefont {Luisa}\ \bibnamefont {Zintgraf}}, \bibinfo {author}
  {\bibfnamefont {Chelsea}\ \bibnamefont {Finn}}, and\ \bibinfo {author}
  {\bibfnamefont {Shimon}\ \bibnamefont {Whiteson}}} (\bibinfo {year} {2025}),\
  \bibfield  {title} {\enquote {\bibinfo {title} {A tutorial on
  meta-reinforcement learning},}\ }\href {https://doi.org/10.1561/2200000080}
  {\bibfield  {journal} {\bibinfo  {journal} {Foundations and Trends in Machine
  Learning}\ }\textbf {\bibinfo {volume} {18}}~(\bibinfo {number} {2-3}),\
  \bibinfo {pages} {224--384}}\BibitemShut {NoStop}%
\bibitem [{\citenamefont {Bellemare}\ \emph {et~al.}(2020)\citenamefont
  {Bellemare}, \citenamefont {Candido}, \citenamefont {Castro}, \citenamefont
  {Gong}, \citenamefont {Machado}, \citenamefont {Moitra}, \citenamefont
  {Ponda},\ and\ \citenamefont {Wang}}]{bellemare2020autonomous}%
  \BibitemOpen
  \bibfield  {author} {\bibinfo {author} {\bibnamefont {Bellemare},
  \bibfnamefont {Marc~G}}, \bibinfo {author} {\bibfnamefont {Salvatore}\
  \bibnamefont {Candido}}, \bibinfo {author} {\bibfnamefont {Pablo~Samuel}\
  \bibnamefont {Castro}}, \bibinfo {author} {\bibfnamefont {Jun}\ \bibnamefont
  {Gong}}, \bibinfo {author} {\bibfnamefont {Marlos~C}\ \bibnamefont
  {Machado}}, \bibinfo {author} {\bibfnamefont {Subhodeep}\ \bibnamefont
  {Moitra}}, \bibinfo {author} {\bibfnamefont {Sameera~S}\ \bibnamefont
  {Ponda}}, and\ \bibinfo {author} {\bibfnamefont {Ziyu}\ \bibnamefont {Wang}}}
  (\bibinfo {year} {2020}),\ \bibfield  {title} {\enquote {\bibinfo {title}
  {Autonomous navigation of stratospheric balloons using reinforcement
  learning},}\ }\href {https://doi.org/10.1038/s41586-020-2939-8} {\bibfield
  {journal} {\bibinfo  {journal} {Nature}\ }\textbf {\bibinfo {volume}
  {588}}~(\bibinfo {number} {7836}),\ \bibinfo {pages} {77--82}}\BibitemShut
  {NoStop}%
\bibitem [{\citenamefont {Belliardo}\ \emph
  {et~al.}(2024{\natexlab{a}})\citenamefont {Belliardo}, \citenamefont
  {Zoratti},\ and\ \citenamefont {Giovannetti}}]{belliardo2024application}%
  \BibitemOpen
  \bibfield  {author} {\bibinfo {author} {\bibnamefont {Belliardo},
  \bibfnamefont {Federico}}, \bibinfo {author} {\bibfnamefont {Fabio}\
  \bibnamefont {Zoratti}}, and\ \bibinfo {author} {\bibfnamefont {Vittorio}\
  \bibnamefont {Giovannetti}}} (\bibinfo {year} {2024}{\natexlab{a}}),\
  \bibfield  {title} {\enquote {\bibinfo {title} {Application of machine
  learning to experimental design in quantum mechanics},}\ }\href
  {https://doi.org/10.1142/s0219749924500023} {\bibfield  {journal} {\bibinfo
  {journal} {International Journal of Quantum Information}\
  }10.1142/s0219749924500023}\BibitemShut {NoStop}%
\bibitem [{\citenamefont {Belliardo}\ \emph
  {et~al.}(2024{\natexlab{b}})\citenamefont {Belliardo}, \citenamefont
  {Zoratti}, \citenamefont {Marquardt},\ and\ \citenamefont
  {Giovannetti}}]{belliardo2024model}%
  \BibitemOpen
  \bibfield  {author} {\bibinfo {author} {\bibnamefont {Belliardo},
  \bibfnamefont {Federico}}, \bibinfo {author} {\bibfnamefont {Fabio}\
  \bibnamefont {Zoratti}}, \bibinfo {author} {\bibfnamefont {Florian}\
  \bibnamefont {Marquardt}}, and\ \bibinfo {author} {\bibfnamefont {Vittorio}\
  \bibnamefont {Giovannetti}}} (\bibinfo {year} {2024}{\natexlab{b}}),\
  \bibfield  {title} {\enquote {\bibinfo {title} {Model-aware reinforcement
  learning for high-performance bayesian experimental design in quantum
  metrology},}\ }\href {https://doi.org/10.22331/q-2024-12-10-1555} {\bibfield
  {journal} {\bibinfo  {journal} {Quantum}\ }\textbf {\bibinfo {volume} {8}},\
  \bibinfo {pages} {1555}}\BibitemShut {NoStop}%
\bibitem [{\citenamefont {Bolens}\ and\ \citenamefont
  {Heyl}(2021)}]{bolens2021reinforcement}%
  \BibitemOpen
  \bibfield  {author} {\bibinfo {author} {\bibnamefont {Bolens}, \bibfnamefont
  {Adrien}}, and\ \bibinfo {author} {\bibfnamefont {Markus}\ \bibnamefont
  {Heyl}}} (\bibinfo {year} {2021}),\ \bibfield  {title} {\enquote {\bibinfo
  {title} {Reinforcement learning for digital quantum simulation},}\ }\href
  {https://doi.org/10.1103/PhysRevLett.127.110502} {\bibfield  {journal}
  {\bibinfo  {journal} {Phys. Rev. Lett.}\ }\textbf {\bibinfo {volume} {127}},\
  \bibinfo {pages} {110502}}\BibitemShut {NoStop}%
\bibitem [{\citenamefont {Bommasani}(2021)}]{bommasani2021opportunities}%
  \BibitemOpen
  \bibfield  {author} {\bibinfo {author} {\bibnamefont {Bommasani},
  \bibfnamefont {Rishi}}} (\bibinfo {year} {2021}),\ \bibfield  {title}
  {\enquote {\bibinfo {title} {On the opportunities and risks of foundation
  models},}\ }\href {https://doi.org/10.48550/arXiv.2108.07258} {\bibinfo
  {journal} {arXiv preprint arXiv:2108.07258}\ }\BibitemShut {NoStop}%
\bibitem [{\citenamefont {Bordoni}\ \emph {et~al.}(2024)\citenamefont
  {Bordoni}, \citenamefont {Papaluca}, \citenamefont {Buttarini}, \citenamefont
  {Sopena}, \citenamefont {Giagu},\ and\ \citenamefont
  {Carrazza}}]{bordoni2024quantum}%
  \BibitemOpen
\bibfield  {journal} {  }\bibfield  {author} {\bibinfo {author} {\bibnamefont
  {Bordoni}, \bibfnamefont {Simone}}, \bibinfo {author} {\bibfnamefont
  {Andrea}\ \bibnamefont {Papaluca}}, \bibinfo {author} {\bibfnamefont
  {Piergiorgio}\ \bibnamefont {Buttarini}}, \bibinfo {author} {\bibfnamefont
  {Alejandro}\ \bibnamefont {Sopena}}, \bibinfo {author} {\bibfnamefont
  {Stefano}\ \bibnamefont {Giagu}}, and\ \bibinfo {author} {\bibfnamefont
  {Stefano}\ \bibnamefont {Carrazza}}} (\bibinfo {year} {2024}),\ \bibfield
  {title} {\enquote {\bibinfo {title} {Quantum noise modeling through
  reinforcement learning},}\ }\href {https://arxiv.org/abs/2408.01506}
  {\bibinfo  {journal} {arXiv preprint arXiv:2408.01506}\ }\BibitemShut
  {NoStop}%
\bibitem [{\citenamefont {Briegel}\ and\ \citenamefont {De~las
  Cuevas}(2012)}]{briegel2012projective}%
  \BibitemOpen
\bibfield  {journal} {  }\bibfield  {author} {\bibinfo {author} {\bibnamefont
  {Briegel}, \bibfnamefont {Hans~J}}, and\ \bibinfo {author} {\bibfnamefont
  {Gemma}\ \bibnamefont {De~las Cuevas}}} (\bibinfo {year} {2012}),\ \bibfield
  {title} {\enquote {\bibinfo {title} {Projective simulation for artificial
  intelligence},}\ }\href {https://doi.org/10.1038/srep00400} {\bibfield
  {journal} {\bibinfo  {journal} {Scientific reports}\ }\textbf {\bibinfo
  {volume} {2}}~(\bibinfo {number} {1}),\ \bibinfo {pages} {400}}\BibitemShut
  {NoStop}%
\bibitem [{\citenamefont {Bronstein}\ \emph {et~al.}(2021)\citenamefont
  {Bronstein}, \citenamefont {Bruna}, \citenamefont {Cohen},\ and\
  \citenamefont {Veli{\v{c}}kovi{\'c}}}]{bronstein2021geometric}%
  \BibitemOpen
  \bibfield  {author} {\bibinfo {author} {\bibnamefont {Bronstein},
  \bibfnamefont {Michael~M}}, \bibinfo {author} {\bibfnamefont {Joan}\
  \bibnamefont {Bruna}}, \bibinfo {author} {\bibfnamefont {Taco}\ \bibnamefont
  {Cohen}}, and\ \bibinfo {author} {\bibfnamefont {Petar}\ \bibnamefont
  {Veli{\v{c}}kovi{\'c}}}} (\bibinfo {year} {2021}),\ \bibfield  {title}
  {\enquote {\bibinfo {title} {Geometric deep learning: Grids, groups, graphs,
  geodesics, and gauges},}\ }\href {https://arxiv.org/abs/2104.13478} {\bibinfo
   {journal} {arXiv preprint arXiv:2104.13478}\ }\BibitemShut {NoStop}%
\bibitem [{\citenamefont {Brown}\ \emph {et~al.}(2021)\citenamefont {Brown},
  \citenamefont {Sgroi}, \citenamefont {Giannelli}, \citenamefont {Paraoanu},
  \citenamefont {Paladino}, \citenamefont {Falci}, \citenamefont
  {Paternostro},\ and\ \citenamefont {Ferraro}}]{brown2021reinforcement}%
  \BibitemOpen
\bibfield  {journal} {  }\bibfield  {author} {\bibinfo {author} {\bibnamefont
  {Brown}, \bibfnamefont {Jonathon}}, \bibinfo {author} {\bibfnamefont
  {Pierpaolo}\ \bibnamefont {Sgroi}}, \bibinfo {author} {\bibfnamefont {Luigi}\
  \bibnamefont {Giannelli}}, \bibinfo {author} {\bibfnamefont {Gheorghe~Sorin}\
  \bibnamefont {Paraoanu}}, \bibinfo {author} {\bibfnamefont {Elisabetta}\
  \bibnamefont {Paladino}}, \bibinfo {author} {\bibfnamefont {Giuseppe}\
  \bibnamefont {Falci}}, \bibinfo {author} {\bibfnamefont {Mauro}\ \bibnamefont
  {Paternostro}}, and\ \bibinfo {author} {\bibfnamefont {Alessandro}\
  \bibnamefont {Ferraro}}} (\bibinfo {year} {2021}),\ \bibfield  {title}
  {\enquote {\bibinfo {title} {Reinforcement learning-enhanced protocols for
  coherent population-transfer in three-level quantum systems},}\ }\href
  {https://doi.org/10.1088/1367-2630/ac2393} {\bibfield  {journal} {\bibinfo
  {journal} {New Journal of Physics}\ }\textbf {\bibinfo {volume}
  {23}}~(\bibinfo {number} {9}),\ \bibinfo {pages} {093035}}\BibitemShut
  {NoStop}%
\bibitem [{\citenamefont {Bukov}(2018)}]{bukov2018kapitza}%
  \BibitemOpen
  \bibfield  {author} {\bibinfo {author} {\bibnamefont {Bukov}, \bibfnamefont
  {Marin}}} (\bibinfo {year} {2018}),\ \bibfield  {title} {\enquote {\bibinfo
  {title} {Reinforcement learning for autonomous preparation of
  floquet-engineered states: Inverting the quantum kapitza oscillator},}\
  }\href {https://doi.org/10.1103/PhysRevB.98.224305} {\bibfield  {journal}
  {\bibinfo  {journal} {Phys. Rev. B}\ }\textbf {\bibinfo {volume} {98}},\
  \bibinfo {pages} {224305}}\BibitemShut {NoStop}%
\bibitem [{\citenamefont {Bukov}\ \emph {et~al.}(2018)\citenamefont {Bukov},
  \citenamefont {Day}, \citenamefont {Sels}, \citenamefont {Weinberg},
  \citenamefont {Polkovnikov},\ and\ \citenamefont
  {Mehta}}]{bukov2018reinforcement}%
  \BibitemOpen
  \bibfield  {author} {\bibinfo {author} {\bibnamefont {Bukov}, \bibfnamefont
  {Marin}}, \bibinfo {author} {\bibfnamefont {Alexandre G.~R.}\ \bibnamefont
  {Day}}, \bibinfo {author} {\bibfnamefont {Dries}\ \bibnamefont {Sels}},
  \bibinfo {author} {\bibfnamefont {Phillip}\ \bibnamefont {Weinberg}},
  \bibinfo {author} {\bibfnamefont {Anatoli}\ \bibnamefont {Polkovnikov}}, and\
  \bibinfo {author} {\bibfnamefont {Pankaj}\ \bibnamefont {Mehta}}} (\bibinfo
  {year} {2018}),\ \bibfield  {title} {\enquote {\bibinfo {title}
  {Reinforcement learning in different phases of quantum control},}\ }\href
  {https://doi.org/10.1103/PhysRevX.8.031086} {\bibfield  {journal} {\bibinfo
  {journal} {Phys. Rev. X}\ }\textbf {\bibinfo {volume} {8}},\ \bibinfo {pages}
  {031086}}\BibitemShut {NoStop}%
\bibitem [{\citenamefont {Cao}\ \emph {et~al.}(2022)\citenamefont {Cao},
  \citenamefont {An}, \citenamefont {Hou}, \citenamefont {Zhou},\ and\
  \citenamefont {Zeng}}]{cao2022quantum}%
  \BibitemOpen
  \bibfield  {author} {\bibinfo {author} {\bibnamefont {Cao}, \bibfnamefont
  {Chenfeng}}, \bibinfo {author} {\bibfnamefont {Zheng}\ \bibnamefont {An}},
  \bibinfo {author} {\bibfnamefont {Shi-Yao}\ \bibnamefont {Hou}}, \bibinfo
  {author} {\bibfnamefont {DL}~\bibnamefont {Zhou}}, and\ \bibinfo {author}
  {\bibfnamefont {Bei}\ \bibnamefont {Zeng}}} (\bibinfo {year} {2022}),\
  \bibfield  {title} {\enquote {\bibinfo {title} {Quantum imaginary time
  evolution steered by reinforcement learning},}\ }\href
  {https://doi.org/10.1038/s42005-022-00837-y} {\bibfield  {journal} {\bibinfo
  {journal} {Communications Physics}\ }\textbf {\bibinfo {volume}
  {5}}~(\bibinfo {number} {1}),\ \bibinfo {pages} {57}}\BibitemShut {NoStop}%
\bibitem [{\citenamefont {Capuano}\ \emph {et~al.}(2025)\citenamefont
  {Capuano}, \citenamefont {Peceli},\ and\ \citenamefont
  {Tiboni}}]{capuano2025shaping}%
  \BibitemOpen
  \bibfield  {author} {\bibinfo {author} {\bibnamefont {Capuano}, \bibfnamefont
  {Francesco}}, \bibinfo {author} {\bibfnamefont {Davorin}\ \bibnamefont
  {Peceli}}, and\ \bibinfo {author} {\bibfnamefont {Gabriele}\ \bibnamefont
  {Tiboni}}} (\bibinfo {year} {2025}),\ \bibfield  {title} {\enquote {\bibinfo
  {title} {Shaping laser pulses with reinforcement learning},}\ }\href
  {https://arxiv.org/abs/2503.00499} {\bibinfo  {journal} {arXiv preprint
  arXiv:2503.00499}\ }\BibitemShut {NoStop}%
\bibitem [{\citenamefont {Carleo}\ \emph {et~al.}(2019)\citenamefont {Carleo},
  \citenamefont {Cirac}, \citenamefont {Cranmer}, \citenamefont {Daudet},
  \citenamefont {Schuld}, \citenamefont {Tishby}, \citenamefont
  {Vogt-Maranto},\ and\ \citenamefont {Zdeborov\'a}}]{carleo2019machine}%
  \BibitemOpen
\bibfield  {journal} {  }\bibfield  {author} {\bibinfo {author} {\bibnamefont
  {Carleo}, \bibfnamefont {Giuseppe}}, \bibinfo {author} {\bibfnamefont
  {Ignacio}\ \bibnamefont {Cirac}}, \bibinfo {author} {\bibfnamefont {Kyle}\
  \bibnamefont {Cranmer}}, \bibinfo {author} {\bibfnamefont {Laurent}\
  \bibnamefont {Daudet}}, \bibinfo {author} {\bibfnamefont {Maria}\
  \bibnamefont {Schuld}}, \bibinfo {author} {\bibfnamefont {Naftali}\
  \bibnamefont {Tishby}}, \bibinfo {author} {\bibfnamefont {Leslie}\
  \bibnamefont {Vogt-Maranto}}, and\ \bibinfo {author} {\bibfnamefont {Lenka}\
  \bibnamefont {Zdeborov\'a}}} (\bibinfo {year} {2019}),\ \bibfield  {title}
  {\enquote {\bibinfo {title} {Machine learning and the physical sciences},}\
  }\href {https://doi.org/10.1103/RevModPhys.91.045002} {\bibfield  {journal}
  {\bibinfo  {journal} {Rev. Mod. Phys.}\ }\textbf {\bibinfo {volume} {91}},\
  \bibinfo {pages} {045002}}\BibitemShut {NoStop}%
\bibitem [{\citenamefont {Carleo}\ and\ \citenamefont
  {Troyer}(2017)}]{carleo2017solving}%
  \BibitemOpen
  \bibfield  {author} {\bibinfo {author} {\bibnamefont {Carleo}, \bibfnamefont
  {Giuseppe}}, and\ \bibinfo {author} {\bibfnamefont {Matthias}\ \bibnamefont
  {Troyer}}} (\bibinfo {year} {2017}),\ \bibfield  {title} {\enquote {\bibinfo
  {title} {Solving the quantum many-body problem with artificial neural
  networks},}\ }\href {https://www.science.org/doi/10.1126/science.aag2302}
  {\bibfield  {journal} {\bibinfo  {journal} {Science}\ }\textbf {\bibinfo
  {volume} {355}}~(\bibinfo {number} {6325}),\ \bibinfo {pages}
  {602--606}}\BibitemShut {NoStop}%
\bibitem [{\citenamefont {Carrasquilla}(2020)}]{carrasquilla2020machine}%
  \BibitemOpen
  \bibfield  {author} {\bibinfo {author} {\bibnamefont {Carrasquilla},
  \bibfnamefont {Juan}}} (\bibinfo {year} {2020}),\ \bibfield  {title}
  {\enquote {\bibinfo {title} {Machine learning for quantum matter},}\ }\href
  {https://doi.org/10.1080/23746149.2020.1797528} {\bibfield  {journal}
  {\bibinfo  {journal} {Advances in Physics: X}\ }\textbf {\bibinfo {volume}
  {5}}~(\bibinfo {number} {1}),\ \bibinfo {pages} {1797528}}\BibitemShut
  {NoStop}%
\bibitem [{\citenamefont {Carrasquilla}\ and\ \citenamefont
  {Melko}(2017)}]{carrasquilla2017machine}%
  \BibitemOpen
  \bibfield  {author} {\bibinfo {author} {\bibnamefont {Carrasquilla},
  \bibfnamefont {Juan}}, and\ \bibinfo {author} {\bibfnamefont {Roger~G}\
  \bibnamefont {Melko}}} (\bibinfo {year} {2017}),\ \bibfield  {title}
  {\enquote {\bibinfo {title} {Machine learning phases of matter},}\ }\href
  {https://doi.org/10.1038/nphys4035} {\bibfield  {journal} {\bibinfo
  {journal} {Nature Physics}\ }\textbf {\bibinfo {volume} {13}}~(\bibinfo
  {number} {5}),\ \bibinfo {pages} {431--434}}\BibitemShut {NoStop}%
\bibitem [{\citenamefont {Cemin}\ \emph {et~al.}(2025)\citenamefont {Cemin},
  \citenamefont {Schmitt},\ and\ \citenamefont {Bukov}}]{cemin2025learning}%
  \BibitemOpen
  \bibfield  {author} {\bibinfo {author} {\bibnamefont {Cemin}, \bibfnamefont
  {Giovanni}}, \bibinfo {author} {\bibfnamefont {Markus}\ \bibnamefont
  {Schmitt}}, and\ \bibinfo {author} {\bibfnamefont {Marin}\ \bibnamefont
  {Bukov}}} (\bibinfo {year} {2025}),\ \bibfield  {title} {\enquote {\bibinfo
  {title} {Learning to stabilize nonequilibrium phases of matter with active
  feedback using partial information},}\ }\href
  {https://doi.org/10.48550/arXiv.2508.06612} {\bibinfo  {journal} {arXiv
  preprint arXiv:2508.06612}\ }\BibitemShut {NoStop}%
\bibitem [{\citenamefont {Chalkiadakis}\ and\ \citenamefont
  {Boutilier}(2003)}]{chalkiadakis2003coordination}%
  \BibitemOpen
\bibfield  {journal} {  }\bibfield  {author} {\bibinfo {author} {\bibnamefont
  {Chalkiadakis}, \bibfnamefont {Georgios}}, and\ \bibinfo {author}
  {\bibfnamefont {Craig}\ \bibnamefont {Boutilier}}} (\bibinfo {year} {2003}),\
  \bibfield  {title} {\enquote {\bibinfo {title} {Coordination in multiagent
  reinforcement learning: A bayesian approach},}\ }in\ \href
  {https://doi.org/10.1145/860575.860689} {\emph {\bibinfo {booktitle}
  {Proceedings of the second international joint conference on Autonomous
  agents and multiagent systems}}},\ pp.\ \bibinfo {pages}
  {709--716}\BibitemShut {NoStop}%
\bibitem [{\citenamefont {Chen}\ and\ \citenamefont
  {Heyl}(2024)}]{chen2024empowering}%
  \BibitemOpen
  \bibfield  {author} {\bibinfo {author} {\bibnamefont {Chen}, \bibfnamefont
  {Ao}}, and\ \bibinfo {author} {\bibfnamefont {Markus}\ \bibnamefont {Heyl}}}
  (\bibinfo {year} {2024}),\ \bibfield  {title} {\enquote {\bibinfo {title}
  {Empowering deep neural quantum states through efficient optimization},}\
  }\href {https://doi.org/10.1038/s41567-024-02566-1} {\bibinfo  {journal}
  {Nature Physics}\ ,\ \bibinfo {pages} {1--6}}\BibitemShut {NoStop}%
\bibitem [{\citenamefont {Chen}\ \emph {et~al.}(2013)\citenamefont {Chen},
  \citenamefont {Dong}, \citenamefont {Li}, \citenamefont {Chu},\ and\
  \citenamefont {Tarn}}]{chen2013fidelity}%
  \BibitemOpen
\bibfield  {journal} {  }\bibfield  {author} {\bibinfo {author} {\bibnamefont
  {Chen}, \bibfnamefont {Chunlin}}, \bibinfo {author} {\bibfnamefont {Daoyi}\
  \bibnamefont {Dong}}, \bibinfo {author} {\bibfnamefont {Han-Xiong}\
  \bibnamefont {Li}}, \bibinfo {author} {\bibfnamefont {Jian}\ \bibnamefont
  {Chu}}, and\ \bibinfo {author} {\bibfnamefont {Tzyh-Jong}\ \bibnamefont
  {Tarn}}} (\bibinfo {year} {2013}),\ \bibfield  {title} {\enquote {\bibinfo
  {title} {Fidelity-based probabilistic q-learning for control of quantum
  systems},}\ }\href {https://doi.org/10.1109/TNNLS.2013.2283574} {\bibfield
  {journal} {\bibinfo  {journal} {IEEE transactions on neural networks and
  learning systems}\ }\textbf {\bibinfo {volume} {25}}~(\bibinfo {number}
  {5}),\ \bibinfo {pages} {920--933}}\BibitemShut {NoStop}%
\bibitem [{\citenamefont {Chen}\ \emph
  {et~al.}(2022{\natexlab{a}})\citenamefont {Chen}, \citenamefont {Aapro},
  \citenamefont {Kipnis}, \citenamefont {Ilin}, \citenamefont {Liljeroth},\
  and\ \citenamefont {Foster}}]{chen2022precise}%
  \BibitemOpen
  \bibfield  {author} {\bibinfo {author} {\bibnamefont {Chen}, \bibfnamefont
  {I-Ju}}, \bibinfo {author} {\bibfnamefont {Markus}\ \bibnamefont {Aapro}},
  \bibinfo {author} {\bibfnamefont {Abraham}\ \bibnamefont {Kipnis}}, \bibinfo
  {author} {\bibfnamefont {Alexander}\ \bibnamefont {Ilin}}, \bibinfo {author}
  {\bibfnamefont {Peter}\ \bibnamefont {Liljeroth}}, and\ \bibinfo {author}
  {\bibfnamefont {Adam~S}\ \bibnamefont {Foster}}} (\bibinfo {year}
  {2022}{\natexlab{a}}),\ \bibfield  {title} {\enquote {\bibinfo {title}
  {Precise atom manipulation through deep reinforcement learning},}\ }\href
  {https://doi.org/10.1038/s41467-022-35149-w} {\bibfield  {journal} {\bibinfo
  {journal} {Nature Communications}\ }\textbf {\bibinfo {volume}
  {13}}~(\bibinfo {number} {1}),\ \bibinfo {pages} {7499}}\BibitemShut
  {NoStop}%
\bibitem [{\citenamefont {Chen}\ and\ \citenamefont
  {Xue}(2019)}]{chen2019manipulation}%
  \BibitemOpen
  \bibfield  {author} {\bibinfo {author} {\bibnamefont {Chen}, \bibfnamefont
  {Jun-Jie}}, and\ \bibinfo {author} {\bibfnamefont {Ming}\ \bibnamefont
  {Xue}}} (\bibinfo {year} {2019}),\ \bibfield  {title} {\enquote {\bibinfo
  {title} {Manipulation of spin dynamics by deep reinforcement learning
  agent},}\ }\href {https://doi.org/10.48550/arXiv.1901.08748} {\bibinfo
  {journal} {arXiv preprint arXiv:1901.08748}\ }\BibitemShut {NoStop}%
\bibitem [{\citenamefont {Chen}\ \emph
  {et~al.}(2022{\natexlab{b}})\citenamefont {Chen}, \citenamefont {Du},
  \citenamefont {Zhao}, \citenamefont {Jiao}, \citenamefont {Lu},\ and\
  \citenamefont {Wu}}]{chen2022efficient}%
  \BibitemOpen
\bibfield  {journal} {  }\bibfield  {author} {\bibinfo {author} {\bibnamefont
  {Chen}, \bibfnamefont {Qiuhao}}, \bibinfo {author} {\bibfnamefont {Yuxuan}\
  \bibnamefont {Du}}, \bibinfo {author} {\bibfnamefont {Qi}~\bibnamefont
  {Zhao}}, \bibinfo {author} {\bibfnamefont {Yuling}\ \bibnamefont {Jiao}},
  \bibinfo {author} {\bibfnamefont {Xiliang}\ \bibnamefont {Lu}}, and\ \bibinfo
  {author} {\bibfnamefont {Xingyao}\ \bibnamefont {Wu}}} (\bibinfo {year}
  {2022}{\natexlab{b}}),\ \bibfield  {title} {\enquote {\bibinfo {title}
  {Efficient and practical quantum compiler towards multi-qubit systems with
  deep reinforcement learning},}\ }\href
  {https://doi.org/10.48550/arXiv.2204.06904} {\bibinfo  {journal} {arXiv
  preprint arXiv:2204.06904}\ }\BibitemShut {NoStop}%
\bibitem [{\citenamefont {Chen}(2024)}]{chen2024introduction}%
  \BibitemOpen
\bibfield  {journal} {  }\bibfield  {author} {\bibinfo {author} {\bibnamefont
  {Chen}, \bibfnamefont {Samuel Yen-Chi}}} (\bibinfo {year} {2024}),\ \bibfield
   {title} {\enquote {\bibinfo {title} {An introduction to quantum
  reinforcement learning (qrl)},}\ }\href {https://arxiv.org/abs/2409.05846}
  {\bibinfo  {journal} {arXiv preprint arXiv:2409.05846}\ }\BibitemShut
  {NoStop}%
\bibitem [{\citenamefont {Chen}\ \emph
  {et~al.}(2022{\natexlab{c}})\citenamefont {Chen}, \citenamefont {Huang},
  \citenamefont {Hsing}, \citenamefont {Goan},\ and\ \citenamefont
  {Kao}}]{chen2022variational}%
  \BibitemOpen
\bibfield  {journal} {  }\bibfield  {author} {\bibinfo {author} {\bibnamefont
  {Chen}, \bibfnamefont {Samuel Yen-Chi}}, \bibinfo {author} {\bibfnamefont
  {Chih-Min}\ \bibnamefont {Huang}}, \bibinfo {author} {\bibfnamefont
  {Chia-Wei}\ \bibnamefont {Hsing}}, \bibinfo {author} {\bibfnamefont
  {Hsi-Sheng}\ \bibnamefont {Goan}}, and\ \bibinfo {author} {\bibfnamefont
  {Ying-Jer}\ \bibnamefont {Kao}}} (\bibinfo {year} {2022}{\natexlab{c}}),\
  \bibfield  {title} {\enquote {\bibinfo {title} {Variational quantum
  reinforcement learning via evolutionary optimization},}\ }\href
  {https://doi.org/10.1088/2632-2153/ac4559} {\bibfield  {journal} {\bibinfo
  {journal} {Machine Learning: Science and Technology}\ }\textbf {\bibinfo
  {volume} {3}}~(\bibinfo {number} {1}),\ \bibinfo {pages}
  {015025}}\BibitemShut {NoStop}%
\bibitem [{\citenamefont {Chih}\ \emph {et~al.}(2022)\citenamefont {Chih},
  \citenamefont {Anderson},\ and\ \citenamefont {Holland}}]{chih2022train}%
  \BibitemOpen
  \bibfield  {author} {\bibinfo {author} {\bibnamefont {Chih}, \bibfnamefont
  {Liang-Ying}}, \bibinfo {author} {\bibfnamefont {Dana~Z}\ \bibnamefont
  {Anderson}}, and\ \bibinfo {author} {\bibfnamefont {Murray}\ \bibnamefont
  {Holland}}} (\bibinfo {year} {2022}),\ \bibfield  {title} {\enquote {\bibinfo
  {title} {How to train your gyro: Reinforcement learning for rotation sensing
  with a shaken optical lattice},}\ }\href {https://arxiv.org/abs/2212.14473}
  {\bibinfo  {journal} {arXiv preprint arXiv:2212.14473}\ }\BibitemShut
  {NoStop}%
\bibitem [{\citenamefont {Colabrese}\ \emph {et~al.}(2017)\citenamefont
  {Colabrese}, \citenamefont {Gustavsson}, \citenamefont {Celani},\ and\
  \citenamefont {Biferale}}]{colabrese2017flow}%
  \BibitemOpen
\bibfield  {journal} {  }\bibfield  {author} {\bibinfo {author} {\bibnamefont
  {Colabrese}, \bibfnamefont {Simona}}, \bibinfo {author} {\bibfnamefont
  {Kristian}\ \bibnamefont {Gustavsson}}, \bibinfo {author} {\bibfnamefont
  {Antonio}\ \bibnamefont {Celani}}, and\ \bibinfo {author} {\bibfnamefont
  {Luca}\ \bibnamefont {Biferale}}} (\bibinfo {year} {2017}),\ \bibfield
  {title} {\enquote {\bibinfo {title} {Flow navigation by smart microswimmers
  via reinforcement learning},}\ }\href
  {https://doi.org/10.1103/PhysRevLett.118.158004} {\bibfield  {journal}
  {\bibinfo  {journal} {Phys. Rev. Lett.}\ }\textbf {\bibinfo {volume} {118}},\
  \bibinfo {pages} {158004}}\BibitemShut {NoStop}%
\bibitem [{\citenamefont {Corli}\ \emph {et~al.}(2024)\citenamefont {Corli},
  \citenamefont {Moro}, \citenamefont {Dragoni}, \citenamefont {Dispenza},\
  and\ \citenamefont {Prati}}]{corli2024quantum}%
  \BibitemOpen
  \bibfield  {author} {\bibinfo {author} {\bibnamefont {Corli}, \bibfnamefont
  {Sebastiano}}, \bibinfo {author} {\bibfnamefont {Lorenzo}\ \bibnamefont
  {Moro}}, \bibinfo {author} {\bibfnamefont {Daniele}\ \bibnamefont {Dragoni}},
  \bibinfo {author} {\bibfnamefont {Massimiliano}\ \bibnamefont {Dispenza}},
  and\ \bibinfo {author} {\bibfnamefont {Enrico}\ \bibnamefont {Prati}}}
  (\bibinfo {year} {2024}),\ \bibfield  {title} {\enquote {\bibinfo {title}
  {Quantum machine learning algorithms for anomaly detection: a survey},}\
  }\href {https://doi.org/10.48550/arXiv.2408.11047} {\bibinfo  {journal}
  {arXiv preprint arXiv:2408.11047}\ }\BibitemShut {NoStop}%
\bibitem [{\citenamefont {Crosta}\ \emph {et~al.}(2024)\citenamefont {Crosta},
  \citenamefont {Rebón}, \citenamefont {Vilariño}, \citenamefont {Matera},\
  and\ \citenamefont {Bilkis}}]{crosta2024automatic}%
  \BibitemOpen
\bibfield  {journal} {  }\bibfield  {author} {\bibinfo {author} {\bibnamefont
  {Crosta}, \bibfnamefont {T}}, \bibinfo {author} {\bibfnamefont
  {L.}~\bibnamefont {Rebón}}, \bibinfo {author} {\bibfnamefont
  {F.}~\bibnamefont {Vilariño}}, \bibinfo {author} {\bibfnamefont {J.~M.}\
  \bibnamefont {Matera}}, and\ \bibinfo {author} {\bibfnamefont
  {M.}~\bibnamefont {Bilkis}}} (\bibinfo {year} {2024}),\ \bibfield  {title}
  {\enquote {\bibinfo {title} {Automatic re-calibration of quantum devices by
  reinforcement learning},}\ }\href {https://arxiv.org/abs/2404.10726}
  {\bibfield  {journal} {\bibinfo  {journal} {arXiv}\ }\textbf {\bibinfo
  {volume} {2404.10726}}}\BibitemShut {NoStop}%
\bibitem [{\citenamefont {Dalgaard}\ \emph {et~al.}(2020)\citenamefont
  {Dalgaard}, \citenamefont {Motzoi}, \citenamefont {S{\o}rensen},\ and\
  \citenamefont {Sherson}}]{dalgaard2020global}%
  \BibitemOpen
  \bibfield  {author} {\bibinfo {author} {\bibnamefont {Dalgaard},
  \bibfnamefont {Mogens}}, \bibinfo {author} {\bibfnamefont {Felix}\
  \bibnamefont {Motzoi}}, \bibinfo {author} {\bibfnamefont {Jens~Jakob}\
  \bibnamefont {S{\o}rensen}}, and\ \bibinfo {author} {\bibfnamefont {Jacob}\
  \bibnamefont {Sherson}}} (\bibinfo {year} {2020}),\ \bibfield  {title}
  {\enquote {\bibinfo {title} {Global optimization of quantum dynamics with
  alphazero deep exploration},}\ }\href
  {https://doi.org/10.1038/s41534-019-0241-0} {\bibfield  {journal} {\bibinfo
  {journal} {NPJ quantum information}\ }\textbf {\bibinfo {volume}
  {6}}~(\bibinfo {number} {1}),\ \bibinfo {pages} {6}}\BibitemShut {NoStop}%
\bibitem [{\citenamefont {Dawid}\ \emph {et~al.}(2022)\citenamefont {Dawid},
  \citenamefont {Arnold}, \citenamefont {Requena}, \citenamefont {Gresch},
  \citenamefont {P{\l}odzie{\'n}}, \citenamefont {Donatella}, \citenamefont
  {Nicoli}, \citenamefont {Stornati}, \citenamefont {Koch}, \citenamefont
  {B{\"u}ttner} \emph {et~al.}}]{dawid2022modern}%
  \BibitemOpen
  \bibfield  {author} {\bibinfo {author} {\bibnamefont {Dawid}, \bibfnamefont
  {Anna}}, \bibinfo {author} {\bibfnamefont {Julian}\ \bibnamefont {Arnold}},
  \bibinfo {author} {\bibfnamefont {Borja}\ \bibnamefont {Requena}}, \bibinfo
  {author} {\bibfnamefont {Alexander}\ \bibnamefont {Gresch}}, \bibinfo
  {author} {\bibfnamefont {Marcin}\ \bibnamefont {P{\l}odzie{\'n}}}, \bibinfo
  {author} {\bibfnamefont {Kaelan}\ \bibnamefont {Donatella}}, \bibinfo
  {author} {\bibfnamefont {Kim~A}\ \bibnamefont {Nicoli}}, \bibinfo {author}
  {\bibfnamefont {Paolo}\ \bibnamefont {Stornati}}, \bibinfo {author}
  {\bibfnamefont {Rouven}\ \bibnamefont {Koch}}, \bibinfo {author}
  {\bibfnamefont {Miriam}\ \bibnamefont {B{\"u}ttner}},  \emph {et~al.}}
  (\bibinfo {year} {2022}),\ \bibfield  {title} {\enquote {\bibinfo {title}
  {Modern applications of machine learning in quantum sciences},}\ }\href
  {https://doi.org/10.48550/arXiv.2204.04198} {\bibinfo  {journal} {arXiv
  preprint arXiv:2204.04198}\ }\BibitemShut {NoStop}%
\bibitem [{\citenamefont {Degrave}\ \emph {et~al.}(2022)\citenamefont
  {Degrave}, \citenamefont {Felici}, \citenamefont {Buchli}, \citenamefont
  {Neunert}, \citenamefont {Tracey}, \citenamefont {Carpanese}, \citenamefont
  {Ewalds}, \citenamefont {Hafner}, \citenamefont {Abdolmaleki}, \citenamefont
  {de~Las~Casas} \emph {et~al.}}]{degrave2022magnetic}%
  \BibitemOpen
\bibfield  {journal} {  }\bibfield  {author} {\bibinfo {author} {\bibnamefont
  {Degrave}, \bibfnamefont {Jonas}}, \bibinfo {author} {\bibfnamefont
  {Federico}\ \bibnamefont {Felici}}, \bibinfo {author} {\bibfnamefont {Jonas}\
  \bibnamefont {Buchli}}, \bibinfo {author} {\bibfnamefont {Michael}\
  \bibnamefont {Neunert}}, \bibinfo {author} {\bibfnamefont {Brendan}\
  \bibnamefont {Tracey}}, \bibinfo {author} {\bibfnamefont {Francesco}\
  \bibnamefont {Carpanese}}, \bibinfo {author} {\bibfnamefont {Timo}\
  \bibnamefont {Ewalds}}, \bibinfo {author} {\bibfnamefont {Roland}\
  \bibnamefont {Hafner}}, \bibinfo {author} {\bibfnamefont {Abbas}\
  \bibnamefont {Abdolmaleki}}, \bibinfo {author} {\bibfnamefont {Diego}\
  \bibnamefont {de~Las~Casas}},  \emph {et~al.}} (\bibinfo {year} {2022}),\
  \bibfield  {title} {\enquote {\bibinfo {title} {Magnetic control of tokamak
  plasmas through deep reinforcement learning},}\ }\href
  {https://doi.org/10.1038/s41586-021-04301-9} {\bibfield  {journal} {\bibinfo
  {journal} {Nature}\ }\textbf {\bibinfo {volume} {602}}~(\bibinfo {number}
  {7897}),\ \bibinfo {pages} {414--419}}\BibitemShut {NoStop}%
\bibitem [{\citenamefont {Ding}\ and\ \citenamefont
  {Englund}(2025)}]{ding2025hardware}%
  \BibitemOpen
  \bibfield  {author} {\bibinfo {author} {\bibnamefont {Ding}, \bibfnamefont
  {Qian}}, and\ \bibinfo {author} {\bibfnamefont {Dirk}\ \bibnamefont
  {Englund}}} (\bibinfo {year} {2025}),\ \bibfield  {title} {\enquote {\bibinfo
  {title} {Hardware co-designed optimal control for programmable atomic quantum
  processors via reinforcement learning},}\ }\href
  {https://arxiv.org/abs/2504.11737} {\bibinfo  {journal} {arXiv preprint
  arXiv:2504.11737}\ }\BibitemShut {NoStop}%
\bibitem [{\citenamefont {Ding}\ \emph {et~al.}(2021)\citenamefont {Ding},
  \citenamefont {Ban}, \citenamefont {Mart\'{\i}n-Guerrero}, \citenamefont
  {Solano}, \citenamefont {Casanova},\ and\ \citenamefont
  {Chen}}]{ding2021breaking}%
  \BibitemOpen
\bibfield  {journal} {  }\bibfield  {author} {\bibinfo {author} {\bibnamefont
  {Ding}, \bibfnamefont {Yongcheng}}, \bibinfo {author} {\bibfnamefont {Yue}\
  \bibnamefont {Ban}}, \bibinfo {author} {\bibfnamefont {Jos\'e~D.}\
  \bibnamefont {Mart\'{\i}n-Guerrero}}, \bibinfo {author} {\bibfnamefont
  {Enrique}\ \bibnamefont {Solano}}, \bibinfo {author} {\bibfnamefont {Jorge}\
  \bibnamefont {Casanova}}, and\ \bibinfo {author} {\bibfnamefont
  {Xi}~\bibnamefont {Chen}}} (\bibinfo {year} {2021}),\ \bibfield  {title}
  {\enquote {\bibinfo {title} {Breaking adiabatic quantum control with deep
  learning},}\ }\href {https://doi.org/10.1103/PhysRevA.103.L040401} {\bibfield
   {journal} {\bibinfo  {journal} {Phys. Rev. A}\ }\textbf {\bibinfo {volume}
  {103}},\ \bibinfo {pages} {L040401}}\BibitemShut {NoStop}%
\bibitem [{\citenamefont {Ding}\ \emph {et~al.}(2023)\citenamefont {Ding},
  \citenamefont {Chen}, \citenamefont {Magdalena-Benedito},\ and\ \citenamefont
  {Martín-Guerrero}}]{ding2023closed}%
  \BibitemOpen
  \bibfield  {author} {\bibinfo {author} {\bibnamefont {Ding}, \bibfnamefont
  {Yongcheng}}, \bibinfo {author} {\bibfnamefont {Xi}~\bibnamefont {Chen}},
  \bibinfo {author} {\bibfnamefont {Rafael}\ \bibnamefont
  {Magdalena-Benedito}}, and\ \bibinfo {author} {\bibfnamefont {José~D}\
  \bibnamefont {Martín-Guerrero}}} (\bibinfo {year} {2023}),\ \bibfield
  {title} {\enquote {\bibinfo {title} {Closed-loop control of a noisy qubit
  with reinforcement learning},}\ }\href
  {https://doi.org/10.1088/2632-2153/acd048} {\bibfield  {journal} {\bibinfo
  {journal} {Machine Learning: Science and Technology}\ }\textbf {\bibinfo
  {volume} {4}}~(\bibinfo {number} {2}),\ \bibinfo {pages}
  {025020}}\BibitemShut {NoStop}%
\bibitem [{\citenamefont {Dong}\ \emph {et~al.}(2008)\citenamefont {Dong},
  \citenamefont {Chen}, \citenamefont {Tarn}, \citenamefont {Pechen},\ and\
  \citenamefont {Rabitz}}]{dong2008incoherent}%
  \BibitemOpen
  \bibfield  {author} {\bibinfo {author} {\bibnamefont {Dong}, \bibfnamefont
  {Daoyi}}, \bibinfo {author} {\bibfnamefont {Chunlin}\ \bibnamefont {Chen}},
  \bibinfo {author} {\bibfnamefont {Tzyh-Jong}\ \bibnamefont {Tarn}}, \bibinfo
  {author} {\bibfnamefont {Alexander}\ \bibnamefont {Pechen}}, and\ \bibinfo
  {author} {\bibfnamefont {Herschel}\ \bibnamefont {Rabitz}}} (\bibinfo {year}
  {2008}),\ \bibfield  {title} {\enquote {\bibinfo {title} {Incoherent control
  of quantum systems with wavefunction-controllable subspaces via quantum
  reinforcement learning},}\ }\href {https://doi.org/10.1109/TSMCB.2008.926603}
  {\bibfield  {journal} {\bibinfo  {journal} {IEEE Transactions on Systems,
  Man, and Cybernetics, Part B (Cybernetics)}\ }\textbf {\bibinfo {volume}
  {38}}~(\bibinfo {number} {4}),\ \bibinfo {pages} {957--962}}\BibitemShut
  {NoStop}%
\bibitem [{\citenamefont {Dubal}\ \emph {et~al.}(2025)\citenamefont {Dubal},
  \citenamefont {Kremer}, \citenamefont {Martiel}, \citenamefont {Villar},
  \citenamefont {Wang},\ and\ \citenamefont {Cruz-Benito}}]{dubal2025pauli}%
  \BibitemOpen
  \bibfield  {author} {\bibinfo {author} {\bibnamefont {Dubal}, \bibfnamefont
  {Ayushi}}, \bibinfo {author} {\bibfnamefont {David}\ \bibnamefont {Kremer}},
  \bibinfo {author} {\bibfnamefont {Simon}\ \bibnamefont {Martiel}}, \bibinfo
  {author} {\bibfnamefont {Victor}\ \bibnamefont {Villar}}, \bibinfo {author}
  {\bibfnamefont {Derek}\ \bibnamefont {Wang}}, and\ \bibinfo {author}
  {\bibfnamefont {Juan}\ \bibnamefont {Cruz-Benito}}} (\bibinfo {year}
  {2025}),\ \bibfield  {title} {\enquote {\bibinfo {title} {Pauli network
  circuit synthesis with reinforcement learning},}\ }\href
  {https://arxiv.org/abs/2503.14448} {\bibinfo  {journal} {arXiv preprint
  arXiv:2503.14448}\ }\BibitemShut {NoStop}%
\bibitem [{\citenamefont {Duncan}\ \emph {et~al.}(2025)\citenamefont {Duncan},
  \citenamefont {Poggi}, \citenamefont {Bukov}, \citenamefont {Zinner},\ and\
  \citenamefont {Campbell}}]{duncan2025taming}%
  \BibitemOpen
\bibfield  {journal} {  }\bibfield  {author} {\bibinfo {author} {\bibnamefont
  {Duncan}, \bibfnamefont {Callum~W}}, \bibinfo {author} {\bibfnamefont
  {Pablo~M}\ \bibnamefont {Poggi}}, \bibinfo {author} {\bibfnamefont {Marin}\
  \bibnamefont {Bukov}}, \bibinfo {author} {\bibfnamefont {Nikolaj~Thomas}\
  \bibnamefont {Zinner}}, and\ \bibinfo {author} {\bibfnamefont {Steve}\
  \bibnamefont {Campbell}}} (\bibinfo {year} {2025}),\ \bibfield  {title}
  {\enquote {\bibinfo {title} {Taming quantum systems: A tutorial for using
  shortcuts-to-adiabaticity, quantum optimal control, and reinforcement
  learning},}\ }\href {https://arxiv.org/abs/2501.16436} {\bibinfo  {journal}
  {arXiv preprint arXiv:2501.16436}\ }\BibitemShut {NoStop}%
\bibitem [{\citenamefont {Dunjko}\ and\ \citenamefont
  {Briegel}(2018)}]{dunjko2018machine}%
  \BibitemOpen
\bibfield  {journal} {  }\bibfield  {author} {\bibinfo {author} {\bibnamefont
  {Dunjko}, \bibfnamefont {Vedran}}, and\ \bibinfo {author} {\bibfnamefont
  {Hans~J}\ \bibnamefont {Briegel}}} (\bibinfo {year} {2018}),\ \bibfield
  {title} {\enquote {\bibinfo {title} {Machine learning \& artificial
  intelligence in the quantum domain: a review of recent progress},}\ }\href
  {https://doi.org/10.1088/1361-6633/aab406} {\bibfield  {journal} {\bibinfo
  {journal} {Reports on Progress in Physics}\ }\textbf {\bibinfo {volume}
  {81}}~(\bibinfo {number} {7}),\ \bibinfo {pages} {074001}}\BibitemShut
  {NoStop}%
\bibitem [{\citenamefont {Dunjko}\ \emph {et~al.}(2016)\citenamefont {Dunjko},
  \citenamefont {Taylor},\ and\ \citenamefont {Briegel}}]{dunko2016quantum}%
  \BibitemOpen
  \bibfield  {author} {\bibinfo {author} {\bibnamefont {Dunjko}, \bibfnamefont
  {Vedran}}, \bibinfo {author} {\bibfnamefont {Jacob~M.}\ \bibnamefont
  {Taylor}}, and\ \bibinfo {author} {\bibfnamefont {Hans~J.}\ \bibnamefont
  {Briegel}}} (\bibinfo {year} {2016}),\ \bibfield  {title} {\enquote {\bibinfo
  {title} {Quantum-enhanced machine learning},}\ }\href
  {https://doi.org/10.1103/PhysRevLett.117.130501} {\bibfield  {journal}
  {\bibinfo  {journal} {Phys. Rev. Lett.}\ }\textbf {\bibinfo {volume} {117}},\
  \bibinfo {pages} {130501}}\BibitemShut {NoStop}%
\bibitem [{\citenamefont {Dunjko}\ \emph {et~al.}(2017)\citenamefont {Dunjko},
  \citenamefont {Taylor},\ and\ \citenamefont {Briegel}}]{dunjko2017advances}%
  \BibitemOpen
  \bibfield  {author} {\bibinfo {author} {\bibnamefont {Dunjko}, \bibfnamefont
  {Vedran}}, \bibinfo {author} {\bibfnamefont {Jacob~M}\ \bibnamefont
  {Taylor}}, and\ \bibinfo {author} {\bibfnamefont {Hans~J}\ \bibnamefont
  {Briegel}}} (\bibinfo {year} {2017}),\ \bibfield  {title} {\enquote {\bibinfo
  {title} {Advances in quantum reinforcement learning},}\ }in\ \href
  {https://doi.org/10.1109/SMC.2017.8122616} {\emph {\bibinfo {booktitle} {2017
  IEEE international conference on systems, man, and cybernetics (SMC)}}}\
  (\bibinfo {organization} {IEEE})\ pp.\ \bibinfo {pages}
  {282--287}\BibitemShut {NoStop}%
\bibitem [{\citenamefont {Eiben}\ and\ \citenamefont
  {Smith}(2015)}]{eiben2015introduction}%
  \BibitemOpen
  \bibfield  {author} {\bibinfo {author} {\bibnamefont {Eiben}, \bibfnamefont
  {Agoston~E}}, and\ \bibinfo {author} {\bibfnamefont {James~E}\ \bibnamefont
  {Smith}}} (\bibinfo {year} {2015}),\ \href
  {https://doi.org/10.1007/978-3-662-05094-1} {\emph {\bibinfo {title}
  {Introduction to evolutionary computing}}}\ (\bibinfo  {publisher}
  {Springer})\BibitemShut {NoStop}%
\bibitem [{\citenamefont {Eisenmann}\ \emph {et~al.}(2024)\citenamefont
  {Eisenmann}, \citenamefont {Hein}, \citenamefont {Udluft},\ and\
  \citenamefont {Runkler}}]{eisenmann2024modelbased}%
  \BibitemOpen
  \bibfield  {author} {\bibinfo {author} {\bibnamefont {Eisenmann},
  \bibfnamefont {Simon}}, \bibinfo {author} {\bibfnamefont {Daniel}\
  \bibnamefont {Hein}}, \bibinfo {author} {\bibfnamefont {Steffen}\
  \bibnamefont {Udluft}}, and\ \bibinfo {author} {\bibfnamefont {Thomas~A.}\
  \bibnamefont {Runkler}}} (\bibinfo {year} {2024}),\ \bibfield  {title}
  {\enquote {\bibinfo {title} {Model-based offline quantum reinforcement
  learning},}\ }\href {https://arxiv.org/abs/2404.10017} {\bibfield  {journal}
  {\bibinfo  {journal} {arXiv}\ }\textbf {\bibinfo {volume}
  {2404.10017}}}\BibitemShut {NoStop}%
\bibitem [{\citenamefont {Erdman}\ and\ \citenamefont
  {No{\'e}}(2023)}]{erdman2023model}%
  \BibitemOpen
  \bibfield  {author} {\bibinfo {author} {\bibnamefont {Erdman}, \bibfnamefont
  {Paolo~A}}, and\ \bibinfo {author} {\bibfnamefont {Frank}\ \bibnamefont
  {No{\'e}}}} (\bibinfo {year} {2023}),\ \bibfield  {title} {\enquote {\bibinfo
  {title} {Model-free optimization of power/efficiency tradeoffs in quantum
  thermal machines using reinforcement learning},}\ }\href
  {https://doi.org/10.1093/pnasnexus/pgad248} {\bibfield  {journal} {\bibinfo
  {journal} {PNAS nexus}\ }\textbf {\bibinfo {volume} {2}}~(\bibinfo {number}
  {8}),\ \bibinfo {pages} {pgad248}}\BibitemShut {NoStop}%
\bibitem [{\citenamefont {Erdman}\ \emph {et~al.}(2023)\citenamefont {Erdman},
  \citenamefont {Rolandi}, \citenamefont {Abiuso}, \citenamefont
  {Perarnau-Llobet},\ and\ \citenamefont {No\'e}}]{erdman2023pareto}%
  \BibitemOpen
  \bibfield  {author} {\bibinfo {author} {\bibnamefont {Erdman}, \bibfnamefont
  {Paolo~A}}, \bibinfo {author} {\bibfnamefont {Alberto}\ \bibnamefont
  {Rolandi}}, \bibinfo {author} {\bibfnamefont {Paolo}\ \bibnamefont {Abiuso}},
  \bibinfo {author} {\bibfnamefont {Mart\'{\i}}\ \bibnamefont
  {Perarnau-Llobet}}, and\ \bibinfo {author} {\bibfnamefont {Frank}\
  \bibnamefont {No\'e}}} (\bibinfo {year} {2023}),\ \bibfield  {title}
  {\enquote {\bibinfo {title} {Pareto-optimal cycles for power, efficiency and
  fluctuations of quantum heat engines using reinforcement learning},}\ }\href
  {https://doi.org/10.1103/PhysRevResearch.5.L022017} {\bibfield  {journal}
  {\bibinfo  {journal} {Phys. Rev. Res.}\ }\textbf {\bibinfo {volume} {5}},\
  \bibinfo {pages} {L022017}}\BibitemShut {NoStop}%
\bibitem [{\citenamefont {Erdman}\ \emph
  {et~al.}(2024{\natexlab{a}})\citenamefont {Erdman}, \citenamefont {Andolina},
  \citenamefont {Giovannetti},\ and\ \citenamefont
  {No\'e}}]{erdman2024reinforcement}%
  \BibitemOpen
  \bibfield  {author} {\bibinfo {author} {\bibnamefont {Erdman}, \bibfnamefont
  {Paolo~Andrea}}, \bibinfo {author} {\bibfnamefont {Gian~Marcello}\
  \bibnamefont {Andolina}}, \bibinfo {author} {\bibfnamefont {Vittorio}\
  \bibnamefont {Giovannetti}}, and\ \bibinfo {author} {\bibfnamefont {Frank}\
  \bibnamefont {No\'e}}} (\bibinfo {year} {2024}{\natexlab{a}}),\ \bibfield
  {title} {\enquote {\bibinfo {title} {Reinforcement learning optimization of
  the charging of a dicke quantum battery},}\ }\href
  {https://doi.org/10.1103/PhysRevLett.133.243602} {\bibfield  {journal}
  {\bibinfo  {journal} {Phys. Rev. Lett.}\ }\textbf {\bibinfo {volume} {133}},\
  \bibinfo {pages} {243602}}\BibitemShut {NoStop}%
\bibitem [{\citenamefont {Erdman}\ \emph
  {et~al.}(2024{\natexlab{b}})\citenamefont {Erdman}, \citenamefont
  {Czupryniak}, \citenamefont {Bhandari}, \citenamefont {Jordan}, \citenamefont
  {Noé}, \citenamefont {Eisert},\ and\ \citenamefont
  {Guarnieri}}]{erdman2024maxwelldemon}%
  \BibitemOpen
  \bibfield  {author} {\bibinfo {author} {\bibnamefont {Erdman}, \bibfnamefont
  {Paolo~Andrea}}, \bibinfo {author} {\bibfnamefont {Robert}\ \bibnamefont
  {Czupryniak}}, \bibinfo {author} {\bibfnamefont {Bibek}\ \bibnamefont
  {Bhandari}}, \bibinfo {author} {\bibfnamefont {Andrew~N.}\ \bibnamefont
  {Jordan}}, \bibinfo {author} {\bibfnamefont {Frank}\ \bibnamefont {Noé}},
  \bibinfo {author} {\bibfnamefont {Jens}\ \bibnamefont {Eisert}}, and\
  \bibinfo {author} {\bibfnamefont {Giacomo}\ \bibnamefont {Guarnieri}}}
  (\bibinfo {year} {2024}{\natexlab{b}}),\ \bibfield  {title} {\enquote
  {\bibinfo {title} {Artificially intelligent maxwell's demon for optimal
  control of open quantum systems},}\ }\href {https://arxiv.org/abs/2408.15328}
  {\bibinfo  {journal} {arXiv preprint arXiv:2408.15328}\ }\BibitemShut
  {NoStop}%
\bibitem [{\citenamefont {Ernst}\ \emph {et~al.}(2025)\citenamefont {Ernst},
  \citenamefont {Chatterjee}, \citenamefont {Franzmeyer},\ and\ \citenamefont
  {Kuhn}}]{ernst2025reinforcement}%
  \BibitemOpen
\bibfield  {journal} {  }\bibfield  {author} {\bibinfo {author} {\bibnamefont
  {Ernst}, \bibfnamefont {Jan~Ole}}, \bibinfo {author} {\bibfnamefont {Aniket}\
  \bibnamefont {Chatterjee}}, \bibinfo {author} {\bibfnamefont {Tim}\
  \bibnamefont {Franzmeyer}}, and\ \bibinfo {author} {\bibfnamefont {Axel}\
  \bibnamefont {Kuhn}}} (\bibinfo {year} {2025}),\ \bibfield  {title} {\enquote
  {\bibinfo {title} {Reinforcement learning for quantum control under physical
  constraints},}\ }\href {https://arxiv.org/abs/2501.14372} {\bibinfo
  {journal} {arXiv preprint arXiv:2501.14372}\ }\BibitemShut {NoStop}%
\bibitem [{\citenamefont {Fallani}\ \emph {et~al.}(2022)\citenamefont
  {Fallani}, \citenamefont {Rossi}, \citenamefont {Tamascelli},\ and\
  \citenamefont {Genoni}}]{fallani2022learning}%
  \BibitemOpen
\bibfield  {journal} {  }\bibfield  {author} {\bibinfo {author} {\bibnamefont
  {Fallani}, \bibfnamefont {Alessio}}, \bibinfo {author} {\bibfnamefont {Matteo
  A.~C.}\ \bibnamefont {Rossi}}, \bibinfo {author} {\bibfnamefont {Dario}\
  \bibnamefont {Tamascelli}}, and\ \bibinfo {author} {\bibfnamefont {Marco~G.}\
  \bibnamefont {Genoni}}} (\bibinfo {year} {2022}),\ \bibfield  {title}
  {\enquote {\bibinfo {title} {Learning feedback control strategies for quantum
  metrology},}\ }\href {https://doi.org/10.1103/PRXQuantum.3.020310} {\bibfield
   {journal} {\bibinfo  {journal} {PRX Quantum}\ }\textbf {\bibinfo {volume}
  {3}},\ \bibinfo {pages} {020310}}\BibitemShut {NoStop}%
\bibitem [{\citenamefont {Fan}\ \emph {et~al.}(2023)\citenamefont {Fan},
  \citenamefont {Shen}, \citenamefont {Nussinov}, \citenamefont {Liu},
  \citenamefont {Sun},\ and\ \citenamefont {Liu}}]{fan2023searching}%
  \BibitemOpen
  \bibfield  {author} {\bibinfo {author} {\bibnamefont {Fan}, \bibfnamefont
  {Changjun}}, \bibinfo {author} {\bibfnamefont {Mutian}\ \bibnamefont {Shen}},
  \bibinfo {author} {\bibfnamefont {Zohar}\ \bibnamefont {Nussinov}}, \bibinfo
  {author} {\bibfnamefont {Zhong}\ \bibnamefont {Liu}}, \bibinfo {author}
  {\bibfnamefont {Yizhou}\ \bibnamefont {Sun}}, and\ \bibinfo {author}
  {\bibfnamefont {Yang-Yu}\ \bibnamefont {Liu}}} (\bibinfo {year} {2023}),\
  \bibfield  {title} {\enquote {\bibinfo {title} {Searching for spin glass
  ground states through deep reinforcement learning},}\ }\href
  {https://doi.org/10.1038/s41467-023-36363-w} {\bibfield  {journal} {\bibinfo
  {journal} {Nature communications}\ }\textbf {\bibinfo {volume}
  {14}}~(\bibinfo {number} {1}),\ \bibinfo {pages} {725}}\BibitemShut {NoStop}%
\bibitem [{\citenamefont {Fauquenot}\ \emph {et~al.}(2024)\citenamefont
  {Fauquenot}, \citenamefont {Sarkar},\ and\ \citenamefont
  {Feld}}]{fauquenot2024eo}%
  \BibitemOpen
  \bibfield  {author} {\bibinfo {author} {\bibnamefont {Fauquenot},
  \bibfnamefont {Sebastiaan}}, \bibinfo {author} {\bibfnamefont {Aritra}\
  \bibnamefont {Sarkar}}, and\ \bibinfo {author} {\bibfnamefont {Sebastian}\
  \bibnamefont {Feld}}} (\bibinfo {year} {2024}),\ \bibfield  {title} {\enquote
  {\bibinfo {title} {Eo-grape and eo-drlpe: Open and closed loop approaches for
  energy efficient quantum optimal control},}\ }\href
  {https://arxiv.org/abs/2411.06556} {\bibinfo  {journal} {arXiv preprint
  arXiv:2411.06556}\ }\BibitemShut {NoStop}%
\bibitem [{\citenamefont {Fitzek}\ \emph {et~al.}(2020)\citenamefont {Fitzek},
  \citenamefont {Eliasson}, \citenamefont {Kockum},\ and\ \citenamefont
  {Granath}}]{fitzek2020deep}%
  \BibitemOpen
\bibfield  {journal} {  }\bibfield  {author} {\bibinfo {author} {\bibnamefont
  {Fitzek}, \bibfnamefont {David}}, \bibinfo {author} {\bibfnamefont {Mattias}\
  \bibnamefont {Eliasson}}, \bibinfo {author} {\bibfnamefont {Anton~Frisk}\
  \bibnamefont {Kockum}}, and\ \bibinfo {author} {\bibfnamefont {Mats}\
  \bibnamefont {Granath}}} (\bibinfo {year} {2020}),\ \bibfield  {title}
  {\enquote {\bibinfo {title} {Deep q-learning decoder for depolarizing noise
  on the toric code},}\ }\href
  {https://doi.org/10.1103/PhysRevResearch.2.023230} {\bibfield  {journal}
  {\bibinfo  {journal} {Phys. Rev. Res.}\ }\textbf {\bibinfo {volume} {2}},\
  \bibinfo {pages} {023230}}\BibitemShut {NoStop}%
\bibitem [{\citenamefont {Foder{\`a}}\ \emph {et~al.}(2024)\citenamefont
  {Foder{\`a}}, \citenamefont {Turati}, \citenamefont {Nembrini}, \citenamefont
  {Dacrema},\ and\ \citenamefont {Cremonesi}}]{fodera2024reinforcement}%
  \BibitemOpen
  \bibfield  {author} {\bibinfo {author} {\bibnamefont {Foder{\`a}},
  \bibfnamefont {Simone}}, \bibinfo {author} {\bibfnamefont {Gloria}\
  \bibnamefont {Turati}}, \bibinfo {author} {\bibfnamefont {Riccardo}\
  \bibnamefont {Nembrini}}, \bibinfo {author} {\bibfnamefont
  {Maurizio~Ferrari}\ \bibnamefont {Dacrema}}, and\ \bibinfo {author}
  {\bibfnamefont {Paolo}\ \bibnamefont {Cremonesi}}} (\bibinfo {year} {2024}),\
  \bibfield  {title} {\enquote {\bibinfo {title} {Reinforcement learning for
  variational quantum circuits design},}\ }\href
  {https://doi.org/10.48550/arXiv.2409.05475} {\bibinfo  {journal} {arXiv
  preprint arXiv:2409.05475}\ }\BibitemShut {NoStop}%
\bibitem [{\citenamefont {Fortunato}\ \emph {et~al.}(2017)\citenamefont
  {Fortunato}, \citenamefont {Azar}, \citenamefont {Piot}, \citenamefont
  {Menick}, \citenamefont {Osband}, \citenamefont {Graves}, \citenamefont
  {Mnih}, \citenamefont {Munos}, \citenamefont {Hassabis}, \citenamefont
  {Pietquin} \emph {et~al.}}]{fortunato2017noisy}%
  \BibitemOpen
\bibfield  {journal} {  }\bibfield  {author} {\bibinfo {author} {\bibnamefont
  {Fortunato}, \bibfnamefont {M}}, \bibinfo {author} {\bibfnamefont
  {MG}~\bibnamefont {Azar}}, \bibinfo {author} {\bibfnamefont {B}~\bibnamefont
  {Piot}}, \bibinfo {author} {\bibfnamefont {J}~\bibnamefont {Menick}},
  \bibinfo {author} {\bibfnamefont {I}~\bibnamefont {Osband}}, \bibinfo
  {author} {\bibfnamefont {A}~\bibnamefont {Graves}}, \bibinfo {author}
  {\bibfnamefont {V}~\bibnamefont {Mnih}}, \bibinfo {author} {\bibfnamefont
  {R}~\bibnamefont {Munos}}, \bibinfo {author} {\bibfnamefont {D}~\bibnamefont
  {Hassabis}}, \bibinfo {author} {\bibfnamefont {O}~\bibnamefont {Pietquin}},
  \emph {et~al.}} (\bibinfo {year} {2017}),\ \bibfield  {title} {\enquote
  {\bibinfo {title} {Noisy networks for exploration. arxiv 2017},}\ }\href
  {https://doi.org/10.48550/arXiv.1706.10295} {\bibinfo  {journal} {arXiv
  preprint arXiv:1706.10295}\ }\BibitemShut {NoStop}%
\bibitem [{\citenamefont {F{\"o}sel}\ \emph {et~al.}(2021)\citenamefont
  {F{\"o}sel}, \citenamefont {Niu}, \citenamefont {Marquardt},\ and\
  \citenamefont {Li}}]{foesel2021quantum}%
  \BibitemOpen
\bibfield  {journal} {  }\bibfield  {author} {\bibinfo {author} {\bibnamefont
  {F{\"o}sel}, \bibfnamefont {Thomas}}, \bibinfo {author} {\bibfnamefont
  {Murphy~Yuezhen}\ \bibnamefont {Niu}}, \bibinfo {author} {\bibfnamefont
  {Florian}\ \bibnamefont {Marquardt}}, and\ \bibinfo {author} {\bibfnamefont
  {Li}~\bibnamefont {Li}}} (\bibinfo {year} {2021}),\ \bibfield  {title}
  {\enquote {\bibinfo {title} {Quantum circuit optimization with deep
  reinforcement learning},}\ }\href {https://doi.org/10.48550/arXiv.2103.07585}
  {\bibinfo  {journal} {arXiv preprint arXiv:2103.07585}\ }\BibitemShut
  {NoStop}%
\bibitem [{\citenamefont {F\"osel}\ \emph {et~al.}(2018)\citenamefont
  {F\"osel}, \citenamefont {Tighineanu}, \citenamefont {Weiss},\ and\
  \citenamefont {Marquardt}}]{foesel2018reinforcement}%
  \BibitemOpen
\bibfield  {journal} {  }\bibfield  {author} {\bibinfo {author} {\bibnamefont
  {F\"osel}, \bibfnamefont {Thomas}}, \bibinfo {author} {\bibfnamefont {Petru}\
  \bibnamefont {Tighineanu}}, \bibinfo {author} {\bibfnamefont {Talitha}\
  \bibnamefont {Weiss}}, and\ \bibinfo {author} {\bibfnamefont {Florian}\
  \bibnamefont {Marquardt}}} (\bibinfo {year} {2018}),\ \bibfield  {title}
  {\enquote {\bibinfo {title} {Reinforcement learning with neural networks for
  quantum feedback},}\ }\href {https://doi.org/10.1103/PhysRevX.8.031084}
  {\bibfield  {journal} {\bibinfo  {journal} {Phys. Rev. X}\ }\textbf {\bibinfo
  {volume} {8}},\ \bibinfo {pages} {031084}}\BibitemShut {NoStop}%
\bibitem [{\citenamefont {Freire}\ \emph {et~al.}(2025)\citenamefont {Freire},
  \citenamefont {Delfosse},\ and\ \citenamefont
  {Leverrier}}]{freire2025optimizing}%
  \BibitemOpen
  \bibfield  {author} {\bibinfo {author} {\bibnamefont {Freire}, \bibfnamefont
  {Bruno~CA}}, \bibinfo {author} {\bibfnamefont {Nicolas}\ \bibnamefont
  {Delfosse}}, and\ \bibinfo {author} {\bibfnamefont {Anthony}\ \bibnamefont
  {Leverrier}}} (\bibinfo {year} {2025}),\ \bibfield  {title} {\enquote
  {\bibinfo {title} {Optimizing hypergraph product codes with random walks,
  simulated annealing and reinforcement learning},}\ }\href
  {https://arxiv.org/abs/2501.09622} {\bibinfo  {journal} {arXiv preprint
  arXiv:2501.09622}\ }\BibitemShut {NoStop}%
\bibitem [{\citenamefont {Fujimoto}\ \emph {et~al.}(2018)\citenamefont
  {Fujimoto}, \citenamefont {Hoof},\ and\ \citenamefont
  {Meger}}]{fujimoto2018addressing}%
  \BibitemOpen
\bibfield  {journal} {  }\bibfield  {author} {\bibinfo {author} {\bibnamefont
  {Fujimoto}, \bibfnamefont {Scott}}, \bibinfo {author} {\bibfnamefont {Herke}\
  \bibnamefont {Hoof}}, and\ \bibinfo {author} {\bibfnamefont {David}\
  \bibnamefont {Meger}}} (\bibinfo {year} {2018}),\ \bibfield  {title}
  {\enquote {\bibinfo {title} {Addressing function approximation error in
  actor-critic methods},}\ }\href {https://arxiv.org/abs/1802.09477} {\bibinfo
  {journal} {PMLR}\ ,\ \bibinfo {pages} {1587--1596}}\BibitemShut {NoStop}%
\bibitem [{\citenamefont {Garcia-Saez}\ and\ \citenamefont
  {Riu}(2019)}]{garcia2019quantum}%
  \BibitemOpen
\bibfield  {journal} {  }\bibfield  {author} {\bibinfo {author} {\bibnamefont
  {Garcia-Saez}, \bibfnamefont {Artur}}, and\ \bibinfo {author} {\bibfnamefont
  {Jordi}\ \bibnamefont {Riu}}} (\bibinfo {year} {2019}),\ \bibfield  {title}
  {\enquote {\bibinfo {title} {Quantum observables for continuous control of
  the quantum approximate optimization algorithm via reinforcement learning},}\
  }\href {https://doi.org/10.48550/arXiv.1911.09682} {\bibinfo  {journal}
  {arXiv preprint arXiv:1911.09682}\ }\BibitemShut {NoStop}%
\bibitem [{\citenamefont {Giannelli}\ \emph {et~al.}(2022)\citenamefont
  {Giannelli}, \citenamefont {Sgroi}, \citenamefont {Brown}, \citenamefont
  {Paraoanu}, \citenamefont {Paternostro}, \citenamefont {Paladino},\ and\
  \citenamefont {Falci}}]{giannelli2022tutorial}%
  \BibitemOpen
\bibfield  {journal} {  }\bibfield  {author} {\bibinfo {author} {\bibnamefont
  {Giannelli}, \bibfnamefont {Luigi}}, \bibinfo {author} {\bibfnamefont
  {Sofia}\ \bibnamefont {Sgroi}}, \bibinfo {author} {\bibfnamefont {Jonathon}\
  \bibnamefont {Brown}}, \bibinfo {author} {\bibfnamefont {Gheorghe~Sorin}\
  \bibnamefont {Paraoanu}}, \bibinfo {author} {\bibfnamefont {Mauro}\
  \bibnamefont {Paternostro}}, \bibinfo {author} {\bibfnamefont {Elisabetta}\
  \bibnamefont {Paladino}}, and\ \bibinfo {author} {\bibfnamefont {Giuseppe}\
  \bibnamefont {Falci}}} (\bibinfo {year} {2022}),\ \bibfield  {title}
  {\enquote {\bibinfo {title} {A tutorial on optimal control and reinforcement
  learning methods for quantum technologies},}\ }\href
  {https://doi.org/10.1016/j.physleta.2022.128054} {\bibfield  {journal}
  {\bibinfo  {journal} {Physics Letters A}\ }\textbf {\bibinfo {volume}
  {434}},\ \bibinfo {pages} {128054}}\BibitemShut {NoStop}%
\bibitem [{\citenamefont {Gillman}\ \emph {et~al.}(2022)\citenamefont
  {Gillman}, \citenamefont {Rose},\ and\ \citenamefont
  {Garrahan}}]{gillman2022reinforcement}%
  \BibitemOpen
  \bibfield  {author} {\bibinfo {author} {\bibnamefont {Gillman}, \bibfnamefont
  {Edward}}, \bibinfo {author} {\bibfnamefont {Dominic~C}\ \bibnamefont
  {Rose}}, and\ \bibinfo {author} {\bibfnamefont {Juan~P}\ \bibnamefont
  {Garrahan}}} (\bibinfo {year} {2022}),\ \bibfield  {title} {\enquote
  {\bibinfo {title} {Reinforcement learning with tensor networks: Application
  to dynamical large deviations},}\ }\href
  {https://doi.org/10.48550/arXiv.2209.14089} {\bibinfo  {journal} {arXiv
  preprint arXiv:2209.14089}\ }\BibitemShut {NoStop}%
\bibitem [{\citenamefont {Giordano}\ and\ \citenamefont
  {Martin-Delgado}(2022)}]{giordano2022reinforcement}%
  \BibitemOpen
\bibfield  {journal} {  }\bibfield  {author} {\bibinfo {author} {\bibnamefont
  {Giordano}, \bibfnamefont {Sara}}, and\ \bibinfo {author} {\bibfnamefont
  {Miguel~A.}\ \bibnamefont {Martin-Delgado}}} (\bibinfo {year} {2022}),\
  \bibfield  {title} {\enquote {\bibinfo {title} {Reinforcement-learning
  generation of four-qubit entangled states},}\ }\href
  {https://doi.org/10.1103/PhysRevResearch.4.043056} {\bibfield  {journal}
  {\bibinfo  {journal} {Phys. Rev. Res.}\ }\textbf {\bibinfo {volume} {4}},\
  \bibinfo {pages} {043056}}\BibitemShut {NoStop}%
\bibitem [{\citenamefont {Guatto}\ \emph {et~al.}(2024)\citenamefont {Guatto},
  \citenamefont {Susto},\ and\ \citenamefont {Ticozzi}}]{guatto2024improving}%
  \BibitemOpen
  \bibfield  {author} {\bibinfo {author} {\bibnamefont {Guatto}, \bibfnamefont
  {Manuel}}, \bibinfo {author} {\bibfnamefont {Gian~Antonio}\ \bibnamefont
  {Susto}}, and\ \bibinfo {author} {\bibfnamefont {Francesco}\ \bibnamefont
  {Ticozzi}}} (\bibinfo {year} {2024}),\ \bibfield  {title} {\enquote {\bibinfo
  {title} {Improving robustness of quantum feedback control with reinforcement
  learning},}\ }\href {https://arxiv.org/abs/2401.17190} {\bibfield  {journal}
  {\bibinfo  {journal} {arXiv}\ }}\Eprint {https://arxiv.org/abs/2401.17190}
  {2401.17190} \BibitemShut {NoStop}%
\bibitem [{\citenamefont {Guo}\ \emph {et~al.}(2021)\citenamefont {Guo},
  \citenamefont {Chen}, \citenamefont {Liu}, \citenamefont {Xue}, \citenamefont
  {Chen}, \citenamefont {Cao}, \citenamefont {Mao}, \citenamefont {Tey},\ and\
  \citenamefont {You}}]{guo2021faster}%
  \BibitemOpen
  \bibfield  {author} {\bibinfo {author} {\bibnamefont {Guo}, \bibfnamefont
  {Shuai-Feng}}, \bibinfo {author} {\bibfnamefont {Feng}\ \bibnamefont {Chen}},
  \bibinfo {author} {\bibfnamefont {Qi}~\bibnamefont {Liu}}, \bibinfo {author}
  {\bibfnamefont {Ming}\ \bibnamefont {Xue}}, \bibinfo {author} {\bibfnamefont
  {Jun-Jie}\ \bibnamefont {Chen}}, \bibinfo {author} {\bibfnamefont {Jia-Hao}\
  \bibnamefont {Cao}}, \bibinfo {author} {\bibfnamefont {Tian-Wei}\
  \bibnamefont {Mao}}, \bibinfo {author} {\bibfnamefont {Meng~Khoon}\
  \bibnamefont {Tey}}, and\ \bibinfo {author} {\bibfnamefont {Li}~\bibnamefont
  {You}}} (\bibinfo {year} {2021}),\ \bibfield  {title} {\enquote {\bibinfo
  {title} {Faster state preparation across quantum phase transition assisted by
  reinforcement learning},}\ }\href
  {https://doi.org/10.1103/PhysRevLett.126.060401} {\bibfield  {journal}
  {\bibinfo  {journal} {Phys. Rev. Lett.}\ }\textbf {\bibinfo {volume} {126}},\
  \bibinfo {pages} {060401}}\BibitemShut {NoStop}%
\bibitem [{\citenamefont {Haarnoja}\ \emph {et~al.}(2018)\citenamefont
  {Haarnoja}, \citenamefont {Zhou}, \citenamefont {Abbeel},\ and\ \citenamefont
  {Levine}}]{haarnoja2018soft}%
  \BibitemOpen
  \bibfield  {author} {\bibinfo {author} {\bibnamefont {Haarnoja},
  \bibfnamefont {Tuomas}}, \bibinfo {author} {\bibfnamefont {Aurick}\
  \bibnamefont {Zhou}}, \bibinfo {author} {\bibfnamefont {Pieter}\ \bibnamefont
  {Abbeel}}, and\ \bibinfo {author} {\bibfnamefont {Sergey}\ \bibnamefont
  {Levine}}} (\bibinfo {year} {2018}),\ \bibfield  {title} {\enquote {\bibinfo
  {title} {Soft actor-critic: Off-policy maximum entropy deep reinforcement
  learning with a stochastic actor},}\ }\bibfield  {booktitle} {\emph {\bibinfo
  {booktitle} {International conference on machine learning}},\ }\href
  {https://doi.org/10.48550/arXiv.1801.01290} {\bibinfo  {journal} {PMLR}\ ,\
  \bibinfo {pages} {1861--1870}}\BibitemShut {NoStop}%
\bibitem [{\citenamefont {Halverson}\ \emph {et~al.}(2019)\citenamefont
  {Halverson}, \citenamefont {Nelson},\ and\ \citenamefont
  {Ruehle}}]{halverson2019branes}%
  \BibitemOpen
\bibfield  {journal} {  }\bibfield  {author} {\bibinfo {author} {\bibnamefont
  {Halverson}, \bibfnamefont {James}}, \bibinfo {author} {\bibfnamefont
  {Brent}\ \bibnamefont {Nelson}}, and\ \bibinfo {author} {\bibfnamefont
  {Fabian}\ \bibnamefont {Ruehle}}} (\bibinfo {year} {2019}),\ \bibfield
  {title} {\enquote {\bibinfo {title} {Branes with brains: exploring string
  vacua with deep reinforcement learning},}\ }\href
  {https://doi.org/10.1007/JHEP06(2019)003} {\bibfield  {journal} {\bibinfo
  {journal} {Journal of High Energy Physics}\ }\textbf {\bibinfo {volume}
  {2019}}~(\bibinfo {number} {6}),\ \bibinfo {pages} {1--60}}\BibitemShut
  {NoStop}%
\bibitem [{\citenamefont {Haug}\ \emph {et~al.}(2021)\citenamefont {Haug},
  \citenamefont {Dumke}, \citenamefont {Kwek}, \citenamefont {Miniatura},\ and\
  \citenamefont {Amico}}]{haug2021machine}%
  \BibitemOpen
  \bibfield  {author} {\bibinfo {author} {\bibnamefont {Haug}, \bibfnamefont
  {Tobias}}, \bibinfo {author} {\bibfnamefont {Rainer}\ \bibnamefont {Dumke}},
  \bibinfo {author} {\bibfnamefont {Leong-Chuan}\ \bibnamefont {Kwek}},
  \bibinfo {author} {\bibfnamefont {Christian}\ \bibnamefont {Miniatura}}, and\
  \bibinfo {author} {\bibfnamefont {Luigi}\ \bibnamefont {Amico}}} (\bibinfo
  {year} {2021}),\ \bibfield  {title} {\enquote {\bibinfo {title}
  {Machine-learning engineering of quantum currents},}\ }\href
  {https://doi.org/10.1103/PhysRevResearch.3.013034} {\bibfield  {journal}
  {\bibinfo  {journal} {Phys. Rev. Res.}\ }\textbf {\bibinfo {volume} {3}},\
  \bibinfo {pages} {013034}}\BibitemShut {NoStop}%
\bibitem [{\citenamefont {Haug}\ \emph {et~al.}(2020)\citenamefont {Haug},
  \citenamefont {Mok}, \citenamefont {You}, \citenamefont {Zhang},
  \citenamefont {Png},\ and\ \citenamefont {Kwek}}]{haug2020classifying}%
  \BibitemOpen
  \bibfield  {author} {\bibinfo {author} {\bibnamefont {Haug}, \bibfnamefont
  {Tobias}}, \bibinfo {author} {\bibfnamefont {Wai-Keong}\ \bibnamefont {Mok}},
  \bibinfo {author} {\bibfnamefont {Jia-Bin}\ \bibnamefont {You}}, \bibinfo
  {author} {\bibfnamefont {Wenzu}\ \bibnamefont {Zhang}}, \bibinfo {author}
  {\bibfnamefont {Ching~Eng}\ \bibnamefont {Png}}, and\ \bibinfo {author}
  {\bibfnamefont {Leong-Chuan}\ \bibnamefont {Kwek}}} (\bibinfo {year}
  {2020}),\ \bibfield  {title} {\enquote {\bibinfo {title} {Classifying global
  state preparation via deep reinforcement learning},}\ }\href
  {https://doi.org/10.1088/2632-2153/abc81f} {\bibfield  {journal} {\bibinfo
  {journal} {Machine Learning: Science and Technology}\ }\textbf {\bibinfo
  {volume} {2}}~(\bibinfo {number} {1}),\ \bibinfo {pages}
  {01LT02}}\BibitemShut {NoStop}%
\bibitem [{\citenamefont {Hausknecht}\ and\ \citenamefont
  {Stone}(2017)}]{hausknecht2015deep}%
  \BibitemOpen
  \bibfield  {author} {\bibinfo {author} {\bibnamefont {Hausknecht},
  \bibfnamefont {Matthew}}, and\ \bibinfo {author} {\bibfnamefont {Peter}\
  \bibnamefont {Stone}}} (\bibinfo {year} {2017}),\ \bibfield  {title}
  {\enquote {\bibinfo {title} {Deep recurrent q-learning for partially
  observable mdps},}\ }\href {https://arxiv.org/abs/1507.06527} {\bibinfo
  {journal} {arXiv:1507.06527}\ }\BibitemShut {NoStop}%
\bibitem [{\citenamefont {He}\ and\ \citenamefont
  {Liu}(2025)}]{he2025discovering}%
  \BibitemOpen
\bibfield  {journal} {  }\bibfield  {author} {\bibinfo {author} {\bibnamefont
  {He}, \bibfnamefont {Austin~Yubo}}, and\ \bibinfo {author} {\bibfnamefont
  {Zi-Wen}\ \bibnamefont {Liu}}} (\bibinfo {year} {2025}),\ \bibfield  {title}
  {\enquote {\bibinfo {title} {Discovering highly efficient low-weight quantum
  error-correcting codes with reinforcement learning},}\ }\href
  {https://doi.org/10.48550/arXiv.2502.14372} {\bibinfo  {journal} {arXiv
  preprint arXiv:2502.14372}\ }\BibitemShut {NoStop}%
\bibitem [{\citenamefont {He}\ \emph {et~al.}(2021{\natexlab{a}})\citenamefont
  {He}, \citenamefont {Wang}, \citenamefont {Nie}, \citenamefont {Wu},
  \citenamefont {Zhang},\ and\ \citenamefont {Wang}}]{he2021deep}%
  \BibitemOpen
\bibfield  {journal} {  }\bibfield  {author} {\bibinfo {author} {\bibnamefont
  {He}, \bibfnamefont {Run-Hong}}, \bibinfo {author} {\bibfnamefont {Rui}\
  \bibnamefont {Wang}}, \bibinfo {author} {\bibfnamefont {Shen-Shuang}\
  \bibnamefont {Nie}}, \bibinfo {author} {\bibfnamefont {Jing}\ \bibnamefont
  {Wu}}, \bibinfo {author} {\bibfnamefont {Jia-Hui}\ \bibnamefont {Zhang}},
  and\ \bibinfo {author} {\bibfnamefont {Zhao-Ming}\ \bibnamefont {Wang}}}
  (\bibinfo {year} {2021}{\natexlab{a}}),\ \bibfield  {title} {\enquote
  {\bibinfo {title} {Deep reinforcement learning for universal quantum state
  preparation via dynamic pulse control},}\ }\href
  {https://doi.org/10.1140/epjqt/s40507-021-00119-6} {\bibfield  {journal}
  {\bibinfo  {journal} {EPJ Quantum Technology}\ }\textbf {\bibinfo {volume}
  {8}}~(\bibinfo {number} {1}),\ \bibinfo {pages} {29}}\BibitemShut {NoStop}%
\bibitem [{\citenamefont {He}\ \emph {et~al.}(2021{\natexlab{b}})\citenamefont
  {He}, \citenamefont {Li}, \citenamefont {Zheng}, \citenamefont {Li},\ and\
  \citenamefont {Situ}}]{he2021variational}%
  \BibitemOpen
  \bibfield  {author} {\bibinfo {author} {\bibnamefont {He}, \bibfnamefont
  {Zhimin}}, \bibinfo {author} {\bibfnamefont {Lvzhou}\ \bibnamefont {Li}},
  \bibinfo {author} {\bibfnamefont {Shenggen}\ \bibnamefont {Zheng}}, \bibinfo
  {author} {\bibfnamefont {Yongyao}\ \bibnamefont {Li}}, and\ \bibinfo {author}
  {\bibfnamefont {Haozhen}\ \bibnamefont {Situ}}} (\bibinfo {year}
  {2021}{\natexlab{b}}),\ \bibfield  {title} {\enquote {\bibinfo {title}
  {Variational quantum compiling with double q-learning},}\ }\href
  {https://doi.org/10.1088/1367-2630/abe0ae} {\bibfield  {journal} {\bibinfo
  {journal} {New Journal of Physics}\ }\textbf {\bibinfo {volume}
  {23}}~(\bibinfo {number} {3}),\ \bibinfo {pages} {033002}}\BibitemShut
  {NoStop}%
\bibitem [{\citenamefont {Herrera-Mart{\'{i}}}(2022)}]{herreramarti2022policy}%
  \BibitemOpen
  \bibfield  {author} {\bibinfo {author} {\bibnamefont {Herrera-Mart{\'{i}}},
  \bibfnamefont {David~A}}} (\bibinfo {year} {2022}),\ \bibfield  {title}
  {\enquote {\bibinfo {title} {Policy {G}radient {A}pproach to {C}ompilation of
  {V}ariational {Q}uantum {C}ircuits},}\ }\href
  {https://doi.org/10.22331/q-2022-09-08-797} {\bibfield  {journal} {\bibinfo
  {journal} {{Quantum}}\ }\textbf {\bibinfo {volume} {6}},\ \bibinfo {pages}
  {797}}\BibitemShut {NoStop}%
\bibitem [{\citenamefont {Hu}\ \emph {et~al.}(2022)\citenamefont {Hu},
  \citenamefont {Chen},\ and\ \citenamefont {Dong}}]{hu2022deep}%
  \BibitemOpen
  \bibfield  {author} {\bibinfo {author} {\bibnamefont {Hu}, \bibfnamefont
  {Shouliang}}, \bibinfo {author} {\bibfnamefont {Chunlin}\ \bibnamefont
  {Chen}}, and\ \bibinfo {author} {\bibfnamefont {Daoyi}\ \bibnamefont {Dong}}}
  (\bibinfo {year} {2022}),\ \bibfield  {title} {\enquote {\bibinfo {title}
  {Deep reinforcement learning for control design of quantum gates},}\
  }\bibfield  {booktitle} {\emph {\bibinfo {booktitle} {2022 13th Asian Control
  Conference (ASCC)}},\ }\href
  {https://doi.org/10.23919/ASCC56756.2022.9828135} {\bibinfo  {journal}
  {IEEE}\ ,\ \bibinfo {pages} {2367--2372}}\BibitemShut {NoStop}%
\bibitem [{\citenamefont {Hutin}\ \emph {et~al.}(2025)\citenamefont {Hutin},
  \citenamefont {Bilous}, \citenamefont {Ye}, \citenamefont {Abdollahi},
  \citenamefont {Cros}, \citenamefont {Dvir}, \citenamefont {Shah},
  \citenamefont {Cohen}, \citenamefont {Bienfait}, \citenamefont {Marquardt}
  \emph {et~al.}}]{hutin2025preparing}%
  \BibitemOpen
\bibfield  {journal} {  }\bibfield  {author} {\bibinfo {author} {\bibnamefont
  {Hutin}, \bibfnamefont {Hector}}, \bibinfo {author} {\bibfnamefont {Pavlo}\
  \bibnamefont {Bilous}}, \bibinfo {author} {\bibfnamefont {Chengzhi}\
  \bibnamefont {Ye}}, \bibinfo {author} {\bibfnamefont {Sepideh}\ \bibnamefont
  {Abdollahi}}, \bibinfo {author} {\bibfnamefont {Loris}\ \bibnamefont {Cros}},
  \bibinfo {author} {\bibfnamefont {Tom}\ \bibnamefont {Dvir}}, \bibinfo
  {author} {\bibfnamefont {Tirth}\ \bibnamefont {Shah}}, \bibinfo {author}
  {\bibfnamefont {Yonatan}\ \bibnamefont {Cohen}}, \bibinfo {author}
  {\bibfnamefont {Audrey}\ \bibnamefont {Bienfait}}, \bibinfo {author}
  {\bibfnamefont {Florian}\ \bibnamefont {Marquardt}},  \emph {et~al.}}
  (\bibinfo {year} {2025}),\ \bibfield  {title} {\enquote {\bibinfo {title}
  {Preparing schr{\"o}dinger cat states in a microwave cavity using a neural
  network},}\ }\href {https://doi.org/10.1103/PRXQuantum.6.010321} {\bibfield
  {journal} {\bibinfo  {journal} {PRX Quantum}\ }\textbf {\bibinfo {volume}
  {6}}~(\bibinfo {number} {1}),\ \bibinfo {pages} {010321}}\BibitemShut
  {NoStop}%
\bibitem [{\citenamefont {Ivanova-Rohling}\ \emph {et~al.}(2024)\citenamefont
  {Ivanova-Rohling}, \citenamefont {Rohling},\ and\ \citenamefont
  {Burkard}}]{ivanova2024discovery}%
  \BibitemOpen
  \bibfield  {author} {\bibinfo {author} {\bibnamefont {Ivanova-Rohling},
  \bibfnamefont {Violeta~N}}, \bibinfo {author} {\bibfnamefont {Niklas}\
  \bibnamefont {Rohling}}, and\ \bibinfo {author} {\bibfnamefont {Guido}\
  \bibnamefont {Burkard}}} (\bibinfo {year} {2024}),\ \bibfield  {title}
  {\enquote {\bibinfo {title} {Discovery of an exchange-only gate sequence for
  cnot with record-low gate time using reinforcement learning},}\ }\href
  {https://arxiv.org/abs/2402.10559} {\bibinfo  {journal} {arXiv preprint
  arXiv:2402.10559}\ }\BibitemShut {NoStop}%
\bibitem [{\citenamefont {Jae}\ \emph {et~al.}(2024)\citenamefont {Jae},
  \citenamefont {Hong}, \citenamefont {Choo},\ and\ \citenamefont
  {Kwon}}]{jae2024reinforcement}%
  \BibitemOpen
\bibfield  {journal} {  }\bibfield  {author} {\bibinfo {author} {\bibnamefont
  {Jae}, \bibfnamefont {Jeongwoo}}, \bibinfo {author} {\bibfnamefont
  {Jeonghoon}\ \bibnamefont {Hong}}, \bibinfo {author} {\bibfnamefont {Jinho}\
  \bibnamefont {Choo}}, and\ \bibinfo {author} {\bibfnamefont {Yeong-Dae}\
  \bibnamefont {Kwon}}} (\bibinfo {year} {2024}),\ \bibfield  {title} {\enquote
  {\bibinfo {title} {Reinforcement learning to learn quantum states for
  heisenberg scaling accuracy},}\ }\href {https://arxiv.org/abs/2412.02334}
  {\bibinfo  {journal} {arXiv preprint arXiv:2412.02334}\ }\BibitemShut
  {NoStop}%
\bibitem [{\citenamefont {Jerbi}\ \emph {et~al.}(2021)\citenamefont {Jerbi},
  \citenamefont {Trenkwalder}, \citenamefont {Poulsen~Nautrup}, \citenamefont
  {Briegel},\ and\ \citenamefont {Dunjko}}]{jerbi2021quantum}%
  \BibitemOpen
\bibfield  {journal} {  }\bibfield  {author} {\bibinfo {author} {\bibnamefont
  {Jerbi}, \bibfnamefont {Sofiene}}, \bibinfo {author} {\bibfnamefont {Lea~M.}\
  \bibnamefont {Trenkwalder}}, \bibinfo {author} {\bibfnamefont {Hendrik}\
  \bibnamefont {Poulsen~Nautrup}}, \bibinfo {author} {\bibfnamefont {Hans~J.}\
  \bibnamefont {Briegel}}, and\ \bibinfo {author} {\bibfnamefont {Vedran}\
  \bibnamefont {Dunjko}}} (\bibinfo {year} {2021}),\ \bibfield  {title}
  {\enquote {\bibinfo {title} {Quantum enhancements for deep reinforcement
  learning in large spaces},}\ }\href
  {https://doi.org/10.1103/PRXQuantum.2.010328} {\bibfield  {journal} {\bibinfo
   {journal} {PRX Quantum}\ }\textbf {\bibinfo {volume} {2}},\ \bibinfo {pages}
  {010328}}\BibitemShut {NoStop}%
\bibitem [{\citenamefont {Jiang}\ \emph {et~al.}(2022)\citenamefont {Jiang},
  \citenamefont {Pan}, \citenamefont {Wu}, \citenamefont {Gao},\ and\
  \citenamefont {Dong}}]{chen2022robust}%
  \BibitemOpen
  \bibfield  {author} {\bibinfo {author} {\bibnamefont {Jiang}, \bibfnamefont
  {Chen}}, \bibinfo {author} {\bibfnamefont {Yu}~\bibnamefont {Pan}}, \bibinfo
  {author} {\bibfnamefont {Zheng-Guang}\ \bibnamefont {Wu}}, \bibinfo {author}
  {\bibfnamefont {Qing}\ \bibnamefont {Gao}}, and\ \bibinfo {author}
  {\bibfnamefont {Daoyi}\ \bibnamefont {Dong}}} (\bibinfo {year} {2022}),\
  \bibfield  {title} {\enquote {\bibinfo {title} {Robust optimization for
  quantum reinforcement learning control using partial observations},}\ }\href
  {https://doi.org/10.1103/PhysRevA.105.062443} {\bibfield  {journal} {\bibinfo
   {journal} {Phys. Rev. A}\ }\textbf {\bibinfo {volume} {105}},\ \bibinfo
  {pages} {062443}}\BibitemShut {NoStop}%
\bibitem [{\citenamefont {Jin}\ \emph {et~al.}(2025)\citenamefont {Jin},
  \citenamefont {Wang}, \citenamefont {Xu}, \citenamefont {Zhuang},
  \citenamefont {Hu},\ and\ \citenamefont {Liu}}]{jin2025ppo}%
  \BibitemOpen
  \bibfield  {author} {\bibinfo {author} {\bibnamefont {Jin}, \bibfnamefont
  {Yu-Xin}}, \bibinfo {author} {\bibfnamefont {Zi-Wei}\ \bibnamefont {Wang}},
  \bibinfo {author} {\bibfnamefont {Hong-Ze}\ \bibnamefont {Xu}}, \bibinfo
  {author} {\bibfnamefont {Wei-Feng}\ \bibnamefont {Zhuang}}, \bibinfo {author}
  {\bibfnamefont {Meng-Jun}\ \bibnamefont {Hu}}, and\ \bibinfo {author}
  {\bibfnamefont {Dong~E}\ \bibnamefont {Liu}}} (\bibinfo {year} {2025}),\
  \bibfield  {title} {\enquote {\bibinfo {title} {Ppo-q: Proximal policy
  optimization with parametrized quantum policies or values},}\ }\href
  {https://arxiv.org/abs/2501.07085} {\bibinfo  {journal} {arXiv preprint
  arXiv:2501.07085}\ }\BibitemShut {NoStop}%
\bibitem [{\citenamefont {Judson}\ and\ \citenamefont
  {Rabitz}(1992)}]{rabitz1992teaching}%
  \BibitemOpen
\bibfield  {journal} {  }\bibfield  {author} {\bibinfo {author} {\bibnamefont
  {Judson}, \bibfnamefont {Richard~S}}, and\ \bibinfo {author} {\bibfnamefont
  {Herschel}\ \bibnamefont {Rabitz}}} (\bibinfo {year} {1992}),\ \bibfield
  {title} {\enquote {\bibinfo {title} {Teaching lasers to control molecules},}\
  }\href {https://doi.org/10.1103/PhysRevLett.68.1500} {\bibfield  {journal}
  {\bibinfo  {journal} {Phys. Rev. Lett.}\ }\textbf {\bibinfo {volume} {68}},\
  \bibinfo {pages} {1500--1503}}\BibitemShut {NoStop}%
\bibitem [{\citenamefont {Kakade}(2001)}]{kakade2001natural}%
  \BibitemOpen
  \bibfield  {author} {\bibinfo {author} {\bibnamefont {Kakade}, \bibfnamefont
  {Sham~M}}} (\bibinfo {year} {2001}),\ \bibfield  {title} {\enquote {\bibinfo
  {title} {A natural policy gradient},}\ }\href
  {https://proceedings.neurips.cc/paper_files/paper/2001/file/4b86abe48d358ecf194c56c69108433e-Paper.pdf}
  {\bibfield  {journal} {\bibinfo  {journal} {Advances in neural information
  processing systems}\ }\textbf {\bibinfo {volume} {14}}}\BibitemShut {NoStop}%
\bibitem [{\citenamefont {Kalita}\ and\ \citenamefont
  {Sarma}(2024)}]{kalita2024domino}%
  \BibitemOpen
  \bibfield  {author} {\bibinfo {author} {\bibnamefont {Kalita}, \bibfnamefont
  {Sampreet}}, and\ \bibinfo {author} {\bibfnamefont {Amarendra~K}\
  \bibnamefont {Sarma}}} (\bibinfo {year} {2024}),\ \bibfield  {title}
  {\enquote {\bibinfo {title} {Domino-cooling oscillator networks with deep
  reinforcement learning},}\ }\href {https://arxiv.org/abs/2408.12271}
  {\bibinfo  {journal} {arXiv preprint arXiv:2408.12271}\ }\BibitemShut
  {NoStop}%
\bibitem [{\citenamefont {Khairy}\ \emph {et~al.}(2019)\citenamefont {Khairy},
  \citenamefont {Shaydulin}, \citenamefont {Cincio}, \citenamefont {Alexeev},\
  and\ \citenamefont {Balaprakash}}]{khairy2019reinforcement}%
  \BibitemOpen
\bibfield  {journal} {  }\bibfield  {author} {\bibinfo {author} {\bibnamefont
  {Khairy}, \bibfnamefont {Sami}}, \bibinfo {author} {\bibfnamefont {Ruslan}\
  \bibnamefont {Shaydulin}}, \bibinfo {author} {\bibfnamefont {Lukasz}\
  \bibnamefont {Cincio}}, \bibinfo {author} {\bibfnamefont {Yuri}\ \bibnamefont
  {Alexeev}}, and\ \bibinfo {author} {\bibfnamefont {Prasanna}\ \bibnamefont
  {Balaprakash}}} (\bibinfo {year} {2019}),\ \bibfield  {title} {\enquote
  {\bibinfo {title} {Reinforcement-learning-based variational quantum circuits
  optimization for combinatorial problems},}\ }\href
  {https://doi.org/10.48550/arXiv.1911.04574} {\bibinfo  {journal} {arXiv
  preprint arXiv:1911.04574}\ }\BibitemShut {NoStop}%
\bibitem [{\citenamefont {Khalid}\ \emph {et~al.}(2023)\citenamefont {Khalid},
  \citenamefont {Weidner}, \citenamefont {Jonckheere}, \citenamefont
  {Schirmer},\ and\ \citenamefont {Langbein}}]{khalid2023sample}%
  \BibitemOpen
\bibfield  {journal} {  }\bibfield  {author} {\bibinfo {author} {\bibnamefont
  {Khalid}, \bibfnamefont {Irtaza}}, \bibinfo {author} {\bibfnamefont
  {Carrie~A.}\ \bibnamefont {Weidner}}, \bibinfo {author} {\bibfnamefont
  {Edmond~A.}\ \bibnamefont {Jonckheere}}, \bibinfo {author} {\bibfnamefont
  {Sophie~G.}\ \bibnamefont {Schirmer}}, and\ \bibinfo {author} {\bibfnamefont
  {Frank~C.}\ \bibnamefont {Langbein}}} (\bibinfo {year} {2023}),\ \bibfield
  {title} {\enquote {\bibinfo {title} {Sample-efficient model-based
  reinforcement learning for quantum control},}\ }\href
  {https://doi.org/10.1103/PhysRevResearch.5.043002} {\bibfield  {journal}
  {\bibinfo  {journal} {Phys. Rev. Res.}\ }\textbf {\bibinfo {volume} {5}},\
  \bibinfo {pages} {043002}}\BibitemShut {NoStop}%
\bibitem [{\citenamefont {Koch}\ \emph {et~al.}(2022)\citenamefont {Koch},
  \citenamefont {Boscain}, \citenamefont {Calarco}, \citenamefont {Dirr},
  \citenamefont {Filipp}, \citenamefont {Glaser}, \citenamefont {Kosloff},
  \citenamefont {Montangero}, \citenamefont {Schulte-Herbr{\"u}ggen},
  \citenamefont {Sugny} \emph {et~al.}}]{koch2022quantum}%
  \BibitemOpen
  \bibfield  {author} {\bibinfo {author} {\bibnamefont {Koch}, \bibfnamefont
  {Christiane~P}}, \bibinfo {author} {\bibfnamefont {Ugo}\ \bibnamefont
  {Boscain}}, \bibinfo {author} {\bibfnamefont {Tommaso}\ \bibnamefont
  {Calarco}}, \bibinfo {author} {\bibfnamefont {Gunther}\ \bibnamefont {Dirr}},
  \bibinfo {author} {\bibfnamefont {Stefan}\ \bibnamefont {Filipp}}, \bibinfo
  {author} {\bibfnamefont {Steffen~J}\ \bibnamefont {Glaser}}, \bibinfo
  {author} {\bibfnamefont {Ronnie}\ \bibnamefont {Kosloff}}, \bibinfo {author}
  {\bibfnamefont {Simone}\ \bibnamefont {Montangero}}, \bibinfo {author}
  {\bibfnamefont {Thomas}\ \bibnamefont {Schulte-Herbr{\"u}ggen}}, \bibinfo
  {author} {\bibfnamefont {Dominique}\ \bibnamefont {Sugny}},  \emph {et~al.}}
  (\bibinfo {year} {2022}),\ \bibfield  {title} {\enquote {\bibinfo {title}
  {Quantum optimal control in quantum technologies. strategic report on current
  status, visions and goals for research in europe},}\ }\href
  {https://doi.org/10.1140/epjqt/s40507-022-00138-x} {\bibfield  {journal}
  {\bibinfo  {journal} {EPJ Quantum Technology}\ }\textbf {\bibinfo {volume}
  {9}}~(\bibinfo {number} {1}),\ \bibinfo {pages} {19}}\BibitemShut {NoStop}%
\bibitem [{\citenamefont {K{\"o}lle}\ \emph
  {et~al.}(2024{\natexlab{a}})\citenamefont {K{\"o}lle}, \citenamefont
  {Schneider}, \citenamefont {Egger}, \citenamefont {Topp}, \citenamefont
  {Phan}, \citenamefont {Altmann}, \citenamefont {N{\"u}{\ss}lein},\ and\
  \citenamefont {Linnhoff-Popien}}]{kolle2024architectural}%
  \BibitemOpen
  \bibfield  {author} {\bibinfo {author} {\bibnamefont {K{\"o}lle},
  \bibfnamefont {Michael}}, \bibinfo {author} {\bibfnamefont {Karola}\
  \bibnamefont {Schneider}}, \bibinfo {author} {\bibfnamefont {Sabrina}\
  \bibnamefont {Egger}}, \bibinfo {author} {\bibfnamefont {Felix}\ \bibnamefont
  {Topp}}, \bibinfo {author} {\bibfnamefont {Thomy}\ \bibnamefont {Phan}},
  \bibinfo {author} {\bibfnamefont {Philipp}\ \bibnamefont {Altmann}}, \bibinfo
  {author} {\bibfnamefont {Jonas}\ \bibnamefont {N{\"u}{\ss}lein}}, and\
  \bibinfo {author} {\bibfnamefont {Claudia}\ \bibnamefont {Linnhoff-Popien}}}
  (\bibinfo {year} {2024}{\natexlab{a}}),\ \bibfield  {title} {\enquote
  {\bibinfo {title} {Architectural influence on variational quantum circuits in
  multi-agent reinforcement learning: Evolutionary strategies for
  optimization},}\ }\href {https://arxiv.org/abs/2407.20739} {\bibinfo
  {journal} {arXiv preprint arXiv:2407.20739}\ }\BibitemShut {NoStop}%
\bibitem [{\citenamefont {K{\"o}lle}\ \emph
  {et~al.}(2024{\natexlab{b}})\citenamefont {K{\"o}lle}, \citenamefont
  {Schubert}, \citenamefont {Altmann}, \citenamefont {Zorn}, \citenamefont
  {Stein},\ and\ \citenamefont {Linnhoff-Popien}}]{kolle2024reinforcement}%
  \BibitemOpen
\bibfield  {journal} {  }\bibfield  {author} {\bibinfo {author} {\bibnamefont
  {K{\"o}lle}, \bibfnamefont {Michael}}, \bibinfo {author} {\bibfnamefont
  {Tom}\ \bibnamefont {Schubert}}, \bibinfo {author} {\bibfnamefont {Philipp}\
  \bibnamefont {Altmann}}, \bibinfo {author} {\bibfnamefont {Maximilian}\
  \bibnamefont {Zorn}}, \bibinfo {author} {\bibfnamefont {Jonas}\ \bibnamefont
  {Stein}}, and\ \bibinfo {author} {\bibfnamefont {Claudia}\ \bibnamefont
  {Linnhoff-Popien}}} (\bibinfo {year} {2024}{\natexlab{b}}),\ \bibfield
  {title} {\enquote {\bibinfo {title} {A reinforcement learning environment for
  directed quantum circuit synthesis},}\ }\href
  {https://doi.org/10.48550/arXiv.2401.07054} {\bibinfo  {journal} {arXiv
  preprint arXiv:2401.07054}\ }\BibitemShut {NoStop}%
\bibitem [{\citenamefont {K{\"o}lle}\ \emph
  {et~al.}(2024{\natexlab{c}})\citenamefont {K{\"o}lle}, \citenamefont {Seidl},
  \citenamefont {Zorn}, \citenamefont {Altmann}, \citenamefont {Stein},\ and\
  \citenamefont {Gabor}}]{kolle2024optimizing}%
  \BibitemOpen
\bibfield  {journal} {  }\bibfield  {author} {\bibinfo {author} {\bibnamefont
  {K{\"o}lle}, \bibfnamefont {Michael}}, \bibinfo {author} {\bibfnamefont
  {Daniel}\ \bibnamefont {Seidl}}, \bibinfo {author} {\bibfnamefont
  {Maximilian}\ \bibnamefont {Zorn}}, \bibinfo {author} {\bibfnamefont
  {Philipp}\ \bibnamefont {Altmann}}, \bibinfo {author} {\bibfnamefont {Jonas}\
  \bibnamefont {Stein}}, and\ \bibinfo {author} {\bibfnamefont {Thomas}\
  \bibnamefont {Gabor}}} (\bibinfo {year} {2024}{\natexlab{c}}),\ \bibfield
  {title} {\enquote {\bibinfo {title} {Optimizing variational quantum circuits
  using metaheuristic strategies in reinforcement learning},}\ }\href
  {https://doi.org/10.48550/arXiv.2408.01187} {\bibinfo  {journal} {arXiv
  preprint arXiv:2408.01187}\ }\BibitemShut {NoStop}%
\bibitem [{\citenamefont {K{\"o}lle}\ \emph
  {et~al.}(2024{\natexlab{d}})\citenamefont {K{\"o}lle}, \citenamefont
  {Witter}, \citenamefont {Rohe}, \citenamefont {Stenzel}, \citenamefont
  {Altmann},\ and\ \citenamefont {Gabor}}]{kolle2024study}%
  \BibitemOpen
\bibfield  {journal} {  }\bibfield  {author} {\bibinfo {author} {\bibnamefont
  {K{\"o}lle}, \bibfnamefont {Michael}}, \bibinfo {author} {\bibfnamefont
  {Timo}\ \bibnamefont {Witter}}, \bibinfo {author} {\bibfnamefont {Tobias}\
  \bibnamefont {Rohe}}, \bibinfo {author} {\bibfnamefont {Gerhard}\
  \bibnamefont {Stenzel}}, \bibinfo {author} {\bibfnamefont {Philipp}\
  \bibnamefont {Altmann}}, and\ \bibinfo {author} {\bibfnamefont {Thomas}\
  \bibnamefont {Gabor}}} (\bibinfo {year} {2024}{\natexlab{d}}),\ \bibfield
  {title} {\enquote {\bibinfo {title} {A study on optimization techniques for
  variational quantum circuits in reinforcement learning},}\ }\href
  {https://doi.org/10.48550/arXiv.2405.12354} {\bibinfo  {journal} {arXiv
  preprint arXiv:2405.12354}\ }\BibitemShut {NoStop}%
\bibitem [{\citenamefont {Konaka}\ \emph {et~al.}(2025)\citenamefont {Konaka},
  \citenamefont {R{\"o}hm}, \citenamefont {Mihana},\ and\ \citenamefont
  {Horisaki}}]{konaka2025scalable}%
  \BibitemOpen
\bibfield  {journal} {  }\bibfield  {author} {\bibinfo {author} {\bibnamefont
  {Konaka}, \bibfnamefont {Kohei}}, \bibinfo {author} {\bibfnamefont
  {Andr{\'e}}\ \bibnamefont {R{\"o}hm}}, \bibinfo {author} {\bibfnamefont
  {Takatomo}\ \bibnamefont {Mihana}}, and\ \bibinfo {author} {\bibfnamefont
  {Ryoichi}\ \bibnamefont {Horisaki}}} (\bibinfo {year} {2025}),\ \bibfield
  {title} {\enquote {\bibinfo {title} {Scalable conflict-free decision making
  with photons},}\ }\href {https://arxiv.org/abs/2504.08331} {\bibinfo
  {journal} {arXiv preprint arXiv:2504.08331}\ }\BibitemShut {NoStop}%
\bibitem [{\citenamefont {Konda}\ and\ \citenamefont
  {Tsitsiklis}(1999)}]{konda1999actor}%
  \BibitemOpen
\bibfield  {journal} {  }\bibfield  {author} {\bibinfo {author} {\bibnamefont
  {Konda}, \bibfnamefont {Vijay}}, and\ \bibinfo {author} {\bibfnamefont
  {John}\ \bibnamefont {Tsitsiklis}}} (\bibinfo {year} {1999}),\ \bibfield
  {title} {\enquote {\bibinfo {title} {Actor-critic algorithms},}\ }\href
  {https://doi.org/10.1137/S0363012901385691} {\bibfield  {journal} {\bibinfo
  {journal} {Advances in neural information processing systems}\ }\textbf
  {\bibinfo {volume} {12}}}\BibitemShut {NoStop}%
\bibitem [{\citenamefont {Kremer}\ \emph {et~al.}(2024)\citenamefont {Kremer},
  \citenamefont {Villar}, \citenamefont {Paik}, \citenamefont {Duran},
  \citenamefont {Faro},\ and\ \citenamefont
  {Cruz-Benito}}]{kremer2024practical}%
  \BibitemOpen
  \bibfield  {author} {\bibinfo {author} {\bibnamefont {Kremer}, \bibfnamefont
  {David}}, \bibinfo {author} {\bibfnamefont {Victor}\ \bibnamefont {Villar}},
  \bibinfo {author} {\bibfnamefont {Hanhee}\ \bibnamefont {Paik}}, \bibinfo
  {author} {\bibfnamefont {Ivan}\ \bibnamefont {Duran}}, \bibinfo {author}
  {\bibfnamefont {Ismael}\ \bibnamefont {Faro}}, and\ \bibinfo {author}
  {\bibfnamefont {Juan}\ \bibnamefont {Cruz-Benito}}} (\bibinfo {year}
  {2024}),\ \bibfield  {title} {\enquote {\bibinfo {title} {Practical and
  efficient quantum circuit synthesis and transpiling with reinforcement
  learning},}\ }\href {https://doi.org/10.48550/arXiv.2405.13196} {\bibinfo
  {journal} {arXiv preprint arXiv:2405.13196}\ }\BibitemShut {NoStop}%
\bibitem [{\citenamefont {Krenn}\ \emph {et~al.}(2023)\citenamefont {Krenn},
  \citenamefont {Landgraf}, \citenamefont {Foesel},\ and\ \citenamefont
  {Marquardt}}]{krenn2023artificial}%
  \BibitemOpen
\bibfield  {journal} {  }\bibfield  {author} {\bibinfo {author} {\bibnamefont
  {Krenn}, \bibfnamefont {Mario}}, \bibinfo {author} {\bibfnamefont {Jonas}\
  \bibnamefont {Landgraf}}, \bibinfo {author} {\bibfnamefont {Thomas}\
  \bibnamefont {Foesel}}, and\ \bibinfo {author} {\bibfnamefont {Florian}\
  \bibnamefont {Marquardt}}} (\bibinfo {year} {2023}),\ \bibfield  {title}
  {\enquote {\bibinfo {title} {Artificial intelligence and machine learning for
  quantum technologies},}\ }\href {https://doi.org/10.1103/PhysRevA.107.010101}
  {\bibfield  {journal} {\bibinfo  {journal} {Phys. Rev. A}\ }\textbf {\bibinfo
  {volume} {107}},\ \bibinfo {pages} {010101}}\BibitemShut {NoStop}%
\bibitem [{\citenamefont {Kruse}\ \emph {et~al.}(2025)\citenamefont {Kruse},
  \citenamefont {Coelho}, \citenamefont {Rosskopf}, \citenamefont {Wille},\
  and\ \citenamefont {Lorenz}}]{kruse2025benchmarking}%
  \BibitemOpen
  \bibfield  {author} {\bibinfo {author} {\bibnamefont {Kruse}, \bibfnamefont
  {Georg}}, \bibinfo {author} {\bibfnamefont {Rodrigo}\ \bibnamefont {Coelho}},
  \bibinfo {author} {\bibfnamefont {Andreas}\ \bibnamefont {Rosskopf}},
  \bibinfo {author} {\bibfnamefont {Robert}\ \bibnamefont {Wille}}, and\
  \bibinfo {author} {\bibfnamefont {Jeanette-Miriam}\ \bibnamefont {Lorenz}}}
  (\bibinfo {year} {2025}),\ \bibfield  {title} {\enquote {\bibinfo {title}
  {Benchmarking quantum reinforcement learning},}\ }\href
  {https://arxiv.org/abs/2502.04909} {\bibinfo  {journal} {arXiv preprint
  arXiv:2502.04909}\ }\BibitemShut {NoStop}%
\bibitem [{\citenamefont {Kundu}(2025)}]{kundu2025improving}%
  \BibitemOpen
\bibfield  {journal} {  }\bibfield  {author} {\bibinfo {author} {\bibnamefont
  {Kundu}, \bibfnamefont {Akash}}} (\bibinfo {year} {2025}),\ \bibfield
  {title} {\enquote {\bibinfo {title} {Improving thermal state preparation of
  sachdev-ye-kitaev model with reinforcement learning on quantum hardware},}\
  }\href {https://arxiv.org/abs/2501.11454} {\bibinfo  {journal} {arXiv
  preprint arXiv:2501.11454}\ }\BibitemShut {NoStop}%
\bibitem [{\citenamefont {Kundu}\ \emph {et~al.}(2024)\citenamefont {Kundu},
  \citenamefont {Sarkar},\ and\ \citenamefont {Sadhu}}]{kundu2024kanqas}%
  \BibitemOpen
\bibfield  {journal} {  }\bibfield  {author} {\bibinfo {author} {\bibnamefont
  {Kundu}, \bibfnamefont {Akash}}, \bibinfo {author} {\bibfnamefont {Aritra}\
  \bibnamefont {Sarkar}}, and\ \bibinfo {author} {\bibfnamefont {Abhishek}\
  \bibnamefont {Sadhu}}} (\bibinfo {year} {2024}),\ \bibfield  {title}
  {\enquote {\bibinfo {title} {Kanqas: Kolmogorov-arnold network for quantum
  architecture search},}\ }\href
  {https://doi.org/10.1140/epjqt/s40507-024-00289-z} {\bibfield  {journal}
  {\bibinfo  {journal} {EPJ Quantum Technology}\ }\textbf {\bibinfo {volume}
  {11}}~(\bibinfo {number} {1}),\ \bibinfo {pages} {76}}\BibitemShut {NoStop}%
\bibitem [{\citenamefont {Kundu}\ and\ \citenamefont
  {Sarra}(2024)}]{kundu2024easy}%
  \BibitemOpen
  \bibfield  {author} {\bibinfo {author} {\bibnamefont {Kundu}, \bibfnamefont
  {Akash}}, and\ \bibinfo {author} {\bibfnamefont {Leopoldo}\ \bibnamefont
  {Sarra}}} (\bibinfo {year} {2024}),\ \bibfield  {title} {\enquote {\bibinfo
  {title} {From easy to hard: Tackling quantum problems with learned gadgets
  for real hardware},}\ }\href {https://arxiv.org/abs/2411.00230} {\bibinfo
  {journal} {arXiv preprint arXiv:2411.00230}\ }\BibitemShut {NoStop}%
\bibitem [{\citenamefont {Kuprikov}\ \emph {et~al.}(2022)\citenamefont
  {Kuprikov}, \citenamefont {Kokhanovskiy}, \citenamefont {Serebrennikov},\
  and\ \citenamefont {Turitsyn}}]{kuprikov2022deep}%
  \BibitemOpen
\bibfield  {journal} {  }\bibfield  {author} {\bibinfo {author} {\bibnamefont
  {Kuprikov}, \bibfnamefont {Evgeny}}, \bibinfo {author} {\bibfnamefont
  {Alexey}\ \bibnamefont {Kokhanovskiy}}, \bibinfo {author} {\bibfnamefont
  {Kirill}\ \bibnamefont {Serebrennikov}}, and\ \bibinfo {author}
  {\bibfnamefont {Sergey}\ \bibnamefont {Turitsyn}}} (\bibinfo {year} {2022}),\
  \bibfield  {title} {\enquote {\bibinfo {title} {Deep reinforcement learning
  for self-tuning laser source of dissipative solitons},}\ }\href
  {https://doi.org/10.1038/s41598-022-11274-w} {\bibfield  {journal} {\bibinfo
  {journal} {Scientific Reports}\ }\textbf {\bibinfo {volume} {12}}~(\bibinfo
  {number} {1}),\ \bibinfo {pages} {7185}}\BibitemShut {NoStop}%
\bibitem [{\citenamefont {Lanore}\ \emph {et~al.}(2024)\citenamefont {Lanore},
  \citenamefont {Grasselli}, \citenamefont {Valcarce}, \citenamefont {Bancal},\
  and\ \citenamefont {Sangouard}}]{lanore2024automated}%
  \BibitemOpen
  \bibfield  {author} {\bibinfo {author} {\bibnamefont {Lanore}, \bibfnamefont
  {Corentin}}, \bibinfo {author} {\bibfnamefont {Federico}\ \bibnamefont
  {Grasselli}}, \bibinfo {author} {\bibfnamefont {Xavier}\ \bibnamefont
  {Valcarce}}, \bibinfo {author} {\bibfnamefont {Jean-Daniel}\ \bibnamefont
  {Bancal}}, and\ \bibinfo {author} {\bibfnamefont {Nicolas}\ \bibnamefont
  {Sangouard}}} (\bibinfo {year} {2024}),\ \bibfield  {title} {\enquote
  {\bibinfo {title} {Automated generation of photonic circuits for bell tests
  with homodyne measurements},}\ }\href {https://arxiv.org/abs/2410.19670}
  {\bibinfo  {journal} {arXiv preprint arXiv:2410.19670}\ }\BibitemShut
  {NoStop}%
\bibitem [{\citenamefont {Levine}(2018)}]{levine2018reinforcement}%
  \BibitemOpen
\bibfield  {journal} {  }\bibfield  {author} {\bibinfo {author} {\bibnamefont
  {Levine}, \bibfnamefont {Sergey}}} (\bibinfo {year} {2018}),\ \bibfield
  {title} {\enquote {\bibinfo {title} {Reinforcement learning and control as
  probabilistic inference: Tutorial and review},}\ }\href
  {https://arxiv.org/abs/1805.00909} {\bibinfo  {journal} {arXiv preprint
  arXiv:1805.00909}\ }\BibitemShut {NoStop}%
\bibitem [{\citenamefont {Li}\ \emph {et~al.}(2025)\citenamefont {Li},
  \citenamefont {Fan}, \citenamefont {Li}, \citenamefont {Ruan}, \citenamefont
  {Zhao}, \citenamefont {Peng}, \citenamefont {Wu}, \citenamefont {Zhang},\
  and\ \citenamefont {Song}}]{li2025robust}%
  \BibitemOpen
\bibfield  {journal} {  }\bibfield  {author} {\bibinfo {author} {\bibnamefont
  {Li}, \bibfnamefont {Shengyong}}, \bibinfo {author} {\bibfnamefont {Yidian}\
  \bibnamefont {Fan}}, \bibinfo {author} {\bibfnamefont {Xiang}\ \bibnamefont
  {Li}}, \bibinfo {author} {\bibfnamefont {Xinhui}\ \bibnamefont {Ruan}},
  \bibinfo {author} {\bibfnamefont {Qianchuan}\ \bibnamefont {Zhao}}, \bibinfo
  {author} {\bibfnamefont {Zhihui}\ \bibnamefont {Peng}}, \bibinfo {author}
  {\bibfnamefont {Re-Bing}\ \bibnamefont {Wu}}, \bibinfo {author}
  {\bibfnamefont {Jing}\ \bibnamefont {Zhang}}, and\ \bibinfo {author}
  {\bibfnamefont {Pengtao}\ \bibnamefont {Song}}} (\bibinfo {year} {2025}),\
  \bibfield  {title} {\enquote {\bibinfo {title} {Robust quantum control using
  reinforcement learning from demonstration},}\ }\href
  {https://arxiv.org/abs/2503.21085} {\bibinfo  {journal} {arXiv preprint
  arXiv:2503.21085}\ }\BibitemShut {NoStop}%
\bibitem [{\citenamefont {Li}\ \emph {et~al.}(2024)\citenamefont {Li},
  \citenamefont {Zhao}, \citenamefont {Wang}, \citenamefont {Liu},
  \citenamefont {Miao},\ and\ \citenamefont {Zhao}}]{li2024reinforcement}%
  \BibitemOpen
\bibfield  {journal} {  }\bibfield  {author} {\bibinfo {author} {\bibnamefont
  {Li}, \bibfnamefont {Tingting}}, \bibinfo {author} {\bibfnamefont {Yiming}\
  \bibnamefont {Zhao}}, \bibinfo {author} {\bibfnamefont {Yong}\ \bibnamefont
  {Wang}}, \bibinfo {author} {\bibfnamefont {Yanping}\ \bibnamefont {Liu}},
  \bibinfo {author} {\bibfnamefont {Yazhuang}\ \bibnamefont {Miao}}, and\
  \bibinfo {author} {\bibfnamefont {Xiaolong}\ \bibnamefont {Zhao}}} (\bibinfo
  {year} {2024}),\ \bibfield  {title} {\enquote {\bibinfo {title}
  {Reinforcement learning enhancing entanglement for two-photon-driven rabi
  model},}\ }\href {https://arxiv.org/abs/2411.15841} {\bibinfo  {journal}
  {arXiv preprint arXiv:2411.15841}\ }\BibitemShut {NoStop}%
\bibitem [{\citenamefont {Li}\ \emph {et~al.}(2023)\citenamefont {Li},
  \citenamefont {Liu}, \citenamefont {Li},\ and\ \citenamefont
  {Li}}]{li2023quantum}%
  \BibitemOpen
\bibfield  {journal} {  }\bibfield  {author} {\bibinfo {author} {\bibnamefont
  {Li}, \bibfnamefont {Yangzhi}}, \bibinfo {author} {\bibfnamefont {Wen}\
  \bibnamefont {Liu}}, \bibinfo {author} {\bibfnamefont {Maoduo}\ \bibnamefont
  {Li}}, and\ \bibinfo {author} {\bibfnamefont {Yugang}\ \bibnamefont {Li}}}
  (\bibinfo {year} {2023}),\ \bibfield  {title} {\enquote {\bibinfo {title}
  {Quantum circuit compilation for nearest-neighbor architecture based on
  reinforcement learning},}\ }\href
  {https://doi.org/10.1007/s11128-023-04050-w} {\bibfield  {journal} {\bibinfo
  {journal} {Quantum Information Processing}\ }\textbf {\bibinfo {volume}
  {22}},\ 10.1007/s11128-023-04050-w}\BibitemShut {NoStop}%
\bibitem [{\citenamefont {Lillicrap}\ \emph {et~al.}(2015)\citenamefont
  {Lillicrap}, \citenamefont {Hunt}, \citenamefont {Pritzel}, \citenamefont
  {Heess}, \citenamefont {Erez}, \citenamefont {Tassa}, \citenamefont
  {Silver},\ and\ \citenamefont {Wierstra}}]{lillicrap2015continuous}%
  \BibitemOpen
  \bibfield  {author} {\bibinfo {author} {\bibnamefont {Lillicrap},
  \bibfnamefont {Timothy~P}}, \bibinfo {author} {\bibfnamefont {Jonathan~J}\
  \bibnamefont {Hunt}}, \bibinfo {author} {\bibfnamefont {Alexander}\
  \bibnamefont {Pritzel}}, \bibinfo {author} {\bibfnamefont {Nicolas}\
  \bibnamefont {Heess}}, \bibinfo {author} {\bibfnamefont {Tom}\ \bibnamefont
  {Erez}}, \bibinfo {author} {\bibfnamefont {Yuval}\ \bibnamefont {Tassa}},
  \bibinfo {author} {\bibfnamefont {David}\ \bibnamefont {Silver}}, and\
  \bibinfo {author} {\bibfnamefont {Daan}\ \bibnamefont {Wierstra}}} (\bibinfo
  {year} {2015}),\ \bibfield  {title} {\enquote {\bibinfo {title} {Continuous
  control with deep reinforcement learning},}\ }\href
  {https://doi.org/10.48550/arXiv.1509.02971} {\bibinfo  {journal} {arXiv
  preprint arXiv:1509.02971}\ }\BibitemShut {NoStop}%
\bibitem [{\citenamefont {Liu}\ \emph {et~al.}(2024)\citenamefont {Liu},
  \citenamefont {Lin}, \citenamefont {Yang}, \citenamefont {Chen},\ and\
  \citenamefont {Hsieh}}]{liu2024qtrl}%
  \BibitemOpen
\bibfield  {journal} {  }\bibfield  {author} {\bibinfo {author} {\bibnamefont
  {Liu}, \bibfnamefont {Chen-Yu}}, \bibinfo {author} {\bibfnamefont
  {Chu-Hsuan~Abraham}\ \bibnamefont {Lin}}, \bibinfo {author} {\bibfnamefont
  {Chao-Han~Huck}\ \bibnamefont {Yang}}, \bibinfo {author} {\bibfnamefont
  {Kuan-Cheng}\ \bibnamefont {Chen}}, and\ \bibinfo {author} {\bibfnamefont
  {Min-Hsiu}\ \bibnamefont {Hsieh}}} (\bibinfo {year} {2024}),\ \bibfield
  {title} {\enquote {\bibinfo {title} {Qtrl: Toward practical quantum
  reinforcement learning via quantum-train},}\ }\href
  {https://arxiv.org/abs/2407.06103} {\bibinfo  {journal} {arXiv preprint
  arXiv:2407.06103}\ }\BibitemShut {NoStop}%
\bibitem [{\citenamefont {Liu}\ \emph {et~al.}(2025)\citenamefont {Liu},
  \citenamefont {Tan}, \citenamefont {Kuang},\ and\ \citenamefont
  {Liao}}]{liu2025deterministic}%
  \BibitemOpen
\bibfield  {journal} {  }\bibfield  {author} {\bibinfo {author} {\bibnamefont
  {Liu}, \bibfnamefont {Yu-Hong}}, \bibinfo {author} {\bibfnamefont
  {Qing-Shou}\ \bibnamefont {Tan}}, \bibinfo {author} {\bibfnamefont {Le-Man}\
  \bibnamefont {Kuang}}, and\ \bibinfo {author} {\bibfnamefont {Jie-Qiao}\
  \bibnamefont {Liao}}} (\bibinfo {year} {2025}),\ \bibfield  {title} {\enquote
  {\bibinfo {title} {Deterministic generation of non-classical mechanical
  states in cavity optomechanics via reinforcement learning},}\ }\href
  {https://arxiv.org/abs/2502.08350} {\bibinfo  {journal} {arXiv preprint
  arXiv:2502.08350}\ }\BibitemShut {NoStop}%
\bibitem [{\citenamefont {Lockwood}\ and\ \citenamefont
  {Si}(2020)}]{lockwood2020reinforcement}%
  \BibitemOpen
\bibfield  {journal} {  }\bibfield  {author} {\bibinfo {author} {\bibnamefont
  {Lockwood}, \bibfnamefont {Owen}}, and\ \bibinfo {author} {\bibfnamefont
  {Mei}\ \bibnamefont {Si}}} (\bibinfo {year} {2020}),\ \bibfield  {title}
  {\enquote {\bibinfo {title} {Reinforcement learning with quantum variational
  circuits},}\ }\href {https://arxiv.org/abs/2008.07524} {\bibinfo  {journal}
  {arXiv:2008.07524}\ }\BibitemShut {NoStop}%
\bibitem [{\citenamefont {Ma}\ \emph {et~al.}(2025)\citenamefont {Ma},
  \citenamefont {Qi}, \citenamefont {Petersen}, \citenamefont {Wu},
  \citenamefont {Rabitz},\ and\ \citenamefont {Dong}}]{ma2025machine}%
  \BibitemOpen
\bibfield  {journal} {  }\bibfield  {author} {\bibinfo {author} {\bibnamefont
  {Ma}, \bibfnamefont {Hailan}}, \bibinfo {author} {\bibfnamefont
  {Bo}~\bibnamefont {Qi}}, \bibinfo {author} {\bibfnamefont {Ian~R}\
  \bibnamefont {Petersen}}, \bibinfo {author} {\bibfnamefont {Re-Bing}\
  \bibnamefont {Wu}}, \bibinfo {author} {\bibfnamefont {Herschel}\ \bibnamefont
  {Rabitz}}, and\ \bibinfo {author} {\bibfnamefont {Daoyi}\ \bibnamefont
  {Dong}}} (\bibinfo {year} {2025}),\ \bibfield  {title} {\enquote {\bibinfo
  {title} {Machine learning for estimation and control of quantum systems},}\
  }\href {https://arxiv.org/abs/2503.03164} {\bibinfo  {journal} {arXiv
  preprint arXiv:2503.03164}\ }\BibitemShut {NoStop}%
\bibitem [{\citenamefont {Mackeprang}\ \emph {et~al.}(2020)\citenamefont
  {Mackeprang}, \citenamefont {Dasari},\ and\ \citenamefont
  {Wrachtrup}}]{mackeprang2020reinforcement}%
  \BibitemOpen
\bibfield  {journal} {  }\bibfield  {author} {\bibinfo {author} {\bibnamefont
  {Mackeprang}, \bibfnamefont {Jelena}}, \bibinfo {author} {\bibfnamefont
  {Durga B~Rao}\ \bibnamefont {Dasari}}, and\ \bibinfo {author} {\bibfnamefont
  {J{\"o}rg}\ \bibnamefont {Wrachtrup}}} (\bibinfo {year} {2020}),\ \bibfield
  {title} {\enquote {\bibinfo {title} {A reinforcement learning approach for
  quantum state engineering},}\ }\href
  {https://doi.org/10.1007/s42484-020-00016-8} {\bibfield  {journal} {\bibinfo
  {journal} {Quantum Machine Intelligence}\ }\textbf {\bibinfo {volume} {2}},\
  \bibinfo {pages} {1--14}}\BibitemShut {NoStop}%
\bibitem [{\citenamefont {Margolis}\ \emph {et~al.}(2024)\citenamefont
  {Margolis}, \citenamefont {Yang}, \citenamefont {Paigwar}, \citenamefont
  {Chen},\ and\ \citenamefont {Agrawal}}]{margolis2024rapid}%
  \BibitemOpen
  \bibfield  {author} {\bibinfo {author} {\bibnamefont {Margolis},
  \bibfnamefont {Gabriel~B}}, \bibinfo {author} {\bibfnamefont
  {Ge}~\bibnamefont {Yang}}, \bibinfo {author} {\bibfnamefont {Kartik}\
  \bibnamefont {Paigwar}}, \bibinfo {author} {\bibfnamefont {Tao}\ \bibnamefont
  {Chen}}, and\ \bibinfo {author} {\bibfnamefont {Pulkit}\ \bibnamefont
  {Agrawal}}} (\bibinfo {year} {2024}),\ \bibfield  {title} {\enquote {\bibinfo
  {title} {Rapid locomotion via reinforcement learning},}\ }\href
  {https://arxiv.org/abs/2205.02824} {\bibfield  {journal} {\bibinfo  {journal}
  {The International Journal of Robotics Research}\ }\textbf {\bibinfo {volume}
  {43}}~(\bibinfo {number} {4}),\ \bibinfo {pages} {572--587}}\BibitemShut
  {NoStop}%
\bibitem [{\citenamefont {Matekole}\ \emph {et~al.}(2022)\citenamefont
  {Matekole}, \citenamefont {Ye}, \citenamefont {Iyer},\ and\ \citenamefont
  {Chen}}]{matekole2022decoding}%
  \BibitemOpen
  \bibfield  {author} {\bibinfo {author} {\bibnamefont {Matekole},
  \bibfnamefont {Elisha~Siddiqui}}, \bibinfo {author} {\bibfnamefont {Esther}\
  \bibnamefont {Ye}}, \bibinfo {author} {\bibfnamefont {Ramya}\ \bibnamefont
  {Iyer}}, and\ \bibinfo {author} {\bibfnamefont {Samuel Yen-Chi}\ \bibnamefont
  {Chen}}} (\bibinfo {year} {2022}),\ \bibfield  {title} {\enquote {\bibinfo
  {title} {Decoding surface codes with deep reinforcement learning and
  probabilistic policy reuse},}\ }\href {https://arxiv.org/abs/2212.11890}
  {\bibinfo  {journal} {arXiv preprint arXiv:2212.11890}\ }\BibitemShut
  {NoStop}%
\bibitem [{\citenamefont {Mattick}\ \emph {et~al.}(2025)\citenamefont
  {Mattick}, \citenamefont {Periyasamy}, \citenamefont {Ufrecht}, \citenamefont
  {Dubey}, \citenamefont {Mutschler}, \citenamefont {Plinge},\ and\
  \citenamefont {Scherer}}]{mattick2025optimizing}%
  \BibitemOpen
\bibfield  {journal} {  }\bibfield  {author} {\bibinfo {author} {\bibnamefont
  {Mattick}, \bibfnamefont {Alexander}}, \bibinfo {author} {\bibfnamefont
  {Maniraman}\ \bibnamefont {Periyasamy}}, \bibinfo {author} {\bibfnamefont
  {Christian}\ \bibnamefont {Ufrecht}}, \bibinfo {author} {\bibfnamefont
  {Abhishek~Y}\ \bibnamefont {Dubey}}, \bibinfo {author} {\bibfnamefont
  {Christopher}\ \bibnamefont {Mutschler}}, \bibinfo {author} {\bibfnamefont
  {Axel}\ \bibnamefont {Plinge}}, and\ \bibinfo {author} {\bibfnamefont
  {Daniel~D}\ \bibnamefont {Scherer}}} (\bibinfo {year} {2025}),\ \bibfield
  {title} {\enquote {\bibinfo {title} {Optimizing quantum circuits via zx
  diagrams using reinforcement learning and graph neural networks},}\ }\href
  {https://arxiv.org/abs/2504.03429} {\bibinfo  {journal} {arXiv preprint
  arXiv:2504.03429}\ }\BibitemShut {NoStop}%
\bibitem [{\citenamefont {Mehta}\ \emph {et~al.}(2019)\citenamefont {Mehta},
  \citenamefont {Bukov}, \citenamefont {Wang}, \citenamefont {Day},
  \citenamefont {Richardson}, \citenamefont {Fisher},\ and\ \citenamefont
  {Schwab}}]{mehta2019high}%
  \BibitemOpen
\bibfield  {journal} {  }\bibfield  {author} {\bibinfo {author} {\bibnamefont
  {Mehta}, \bibfnamefont {Pankaj}}, \bibinfo {author} {\bibfnamefont {Marin}\
  \bibnamefont {Bukov}}, \bibinfo {author} {\bibfnamefont {Ching-Hao}\
  \bibnamefont {Wang}}, \bibinfo {author} {\bibfnamefont {Alexandre~GR}\
  \bibnamefont {Day}}, \bibinfo {author} {\bibfnamefont {Clint}\ \bibnamefont
  {Richardson}}, \bibinfo {author} {\bibfnamefont {Charles~K}\ \bibnamefont
  {Fisher}}, and\ \bibinfo {author} {\bibfnamefont {David~J}\ \bibnamefont
  {Schwab}}} (\bibinfo {year} {2019}),\ \bibfield  {title} {\enquote {\bibinfo
  {title} {A high-bias, low-variance introduction to machine learning for
  physicists},}\ }\href {https://doi.org/10.1016/j.physrep.2019.03.001}
  {\bibfield  {journal} {\bibinfo  {journal} {Physics reports}\ }\textbf
  {\bibinfo {volume} {810}},\ \bibinfo {pages} {1--124}}\BibitemShut {NoStop}%
\bibitem [{\citenamefont {Melnikov}\ \emph {et~al.}(2018)\citenamefont
  {Melnikov}, \citenamefont {Poulsen~Nautrup}, \citenamefont {Krenn},
  \citenamefont {Dunjko}, \citenamefont {Tiersch}, \citenamefont {Zeilinger},\
  and\ \citenamefont {Briegel}}]{melnikov2018active}%
  \BibitemOpen
  \bibfield  {author} {\bibinfo {author} {\bibnamefont {Melnikov},
  \bibfnamefont {Alexey~A}}, \bibinfo {author} {\bibfnamefont {Hendrik}\
  \bibnamefont {Poulsen~Nautrup}}, \bibinfo {author} {\bibfnamefont {Mario}\
  \bibnamefont {Krenn}}, \bibinfo {author} {\bibfnamefont {Vedran}\
  \bibnamefont {Dunjko}}, \bibinfo {author} {\bibfnamefont {Markus}\
  \bibnamefont {Tiersch}}, \bibinfo {author} {\bibfnamefont {Anton}\
  \bibnamefont {Zeilinger}}, and\ \bibinfo {author} {\bibfnamefont {Hans~J}\
  \bibnamefont {Briegel}}} (\bibinfo {year} {2018}),\ \bibfield  {title}
  {\enquote {\bibinfo {title} {Active learning machine learns to create new
  quantum experiments},}\ }\href {https://doi.org/10.1073/pnas.1714936115}
  {\bibfield  {journal} {\bibinfo  {journal} {Proceedings of the National
  Academy of Sciences}\ }\textbf {\bibinfo {volume} {115}}~(\bibinfo {number}
  {6}),\ \bibinfo {pages} {1221--1226}}\BibitemShut {NoStop}%
\bibitem [{\citenamefont {Metz}\ and\ \citenamefont
  {Bukov}(2023)}]{metz2023self}%
  \BibitemOpen
  \bibfield  {author} {\bibinfo {author} {\bibnamefont {Metz}, \bibfnamefont
  {Friederike}}, and\ \bibinfo {author} {\bibfnamefont {Marin}\ \bibnamefont
  {Bukov}}} (\bibinfo {year} {2023}),\ \bibfield  {title} {\enquote {\bibinfo
  {title} {Self-correcting quantum many-body control using reinforcement
  learning with tensor networks},}\ }\href
  {https://doi.org/10.1038/s42256-023-00687-5} {\bibfield  {journal} {\bibinfo
  {journal} {Nature Machine Intelligence}\ }\textbf {\bibinfo {volume}
  {5}}~(\bibinfo {number} {7}),\ \bibinfo {pages} {780--791}}\BibitemShut
  {NoStop}%
\bibitem [{\citenamefont {Mills}\ \emph {et~al.}(2020)\citenamefont {Mills},
  \citenamefont {Ronagh},\ and\ \citenamefont {Tamblyn}}]{mills2020finding}%
  \BibitemOpen
  \bibfield  {author} {\bibinfo {author} {\bibnamefont {Mills}, \bibfnamefont
  {Kyle}}, \bibinfo {author} {\bibfnamefont {Pooya}\ \bibnamefont {Ronagh}},
  and\ \bibinfo {author} {\bibfnamefont {Isaac}\ \bibnamefont {Tamblyn}}}
  (\bibinfo {year} {2020}),\ \bibfield  {title} {\enquote {\bibinfo {title}
  {Finding the ground state of spin hamiltonians with reinforcement
  learning},}\ }\href {https://doi.org/10.1038/s42256-020-0226-x} {\bibfield
  {journal} {\bibinfo  {journal} {Nature Machine Intelligence}\ }\textbf
  {\bibinfo {volume} {2}}~(\bibinfo {number} {9}),\ \bibinfo {pages}
  {509--517}}\BibitemShut {NoStop}%
\bibitem [{\citenamefont {Mirhoseini}\ \emph {et~al.}(2021)\citenamefont
  {Mirhoseini}, \citenamefont {Goldie}, \citenamefont {Yazgan}, \citenamefont
  {Jiang}, \citenamefont {Songhori}, \citenamefont {Wang}, \citenamefont {Lee},
  \citenamefont {Johnson}, \citenamefont {Pathak}, \citenamefont {Nazi} \emph
  {et~al.}}]{mirhoseini2021graph}%
  \BibitemOpen
  \bibfield  {author} {\bibinfo {author} {\bibnamefont {Mirhoseini},
  \bibfnamefont {Azalia}}, \bibinfo {author} {\bibfnamefont {Anna}\
  \bibnamefont {Goldie}}, \bibinfo {author} {\bibfnamefont {Mustafa}\
  \bibnamefont {Yazgan}}, \bibinfo {author} {\bibfnamefont {Joe~Wenjie}\
  \bibnamefont {Jiang}}, \bibinfo {author} {\bibfnamefont {Ebrahim}\
  \bibnamefont {Songhori}}, \bibinfo {author} {\bibfnamefont {Shen}\
  \bibnamefont {Wang}}, \bibinfo {author} {\bibfnamefont {Young-Joon}\
  \bibnamefont {Lee}}, \bibinfo {author} {\bibfnamefont {Eric}\ \bibnamefont
  {Johnson}}, \bibinfo {author} {\bibfnamefont {Omkar}\ \bibnamefont {Pathak}},
  \bibinfo {author} {\bibfnamefont {Azade}\ \bibnamefont {Nazi}},  \emph
  {et~al.}} (\bibinfo {year} {2021}),\ \bibfield  {title} {\enquote {\bibinfo
  {title} {A graph placement methodology for fast chip design},}\ }\href
  {https://www.nature.com/articles/s41586-021-03544-w} {\bibfield  {journal}
  {\bibinfo  {journal} {Nature}\ }\textbf {\bibinfo {volume} {594}}~(\bibinfo
  {number} {7862}),\ \bibinfo {pages} {207--212}}\BibitemShut {NoStop}%
\bibitem [{\citenamefont {Mnih}\ \emph {et~al.}(2016)\citenamefont {Mnih},
  \citenamefont {Badia}, \citenamefont {Mirza}, \citenamefont {Graves},
  \citenamefont {Lillicrap}, \citenamefont {Harley}, \citenamefont {Silver},\
  and\ \citenamefont {Kavukcuoglu}}]{mnih2016asynchronous}%
  \BibitemOpen
  \bibfield  {author} {\bibinfo {author} {\bibnamefont {Mnih}, \bibfnamefont
  {Volodymyr}}, \bibinfo {author} {\bibfnamefont {Adria~Puigdomenech}\
  \bibnamefont {Badia}}, \bibinfo {author} {\bibfnamefont {Mehdi}\ \bibnamefont
  {Mirza}}, \bibinfo {author} {\bibfnamefont {Alex}\ \bibnamefont {Graves}},
  \bibinfo {author} {\bibfnamefont {Timothy}\ \bibnamefont {Lillicrap}},
  \bibinfo {author} {\bibfnamefont {Tim}\ \bibnamefont {Harley}}, \bibinfo
  {author} {\bibfnamefont {David}\ \bibnamefont {Silver}}, and\ \bibinfo
  {author} {\bibfnamefont {Koray}\ \bibnamefont {Kavukcuoglu}}} (\bibinfo
  {year} {2016}),\ \bibfield  {title} {\enquote {\bibinfo {title} {Asynchronous
  methods for deep reinforcement learning},}\ }\bibfield  {booktitle} {\emph
  {\bibinfo {booktitle} {International conference on machine learning}},\
  }\href {https://doi.org/10.48550/arXiv.1602.01783} {\bibinfo  {journal}
  {PMLR}\ ,\ \bibinfo {pages} {1928--1937}}\BibitemShut {NoStop}%
\bibitem [{\citenamefont {Mnih}\ \emph {et~al.}(2013)\citenamefont {Mnih},
  \citenamefont {Kavukcuoglu}, \citenamefont {Silver}, \citenamefont {Graves},
  \citenamefont {Antonoglou}, \citenamefont {Wierstra},\ and\ \citenamefont
  {Riedmiller}}]{mnih2013playing}%
  \BibitemOpen
\bibfield  {journal} {  }\bibfield  {author} {\bibinfo {author} {\bibnamefont
  {Mnih}, \bibfnamefont {Volodymyr}}, \bibinfo {author} {\bibfnamefont {Koray}\
  \bibnamefont {Kavukcuoglu}}, \bibinfo {author} {\bibfnamefont {David}\
  \bibnamefont {Silver}}, \bibinfo {author} {\bibfnamefont {Alex}\ \bibnamefont
  {Graves}}, \bibinfo {author} {\bibfnamefont {Ioannis}\ \bibnamefont
  {Antonoglou}}, \bibinfo {author} {\bibfnamefont {Daan}\ \bibnamefont
  {Wierstra}}, and\ \bibinfo {author} {\bibfnamefont {Martin}\ \bibnamefont
  {Riedmiller}}} (\bibinfo {year} {2013}),\ \bibfield  {title} {\enquote
  {\bibinfo {title} {Playing atari with deep reinforcement learning},}\ }\href
  {https://arxiv.org/abs/1312.5602} {\bibinfo  {journal} {arXiv preprint
  arXiv:1312.5602}\ }\BibitemShut {NoStop}%
\bibitem [{\citenamefont {Mnih}\ \emph {et~al.}(2015)\citenamefont {Mnih},
  \citenamefont {Kavukcuoglu}, \citenamefont {Silver}, \citenamefont {Rusu},
  \citenamefont {Veness}, \citenamefont {Bellemare}, \citenamefont {Graves},
  \citenamefont {Riedmiller}, \citenamefont {Fidjeland}, \citenamefont
  {Ostrovski} \emph {et~al.}}]{mnih2015human}%
  \BibitemOpen
\bibfield  {journal} {  }\bibfield  {author} {\bibinfo {author} {\bibnamefont
  {Mnih}, \bibfnamefont {Volodymyr}}, \bibinfo {author} {\bibfnamefont {Koray}\
  \bibnamefont {Kavukcuoglu}}, \bibinfo {author} {\bibfnamefont {David}\
  \bibnamefont {Silver}}, \bibinfo {author} {\bibfnamefont {Andrei~A}\
  \bibnamefont {Rusu}}, \bibinfo {author} {\bibfnamefont {Joel}\ \bibnamefont
  {Veness}}, \bibinfo {author} {\bibfnamefont {Marc~G}\ \bibnamefont
  {Bellemare}}, \bibinfo {author} {\bibfnamefont {Alex}\ \bibnamefont
  {Graves}}, \bibinfo {author} {\bibfnamefont {Martin}\ \bibnamefont
  {Riedmiller}}, \bibinfo {author} {\bibfnamefont {Andreas~K}\ \bibnamefont
  {Fidjeland}}, \bibinfo {author} {\bibfnamefont {Georg}\ \bibnamefont
  {Ostrovski}},  \emph {et~al.}} (\bibinfo {year} {2015}),\ \bibfield  {title}
  {\enquote {\bibinfo {title} {Human-level control through deep reinforcement
  learning},}\ }\href {https://doi.org/10.1038/nature14236} {\bibfield
  {journal} {\bibinfo  {journal} {nature}\ }\textbf {\bibinfo {volume}
  {518}}~(\bibinfo {number} {7540}),\ \bibinfo {pages} {529--533}}\BibitemShut
  {NoStop}%
\bibitem [{\citenamefont {Moro}\ \emph {et~al.}(2021)\citenamefont {Moro},
  \citenamefont {Paris}, \citenamefont {Restelli},\ and\ \citenamefont
  {Prati}}]{moro2021quantum}%
  \BibitemOpen
  \bibfield  {author} {\bibinfo {author} {\bibnamefont {Moro}, \bibfnamefont
  {Lorenzo}}, \bibinfo {author} {\bibfnamefont {Matteo~GA}\ \bibnamefont
  {Paris}}, \bibinfo {author} {\bibfnamefont {Marcello}\ \bibnamefont
  {Restelli}}, and\ \bibinfo {author} {\bibfnamefont {Enrico}\ \bibnamefont
  {Prati}}} (\bibinfo {year} {2021}),\ \bibfield  {title} {\enquote {\bibinfo
  {title} {Quantum compiling by deep reinforcement learning},}\ }\href
  {https://doi.org/10.1038/s42005-021-00684-3} {\bibfield  {journal} {\bibinfo
  {journal} {Communications Physics}\ }\textbf {\bibinfo {volume}
  {4}}~(\bibinfo {number} {1}),\ \bibinfo {pages} {178}}\BibitemShut {NoStop}%
\bibitem [{\citenamefont {Morral-Yepes}\ \emph {et~al.}(2024)\citenamefont
  {Morral-Yepes}, \citenamefont {Smith}, \citenamefont {Sondhi},\ and\
  \citenamefont {Pollmann}}]{morral-yepes2024entanglement}%
  \BibitemOpen
  \bibfield  {author} {\bibinfo {author} {\bibnamefont {Morral-Yepes},
  \bibfnamefont {Ra\'ul}}, \bibinfo {author} {\bibfnamefont {Adam}\
  \bibnamefont {Smith}}, \bibinfo {author} {\bibfnamefont {S.L.}\ \bibnamefont
  {Sondhi}}, and\ \bibinfo {author} {\bibfnamefont {Frank}\ \bibnamefont
  {Pollmann}}} (\bibinfo {year} {2024}),\ \bibfield  {title} {\enquote
  {\bibinfo {title} {Entanglement transitions in unitary circuit games},}\
  }\href {https://doi.org/10.1103/PRXQuantum.5.010309} {\bibfield  {journal}
  {\bibinfo  {journal} {PRX Quantum}\ }\textbf {\bibinfo {volume} {5}},\
  \bibinfo {pages} {010309}}\BibitemShut {NoStop}%
\bibitem [{\citenamefont {N{\"a}gele}\ and\ \citenamefont
  {Marquardt}(2024)}]{nagele2024optimizing}%
  \BibitemOpen
  \bibfield  {author} {\bibinfo {author} {\bibnamefont {N{\"a}gele},
  \bibfnamefont {Maximilian}}, and\ \bibinfo {author} {\bibfnamefont {Florian}\
  \bibnamefont {Marquardt}}} (\bibinfo {year} {2024}),\ \bibfield  {title}
  {\enquote {\bibinfo {title} {Optimizing zx-diagrams with deep reinforcement
  learning},}\ }\href {https://doi.org/10.1088/2632-2153/ad76f7} {\bibfield
  {journal} {\bibinfo  {journal} {Machine Learning: Science and Technology}\
  }\textbf {\bibinfo {volume} {5}}~(\bibinfo {number} {3}),\ \bibinfo {pages}
  {035077}}\BibitemShut {NoStop}%
\bibitem [{\citenamefont {Nautrup}\ \emph {et~al.}(2019)\citenamefont
  {Nautrup}, \citenamefont {Delfosse}, \citenamefont {Dunjko}, \citenamefont
  {Briegel},\ and\ \citenamefont {Friis}}]{nautrup2019optimizing}%
  \BibitemOpen
  \bibfield  {author} {\bibinfo {author} {\bibnamefont {Nautrup}, \bibfnamefont
  {Hendrik~Poulsen}}, \bibinfo {author} {\bibfnamefont {Nicolas}\ \bibnamefont
  {Delfosse}}, \bibinfo {author} {\bibfnamefont {Vedran}\ \bibnamefont
  {Dunjko}}, \bibinfo {author} {\bibfnamefont {Hans~J}\ \bibnamefont
  {Briegel}}, and\ \bibinfo {author} {\bibfnamefont {Nicolai}\ \bibnamefont
  {Friis}}} (\bibinfo {year} {2019}),\ \bibfield  {title} {\enquote {\bibinfo
  {title} {Optimizing quantum error correction codes with reinforcement
  learning},}\ }\href {https://doi.org/10.22331/q-2019-12-16-215} {\bibfield
  {journal} {\bibinfo  {journal} {Quantum}\ }\textbf {\bibinfo {volume} {3}},\
  \bibinfo {pages} {215}}\BibitemShut {NoStop}%
\bibitem [{\citenamefont {Nguyen}\ \emph {et~al.}(2024)\citenamefont {Nguyen},
  \citenamefont {Motzoi}, \citenamefont {Metcalf}, \citenamefont {Whaley},
  \citenamefont {Bukov},\ and\ \citenamefont
  {Schmitt}}]{nguyen2024reinforcement}%
  \BibitemOpen
  \bibfield  {author} {\bibinfo {author} {\bibnamefont {Nguyen}, \bibfnamefont
  {Ho~Nam}}, \bibinfo {author} {\bibfnamefont {Felix}\ \bibnamefont {Motzoi}},
  \bibinfo {author} {\bibfnamefont {Mekena}\ \bibnamefont {Metcalf}}, \bibinfo
  {author} {\bibfnamefont {K~Birgitta}\ \bibnamefont {Whaley}}, \bibinfo
  {author} {\bibfnamefont {Marin}\ \bibnamefont {Bukov}}, and\ \bibinfo
  {author} {\bibfnamefont {Markus}\ \bibnamefont {Schmitt}}} (\bibinfo {year}
  {2024}),\ \bibfield  {title} {\enquote {\bibinfo {title} {Reinforcement
  learning pulses for transmon qubit entangling gates},}\ }\href
  {https://doi.org/10.1088/2632-2153/ad4f4d} {\bibfield  {journal} {\bibinfo
  {journal} {Machine Learning: Science and Technology}\ }\textbf {\bibinfo
  {volume} {5}}~(\bibinfo {number} {2}),\ \bibinfo {pages}
  {025066}}\BibitemShut {NoStop}%
\bibitem [{\citenamefont {Nguyen}\ \emph {et~al.}(2021)\citenamefont {Nguyen},
  \citenamefont {Orbell}, \citenamefont {Lennon}, \citenamefont {Moon},
  \citenamefont {Vigneau}, \citenamefont {Camenzind}, \citenamefont {Yu},
  \citenamefont {Zumb{\"u}hl}, \citenamefont {Briggs}, \citenamefont {Osborne}
  \emph {et~al.}}]{nguyen2021deep}%
  \BibitemOpen
  \bibfield  {author} {\bibinfo {author} {\bibnamefont {Nguyen}, \bibfnamefont
  {V}}, \bibinfo {author} {\bibfnamefont {SB}~\bibnamefont {Orbell}}, \bibinfo
  {author} {\bibfnamefont {Dominic~T}\ \bibnamefont {Lennon}}, \bibinfo
  {author} {\bibfnamefont {Hyungil}\ \bibnamefont {Moon}}, \bibinfo {author}
  {\bibfnamefont {Florian}\ \bibnamefont {Vigneau}}, \bibinfo {author}
  {\bibfnamefont {Leon~C}\ \bibnamefont {Camenzind}}, \bibinfo {author}
  {\bibfnamefont {Liuqi}\ \bibnamefont {Yu}}, \bibinfo {author} {\bibfnamefont
  {Dominik~M}\ \bibnamefont {Zumb{\"u}hl}}, \bibinfo {author} {\bibfnamefont
  {G~Andrew~D}\ \bibnamefont {Briggs}}, \bibinfo {author} {\bibfnamefont
  {Michael~A}\ \bibnamefont {Osborne}},  \emph {et~al.}} (\bibinfo {year}
  {2021}),\ \bibfield  {title} {\enquote {\bibinfo {title} {Deep reinforcement
  learning for efficient measurement of quantum devices},}\ }\href
  {https://doi.org/10.1038/s41534-021-00434-x} {\bibfield  {journal} {\bibinfo
  {journal} {npj Quantum Information}\ }\textbf {\bibinfo {volume}
  {7}}~(\bibinfo {number} {1}),\ \bibinfo {pages} {100}}\BibitemShut {NoStop}%
\bibitem [{\citenamefont {Nielsen}\ and\ \citenamefont
  {Chuang}(2010)}]{nielsen2010quantum}%
  \BibitemOpen
  \bibfield  {author} {\bibinfo {author} {\bibnamefont {Nielsen}, \bibfnamefont
  {Michael~A}}, and\ \bibinfo {author} {\bibfnamefont {Isaac~L}\ \bibnamefont
  {Chuang}}} (\bibinfo {year} {2010}),\ \href@noop {} {\emph {\bibinfo {title}
  {Quantum computation and quantum information}}}\ (\bibinfo  {publisher}
  {Cambridge university press})\BibitemShut {NoStop}%
\bibitem [{\citenamefont {Niu}\ \emph {et~al.}(2019)\citenamefont {Niu},
  \citenamefont {Boixo}, \citenamefont {Smelyanskiy},\ and\ \citenamefont
  {Neven}}]{niu2019universal}%
  \BibitemOpen
  \bibfield  {author} {\bibinfo {author} {\bibnamefont {Niu}, \bibfnamefont
  {Murphy~Yuezhen}}, \bibinfo {author} {\bibfnamefont {Sergio}\ \bibnamefont
  {Boixo}}, \bibinfo {author} {\bibfnamefont {Vadim~N}\ \bibnamefont
  {Smelyanskiy}}, and\ \bibinfo {author} {\bibfnamefont {Hartmut}\ \bibnamefont
  {Neven}}} (\bibinfo {year} {2019}),\ \bibfield  {title} {\enquote {\bibinfo
  {title} {Universal quantum control through deep reinforcement learning},}\
  }\href {https://doi.org/10.1038/s41534-019-0141-3} {\bibfield  {journal}
  {\bibinfo  {journal} {npj Quantum Information}\ }\textbf {\bibinfo {volume}
  {5}}~(\bibinfo {number} {1}),\ \bibinfo {pages} {33}}\BibitemShut {NoStop}%
\bibitem [{\citenamefont {O'Connor}\ \emph {et~al.}(2025)\citenamefont
  {O'Connor}, \citenamefont {Ma},\ and\ \citenamefont
  {Genoni}}]{o2025bounding}%
  \BibitemOpen
  \bibfield  {author} {\bibinfo {author} {\bibnamefont {O'Connor},
  \bibfnamefont {Eoin}}, \bibinfo {author} {\bibfnamefont {Hailan}\
  \bibnamefont {Ma}}, and\ \bibinfo {author} {\bibfnamefont {Marco~G}\
  \bibnamefont {Genoni}}} (\bibinfo {year} {2025}),\ \bibfield  {title}
  {\enquote {\bibinfo {title} {Bounding fidelity in quantum feedback control:
  Theory and applications to dicke state preparation},}\ }\href
  {https://doi.org/10.48550/arXiv.2503.19151} {\bibinfo  {journal} {arXiv
  preprint arXiv:2503.19151}\ }\BibitemShut {NoStop}%
\bibitem [{\citenamefont {Oh}\ \emph {et~al.}(2025)\citenamefont {Oh},
  \citenamefont {Farquhar}, \citenamefont {Kemaev}, \citenamefont {Calian},
  \citenamefont {Hessel}, \citenamefont {Zintgraf}, \citenamefont {Singh},
  \citenamefont {Van~Hasselt},\ and\ \citenamefont
  {Silver}}]{oh2025discovering}%
  \BibitemOpen
\bibfield  {journal} {  }\bibfield  {author} {\bibinfo {author} {\bibnamefont
  {Oh}, \bibfnamefont {Junhyuk}}, \bibinfo {author} {\bibfnamefont {Greg}\
  \bibnamefont {Farquhar}}, \bibinfo {author} {\bibfnamefont {Iurii}\
  \bibnamefont {Kemaev}}, \bibinfo {author} {\bibfnamefont {Dan~A}\
  \bibnamefont {Calian}}, \bibinfo {author} {\bibfnamefont {Matteo}\
  \bibnamefont {Hessel}}, \bibinfo {author} {\bibfnamefont {Luisa}\
  \bibnamefont {Zintgraf}}, \bibinfo {author} {\bibfnamefont {Satinder}\
  \bibnamefont {Singh}}, \bibinfo {author} {\bibfnamefont {Hado}\ \bibnamefont
  {Van~Hasselt}}, and\ \bibinfo {author} {\bibfnamefont {David}\ \bibnamefont
  {Silver}}} (\bibinfo {year} {2025}),\ \bibfield  {title} {\enquote {\bibinfo
  {title} {Discovering state-of-the-art reinforcement learning algorithms},}\
  }\href {https://doi.org/10.1038/s41586-025-09761-x} {\bibinfo  {journal}
  {Nature}\ ,\ \bibinfo {pages} {1--2}}\BibitemShut {NoStop}%
\bibitem [{\citenamefont {Olle}\ \emph {et~al.}(2025)\citenamefont {Olle},
  \citenamefont {Yevtushenko},\ and\ \citenamefont
  {Marquardt}}]{olle2025scaling}%
  \BibitemOpen
\bibfield  {journal} {  }\bibfield  {author} {\bibinfo {author} {\bibnamefont
  {Olle}, \bibfnamefont {Jan}}, \bibinfo {author} {\bibfnamefont {Oleg~M}\
  \bibnamefont {Yevtushenko}}, and\ \bibinfo {author} {\bibfnamefont {Florian}\
  \bibnamefont {Marquardt}}} (\bibinfo {year} {2025}),\ \bibfield  {title}
  {\enquote {\bibinfo {title} {Scaling the automated discovery of quantum
  circuits via reinforcement learning with gadgets},}\ }\href
  {https://doi.org/10.48550/arXiv.2503.11638} {\bibinfo  {journal} {arXiv
  preprint arXiv:2503.11638}\ }\BibitemShut {NoStop}%
\bibitem [{\citenamefont {Olle}\ \emph {et~al.}(2024)\citenamefont {Olle},
  \citenamefont {Zen}, \citenamefont {Puviani},\ and\ \citenamefont
  {Marquardt}}]{olle2023simultaneous}%
  \BibitemOpen
\bibfield  {journal} {  }\bibfield  {author} {\bibinfo {author} {\bibnamefont
  {Olle}, \bibfnamefont {Jan}}, \bibinfo {author} {\bibfnamefont {Remmy}\
  \bibnamefont {Zen}}, \bibinfo {author} {\bibfnamefont {Matteo}\ \bibnamefont
  {Puviani}}, and\ \bibinfo {author} {\bibfnamefont {Florian}\ \bibnamefont
  {Marquardt}}} (\bibinfo {year} {2024}),\ \bibfield  {title} {\enquote
  {\bibinfo {title} {Simultaneous discovery of quantum error correction codes
  and encoders with a noise-aware reinforcement learning agent},}\ }\href
  {https://doi.org/10.1038/s41534-024-00920-y} {\bibfield  {journal} {\bibinfo
  {journal} {npj Quantum Information}\ }\textbf {\bibinfo {volume}
  {10}}~(\bibinfo {number} {1}),\ \bibinfo {pages} {126}}\BibitemShut {NoStop}%
\bibitem [{\citenamefont {Ouyang}\ \emph {et~al.}(2022)\citenamefont {Ouyang},
  \citenamefont {Wu}, \citenamefont {Jiang}, \citenamefont {Almeida},
  \citenamefont {Wainwright}, \citenamefont {Mishkin}, \citenamefont {Zhang},
  \citenamefont {Agarwal}, \citenamefont {Slama}, \citenamefont {Ray} \emph
  {et~al.}}]{ouyang2022training}%
  \BibitemOpen
  \bibfield  {author} {\bibinfo {author} {\bibnamefont {Ouyang}, \bibfnamefont
  {Long}}, \bibinfo {author} {\bibfnamefont {Jeffrey}\ \bibnamefont {Wu}},
  \bibinfo {author} {\bibfnamefont {Xu}~\bibnamefont {Jiang}}, \bibinfo
  {author} {\bibfnamefont {Diogo}\ \bibnamefont {Almeida}}, \bibinfo {author}
  {\bibfnamefont {Carroll}\ \bibnamefont {Wainwright}}, \bibinfo {author}
  {\bibfnamefont {Pamela}\ \bibnamefont {Mishkin}}, \bibinfo {author}
  {\bibfnamefont {Chong}\ \bibnamefont {Zhang}}, \bibinfo {author}
  {\bibfnamefont {Sandhini}\ \bibnamefont {Agarwal}}, \bibinfo {author}
  {\bibfnamefont {Katarina}\ \bibnamefont {Slama}}, \bibinfo {author}
  {\bibfnamefont {Alex}\ \bibnamefont {Ray}},  \emph {et~al.}} (\bibinfo {year}
  {2022}),\ \bibfield  {title} {\enquote {\bibinfo {title} {Training language
  models to follow instructions with human feedback},}\ }\href
  {https://doi.org/10.48550/arXiv.2203.02155} {\bibfield  {journal} {\bibinfo
  {journal} {Advances in neural information processing systems}\ }\textbf
  {\bibinfo {volume} {35}},\ \bibinfo {pages} {27730--27744}}\BibitemShut
  {NoStop}%
\bibitem [{\citenamefont {Paparelle}\ \emph {et~al.}(2020)\citenamefont
  {Paparelle}, \citenamefont {Moro},\ and\ \citenamefont
  {Prati}}]{paparelle2020digitally}%
  \BibitemOpen
  \bibfield  {author} {\bibinfo {author} {\bibnamefont {Paparelle},
  \bibfnamefont {Iris}}, \bibinfo {author} {\bibfnamefont {Lorenzo}\
  \bibnamefont {Moro}}, and\ \bibinfo {author} {\bibfnamefont {Enrico}\
  \bibnamefont {Prati}}} (\bibinfo {year} {2020}),\ \bibfield  {title}
  {\enquote {\bibinfo {title} {Digitally stimulated raman passage by deep
  reinforcement learning},}\ }\href
  {https://doi.org/https://doi.org/10.1016/j.physleta.2020.126266} {\bibfield
  {journal} {\bibinfo  {journal} {Physics Letters A}\ }\textbf {\bibinfo
  {volume} {384}}~(\bibinfo {number} {14}),\ \bibinfo {pages}
  {126266}}\BibitemShut {NoStop}%
\bibitem [{\citenamefont {Paparo}\ \emph {et~al.}(2014)\citenamefont {Paparo},
  \citenamefont {Dunjko}, \citenamefont {Makmal}, \citenamefont
  {Martin-Delgado},\ and\ \citenamefont {Briegel}}]{paparo2014quantum}%
  \BibitemOpen
  \bibfield  {author} {\bibinfo {author} {\bibnamefont {Paparo}, \bibfnamefont
  {Giuseppe~Davide}}, \bibinfo {author} {\bibfnamefont {Vedran}\ \bibnamefont
  {Dunjko}}, \bibinfo {author} {\bibfnamefont {Adi}\ \bibnamefont {Makmal}},
  \bibinfo {author} {\bibfnamefont {Miguel~Angel}\ \bibnamefont
  {Martin-Delgado}}, and\ \bibinfo {author} {\bibfnamefont {Hans~J.}\
  \bibnamefont {Briegel}}} (\bibinfo {year} {2014}),\ \bibfield  {title}
  {\enquote {\bibinfo {title} {Quantum speedup for active learning agents},}\
  }\href {https://doi.org/10.1103/PhysRevX.4.031002} {\bibfield  {journal}
  {\bibinfo  {journal} {Phys. Rev. X}\ }\textbf {\bibinfo {volume} {4}},\
  \bibinfo {pages} {031002}}\BibitemShut {NoStop}%
\bibitem [{\citenamefont {Park}\ \emph {et~al.}(2024)\citenamefont {Park},
  \citenamefont {Maskara}, \citenamefont {Kalinowski},\ and\ \citenamefont
  {Lukin}}]{park2024enhancing}%
  \BibitemOpen
  \bibfield  {author} {\bibinfo {author} {\bibnamefont {Park}, \bibfnamefont
  {Mincheol}}, \bibinfo {author} {\bibfnamefont {Nishad}\ \bibnamefont
  {Maskara}}, \bibinfo {author} {\bibfnamefont {Marcin}\ \bibnamefont
  {Kalinowski}}, and\ \bibinfo {author} {\bibfnamefont {Mikhail~D}\
  \bibnamefont {Lukin}}} (\bibinfo {year} {2024}),\ \bibfield  {title}
  {\enquote {\bibinfo {title} {Enhancing quantum memory lifetime with
  measurement-free local error correction and reinforcement learning},}\ }\href
  {https://arxiv.org/abs/2408.09524} {\bibinfo  {journal} {arXiv preprint
  arXiv:2408.09524}\ }\BibitemShut {NoStop}%
\bibitem [{\citenamefont {Pastor}\ \emph {et~al.}(2024)\citenamefont {Pastor},
  \citenamefont {Escofet}, \citenamefont {Rached}, \citenamefont {Alarcón},
  \citenamefont {Barlet-Ros},\ and\ \citenamefont
  {Abadal}}]{pastor2024circuit}%
  \BibitemOpen
\bibfield  {journal} {  }\bibfield  {author} {\bibinfo {author} {\bibnamefont
  {Pastor}, \bibfnamefont {Arnau}}, \bibinfo {author} {\bibfnamefont {Pau}\
  \bibnamefont {Escofet}}, \bibinfo {author} {\bibfnamefont {Sahar~Ben}\
  \bibnamefont {Rached}}, \bibinfo {author} {\bibfnamefont {Eduard}\
  \bibnamefont {Alarcón}}, \bibinfo {author} {\bibfnamefont {Pere}\
  \bibnamefont {Barlet-Ros}}, and\ \bibinfo {author} {\bibfnamefont {Sergi}\
  \bibnamefont {Abadal}}} (\bibinfo {year} {2024}),\ \bibfield  {title}
  {\enquote {\bibinfo {title} {Circuit partitioning for multi-core quantum
  architectures with deep reinforcement learning},}\ }\href
  {https://arxiv.org/abs/2401.17976} {\bibfield  {journal} {\bibinfo  {journal}
  {arXiv}\ }}\Eprint {https://arxiv.org/abs/2401.17976} {2401.17976}
  \BibitemShut {NoStop}%
\bibitem [{\citenamefont {Patel}\ \emph {et~al.}(2024)\citenamefont {Patel},
  \citenamefont {Kundu}, \citenamefont {Ostaszewski}, \citenamefont
  {Bonet-Monroig}, \citenamefont {Dunjko},\ and\ \citenamefont
  {Danaci}}]{patel2024curriculum}%
  \BibitemOpen
  \bibfield  {author} {\bibinfo {author} {\bibnamefont {Patel}, \bibfnamefont
  {Yash~J}}, \bibinfo {author} {\bibfnamefont {Akash}\ \bibnamefont {Kundu}},
  \bibinfo {author} {\bibfnamefont {Mateusz}\ \bibnamefont {Ostaszewski}},
  \bibinfo {author} {\bibfnamefont {Xavier}\ \bibnamefont {Bonet-Monroig}},
  \bibinfo {author} {\bibfnamefont {Vedran}\ \bibnamefont {Dunjko}}, and\
  \bibinfo {author} {\bibfnamefont {Onur}\ \bibnamefont {Danaci}}} (\bibinfo
  {year} {2024}),\ \bibfield  {title} {\enquote {\bibinfo {title} {Curriculum
  reinforcement learning for quantum architecture search under hardware
  errors},}\ }\href {https://arxiv.org/abs/2402.03500} {\bibfield  {journal}
  {\bibinfo  {journal} {arXiv}\ }}\Eprint {https://arxiv.org/abs/2402.03500}
  {2402.03500} \BibitemShut {NoStop}%
\bibitem [{\citenamefont {Piera}\ \emph {et~al.}(2024)\citenamefont {Piera},
  \citenamefont {DeBrota}, \citenamefont {Weiss}, \citenamefont {Lemos},
  \citenamefont {Ara{\'u}jo}, \citenamefont {Aguilar},\ and\ \citenamefont
  {Pienaar}}]{piera2024synthesizing}%
  \BibitemOpen
  \bibfield  {author} {\bibinfo {author} {\bibnamefont {Piera}, \bibfnamefont
  {Rodrigo~S}}, \bibinfo {author} {\bibfnamefont {John~B}\ \bibnamefont
  {DeBrota}}, \bibinfo {author} {\bibfnamefont {Matthew~B}\ \bibnamefont
  {Weiss}}, \bibinfo {author} {\bibfnamefont {Gabriela~B}\ \bibnamefont
  {Lemos}}, \bibinfo {author} {\bibfnamefont {Jailson~Sales}\ \bibnamefont
  {Ara{\'u}jo}}, \bibinfo {author} {\bibfnamefont {Gabriel~H}\ \bibnamefont
  {Aguilar}}, and\ \bibinfo {author} {\bibfnamefont {Jacques~L}\ \bibnamefont
  {Pienaar}}} (\bibinfo {year} {2024}),\ \bibfield  {title} {\enquote {\bibinfo
  {title} {Synthesizing the born rule with reinforcement learning},}\ }\href
  {https://doi.org/10.48550/arXiv.2404.19011} {\bibinfo  {journal} {arXiv
  preprint arXiv:2404.19011}\ }\BibitemShut {NoStop}%
\bibitem [{\citenamefont {Popova}\ \emph {et~al.}(2018)\citenamefont {Popova},
  \citenamefont {Isayev},\ and\ \citenamefont {Tropsha}}]{popova2018deep}%
  \BibitemOpen
\bibfield  {journal} {  }\bibfield  {author} {\bibinfo {author} {\bibnamefont
  {Popova}, \bibfnamefont {Mariya}}, \bibinfo {author} {\bibfnamefont
  {Olexandr}\ \bibnamefont {Isayev}}, and\ \bibinfo {author} {\bibfnamefont
  {Alexander}\ \bibnamefont {Tropsha}}} (\bibinfo {year} {2018}),\ \bibfield
  {title} {\enquote {\bibinfo {title} {Deep reinforcement learning for de novo
  drug design},}\ }\href {https://doi.org/10.1126/sciadv.aap7885} {\bibfield
  {journal} {\bibinfo  {journal} {Science advances}\ }\textbf {\bibinfo
  {volume} {4}}~(\bibinfo {number} {7}),\ \bibinfo {pages}
  {eaap7885}}\BibitemShut {NoStop}%
\bibitem [{\citenamefont {Porotti}\ \emph {et~al.}(2022)\citenamefont
  {Porotti}, \citenamefont {Essig}, \citenamefont {Huard},\ and\ \citenamefont
  {Marquardt}}]{porotti2022deep}%
  \BibitemOpen
  \bibfield  {author} {\bibinfo {author} {\bibnamefont {Porotti}, \bibfnamefont
  {Riccardo}}, \bibinfo {author} {\bibfnamefont {Antoine}\ \bibnamefont
  {Essig}}, \bibinfo {author} {\bibfnamefont {Benjamin}\ \bibnamefont {Huard}},
  and\ \bibinfo {author} {\bibfnamefont {Florian}\ \bibnamefont {Marquardt}}}
  (\bibinfo {year} {2022}),\ \bibfield  {title} {\enquote {\bibinfo {title}
  {Deep reinforcement learning for quantum state preparation with weak
  nonlinear measurements},}\ }\href {https://doi.org/10.22331/q-2022-06-28-747}
  {\bibfield  {journal} {\bibinfo  {journal} {Quantum}\ }\textbf {\bibinfo
  {volume} {6}},\ \bibinfo {pages} {747}}\BibitemShut {NoStop}%
\bibitem [{\citenamefont {Porotti}\ \emph {et~al.}(2023)\citenamefont
  {Porotti}, \citenamefont {Peano},\ and\ \citenamefont
  {Marquardt}}]{porotti2023gradient}%
  \BibitemOpen
  \bibfield  {author} {\bibinfo {author} {\bibnamefont {Porotti}, \bibfnamefont
  {Riccardo}}, \bibinfo {author} {\bibfnamefont {Vittorio}\ \bibnamefont
  {Peano}}, and\ \bibinfo {author} {\bibfnamefont {Florian}\ \bibnamefont
  {Marquardt}}} (\bibinfo {year} {2023}),\ \bibfield  {title} {\enquote
  {\bibinfo {title} {Gradient-ascent pulse engineering with feedback},}\ }\href
  {https://doi.org/10.1103/PRXQuantum.4.030305} {\bibfield  {journal} {\bibinfo
   {journal} {PRX Quantum}\ }\textbf {\bibinfo {volume} {4}}~(\bibinfo {number}
  {3}),\ \bibinfo {pages} {030305}}\BibitemShut {NoStop}%
\bibitem [{\citenamefont {Porotti}\ \emph {et~al.}(2019)\citenamefont
  {Porotti}, \citenamefont {Tamascelli}, \citenamefont {Restelli},\ and\
  \citenamefont {Prati}}]{porotti2019coherent}%
  \BibitemOpen
  \bibfield  {author} {\bibinfo {author} {\bibnamefont {Porotti}, \bibfnamefont
  {Riccardo}}, \bibinfo {author} {\bibfnamefont {Dario}\ \bibnamefont
  {Tamascelli}}, \bibinfo {author} {\bibfnamefont {Marcello}\ \bibnamefont
  {Restelli}}, and\ \bibinfo {author} {\bibfnamefont {Enrico}\ \bibnamefont
  {Prati}}} (\bibinfo {year} {2019}),\ \bibfield  {title} {\enquote {\bibinfo
  {title} {Coherent transport of quantum states by deep reinforcement
  learning},}\ }\href {https://doi.org/10.1038/s42005-019-0169-x} {\bibfield
  {journal} {\bibinfo  {journal} {Communications Physics}\ }\textbf {\bibinfo
  {volume} {2}}~(\bibinfo {number} {1}),\ \bibinfo {pages} {61}}\BibitemShut
  {NoStop}%
\bibitem [{\citenamefont {Pozzi}\ \emph {et~al.}(2022)\citenamefont {Pozzi},
  \citenamefont {Herbert}, \citenamefont {Sengupta},\ and\ \citenamefont
  {Mullins}}]{pozzi2022using}%
  \BibitemOpen
  \bibfield  {author} {\bibinfo {author} {\bibnamefont {Pozzi}, \bibfnamefont
  {Matteo~G}}, \bibinfo {author} {\bibfnamefont {Steven~J}\ \bibnamefont
  {Herbert}}, \bibinfo {author} {\bibfnamefont {Akash}\ \bibnamefont
  {Sengupta}}, and\ \bibinfo {author} {\bibfnamefont {Robert~D}\ \bibnamefont
  {Mullins}}} (\bibinfo {year} {2022}),\ \bibfield  {title} {\enquote {\bibinfo
  {title} {Using reinforcement learning to perform qubit routing in quantum
  compilers},}\ }\href {https://doi.org/10.1145/3520434} {\bibfield  {journal}
  {\bibinfo  {journal} {ACM Transactions on Quantum Computing}\ }\textbf
  {\bibinfo {volume} {3}}~(\bibinfo {number} {2}),\ \bibinfo {pages}
  {1--25}}\BibitemShut {NoStop}%
\bibitem [{\citenamefont {Praeger}\ \emph {et~al.}(2021)\citenamefont
  {Praeger}, \citenamefont {Xie}, \citenamefont {Grant-Jacob}, \citenamefont
  {Eason},\ and\ \citenamefont {Mills}}]{praeger2021playing}%
  \BibitemOpen
  \bibfield  {author} {\bibinfo {author} {\bibnamefont {Praeger}, \bibfnamefont
  {Matthew}}, \bibinfo {author} {\bibfnamefont {Yunhui}\ \bibnamefont {Xie}},
  \bibinfo {author} {\bibfnamefont {James~A}\ \bibnamefont {Grant-Jacob}},
  \bibinfo {author} {\bibfnamefont {Robert~W}\ \bibnamefont {Eason}}, and\
  \bibinfo {author} {\bibfnamefont {Ben}\ \bibnamefont {Mills}}} (\bibinfo
  {year} {2021}),\ \bibfield  {title} {\enquote {\bibinfo {title} {Playing
  optical tweezers with deep reinforcement learning: in virtual, physical and
  augmented environments},}\ }\href
  {https://iopscience.iop.org/article/10.1088/2632-2153/abf0f6} {\bibfield
  {journal} {\bibinfo  {journal} {Machine Learning: Science and Technology}\
  }\textbf {\bibinfo {volume} {2}}~(\bibinfo {number} {3}),\ \bibinfo {pages}
  {035024}}\BibitemShut {NoStop}%
\bibitem [{\citenamefont {Preti}\ \emph {et~al.}(2024)\citenamefont {Preti},
  \citenamefont {Schilling}, \citenamefont {Jerbi}, \citenamefont
  {Trenkwalder}, \citenamefont {Nautrup}, \citenamefont {Motzoi},\ and\
  \citenamefont {Briegel}}]{preti2024hybrid}%
  \BibitemOpen
  \bibfield  {author} {\bibinfo {author} {\bibnamefont {Preti}, \bibfnamefont
  {Francesco}}, \bibinfo {author} {\bibfnamefont {Michael}\ \bibnamefont
  {Schilling}}, \bibinfo {author} {\bibfnamefont {Sofiene}\ \bibnamefont
  {Jerbi}}, \bibinfo {author} {\bibfnamefont {Lea~M}\ \bibnamefont
  {Trenkwalder}}, \bibinfo {author} {\bibfnamefont {Hendrik~Poulsen}\
  \bibnamefont {Nautrup}}, \bibinfo {author} {\bibfnamefont {Felix}\
  \bibnamefont {Motzoi}}, and\ \bibinfo {author} {\bibfnamefont {Hans~J}\
  \bibnamefont {Briegel}}} (\bibinfo {year} {2024}),\ \bibfield  {title}
  {\enquote {\bibinfo {title} {Hybrid discrete-continuous compilation of
  trapped-ion quantum circuits with deep reinforcement learning},}\ }\href
  {https://doi.org/10.22331/q-2024-05-14-1343} {\bibfield  {journal} {\bibinfo
  {journal} {Quantum}\ }\textbf {\bibinfo {volume} {8}},\ \bibinfo {pages}
  {1343}}\BibitemShut {NoStop}%
\bibitem [{\citenamefont {Promponas}\ \emph {et~al.}(2024)\citenamefont
  {Promponas}, \citenamefont {Mudvari}, \citenamefont {Chiesa}, \citenamefont
  {Polakos}, \citenamefont {Samuel},\ and\ \citenamefont
  {Tassiulas}}]{promponas2024compiler}%
  \BibitemOpen
  \bibfield  {author} {\bibinfo {author} {\bibnamefont {Promponas},
  \bibfnamefont {Panagiotis}}, \bibinfo {author} {\bibfnamefont {Akrit}\
  \bibnamefont {Mudvari}}, \bibinfo {author} {\bibfnamefont {Luca~Della}\
  \bibnamefont {Chiesa}}, \bibinfo {author} {\bibfnamefont {Paul}\ \bibnamefont
  {Polakos}}, \bibinfo {author} {\bibfnamefont {Louis}\ \bibnamefont {Samuel}},
  and\ \bibinfo {author} {\bibfnamefont {Leandros}\ \bibnamefont {Tassiulas}}}
  (\bibinfo {year} {2024}),\ \href {https://arxiv.org/abs/2404.17077} {\enquote
  {\bibinfo {title} {Compiler for distributed quantum computing: a
  reinforcement learning approach},}\ }\BibitemShut {NoStop}%
\bibitem [{\citenamefont {Puviani}\ \emph {et~al.}(2025)\citenamefont
  {Puviani}, \citenamefont {Borah}, \citenamefont {Zen}, \citenamefont {Olle},\
  and\ \citenamefont {Marquardt}}]{puviani2025non}%
  \BibitemOpen
  \bibfield  {author} {\bibinfo {author} {\bibnamefont {Puviani}, \bibfnamefont
  {Matteo}}, \bibinfo {author} {\bibfnamefont {Sangkha}\ \bibnamefont {Borah}},
  \bibinfo {author} {\bibfnamefont {Remmy}\ \bibnamefont {Zen}}, \bibinfo
  {author} {\bibfnamefont {Jan}\ \bibnamefont {Olle}}, and\ \bibinfo {author}
  {\bibfnamefont {Florian}\ \bibnamefont {Marquardt}}} (\bibinfo {year}
  {2025}),\ \bibfield  {title} {\enquote {\bibinfo {title} {Non-markovian
  feedback for optimized quantum error correction},}\ }\href
  {https://doi.org/10.1103/PhysRevLett.134.020601} {\bibfield  {journal}
  {\bibinfo  {journal} {Physical Review Letters}\ }\textbf {\bibinfo {volume}
  {134}}~(\bibinfo {number} {2}),\ \bibinfo {pages} {020601}}\BibitemShut
  {NoStop}%
\bibitem [{\citenamefont {Qiu}\ \emph {et~al.}(2022)\citenamefont {Qiu},
  \citenamefont {Zhuang}, \citenamefont {Huang},\ and\ \citenamefont
  {Lee}}]{qiu2022efficient}%
  \BibitemOpen
  \bibfield  {author} {\bibinfo {author} {\bibnamefont {Qiu}, \bibfnamefont
  {Yuxiang}}, \bibinfo {author} {\bibfnamefont {Min}\ \bibnamefont {Zhuang}},
  \bibinfo {author} {\bibfnamefont {Jiahao}\ \bibnamefont {Huang}}, and\
  \bibinfo {author} {\bibfnamefont {Chaohong}\ \bibnamefont {Lee}}} (\bibinfo
  {year} {2022}),\ \bibfield  {title} {\enquote {\bibinfo {title} {Efficient
  and robust entanglement generation with deep reinforcement learning for
  quantum metrology},}\ }\href {https://doi.org/10.1088/1367-2630/ac8285}
  {\bibfield  {journal} {\bibinfo  {journal} {New Journal of Physics}\ }\textbf
  {\bibinfo {volume} {24}}~(\bibinfo {number} {8}),\ \bibinfo {pages}
  {083011}}\BibitemShut {NoStop}%
\bibitem [{\citenamefont {Quetschlich}\ \emph {et~al.}(2023)\citenamefont
  {Quetschlich}, \citenamefont {Burgholzer},\ and\ \citenamefont
  {Wille}}]{quetschlich2023compiler}%
  \BibitemOpen
  \bibfield  {author} {\bibinfo {author} {\bibnamefont {Quetschlich},
  \bibfnamefont {Nils}}, \bibinfo {author} {\bibfnamefont {Lukas}\ \bibnamefont
  {Burgholzer}}, and\ \bibinfo {author} {\bibfnamefont {Robert}\ \bibnamefont
  {Wille}}} (\bibinfo {year} {2023}),\ \bibfield  {title} {\enquote {\bibinfo
  {title} {Compiler optimization for quantum computing using reinforcement
  learning},}\ }\href {https://doi.org/10.1109/DAC56929.2023.10248002}
  {\bibinfo  {journal} {IEEE}\ ,\ \bibinfo {pages} {1--6}}\BibitemShut
  {NoStop}%
\bibitem [{\citenamefont {Rapp}\ \emph {et~al.}(2024)\citenamefont {Rapp},
  \citenamefont {Kreplin},\ and\ \citenamefont {Roth}}]{rapp2024reinforcement}%
  \BibitemOpen
\bibfield  {journal} {  }\bibfield  {author} {\bibinfo {author} {\bibnamefont
  {Rapp}, \bibfnamefont {Frederic}}, \bibinfo {author} {\bibfnamefont
  {David~A}\ \bibnamefont {Kreplin}}, and\ \bibinfo {author} {\bibfnamefont
  {Marco}\ \bibnamefont {Roth}}} (\bibinfo {year} {2024}),\ \bibfield  {title}
  {\enquote {\bibinfo {title} {Reinforcement learning-based architecture search
  for quantum machine learning},}\ }\href {https://arxiv.org/abs/2406.02717}
  {\bibinfo  {journal} {arXiv preprint arXiv:2406.02717}\ }\BibitemShut
  {NoStop}%
\bibitem [{\citenamefont {Reddy}\ \emph {et~al.}(2016)\citenamefont {Reddy},
  \citenamefont {Celani}, \citenamefont {Sejnowski},\ and\ \citenamefont
  {Vergassola}}]{reddy2016learning}%
  \BibitemOpen
\bibfield  {journal} {  }\bibfield  {author} {\bibinfo {author} {\bibnamefont
  {Reddy}, \bibfnamefont {Gautam}}, \bibinfo {author} {\bibfnamefont {Antonio}\
  \bibnamefont {Celani}}, \bibinfo {author} {\bibfnamefont {Terrence~J}\
  \bibnamefont {Sejnowski}}, and\ \bibinfo {author} {\bibfnamefont {Massimo}\
  \bibnamefont {Vergassola}}} (\bibinfo {year} {2016}),\ \bibfield  {title}
  {\enquote {\bibinfo {title} {Learning to soar in turbulent environments},}\
  }\href {https://doi.org/10.1073/pnas.1606075113} {\bibfield  {journal}
  {\bibinfo  {journal} {Proceedings of the National Academy of Sciences}\
  }\textbf {\bibinfo {volume} {113}}~(\bibinfo {number} {33}),\ \bibinfo
  {pages} {E4877--E4884}}\BibitemShut {NoStop}%
\bibitem [{\citenamefont {Reinschmidt}\ \emph {et~al.}(2023)\citenamefont
  {Reinschmidt}, \citenamefont {Fortágh}, \citenamefont {Günther},\ and\
  \citenamefont {Volchkov}}]{reinschmidt2023reinforcement}%
  \BibitemOpen
  \bibfield  {author} {\bibinfo {author} {\bibnamefont {Reinschmidt},
  \bibfnamefont {Malte}}, \bibinfo {author} {\bibfnamefont {József}\
  \bibnamefont {Fortágh}}, \bibinfo {author} {\bibfnamefont {Andreas}\
  \bibnamefont {Günther}}, and\ \bibinfo {author} {\bibfnamefont {Valentin}\
  \bibnamefont {Volchkov}}} (\bibinfo {year} {2023}),\ \bibfield  {title}
  {\enquote {\bibinfo {title} {Reinforcement learning in ultracold atom
  experiments},}\ }\href {https://arxiv.org/abs/2306.16764} {\bibinfo
  {journal} {arXiv preprint arXiv:2306.16764}\ }\BibitemShut {NoStop}%
\bibitem [{\citenamefont {Rende}\ \emph {et~al.}(2025)\citenamefont {Rende},
  \citenamefont {Viteritti}, \citenamefont {Becca}, \citenamefont
  {Scardicchio}, \citenamefont {Laio},\ and\ \citenamefont
  {Carleo}}]{rende2025foundation}%
  \BibitemOpen
\bibfield  {journal} {  }\bibfield  {author} {\bibinfo {author} {\bibnamefont
  {Rende}, \bibfnamefont {Riccardo}}, \bibinfo {author} {\bibfnamefont
  {Luciano~Loris}\ \bibnamefont {Viteritti}}, \bibinfo {author} {\bibfnamefont
  {Federico}\ \bibnamefont {Becca}}, \bibinfo {author} {\bibfnamefont
  {Antonello}\ \bibnamefont {Scardicchio}}, \bibinfo {author} {\bibfnamefont
  {Alessandro}\ \bibnamefont {Laio}}, and\ \bibinfo {author} {\bibfnamefont
  {Giuseppe}\ \bibnamefont {Carleo}}} (\bibinfo {year} {2025}),\ \bibfield
  {title} {\enquote {\bibinfo {title} {Foundation neural-networks quantum
  states as a unified ansatz for multiple hamiltonians},}\ }\href
  {https://doi.org/10.1038/s41467-025-62098-x} {\bibfield  {journal} {\bibinfo
  {journal} {Nature communications}\ }\textbf {\bibinfo {volume}
  {16}}~(\bibinfo {number} {1}),\ \bibinfo {pages} {7213}}\BibitemShut
  {NoStop}%
\bibitem [{\citenamefont {Reuer}\ \emph {et~al.}(2023)\citenamefont {Reuer},
  \citenamefont {Landgraf}, \citenamefont {Fösel}, \citenamefont
  {O’Sullivan}, \citenamefont {Beltrán}, \citenamefont {Akin}, \citenamefont
  {Norris}, \citenamefont {Remm}, \citenamefont {Kerschbaum}, \citenamefont
  {Besse}, \citenamefont {Marquardt}, \citenamefont {Wallraff},\ and\
  \citenamefont {Eichler}}]{reuer2023realizing}%
  \BibitemOpen
  \bibfield  {author} {\bibinfo {author} {\bibnamefont {Reuer}, \bibfnamefont
  {Kevin}}, \bibinfo {author} {\bibfnamefont {Jonas}\ \bibnamefont {Landgraf}},
  \bibinfo {author} {\bibfnamefont {Thomas}\ \bibnamefont {Fösel}}, \bibinfo
  {author} {\bibfnamefont {James}\ \bibnamefont {O’Sullivan}}, \bibinfo
  {author} {\bibfnamefont {Liberto}\ \bibnamefont {Beltrán}}, \bibinfo
  {author} {\bibfnamefont {Abdulkadir}\ \bibnamefont {Akin}}, \bibinfo {author}
  {\bibfnamefont {Graham~J.}\ \bibnamefont {Norris}}, \bibinfo {author}
  {\bibfnamefont {Ants}\ \bibnamefont {Remm}}, \bibinfo {author} {\bibfnamefont
  {Michael}\ \bibnamefont {Kerschbaum}}, \bibinfo {author} {\bibfnamefont
  {Jean-Claude}\ \bibnamefont {Besse}}, \bibinfo {author} {\bibfnamefont
  {Florian}\ \bibnamefont {Marquardt}}, \bibinfo {author} {\bibfnamefont
  {Andreas}\ \bibnamefont {Wallraff}}, and\ \bibinfo {author} {\bibfnamefont
  {Christopher}\ \bibnamefont {Eichler}}} (\bibinfo {year} {2023}),\ \bibfield
  {title} {\enquote {\bibinfo {title} {Realizing a deep reinforcement learning
  agent for real-time quantum feedback},}\ }\href
  {https://doi.org/10.1038/s41467-023-42901-3} {\bibfield  {journal} {\bibinfo
  {journal} {Nature Communications}\ }\textbf {\bibinfo {volume}
  {14}}~(\bibinfo {number} {1}),\ \bibinfo {pages} {7138}}\BibitemShut
  {NoStop}%
\bibitem [{\citenamefont {Rietsch}\ \emph {et~al.}(2024)\citenamefont
  {Rietsch}, \citenamefont {Dubey}, \citenamefont {Ufrecht}, \citenamefont
  {Periyasamy}, \citenamefont {Plinge}, \citenamefont {Mutschler},\ and\
  \citenamefont {Scherer}}]{rietsch2024unitary}%
  \BibitemOpen
  \bibfield  {author} {\bibinfo {author} {\bibnamefont {Rietsch}, \bibfnamefont
  {Sebastian}}, \bibinfo {author} {\bibfnamefont {Abhishek~Y}\ \bibnamefont
  {Dubey}}, \bibinfo {author} {\bibfnamefont {Christian}\ \bibnamefont
  {Ufrecht}}, \bibinfo {author} {\bibfnamefont {Maniraman}\ \bibnamefont
  {Periyasamy}}, \bibinfo {author} {\bibfnamefont {Axel}\ \bibnamefont
  {Plinge}}, \bibinfo {author} {\bibfnamefont {Christopher}\ \bibnamefont
  {Mutschler}}, and\ \bibinfo {author} {\bibfnamefont {Daniel~D}\ \bibnamefont
  {Scherer}}} (\bibinfo {year} {2024}),\ \bibfield  {title} {\enquote {\bibinfo
  {title} {Unitary synthesis of clifford+ t circuits with reinforcement
  learning},}\ }\href {https://arxiv.org/abs/2404.14865} {\bibinfo  {journal}
  {arXiv preprint arXiv:2404.14865}\ }\BibitemShut {NoStop}%
\bibitem [{\citenamefont {Riu}\ \emph {et~al.}(2023)\citenamefont {Riu},
  \citenamefont {Nogu{\'e}}, \citenamefont {Vilaplana}, \citenamefont
  {Garcia-Saez},\ and\ \citenamefont {Estarellas}}]{riu2023reinforcement}%
  \BibitemOpen
\bibfield  {journal} {  }\bibfield  {author} {\bibinfo {author} {\bibnamefont
  {Riu}, \bibfnamefont {Jordi}}, \bibinfo {author} {\bibfnamefont {Jan}\
  \bibnamefont {Nogu{\'e}}}, \bibinfo {author} {\bibfnamefont {Gerard}\
  \bibnamefont {Vilaplana}}, \bibinfo {author} {\bibfnamefont {Artur}\
  \bibnamefont {Garcia-Saez}}, and\ \bibinfo {author} {\bibfnamefont {Marta~P}\
  \bibnamefont {Estarellas}}} (\bibinfo {year} {2023}),\ \bibfield  {title}
  {\enquote {\bibinfo {title} {Reinforcement learning based quantum circuit
  optimization via zx-calculus},}\ }\href {https://arxiv.org/abs/2312.11597}
  {\bibinfo  {journal} {arXiv preprint arXiv:2312.11597}\ }\BibitemShut
  {NoStop}%
\bibitem [{\citenamefont {Rose}\ \emph {et~al.}(2021)\citenamefont {Rose},
  \citenamefont {Mair},\ and\ \citenamefont
  {Garrahan}}]{rose2021reinforcement}%
  \BibitemOpen
\bibfield  {journal} {  }\bibfield  {author} {\bibinfo {author} {\bibnamefont
  {Rose}, \bibfnamefont {Dominic~C}}, \bibinfo {author} {\bibfnamefont
  {Jamie~F}\ \bibnamefont {Mair}}, and\ \bibinfo {author} {\bibfnamefont
  {Juan~P}\ \bibnamefont {Garrahan}}} (\bibinfo {year} {2021}),\ \bibfield
  {title} {\enquote {\bibinfo {title} {A reinforcement learning approach to
  rare trajectory sampling},}\ }\href
  {https://doi.org/10.1088/1367-2630/abd7bd} {\bibfield  {journal} {\bibinfo
  {journal} {New Journal of Physics}\ }\textbf {\bibinfo {volume}
  {23}}~(\bibinfo {number} {1}),\ \bibinfo {pages} {013013}}\BibitemShut
  {NoStop}%
\bibitem [{\citenamefont {Ruiz}\ \emph {et~al.}(2024)\citenamefont {Ruiz},
  \citenamefont {Laakkonen}, \citenamefont {Bausch}, \citenamefont {Balog},
  \citenamefont {Barekatain}, \citenamefont {Heras}, \citenamefont {Novikov},
  \citenamefont {Fitzpatrick}, \citenamefont {Romera-Paredes}, \citenamefont
  {van~de Wetering} \emph {et~al.}}]{ruiz2024quantum}%
  \BibitemOpen
  \bibfield  {author} {\bibinfo {author} {\bibnamefont {Ruiz}, \bibfnamefont
  {Francisco~JR}}, \bibinfo {author} {\bibfnamefont {Tuomas}\ \bibnamefont
  {Laakkonen}}, \bibinfo {author} {\bibfnamefont {Johannes}\ \bibnamefont
  {Bausch}}, \bibinfo {author} {\bibfnamefont {Matej}\ \bibnamefont {Balog}},
  \bibinfo {author} {\bibfnamefont {Mohammadamin}\ \bibnamefont {Barekatain}},
  \bibinfo {author} {\bibfnamefont {Francisco~JH}\ \bibnamefont {Heras}},
  \bibinfo {author} {\bibfnamefont {Alexander}\ \bibnamefont {Novikov}},
  \bibinfo {author} {\bibfnamefont {Nathan}\ \bibnamefont {Fitzpatrick}},
  \bibinfo {author} {\bibfnamefont {Bernardino}\ \bibnamefont
  {Romera-Paredes}}, \bibinfo {author} {\bibfnamefont {John}\ \bibnamefont
  {van~de Wetering}},  \emph {et~al.}} (\bibinfo {year} {2024}),\ \bibfield
  {title} {\enquote {\bibinfo {title} {Quantum circuit optimization with
  alphatensor},}\ }\href {https://arxiv.org/abs/2402.14396} {\bibinfo
  {journal} {arXiv preprint arXiv:2402.14396}\ }\BibitemShut {NoStop}%
\bibitem [{\citenamefont {Russell}\ and\ \citenamefont
  {Norvig}(2016)}]{russell-norvig_book}%
  \BibitemOpen
\bibfield  {journal} {  }\bibfield  {author} {\bibinfo {author} {\bibnamefont
  {Russell}, \bibfnamefont {Stuart~J}}, and\ \bibinfo {author} {\bibfnamefont
  {Peter}\ \bibnamefont {Norvig}}} (\bibinfo {year} {2016}),\ \href@noop {}
  {\emph {\bibinfo {title} {Artificial intelligence: a modern approach}}}\
  (\bibinfo  {publisher} {Malaysia; Pearson Education Limited})\ \bibinfo
  {note} {{ISBN:} 978-0-13-604259-4}\BibitemShut {NoStop}%
\bibitem [{\citenamefont {Russo}\ \emph {et~al.}(2024)\citenamefont {Russo},
  \citenamefont {Palesi}, \citenamefont {Patti}, \citenamefont {Ascia},\ and\
  \citenamefont {Catania}}]{russo2024attention}%
  \BibitemOpen
  \bibfield  {author} {\bibinfo {author} {\bibnamefont {Russo}, \bibfnamefont
  {Enrico}}, \bibinfo {author} {\bibfnamefont {Maurizio}\ \bibnamefont
  {Palesi}}, \bibinfo {author} {\bibfnamefont {Davide}\ \bibnamefont {Patti}},
  \bibinfo {author} {\bibfnamefont {Giuseppe}\ \bibnamefont {Ascia}}, and\
  \bibinfo {author} {\bibfnamefont {Vincenzo}\ \bibnamefont {Catania}}}
  (\bibinfo {year} {2024}),\ \bibfield  {title} {\enquote {\bibinfo {title}
  {Attention-based deep reinforcement learning for qubit allocation in modular
  quantum architectures},}\ }\href {https://arxiv.org/abs/2406.11452}
  {\bibfield  {journal} {\bibinfo  {journal} {arXiv}\ }}\Eprint
  {https://arxiv.org/abs/2406.11452} {2406.11452} \BibitemShut {NoStop}%
\bibitem [{\citenamefont {Sadhu}\ \emph {et~al.}(2024)\citenamefont {Sadhu},
  \citenamefont {Sarkar},\ and\ \citenamefont {Kundu}}]{sadhu2024quantum}%
  \BibitemOpen
  \bibfield  {author} {\bibinfo {author} {\bibnamefont {Sadhu}, \bibfnamefont
  {Abhishek}}, \bibinfo {author} {\bibfnamefont {Aritra}\ \bibnamefont
  {Sarkar}}, and\ \bibinfo {author} {\bibfnamefont {Akash}\ \bibnamefont
  {Kundu}}} (\bibinfo {year} {2024}),\ \bibfield  {title} {\enquote {\bibinfo
  {title} {A quantum information theoretic analysis of reinforcement
  learning-assisted quantum architecture search},}\ }\href
  {https://doi.org/10.1007/s42484-024-00181-0} {\bibfield  {journal} {\bibinfo
  {journal} {Quantum Machine Intelligence}\ }\textbf {\bibinfo {volume}
  {6}}~(\bibinfo {number} {2}),\ \bibinfo {pages} {49}}\BibitemShut {NoStop}%
\bibitem [{\citenamefont {Saggio}\ \emph {et~al.}(2021)\citenamefont {Saggio},
  \citenamefont {Asenbeck}, \citenamefont {Hamann}, \citenamefont
  {Str{\"o}mberg}, \citenamefont {Schiansky}, \citenamefont {Dunjko},
  \citenamefont {Friis}, \citenamefont {Harris}, \citenamefont {Hochberg},
  \citenamefont {Englund} \emph {et~al.}}]{saggio2021experimental}%
  \BibitemOpen
  \bibfield  {author} {\bibinfo {author} {\bibnamefont {Saggio}, \bibfnamefont
  {Valeria}}, \bibinfo {author} {\bibfnamefont {Beate~E}\ \bibnamefont
  {Asenbeck}}, \bibinfo {author} {\bibfnamefont {Arne}\ \bibnamefont {Hamann}},
  \bibinfo {author} {\bibfnamefont {Teodor}\ \bibnamefont {Str{\"o}mberg}},
  \bibinfo {author} {\bibfnamefont {Peter}\ \bibnamefont {Schiansky}}, \bibinfo
  {author} {\bibfnamefont {Vedran}\ \bibnamefont {Dunjko}}, \bibinfo {author}
  {\bibfnamefont {Nicolai}\ \bibnamefont {Friis}}, \bibinfo {author}
  {\bibfnamefont {Nicholas~C}\ \bibnamefont {Harris}}, \bibinfo {author}
  {\bibfnamefont {Michael}\ \bibnamefont {Hochberg}}, \bibinfo {author}
  {\bibfnamefont {Dirk}\ \bibnamefont {Englund}},  \emph {et~al.}} (\bibinfo
  {year} {2021}),\ \bibfield  {title} {\enquote {\bibinfo {title} {Experimental
  quantum speed-up in reinforcement learning agents},}\ }\href
  {https://doi.org/10.1038/s41586-021-03242-7} {\bibfield  {journal} {\bibinfo
  {journal} {Nature}\ }\textbf {\bibinfo {volume} {591}}~(\bibinfo {number}
  {7849}),\ \bibinfo {pages} {229--233}}\BibitemShut {NoStop}%
\bibitem [{\citenamefont {Sch{\"a}fer}\ \emph {et~al.}(2020)\citenamefont
  {Sch{\"a}fer}, \citenamefont {Kloc}, \citenamefont {Bruder},\ and\
  \citenamefont {L{\"o}rch}}]{schafer2020differentiable}%
  \BibitemOpen
  \bibfield  {author} {\bibinfo {author} {\bibnamefont {Sch{\"a}fer},
  \bibfnamefont {Frank}}, \bibinfo {author} {\bibfnamefont {Michal}\
  \bibnamefont {Kloc}}, \bibinfo {author} {\bibfnamefont {Christoph}\
  \bibnamefont {Bruder}}, and\ \bibinfo {author} {\bibfnamefont {Niels}\
  \bibnamefont {L{\"o}rch}}} (\bibinfo {year} {2020}),\ \bibfield  {title}
  {\enquote {\bibinfo {title} {A differentiable programming method for quantum
  control},}\ }\href {https://doi.org/10.1088/2632-2153/ab9802} {\bibfield
  {journal} {\bibinfo  {journal} {Machine Learning: Science and Technology}\
  }\textbf {\bibinfo {volume} {1}}~(\bibinfo {number} {3}),\ \bibinfo {pages}
  {035009}}\BibitemShut {NoStop}%
\bibitem [{\citenamefont {Sch{\"a}fer}\ \emph {et~al.}(2021)\citenamefont
  {Sch{\"a}fer}, \citenamefont {Sekatski}, \citenamefont {Koppenh{\"o}fer},
  \citenamefont {Bruder},\ and\ \citenamefont {Kloc}}]{schafer2021control}%
  \BibitemOpen
  \bibfield  {author} {\bibinfo {author} {\bibnamefont {Sch{\"a}fer},
  \bibfnamefont {Frank}}, \bibinfo {author} {\bibfnamefont {Pavel}\
  \bibnamefont {Sekatski}}, \bibinfo {author} {\bibfnamefont {Martin}\
  \bibnamefont {Koppenh{\"o}fer}}, \bibinfo {author} {\bibfnamefont
  {Christoph}\ \bibnamefont {Bruder}}, and\ \bibinfo {author} {\bibfnamefont
  {Michal}\ \bibnamefont {Kloc}}} (\bibinfo {year} {2021}),\ \bibfield  {title}
  {\enquote {\bibinfo {title} {Control of stochastic quantum dynamics by
  differentiable programming},}\ }\href
  {https://iopscience.iop.org/article/10.1088/2632-2153/abec22} {\bibfield
  {journal} {\bibinfo  {journal} {Machine Learning: Science and Technology}\
  }\textbf {\bibinfo {volume} {2}}~(\bibinfo {number} {3}),\ \bibinfo {pages}
  {035004}}\BibitemShut {NoStop}%
\bibitem [{\citenamefont {Schenk}\ \emph {et~al.}(2024)\citenamefont {Schenk},
  \citenamefont {Vasudevan}, \citenamefont {Haranczyk},\ and\ \citenamefont
  {Romero}}]{schenk2024model}%
  \BibitemOpen
  \bibfield  {author} {\bibinfo {author} {\bibnamefont {Schenk}, \bibfnamefont
  {Christina}}, \bibinfo {author} {\bibfnamefont {Aditya}\ \bibnamefont
  {Vasudevan}}, \bibinfo {author} {\bibfnamefont {Maciej}\ \bibnamefont
  {Haranczyk}}, and\ \bibinfo {author} {\bibfnamefont {Ignacio}\ \bibnamefont
  {Romero}}} (\bibinfo {year} {2024}),\ \bibfield  {title} {\enquote {\bibinfo
  {title} {Model-based reinforcement learning control of reaction-diffusion
  problems},}\ }\href {https://doi.org/10.48550/arXiv.2402.14446} {\bibinfo
  {journal} {arXiv preprint arXiv:2402.14446}\ }\BibitemShut {NoStop}%
\bibitem [{\citenamefont {Schmitt}\ and\ \citenamefont
  {Heyl}(2020)}]{schmitt2020quantum}%
  \BibitemOpen
\bibfield  {journal} {  }\bibfield  {author} {\bibinfo {author} {\bibnamefont
  {Schmitt}, \bibfnamefont {Markus}}, and\ \bibinfo {author} {\bibfnamefont
  {Markus}\ \bibnamefont {Heyl}}} (\bibinfo {year} {2020}),\ \bibfield  {title}
  {\enquote {\bibinfo {title} {Quantum many-body dynamics in two dimensions
  with artificial neural networks},}\ }\href
  {https://doi.org/10.1103/PhysRevLett.125.100503} {\bibfield  {journal}
  {\bibinfo  {journal} {Phys. Rev. Lett.}\ }\textbf {\bibinfo {volume} {125}},\
  \bibinfo {pages} {100503}}\BibitemShut {NoStop}%
\bibitem [{\citenamefont {Schuff}\ \emph {et~al.}(2020)\citenamefont {Schuff},
  \citenamefont {Fiderer},\ and\ \citenamefont {Braun}}]{schuff2020improving}%
  \BibitemOpen
  \bibfield  {author} {\bibinfo {author} {\bibnamefont {Schuff}, \bibfnamefont
  {Jonas}}, \bibinfo {author} {\bibfnamefont {Lukas~J}\ \bibnamefont
  {Fiderer}}, and\ \bibinfo {author} {\bibfnamefont {Daniel}\ \bibnamefont
  {Braun}}} (\bibinfo {year} {2020}),\ \bibfield  {title} {\enquote {\bibinfo
  {title} {Improving the dynamics of quantum sensors with reinforcement
  learning},}\ }\href {https://doi.org/10.1088/1367-2630/ab6f1f} {\bibfield
  {journal} {\bibinfo  {journal} {New Journal of Physics}\ }\textbf {\bibinfo
  {volume} {22}}~(\bibinfo {number} {3}),\ \bibinfo {pages}
  {035001}}\BibitemShut {NoStop}%
\bibitem [{\citenamefont {Schulman}\ \emph {et~al.}(2015)\citenamefont
  {Schulman}, \citenamefont {Levine}, \citenamefont {Abbeel}, \citenamefont
  {Jordan},\ and\ \citenamefont {Moritz}}]{schulman2015trust}%
  \BibitemOpen
  \bibfield  {author} {\bibinfo {author} {\bibnamefont {Schulman},
  \bibfnamefont {John}}, \bibinfo {author} {\bibfnamefont {Sergey}\
  \bibnamefont {Levine}}, \bibinfo {author} {\bibfnamefont {Pieter}\
  \bibnamefont {Abbeel}}, \bibinfo {author} {\bibfnamefont {Michael}\
  \bibnamefont {Jordan}}, and\ \bibinfo {author} {\bibfnamefont {Philipp}\
  \bibnamefont {Moritz}}} (\bibinfo {year} {2015}),\ \bibfield  {title}
  {\enquote {\bibinfo {title} {Trust region policy optimization},}\ }\href
  {https://doi.org/10.48550/arXiv.1502.05477} {\bibinfo  {journal} {PMLR}\ ,\
  \bibinfo {pages} {1889--1897}}\BibitemShut {NoStop}%
\bibitem [{\citenamefont {Schulman}\ \emph {et~al.}(2017)\citenamefont
  {Schulman}, \citenamefont {Wolski}, \citenamefont {Dhariwal}, \citenamefont
  {Radford},\ and\ \citenamefont {Klimov}}]{schulman2017proximal}%
  \BibitemOpen
\bibfield  {journal} {  }\bibfield  {author} {\bibinfo {author} {\bibnamefont
  {Schulman}, \bibfnamefont {John}}, \bibinfo {author} {\bibfnamefont {Filip}\
  \bibnamefont {Wolski}}, \bibinfo {author} {\bibfnamefont {Prafulla}\
  \bibnamefont {Dhariwal}}, \bibinfo {author} {\bibfnamefont {Alec}\
  \bibnamefont {Radford}}, and\ \bibinfo {author} {\bibfnamefont {Oleg}\
  \bibnamefont {Klimov}}} (\bibinfo {year} {2017}),\ \bibfield  {title}
  {\enquote {\bibinfo {title} {Proximal policy optimization algorithms},}\
  }\href {https://doi.org/10.48550/arXiv.1707.06347} {\bibinfo  {journal}
  {arXiv preprint arXiv:1707.06347}\ }\BibitemShut {NoStop}%
\bibitem [{\citenamefont {Settaluri}\ \emph {et~al.}(2020)\citenamefont
  {Settaluri}, \citenamefont {Haj-Ali}, \citenamefont {Huang}, \citenamefont
  {Hakhamaneshi},\ and\ \citenamefont {Nikolic}}]{settaluri2020autockt}%
  \BibitemOpen
\bibfield  {journal} {  }\bibfield  {author} {\bibinfo {author} {\bibnamefont
  {Settaluri}, \bibfnamefont {Keertana}}, \bibinfo {author} {\bibfnamefont
  {Ameer}\ \bibnamefont {Haj-Ali}}, \bibinfo {author} {\bibfnamefont {Qijing}\
  \bibnamefont {Huang}}, \bibinfo {author} {\bibfnamefont {Kourosh}\
  \bibnamefont {Hakhamaneshi}}, and\ \bibinfo {author} {\bibfnamefont
  {Borivoje}\ \bibnamefont {Nikolic}}} (\bibinfo {year} {2020}),\ \bibfield
  {title} {\enquote {\bibinfo {title} {Autockt: Deep reinforcement learning of
  analog circuit designs},}\ }\bibfield  {booktitle} {\emph {\bibinfo
  {booktitle} {2020 Design, Automation \& Test in Europe Conference \&
  Exhibition (DATE)}},\ }\href
  {https://doi.org/10.23919/DATE48585.2020.9116200} {\bibinfo  {journal}
  {IEEE}\ ,\ \bibinfo {pages} {490--495}}\BibitemShut {NoStop}%
\bibitem [{\citenamefont {Sgroi}\ \emph {et~al.}(2024)\citenamefont {Sgroi},
  \citenamefont {Zicari}, \citenamefont {Imparato},\ and\ \citenamefont
  {Paternostro}}]{sgroi2024reinforcement}%
  \BibitemOpen
\bibfield  {journal} {  }\bibfield  {author} {\bibinfo {author} {\bibnamefont
  {Sgroi}, \bibfnamefont {S}}, \bibinfo {author} {\bibfnamefont
  {G}~\bibnamefont {Zicari}}, \bibinfo {author} {\bibfnamefont {A}~\bibnamefont
  {Imparato}}, and\ \bibinfo {author} {\bibfnamefont {M}~\bibnamefont
  {Paternostro}}} (\bibinfo {year} {2024}),\ \bibfield  {title} {\enquote
  {\bibinfo {title} {A reinforcement learning approach to the design of quantum
  chains for optimal energy transfer},}\ }\href
  {https://arxiv.org/abs/2402.07561} {\bibinfo  {journal} {arXiv preprint
  arXiv:2402.07561}\ }\BibitemShut {NoStop}%
\bibitem [{\citenamefont {Sgroi}\ \emph {et~al.}(2021)\citenamefont {Sgroi},
  \citenamefont {Palma},\ and\ \citenamefont
  {Paternostro}}]{sgroi2021reinforcement}%
  \BibitemOpen
\bibfield  {journal} {  }\bibfield  {author} {\bibinfo {author} {\bibnamefont
  {Sgroi}, \bibfnamefont {Sofia}}, \bibinfo {author} {\bibfnamefont
  {G.~Massimo}\ \bibnamefont {Palma}}, and\ \bibinfo {author} {\bibfnamefont
  {Mauro}\ \bibnamefont {Paternostro}}} (\bibinfo {year} {2021}),\ \bibfield
  {title} {\enquote {\bibinfo {title} {Reinforcement learning approach to
  nonequilibrium quantum thermodynamics},}\ }\href
  {https://doi.org/10.1103/PhysRevLett.126.020601} {\bibfield  {journal}
  {\bibinfo  {journal} {Phys. Rev. Lett.}\ }\textbf {\bibinfo {volume} {126}},\
  \bibinfo {pages} {020601}}\BibitemShut {NoStop}%
\bibitem [{\citenamefont {Shindi}\ \emph {et~al.}(2023)\citenamefont {Shindi},
  \citenamefont {Yu}, \citenamefont {Girdhar},\ and\ \citenamefont
  {Dong}}]{shindi2023model}%
  \BibitemOpen
  \bibfield  {author} {\bibinfo {author} {\bibnamefont {Shindi}, \bibfnamefont
  {Omar}}, \bibinfo {author} {\bibfnamefont {Qi}~\bibnamefont {Yu}}, \bibinfo
  {author} {\bibfnamefont {Parth}\ \bibnamefont {Girdhar}}, and\ \bibinfo
  {author} {\bibfnamefont {Daoyi}\ \bibnamefont {Dong}}} (\bibinfo {year}
  {2023}),\ \bibfield  {title} {\enquote {\bibinfo {title} {Model-free quantum
  gate design and calibration using deep reinforcement learning},}\ }\href
  {https://doi.org/10.1109/TAI.2023.3243187} {\bibinfo  {journal} {IEEE
  Transactions on Artificial Intelligence}\ }\BibitemShut {NoStop}%
\bibitem [{\citenamefont {Silver}\ \emph {et~al.}(2014)\citenamefont {Silver},
  \citenamefont {Lever}, \citenamefont {Heess}, \citenamefont {Degris},
  \citenamefont {Wierstra},\ and\ \citenamefont
  {Riedmiller}}]{silver2014deterministic}%
  \BibitemOpen
\bibfield  {journal} {  }\bibfield  {author} {\bibinfo {author} {\bibnamefont
  {Silver}, \bibfnamefont {David}}, \bibinfo {author} {\bibfnamefont {Guy}\
  \bibnamefont {Lever}}, \bibinfo {author} {\bibfnamefont {Nicolas}\
  \bibnamefont {Heess}}, \bibinfo {author} {\bibfnamefont {Thomas}\
  \bibnamefont {Degris}}, \bibinfo {author} {\bibfnamefont {Daan}\ \bibnamefont
  {Wierstra}}, and\ \bibinfo {author} {\bibfnamefont {Martin}\ \bibnamefont
  {Riedmiller}}} (\bibinfo {year} {2014}),\ \bibfield  {title} {\enquote
  {\bibinfo {title} {Deterministic policy gradient algorithms},}\ }\bibfield
  {booktitle} {\emph {\bibinfo {booktitle} {International conference on machine
  learning}},\ }\href {https://proceedings.mlr.press/v32/silver14.pdf}
  {\bibinfo  {journal} {PMLR}\ ,\ \bibinfo {pages} {387--395}}\BibitemShut
  {NoStop}%
\bibitem [{\citenamefont {Silver}\ \emph {et~al.}(2017)\citenamefont {Silver},
  \citenamefont {Schrittwieser}, \citenamefont {Simonyan}, \citenamefont
  {Antonoglou}, \citenamefont {Huang}, \citenamefont {Guez}, \citenamefont
  {Hubert}, \citenamefont {Baker}, \citenamefont {Lai}, \citenamefont {Bolton}
  \emph {et~al.}}]{silver2017mastering}%
  \BibitemOpen
\bibfield  {journal} {  }\bibfield  {author} {\bibinfo {author} {\bibnamefont
  {Silver}, \bibfnamefont {David}}, \bibinfo {author} {\bibfnamefont {Julian}\
  \bibnamefont {Schrittwieser}}, \bibinfo {author} {\bibfnamefont {Karen}\
  \bibnamefont {Simonyan}}, \bibinfo {author} {\bibfnamefont {Ioannis}\
  \bibnamefont {Antonoglou}}, \bibinfo {author} {\bibfnamefont {Aja}\
  \bibnamefont {Huang}}, \bibinfo {author} {\bibfnamefont {Arthur}\
  \bibnamefont {Guez}}, \bibinfo {author} {\bibfnamefont {Thomas}\ \bibnamefont
  {Hubert}}, \bibinfo {author} {\bibfnamefont {Lucas}\ \bibnamefont {Baker}},
  \bibinfo {author} {\bibfnamefont {Matthew}\ \bibnamefont {Lai}}, \bibinfo
  {author} {\bibfnamefont {Adrian}\ \bibnamefont {Bolton}},  \emph {et~al.}}
  (\bibinfo {year} {2017}),\ \bibfield  {title} {\enquote {\bibinfo {title}
  {Mastering the game of go without human knowledge},}\ }\href
  {https://doi.org/10.1038/nature24270} {\bibfield  {journal} {\bibinfo
  {journal} {nature}\ }\textbf {\bibinfo {volume} {550}}~(\bibinfo {number}
  {7676}),\ \bibinfo {pages} {354--359}}\BibitemShut {NoStop}%
\bibitem [{\citenamefont {Singh}\ \emph {et~al.}(2019)\citenamefont {Singh},
  \citenamefont {Yang}, \citenamefont {Hartikainen}, \citenamefont {Finn},\
  and\ \citenamefont {Levine}}]{singh2019end}%
  \BibitemOpen
  \bibfield  {author} {\bibinfo {author} {\bibnamefont {Singh}, \bibfnamefont
  {Avi}}, \bibinfo {author} {\bibfnamefont {Larry}\ \bibnamefont {Yang}},
  \bibinfo {author} {\bibfnamefont {Kristian}\ \bibnamefont {Hartikainen}},
  \bibinfo {author} {\bibfnamefont {Chelsea}\ \bibnamefont {Finn}}, and\
  \bibinfo {author} {\bibfnamefont {Sergey}\ \bibnamefont {Levine}}} (\bibinfo
  {year} {2019}),\ \bibfield  {title} {\enquote {\bibinfo {title} {End-to-end
  robotic reinforcement learning without reward engineering},}\ }\href
  {https://arxiv.org/abs/1904.07854} {\bibinfo  {journal} {arXiv preprint
  arXiv:1904.07854}\ }\BibitemShut {NoStop}%
\bibitem [{\citenamefont {Sivak}\ \emph {et~al.}(2022)\citenamefont {Sivak},
  \citenamefont {Eickbusch}, \citenamefont {Liu}, \citenamefont {Royer},
  \citenamefont {Tsioutsios},\ and\ \citenamefont
  {Devoret}}]{sivak2022model-free}%
  \BibitemOpen
\bibfield  {journal} {  }\bibfield  {author} {\bibinfo {author} {\bibnamefont
  {Sivak}, \bibfnamefont {V~V}}, \bibinfo {author} {\bibfnamefont
  {A.}~\bibnamefont {Eickbusch}}, \bibinfo {author} {\bibfnamefont
  {H.}~\bibnamefont {Liu}}, \bibinfo {author} {\bibfnamefont {B.}~\bibnamefont
  {Royer}}, \bibinfo {author} {\bibfnamefont {I.}~\bibnamefont {Tsioutsios}},
  and\ \bibinfo {author} {\bibfnamefont {M.~H.}\ \bibnamefont {Devoret}}}
  (\bibinfo {year} {2022}),\ \bibfield  {title} {\enquote {\bibinfo {title}
  {Model-free quantum control with reinforcement learning},}\ }\href
  {https://doi.org/10.1103/PhysRevX.12.011059} {\bibfield  {journal} {\bibinfo
  {journal} {Phys. Rev. X}\ }\textbf {\bibinfo {volume} {12}},\ \bibinfo
  {pages} {011059}}\BibitemShut {NoStop}%
\bibitem [{\citenamefont {Sivak}\ \emph {et~al.}(2025)\citenamefont {Sivak},
  \citenamefont {Morvan}, \citenamefont {Broughton}, \citenamefont {Neeley},
  \citenamefont {Eickbusch}, \citenamefont {Abanin}, \citenamefont {Abbas},
  \citenamefont {Acharya}, \citenamefont {Beni}, \citenamefont {Aigeldinger}
  \emph {et~al.}}]{sivak2025reinforcement}%
  \BibitemOpen
  \bibfield  {author} {\bibinfo {author} {\bibnamefont {Sivak}, \bibfnamefont
  {Volodymyr}}, \bibinfo {author} {\bibfnamefont {Alexis}\ \bibnamefont
  {Morvan}}, \bibinfo {author} {\bibfnamefont {Michael}\ \bibnamefont
  {Broughton}}, \bibinfo {author} {\bibfnamefont {Matthew}\ \bibnamefont
  {Neeley}}, \bibinfo {author} {\bibfnamefont {Alec}\ \bibnamefont
  {Eickbusch}}, \bibinfo {author} {\bibfnamefont {Dmitry}\ \bibnamefont
  {Abanin}}, \bibinfo {author} {\bibfnamefont {Amira}\ \bibnamefont {Abbas}},
  \bibinfo {author} {\bibfnamefont {Rajeev}\ \bibnamefont {Acharya}}, \bibinfo
  {author} {\bibfnamefont {Laleh~Aghababaie}\ \bibnamefont {Beni}}, \bibinfo
  {author} {\bibfnamefont {Georg}\ \bibnamefont {Aigeldinger}},  \emph
  {et~al.}} (\bibinfo {year} {2025}),\ \bibfield  {title} {\enquote {\bibinfo
  {title} {Reinforcement learning control of quantum error correction},}\
  }\href {https://doi.org/10.48550/arXiv.2511.08493} {\bibinfo  {journal}
  {arXiv preprint arXiv:2511.08493}\ }\BibitemShut {NoStop}%
\bibitem [{\citenamefont {Sivak}\ \emph {et~al.}(2024)\citenamefont {Sivak},
  \citenamefont {Newman},\ and\ \citenamefont
  {Klimov}}]{sivak2024optimization}%
  \BibitemOpen
\bibfield  {journal} {  }\bibfield  {author} {\bibinfo {author} {\bibnamefont
  {Sivak}, \bibfnamefont {Volodymyr}}, \bibinfo {author} {\bibfnamefont
  {Michael}\ \bibnamefont {Newman}}, and\ \bibinfo {author} {\bibfnamefont
  {Paul}\ \bibnamefont {Klimov}}} (\bibinfo {year} {2024}),\ \bibfield  {title}
  {\enquote {\bibinfo {title} {Optimization of decoder priors for accurate
  quantum error correction},}\ }\href
  {https://doi.org/10.1103/PhysRevLett.133.150603} {\bibfield  {journal}
  {\bibinfo  {journal} {Phys. Rev. Lett.}\ }\textbf {\bibinfo {volume} {133}},\
  \bibinfo {pages} {150603}}\BibitemShut {NoStop}%
\bibitem [{\citenamefont {Sivak}\ \emph {et~al.}(2023)\citenamefont {Sivak},
  \citenamefont {Eickbusch}, \citenamefont {Royer}, \citenamefont {Singh},
  \citenamefont {Tsioutsios}, \citenamefont {Ganjam}, \citenamefont {Miano},
  \citenamefont {Brock}, \citenamefont {Ding}, \citenamefont {Frunzio},
  \citenamefont {Girvin}, \citenamefont {Schoelkopf},\ and\ \citenamefont
  {Devoret}}]{sivak2023real}%
  \BibitemOpen
  \bibfield  {author} {\bibinfo {author} {\bibnamefont {Sivak}, \bibfnamefont
  {VV}}, \bibinfo {author} {\bibfnamefont {Alec}\ \bibnamefont {Eickbusch}},
  \bibinfo {author} {\bibfnamefont {Baptiste}\ \bibnamefont {Royer}}, \bibinfo
  {author} {\bibfnamefont {Shraddha}\ \bibnamefont {Singh}}, \bibinfo {author}
  {\bibfnamefont {Ioannis}\ \bibnamefont {Tsioutsios}}, \bibinfo {author}
  {\bibfnamefont {Suhas}\ \bibnamefont {Ganjam}}, \bibinfo {author}
  {\bibfnamefont {Alessandro}\ \bibnamefont {Miano}}, \bibinfo {author}
  {\bibfnamefont {BL}~\bibnamefont {Brock}}, \bibinfo {author} {\bibfnamefont
  {AZ}~\bibnamefont {Ding}}, \bibinfo {author} {\bibfnamefont {Luigi}\
  \bibnamefont {Frunzio}}, \bibinfo {author} {\bibfnamefont {Steve}\
  \bibnamefont {Girvin}}, \bibinfo {author} {\bibfnamefont {R.~J.}\
  \bibnamefont {Schoelkopf}}, and\ \bibinfo {author} {\bibfnamefont {M.~H.}\
  \bibnamefont {Devoret}}} (\bibinfo {year} {2023}),\ \bibfield  {title}
  {\enquote {\bibinfo {title} {Real-time quantum error correction beyond
  break-even},}\ }\href {https://doi.org/10.1038/s41586-023-05782-6} {\bibfield
   {journal} {\bibinfo  {journal} {Nature}\ }\textbf {\bibinfo {volume}
  {616}}~(\bibinfo {number} {7955}),\ \bibinfo {pages} {50--55}}\BibitemShut
  {NoStop}%
\bibitem [{\citenamefont {Sommer}\ \emph {et~al.}(2020)\citenamefont {Sommer},
  \citenamefont {Asjad},\ and\ \citenamefont {Genes}}]{sommer2020prospects}%
  \BibitemOpen
  \bibfield  {author} {\bibinfo {author} {\bibnamefont {Sommer}, \bibfnamefont
  {Christian}}, \bibinfo {author} {\bibfnamefont {Muhammad}\ \bibnamefont
  {Asjad}}, and\ \bibinfo {author} {\bibfnamefont {Claudiu}\ \bibnamefont
  {Genes}}} (\bibinfo {year} {2020}),\ \bibfield  {title} {\enquote {\bibinfo
  {title} {Prospects of reinforcement learning for the simultaneous damping of
  many mechanical modes},}\ }\href
  {https://www.nature.com/articles/s41598-020-59435-z.pdf} {\bibfield
  {journal} {\bibinfo  {journal} {Scientific Reports}\ }\textbf {\bibinfo
  {volume} {10}}~(\bibinfo {number} {1}),\ \bibinfo {pages} {2623}}\BibitemShut
  {NoStop}%
\bibitem [{\citenamefont {S\o{}rdal}\ and\ \citenamefont
  {Bergli}(2019)}]{sordal2019deep}%
  \BibitemOpen
  \bibfield  {author} {\bibinfo {author} {\bibnamefont {S\o{}rdal},
  \bibfnamefont {Vegard~B}}, and\ \bibinfo {author} {\bibfnamefont {Joakim}\
  \bibnamefont {Bergli}}} (\bibinfo {year} {2019}),\ \bibfield  {title}
  {\enquote {\bibinfo {title} {Deep reinforcement learning for quantum szilard
  engine optimization},}\ }\href {https://doi.org/10.1103/PhysRevA.100.042314}
  {\bibfield  {journal} {\bibinfo  {journal} {Phys. Rev. A}\ }\textbf {\bibinfo
  {volume} {100}},\ \bibinfo {pages} {042314}}\BibitemShut {NoStop}%
\bibitem [{\citenamefont {Sriarunothai}\ \emph {et~al.}(2018)\citenamefont
  {Sriarunothai}, \citenamefont {W{\"o}lk}, \citenamefont {Giri}, \citenamefont
  {Friis}, \citenamefont {Dunjko}, \citenamefont {Briegel},\ and\ \citenamefont
  {Wunderlich}}]{sriarunothai2018speeding}%
  \BibitemOpen
  \bibfield  {author} {\bibinfo {author} {\bibnamefont {Sriarunothai},
  \bibfnamefont {Th}}, \bibinfo {author} {\bibfnamefont {Sabine}\ \bibnamefont
  {W{\"o}lk}}, \bibinfo {author} {\bibfnamefont {Gouri~Shankar}\ \bibnamefont
  {Giri}}, \bibinfo {author} {\bibfnamefont {Nicolai}\ \bibnamefont {Friis}},
  \bibinfo {author} {\bibfnamefont {Vedran}\ \bibnamefont {Dunjko}}, \bibinfo
  {author} {\bibfnamefont {Hans~J}\ \bibnamefont {Briegel}}, and\ \bibinfo
  {author} {\bibfnamefont {Ch}~\bibnamefont {Wunderlich}}} (\bibinfo {year}
  {2018}),\ \bibfield  {title} {\enquote {\bibinfo {title} {Speeding-up the
  decision making of a learning agent using an ion trap quantum processor},}\
  }\href {https://doi.org/10.1088/2058-9565/aaef5e} {\bibfield  {journal}
  {\bibinfo  {journal} {Quantum Science and Technology}\ }\textbf {\bibinfo
  {volume} {4}}~(\bibinfo {number} {1}),\ \bibinfo {pages}
  {015014}}\BibitemShut {NoStop}%
\bibitem [{\citenamefont {Sutton}\ and\ \citenamefont
  {Barto}(2018)}]{sutton_barto_book}%
  \BibitemOpen
  \bibfield  {author} {\bibinfo {author} {\bibnamefont {Sutton}, \bibfnamefont
  {Richard~S}}, and\ \bibinfo {author} {\bibfnamefont {Andrew~G}\ \bibnamefont
  {Barto}}} (\bibinfo {year} {2018}),\ \href@noop {} {\emph {\bibinfo {title}
  {Reinforcement learning: An introduction}}}\ (\bibinfo  {publisher} {MIT
  Press})\ \bibinfo {note} {{ISBN:} 978-0-262-03924-6}\BibitemShut {NoStop}%
\bibitem [{\citenamefont {Sweke}\ \emph {et~al.}(2020)\citenamefont {Sweke},
  \citenamefont {Kesselring}, \citenamefont {van Nieuwenburg},\ and\
  \citenamefont {Eisert}}]{sweke2020reinforcement}%
  \BibitemOpen
  \bibfield  {author} {\bibinfo {author} {\bibnamefont {Sweke}, \bibfnamefont
  {Ryan}}, \bibinfo {author} {\bibfnamefont {Markus~S}\ \bibnamefont
  {Kesselring}}, \bibinfo {author} {\bibfnamefont {Evert~PL}\ \bibnamefont {van
  Nieuwenburg}}, and\ \bibinfo {author} {\bibfnamefont {Jens}\ \bibnamefont
  {Eisert}}} (\bibinfo {year} {2020}),\ \bibfield  {title} {\enquote {\bibinfo
  {title} {Reinforcement learning decoders for fault-tolerant quantum
  computation},}\ }\href {https://doi.org/10.1088/2632-2153/abc609} {\bibfield
  {journal} {\bibinfo  {journal} {Machine Learning: Science and Technology}\
  }\textbf {\bibinfo {volume} {2}}~(\bibinfo {number} {2}),\ \bibinfo {pages}
  {025005}}\BibitemShut {NoStop}%
\bibitem [{\citenamefont {Tashev}(2022)}]{tashev2022developing}%
  \BibitemOpen
  \bibfield  {author} {\bibinfo {author} {\bibnamefont {Tashev}, \bibfnamefont
  {Pavel}}} (\bibinfo {year} {2022}),\ \href
  {https://www.pks.mpg.de/fileadmin/user_upload/MPIPKS/group_pages/NQD/master_thesis_tashev.pdf}
  {\emph {\bibinfo {title} {Developing Artificial Intelligence Agents to
  Manipulate Quantum Entanglement}}}\ (\bibinfo  {publisher} {Master thesis,
  Sofia University})\BibitemShut {NoStop}%
\bibitem [{\citenamefont {Tashev}\ \emph {et~al.}(2024)\citenamefont {Tashev},
  \citenamefont {Petrov}, \citenamefont {Metz},\ and\ \citenamefont
  {Bukov}}]{tashev2024reinforcement}%
  \BibitemOpen
  \bibfield  {author} {\bibinfo {author} {\bibnamefont {Tashev}, \bibfnamefont
  {Pavel}}, \bibinfo {author} {\bibfnamefont {Stefan}\ \bibnamefont {Petrov}},
  \bibinfo {author} {\bibfnamefont {Friederike}\ \bibnamefont {Metz}}, and\
  \bibinfo {author} {\bibfnamefont {Marin}\ \bibnamefont {Bukov}}} (\bibinfo
  {year} {2024}),\ \bibfield  {title} {\enquote {\bibinfo {title}
  {Reinforcement learning to disentangle multiqubit quantum states from partial
  observations},}\ }\href {https://arxiv.org/abs/2406.07884} {\bibinfo
  {journal} {arXiv e-prints}\ ,\ \bibinfo {pages} {2406.07884}}\BibitemShut
  {NoStop}%
\bibitem [{\citenamefont {Tesauro}\ \emph {et~al.}(1995)\citenamefont {Tesauro}
  \emph {et~al.}}]{tesauro1995temporal}%
  \BibitemOpen
\bibfield  {journal} {  }\bibfield  {author} {\bibinfo {author} {\bibnamefont
  {Tesauro}, \bibfnamefont {Gerald}},  \emph {et~al.}} (\bibinfo {year}
  {1995}),\ \bibfield  {title} {\enquote {\bibinfo {title} {Temporal difference
  learning and td-gammon},}\ }\href {https://doi.org/10.1145/203330.203343}
  {\bibfield  {journal} {\bibinfo  {journal} {Communications of the ACM}\
  }\textbf {\bibinfo {volume} {38}}~(\bibinfo {number} {3}),\ \bibinfo {pages}
  {58--68}}\BibitemShut {NoStop}%
\bibitem [{\citenamefont {Todorov}(2006)}]{todorov2006optimal}%
  \BibitemOpen
  \bibfield  {author} {\bibinfo {author} {\bibnamefont {Todorov}, \bibfnamefont
  {Emanuel}}} (\bibinfo {year} {2006}),\ \bibfield  {title} {\enquote {\bibinfo
  {title} {Optimal control theory},}\ }\href
  {https://doi.org/10.7551/mitpress/1535.003.0018} {\bibinfo  {journal}
  {Bayesian brain: probabilistic approaches to neural coding}\ ,\ \bibinfo
  {pages} {269--298}}\BibitemShut {NoStop}%
\bibitem [{\citenamefont {Van~Nieuwenburg}\ \emph {et~al.}(2017)\citenamefont
  {Van~Nieuwenburg}, \citenamefont {Liu},\ and\ \citenamefont
  {Huber}}]{van2017learning}%
  \BibitemOpen
\bibfield  {journal} {  }\bibfield  {author} {\bibinfo {author} {\bibnamefont
  {Van~Nieuwenburg}, \bibfnamefont {Evert~PL}}, \bibinfo {author}
  {\bibfnamefont {Ye-Hua}\ \bibnamefont {Liu}}, and\ \bibinfo {author}
  {\bibfnamefont {Sebastian~D}\ \bibnamefont {Huber}}} (\bibinfo {year}
  {2017}),\ \bibfield  {title} {\enquote {\bibinfo {title} {Learning phase
  transitions by confusion},}\ }\href {https://doi.org/10.1038/nphys4037}
  {\bibfield  {journal} {\bibinfo  {journal} {Nature Physics}\ }\textbf
  {\bibinfo {volume} {13}}~(\bibinfo {number} {5}),\ \bibinfo {pages}
  {435--439}}\BibitemShut {NoStop}%
\bibitem [{\citenamefont {Walln\"ofer}\ \emph {et~al.}(2020)\citenamefont
  {Walln\"ofer}, \citenamefont {Melnikov}, \citenamefont {D\"ur},\ and\
  \citenamefont {Briegel}}]{wallnofer2020machine}%
  \BibitemOpen
  \bibfield  {author} {\bibinfo {author} {\bibnamefont {Walln\"ofer},
  \bibfnamefont {Julius}}, \bibinfo {author} {\bibfnamefont {Alexey~A.}\
  \bibnamefont {Melnikov}}, \bibinfo {author} {\bibfnamefont {Wolfgang}\
  \bibnamefont {D\"ur}}, and\ \bibinfo {author} {\bibfnamefont {Hans~J.}\
  \bibnamefont {Briegel}}} (\bibinfo {year} {2020}),\ \bibfield  {title}
  {\enquote {\bibinfo {title} {Machine learning for long-distance quantum
  communication},}\ }\href {https://doi.org/10.1103/PRXQuantum.1.010301}
  {\bibfield  {journal} {\bibinfo  {journal} {PRX Quantum}\ }\textbf {\bibinfo
  {volume} {1}},\ \bibinfo {pages} {010301}}\BibitemShut {NoStop}%
\bibitem [{\citenamefont {Wang}\ \emph
  {et~al.}(2024{\natexlab{a}})\citenamefont {Wang}, \citenamefont {Wang},\ and\
  \citenamefont {Fei}}]{wang2024rnn}%
  \BibitemOpen
  \bibfield  {author} {\bibinfo {author} {\bibnamefont {Wang}, \bibfnamefont
  {Gang}}, \bibinfo {author} {\bibfnamefont {Bang-Hai}\ \bibnamefont {Wang}},
  and\ \bibinfo {author} {\bibfnamefont {Shao-Ming}\ \bibnamefont {Fei}}}
  (\bibinfo {year} {2024}{\natexlab{a}}),\ \bibfield  {title} {\enquote
  {\bibinfo {title} {An rnn--policy gradient approach for quantum architecture
  search},}\ }\href {https://doi.org/10.1007/s11128-024-04393-y} {\bibfield
  {journal} {\bibinfo  {journal} {Quantum Information Processing}\ }\textbf
  {\bibinfo {volume} {23}}~(\bibinfo {number} {5}),\ \bibinfo {pages}
  {184}}\BibitemShut {NoStop}%
\bibitem [{\citenamefont {Wang}\ \emph {et~al.}(2023)\citenamefont {Wang},
  \citenamefont {Gleave}, \citenamefont {Tseng}, \citenamefont {Pelrine},
  \citenamefont {Belrose}, \citenamefont {Miller}, \citenamefont {Dennis},
  \citenamefont {Duan}, \citenamefont {Pogrebniak}, \citenamefont {Levine},\
  and\ \citenamefont {Russell}}]{wang2023adversarial}%
  \BibitemOpen
  \bibfield  {author} {\bibinfo {author} {\bibnamefont {Wang}, \bibfnamefont
  {Tony~T}}, \bibinfo {author} {\bibfnamefont {Adam}\ \bibnamefont {Gleave}},
  \bibinfo {author} {\bibfnamefont {Tom}\ \bibnamefont {Tseng}}, \bibinfo
  {author} {\bibfnamefont {Kellin}\ \bibnamefont {Pelrine}}, \bibinfo {author}
  {\bibfnamefont {Nora}\ \bibnamefont {Belrose}}, \bibinfo {author}
  {\bibfnamefont {Joseph}\ \bibnamefont {Miller}}, \bibinfo {author}
  {\bibfnamefont {Michael~D.}\ \bibnamefont {Dennis}}, \bibinfo {author}
  {\bibfnamefont {Yawen}\ \bibnamefont {Duan}}, \bibinfo {author}
  {\bibfnamefont {Viktor}\ \bibnamefont {Pogrebniak}}, \bibinfo {author}
  {\bibfnamefont {Sergey}\ \bibnamefont {Levine}}, and\ \bibinfo {author}
  {\bibfnamefont {Stuart}\ \bibnamefont {Russell}}} (\bibinfo {year} {2023}),\
  \bibfield  {title} {\enquote {\bibinfo {title} {Adversarial policies beat
  superhuman go ais},}\ }\href {https://arxiv.org/abs/2211.00241} {\bibinfo
  {journal} {arXiv:2211.00241}\ }\BibitemShut {NoStop}%
\bibitem [{\citenamefont {Wang}\ and\ \citenamefont
  {Wang}(2024)}]{wang2024arbitrary}%
  \BibitemOpen
\bibfield  {journal} {  }\bibfield  {author} {\bibinfo {author} {\bibnamefont
  {Wang}, \bibfnamefont {Zhao-Wei}}, and\ \bibinfo {author} {\bibfnamefont
  {Zhao-Ming}\ \bibnamefont {Wang}}} (\bibinfo {year} {2024}),\ \bibfield
  {title} {\enquote {\bibinfo {title} {Arbitrary quantum states preparation
  aided by deep reinforcement learning},}\ }\href
  {https://arxiv.org/abs/2407.16368} {\bibinfo  {journal} {arXiv preprint
  arXiv:2407.16368}\ }\BibitemShut {NoStop}%
\bibitem [{\citenamefont {Wang}\ \emph
  {et~al.}(2024{\natexlab{b}})\citenamefont {Wang}, \citenamefont {Chen},
  \citenamefont {Du}, \citenamefont {Yang}, \citenamefont {Cai}, \citenamefont
  {Huang}, \citenamefont {Zhang}, \citenamefont {Xu}, \citenamefont {Du},
  \citenamefont {Li} \emph {et~al.}}]{wang2024quantum}%
  \BibitemOpen
\bibfield  {journal} {  }\bibfield  {author} {\bibinfo {author} {\bibnamefont
  {Wang}, \bibfnamefont {ZT}}, \bibinfo {author} {\bibfnamefont {Qiuhao}\
  \bibnamefont {Chen}}, \bibinfo {author} {\bibfnamefont {Yuxuan}\ \bibnamefont
  {Du}}, \bibinfo {author} {\bibfnamefont {ZH}~\bibnamefont {Yang}}, \bibinfo
  {author} {\bibfnamefont {Xiaoxia}\ \bibnamefont {Cai}}, \bibinfo {author}
  {\bibfnamefont {Kaixuan}\ \bibnamefont {Huang}}, \bibinfo {author}
  {\bibfnamefont {Jingning}\ \bibnamefont {Zhang}}, \bibinfo {author}
  {\bibfnamefont {Kai}\ \bibnamefont {Xu}}, \bibinfo {author} {\bibfnamefont
  {Jun}\ \bibnamefont {Du}}, \bibinfo {author} {\bibfnamefont {Yinan}\
  \bibnamefont {Li}},  \emph {et~al.}} (\bibinfo {year} {2024}{\natexlab{b}}),\
  \bibfield  {title} {\enquote {\bibinfo {title} {Quantum compiling with
  reinforcement learning on a superconducting processor},}\ }\href
  {https://arxiv.org/abs/2406.12195} {\bibfield  {journal} {\bibinfo  {journal}
  {arXiv}\ }}\Eprint {https://arxiv.org/abs/2406.12195} {2406.12195}
  \BibitemShut {NoStop}%
\bibitem [{\citenamefont {Watkins}\ and\ \citenamefont
  {Dayan}(1992)}]{watkins1992q-learning}%
  \BibitemOpen
  \bibfield  {author} {\bibinfo {author} {\bibnamefont {Watkins}, \bibfnamefont
  {Christopher~JCH}}, and\ \bibinfo {author} {\bibfnamefont {Peter}\
  \bibnamefont {Dayan}}} (\bibinfo {year} {1992}),\ \bibfield  {title}
  {\enquote {\bibinfo {title} {Q-learning},}\ }\href
  {https://doi.org/10.1007/BF00992698} {\bibfield  {journal} {\bibinfo
  {journal} {Machine learning}\ }\textbf {\bibinfo {volume} {8}},\ \bibinfo
  {pages} {279--292}}\BibitemShut {NoStop}%
\bibitem [{\citenamefont {Wauters}\ \emph {et~al.}(2020)\citenamefont
  {Wauters}, \citenamefont {Panizon}, \citenamefont {Mbeng},\ and\
  \citenamefont {Santoro}}]{wauters2020reinforcement}%
  \BibitemOpen
  \bibfield  {author} {\bibinfo {author} {\bibnamefont {Wauters}, \bibfnamefont
  {Matteo~M}}, \bibinfo {author} {\bibfnamefont {Emanuele}\ \bibnamefont
  {Panizon}}, \bibinfo {author} {\bibfnamefont {Glen~B.}\ \bibnamefont
  {Mbeng}}, and\ \bibinfo {author} {\bibfnamefont {Giuseppe~E.}\ \bibnamefont
  {Santoro}}} (\bibinfo {year} {2020}),\ \bibfield  {title} {\enquote {\bibinfo
  {title} {Reinforcement-learning-assisted quantum optimization},}\ }\href
  {https://doi.org/10.1103/PhysRevResearch.2.033446} {\bibfield  {journal}
  {\bibinfo  {journal} {Phys. Rev. Res.}\ }\textbf {\bibinfo {volume} {2}},\
  \bibinfo {pages} {033446}}\BibitemShut {NoStop}%
\bibitem [{\citenamefont {Weiden}\ \emph {et~al.}(2024)\citenamefont {Weiden},
  \citenamefont {Kalloor}, \citenamefont {Younis}, \citenamefont
  {Kubiatowicz},\ and\ \citenamefont {Iancu}}]{weiden2024learning}%
  \BibitemOpen
  \bibfield  {author} {\bibinfo {author} {\bibnamefont {Weiden}, \bibfnamefont
  {Mathias}}, \bibinfo {author} {\bibfnamefont {Justin}\ \bibnamefont
  {Kalloor}}, \bibinfo {author} {\bibfnamefont {Ed}~\bibnamefont {Younis}},
  \bibinfo {author} {\bibfnamefont {John}\ \bibnamefont {Kubiatowicz}}, and\
  \bibinfo {author} {\bibfnamefont {Costin}\ \bibnamefont {Iancu}}} (\bibinfo
  {year} {2024}),\ \bibfield  {title} {\enquote {\bibinfo {title} {High
  precision fault-tolerant quantum circuit synthesis by diagonalization using
  reinforcement learning},}\ }\href {https://arxiv.org/abs/2409.00433}
  {\bibinfo  {journal} {arXiv preprint arXiv:2409.00433}\ }\BibitemShut
  {NoStop}%
\bibitem [{\citenamefont {White}(2021)}]{white2021deep}%
  \BibitemOpen
\bibfield  {journal} {  }\bibfield  {author} {\bibinfo {author} {\bibnamefont
  {White}, \bibfnamefont {Andrew~D}}} (\bibinfo {year} {2021}),\ \bibfield
  {title} {\enquote {\bibinfo {title} {Deep learning for molecules and
  materials},}\ }\href {https://doi.org/10.33011/livecoms.3.1.1499} {\bibfield
  {journal} {\bibinfo  {journal} {Living Journal of Computational Molecular
  Science}\ }\textbf {\bibinfo {volume} {3}}~(\bibinfo {number} {1}),\ \bibinfo
  {pages} {1499}}\BibitemShut {NoStop}%
\bibitem [{\citenamefont {Williams}(1992)}]{williams1992simple}%
  \BibitemOpen
  \bibfield  {author} {\bibinfo {author} {\bibnamefont {Williams},
  \bibfnamefont {Ronald~J}}} (\bibinfo {year} {1992}),\ \bibfield  {title}
  {\enquote {\bibinfo {title} {Simple statistical gradient-following algorithms
  for connectionist reinforcement learning},}\ }\href
  {https://doi.org/10.1007/BF00992696} {\bibfield  {journal} {\bibinfo
  {journal} {Machine learning}\ }\textbf {\bibinfo {volume} {8}},\ \bibinfo
  {pages} {229--256}}\BibitemShut {NoStop}%
\bibitem [{\citenamefont {Wright}\ and\ \citenamefont
  {De~Sousa}(2023)}]{wright2023fast}%
  \BibitemOpen
  \bibfield  {author} {\bibinfo {author} {\bibnamefont {Wright}, \bibfnamefont
  {Emily}}, and\ \bibinfo {author} {\bibfnamefont {Rog{\'e}rio}\ \bibnamefont
  {De~Sousa}}} (\bibinfo {year} {2023}),\ \bibfield  {title} {\enquote
  {\bibinfo {title} {Fast quantum gate design with deep reinforcement learning
  using real-time feedback on readout signals},}\ }\bibfield  {booktitle}
  {\emph {\bibinfo {booktitle} {2023 IEEE International Conference on Quantum
  Computing and Engineering (QCE)}},\ }\href
  {https://doi.org/10.1109/QCE57702.2023.00146} {\bibfield  {journal} {\bibinfo
   {journal} {IEEE}\ }\textbf {\bibinfo {volume} {1}},\ \bibinfo {pages}
  {1295--1303}}\BibitemShut {NoStop}%
\bibitem [{\citenamefont {Wu}\ \emph {et~al.}(2019)\citenamefont {Wu},
  \citenamefont {Wang},\ and\ \citenamefont {Zhang}}]{wu2019solving}%
  \BibitemOpen
  \bibfield  {author} {\bibinfo {author} {\bibnamefont {Wu}, \bibfnamefont
  {Dian}}, \bibinfo {author} {\bibfnamefont {Lei}\ \bibnamefont {Wang}}, and\
  \bibinfo {author} {\bibfnamefont {Pan}\ \bibnamefont {Zhang}}} (\bibinfo
  {year} {2019}),\ \bibfield  {title} {\enquote {\bibinfo {title} {Solving
  statistical mechanics using variational autoregressive networks},}\ }\href
  {https://doi.org/10.1103/PhysRevLett.122.080602} {\bibfield  {journal}
  {\bibinfo  {journal} {Phys. Rev. Lett.}\ }\textbf {\bibinfo {volume} {122}},\
  \bibinfo {pages} {080602}}\BibitemShut {NoStop}%
\bibitem [{\citenamefont {Wu}\ \emph {et~al.}(2025)\citenamefont {Wu},
  \citenamefont {Jin}, \citenamefont {Wen}, \citenamefont {Han},\ and\
  \citenamefont {Wang}}]{wu2025quantum}%
  \BibitemOpen
  \bibfield  {author} {\bibinfo {author} {\bibnamefont {Wu}, \bibfnamefont
  {Shaojun}}, \bibinfo {author} {\bibfnamefont {Shan}\ \bibnamefont {Jin}},
  \bibinfo {author} {\bibfnamefont {Dingding}\ \bibnamefont {Wen}}, \bibinfo
  {author} {\bibfnamefont {Donghong}\ \bibnamefont {Han}}, and\ \bibinfo
  {author} {\bibfnamefont {Xiaoting}\ \bibnamefont {Wang}}} (\bibinfo {year}
  {2025}),\ \bibfield  {title} {\enquote {\bibinfo {title} {Quantum
  reinforcement learning in continuous action space},}\ }\href
  {https://doi.org/10.22331/q-2025-03-12-1660} {\bibfield  {journal} {\bibinfo
  {journal} {Quantum}\ }\textbf {\bibinfo {volume} {9}},\ \bibinfo {pages}
  {1660}}\BibitemShut {NoStop}%
\bibitem [{\citenamefont {Xiao}\ \emph {et~al.}(2022)\citenamefont {Xiao},
  \citenamefont {Fan},\ and\ \citenamefont {Zeng}}]{xiao2022parameter}%
  \BibitemOpen
  \bibfield  {author} {\bibinfo {author} {\bibnamefont {Xiao}, \bibfnamefont
  {Tailong}}, \bibinfo {author} {\bibfnamefont {Jianping}\ \bibnamefont {Fan}},
  and\ \bibinfo {author} {\bibfnamefont {Guihua}\ \bibnamefont {Zeng}}}
  (\bibinfo {year} {2022}),\ \bibfield  {title} {\enquote {\bibinfo {title}
  {Parameter estimation in quantum sensing based on deep reinforcement
  learning},}\ }\href {https://doi.org/10.1038/s41534-021-00513-z} {\bibfield
  {journal} {\bibinfo  {journal} {npj Quantum Information}\ }\textbf {\bibinfo
  {volume} {8}}~(\bibinfo {number} {1}),\ \bibinfo {pages} {2}}\BibitemShut
  {NoStop}%
\bibitem [{\citenamefont {Xu}\ \emph {et~al.}(2019)\citenamefont {Xu},
  \citenamefont {Li}, \citenamefont {Liu}, \citenamefont {Wang}, \citenamefont
  {Yuan},\ and\ \citenamefont {Wang}}]{xu2019generalizable}%
  \BibitemOpen
  \bibfield  {author} {\bibinfo {author} {\bibnamefont {Xu}, \bibfnamefont
  {Han}}, \bibinfo {author} {\bibfnamefont {Junning}\ \bibnamefont {Li}},
  \bibinfo {author} {\bibfnamefont {Liqiang}\ \bibnamefont {Liu}}, \bibinfo
  {author} {\bibfnamefont {Yu}~\bibnamefont {Wang}}, \bibinfo {author}
  {\bibfnamefont {Haidong}\ \bibnamefont {Yuan}}, and\ \bibinfo {author}
  {\bibfnamefont {Xin}\ \bibnamefont {Wang}}} (\bibinfo {year} {2019}),\
  \bibfield  {title} {\enquote {\bibinfo {title} {Generalizable control for
  quantum parameter estimation through reinforcement learning},}\ }\href
  {https://doi.org/10.1038/s41534-019-0198-z} {\bibfield  {journal} {\bibinfo
  {journal} {npj Quantum Information}\ }\textbf {\bibinfo {volume}
  {5}}~(\bibinfo {number} {1}),\ \bibinfo {pages} {82}}\BibitemShut {NoStop}%
\bibitem [{\citenamefont {Xu}\ \emph {et~al.}(2025)\citenamefont {Xu},
  \citenamefont {Xiao}, \citenamefont {Huang}, \citenamefont {He},
  \citenamefont {Fan},\ and\ \citenamefont {Zeng}}]{xu2025towards}%
  \BibitemOpen
  \bibfield  {author} {\bibinfo {author} {\bibnamefont {Xu}, \bibfnamefont
  {Hang}}, \bibinfo {author} {\bibfnamefont {Tailong}\ \bibnamefont {Xiao}},
  \bibinfo {author} {\bibfnamefont {Jingzheng}\ \bibnamefont {Huang}}, \bibinfo
  {author} {\bibfnamefont {Ming}\ \bibnamefont {He}}, \bibinfo {author}
  {\bibfnamefont {Jianping}\ \bibnamefont {Fan}}, and\ \bibinfo {author}
  {\bibfnamefont {Guihua}\ \bibnamefont {Zeng}}} (\bibinfo {year} {2025}),\
  \bibfield  {title} {\enquote {\bibinfo {title} {Towards heisenberg limit
  without critical slowing down via quantum reinforcement learning},}\ }\href
  {https://arxiv.org/abs/2503.02210} {\bibinfo  {journal} {arXiv preprint
  arXiv:2503.02210}\ }\BibitemShut {NoStop}%
\bibitem [{\citenamefont {Xu}\ \emph {et~al.}(2024)\citenamefont {Xu},
  \citenamefont {Ding}, \citenamefont {Mart\'{\i}n-Guerrero},\ and\
  \citenamefont {Chen}}]{xu2024robust}%
  \BibitemOpen
\bibfield  {journal} {  }\bibfield  {author} {\bibinfo {author} {\bibnamefont
  {Xu}, \bibfnamefont {Tian-Niu}}, \bibinfo {author} {\bibfnamefont
  {Yongcheng}\ \bibnamefont {Ding}}, \bibinfo {author} {\bibfnamefont
  {Jos\'e~D.}\ \bibnamefont {Mart\'{\i}n-Guerrero}}, and\ \bibinfo {author}
  {\bibfnamefont {Xi}~\bibnamefont {Chen}}} (\bibinfo {year} {2024}),\
  \bibfield  {title} {\enquote {\bibinfo {title} {Robust two-qubit gate with
  reinforcement learning and dropout},}\ }\href
  {https://doi.org/10.1103/PhysRevA.110.032614} {\bibfield  {journal} {\bibinfo
   {journal} {Phys. Rev. A}\ }\textbf {\bibinfo {volume} {110}},\ \bibinfo
  {pages} {032614}}\BibitemShut {NoStop}%
\bibitem [{\citenamefont {Yao}\ \emph {et~al.}(2020)\citenamefont {Yao},
  \citenamefont {Bukov},\ and\ \citenamefont {Lin}}]{yao2020policy}%
  \BibitemOpen
  \bibfield  {author} {\bibinfo {author} {\bibnamefont {Yao}, \bibfnamefont
  {Jiahao}}, \bibinfo {author} {\bibfnamefont {Marin}\ \bibnamefont {Bukov}},
  and\ \bibinfo {author} {\bibfnamefont {Lin}\ \bibnamefont {Lin}}} (\bibinfo
  {year} {2020}),\ \bibfield  {title} {\enquote {\bibinfo {title} {Policy
  gradient based quantum approximate optimization algorithm},}\ }\bibfield
  {booktitle} {\emph {\bibinfo {booktitle} {Mathematical and scientific machine
  learning}},\ }\href {https://doi.org/10.48550/arXiv.2002.01068} {\bibinfo
  {journal} {PMLR}\ ,\ \bibinfo {pages} {605--634}}\BibitemShut {NoStop}%
\bibitem [{\citenamefont {Yao}\ \emph {et~al.}(2022{\natexlab{a}})\citenamefont
  {Yao}, \citenamefont {Kottering}, \citenamefont {Gundlach}, \citenamefont
  {Lin},\ and\ \citenamefont {Bukov}}]{yao2022noise}%
  \BibitemOpen
\bibfield  {journal} {  }\bibfield  {author} {\bibinfo {author} {\bibnamefont
  {Yao}, \bibfnamefont {Jiahao}}, \bibinfo {author} {\bibfnamefont {Paul}\
  \bibnamefont {Kottering}}, \bibinfo {author} {\bibfnamefont {Hans}\
  \bibnamefont {Gundlach}}, \bibinfo {author} {\bibfnamefont {Lin}\
  \bibnamefont {Lin}}, and\ \bibinfo {author} {\bibfnamefont {Marin}\
  \bibnamefont {Bukov}}} (\bibinfo {year} {2022}{\natexlab{a}}),\ \bibfield
  {title} {\enquote {\bibinfo {title} {Noise-robust end-to-end quantum control
  using deep autoregressive policy networks},}\ }\bibfield  {booktitle} {\emph
  {\bibinfo {booktitle} {Mathematical and scientific machine learning}},\
  }\href {https://doi.org/10.48550/arXiv.2012.06701} {\bibinfo  {journal}
  {PMLR}\ ,\ \bibinfo {pages} {1044--1081}}\BibitemShut {NoStop}%
\bibitem [{\citenamefont {Yao}\ \emph {et~al.}(2022{\natexlab{b}})\citenamefont
  {Yao}, \citenamefont {Li}, \citenamefont {Bukov}, \citenamefont {Lin},\ and\
  \citenamefont {Ying}}]{yao2022monte}%
  \BibitemOpen
\bibfield  {journal} {  }\bibfield  {author} {\bibinfo {author} {\bibnamefont
  {Yao}, \bibfnamefont {Jiahao}}, \bibinfo {author} {\bibfnamefont {Haoya}\
  \bibnamefont {Li}}, \bibinfo {author} {\bibfnamefont {Marin}\ \bibnamefont
  {Bukov}}, \bibinfo {author} {\bibfnamefont {Lin}\ \bibnamefont {Lin}}, and\
  \bibinfo {author} {\bibfnamefont {Lexing}\ \bibnamefont {Ying}}} (\bibinfo
  {year} {2022}{\natexlab{b}}),\ \bibfield  {title} {\enquote {\bibinfo {title}
  {Monte carlo tree search based hybrid optimization of variational quantum
  circuits},}\ }\bibfield  {booktitle} {\emph {\bibinfo {booktitle}
  {Mathematical and Scientific Machine Learning}},\ }\href
  {https://doi.org/10.48550/arXiv.2203.16707} {\bibinfo  {journal} {PMLR}\ ,\
  \bibinfo {pages} {49--64}}\BibitemShut {NoStop}%
\bibitem [{\citenamefont {Yao}\ \emph {et~al.}(2021)\citenamefont {Yao},
  \citenamefont {Lin},\ and\ \citenamefont {Bukov}}]{yao2021reinforcement}%
  \BibitemOpen
\bibfield  {journal} {  }\bibfield  {author} {\bibinfo {author} {\bibnamefont
  {Yao}, \bibfnamefont {Jiahao}}, \bibinfo {author} {\bibfnamefont {Lin}\
  \bibnamefont {Lin}}, and\ \bibinfo {author} {\bibfnamefont {Marin}\
  \bibnamefont {Bukov}}} (\bibinfo {year} {2021}),\ \bibfield  {title}
  {\enquote {\bibinfo {title} {Reinforcement learning for many-body
  ground-state preparation inspired by counterdiabatic driving},}\ }\href
  {https://doi.org/10.1103/PhysRevX.11.031070} {\bibfield  {journal} {\bibinfo
  {journal} {Phys. Rev. X}\ }\textbf {\bibinfo {volume} {11}},\ \bibinfo
  {pages} {031070}}\BibitemShut {NoStop}%
\bibitem [{\citenamefont {Ye}\ \emph {et~al.}(2024)\citenamefont {Ye},
  \citenamefont {Arenz}, \citenamefont {Sinha}, \citenamefont {Lukens},\ and\
  \citenamefont {Lai}}]{ye2024entanglement}%
  \BibitemOpen
  \bibfield  {author} {\bibinfo {author} {\bibnamefont {Ye}, \bibfnamefont
  {Li-Li}}, \bibinfo {author} {\bibfnamefont {Christian}\ \bibnamefont
  {Arenz}}, \bibinfo {author} {\bibfnamefont {Kanu}\ \bibnamefont {Sinha}},
  \bibinfo {author} {\bibfnamefont {Joseph~M}\ \bibnamefont {Lukens}}, and\
  \bibinfo {author} {\bibfnamefont {Ying-Cheng}\ \bibnamefont {Lai}}} (\bibinfo
  {year} {2024}),\ \bibfield  {title} {\enquote {\bibinfo {title} {Entanglement
  engineering of optomechanical systems by reinforcement learning},}\ }\href
  {https://arxiv.org/abs/2406.04550} {\bibinfo  {journal} {arXiv e-prints}\ ,\
  \bibinfo {pages} {2406.04550}}\BibitemShut {NoStop}%
\bibitem [{\citenamefont {Ye}\ and\ \citenamefont
  {Lai}(2024)}]{ye2024controlling}%
  \BibitemOpen
\bibfield  {journal} {  }\bibfield  {author} {\bibinfo {author} {\bibnamefont
  {Ye}, \bibfnamefont {Li-Li}}, and\ \bibinfo {author} {\bibfnamefont
  {Ying-Cheng}\ \bibnamefont {Lai}}} (\bibinfo {year} {2024}),\ \bibfield
  {title} {\enquote {\bibinfo {title} {Controlling nonergodicity in quantum
  many-body systems by reinforcement learning},}\ }\href
  {https://arxiv.org/abs/2408.11989} {\bibinfo  {journal} {arXiv preprint
  arXiv:2408.11989}\ }\BibitemShut {NoStop}%
\bibitem [{\citenamefont {Zaklama}\ \emph {et~al.}(2025)\citenamefont
  {Zaklama}, \citenamefont {Guerci},\ and\ \citenamefont
  {Fu}}]{zaklama2025attention}%
  \BibitemOpen
\bibfield  {journal} {  }\bibfield  {author} {\bibinfo {author} {\bibnamefont
  {Zaklama}, \bibfnamefont {Timothy}}, \bibinfo {author} {\bibfnamefont
  {Daniele}\ \bibnamefont {Guerci}}, and\ \bibinfo {author} {\bibfnamefont
  {Liang}\ \bibnamefont {Fu}}} (\bibinfo {year} {2025}),\ \bibfield  {title}
  {\enquote {\bibinfo {title} {Attention-based foundation model for quantum
  states},}\ }\href {https://doi.org/10.48550/arXiv.2512.11962} {\bibinfo
  {journal} {arXiv preprint arXiv:2512.11962}\ }\BibitemShut {NoStop}%
\bibitem [{\citenamefont {Zen}\ \emph {et~al.}(2025)\citenamefont {Zen},
  \citenamefont {Olle}, \citenamefont {Colmenarez}, \citenamefont {Puviani},
  \citenamefont {M{\"u}ller},\ and\ \citenamefont
  {Marquardt}}]{zen2024quantum}%
  \BibitemOpen
\bibfield  {journal} {  }\bibfield  {author} {\bibinfo {author} {\bibnamefont
  {Zen}, \bibfnamefont {Remmy}}, \bibinfo {author} {\bibfnamefont {Jan}\
  \bibnamefont {Olle}}, \bibinfo {author} {\bibfnamefont {Luis}\ \bibnamefont
  {Colmenarez}}, \bibinfo {author} {\bibfnamefont {Matteo}\ \bibnamefont
  {Puviani}}, \bibinfo {author} {\bibfnamefont {Markus}\ \bibnamefont
  {M{\"u}ller}}, and\ \bibinfo {author} {\bibfnamefont {Florian}\ \bibnamefont
  {Marquardt}}} (\bibinfo {year} {2025}),\ \bibfield  {title} {\enquote
  {\bibinfo {title} {Quantum circuit discovery for fault-tolerant logical state
  preparation with reinforcement learning},}\ }\href
  {https://doi.org/10.1103/gqpr-dgz7} {\bibfield  {journal} {\bibinfo
  {journal} {Physical Review X}\ }\textbf {\bibinfo {volume} {15}}~(\bibinfo
  {number} {4}),\ \bibinfo {pages} {041012}}\BibitemShut {NoStop}%
\bibitem [{\citenamefont {Zhang}\ \emph {et~al.}(2025)\citenamefont {Zhang},
  \citenamefont {Miao}, \citenamefont {Pan}, \citenamefont {Tao},\ and\
  \citenamefont {Chen}}]{zhang2025meta}%
  \BibitemOpen
  \bibfield  {author} {\bibinfo {author} {\bibnamefont {Zhang}, \bibfnamefont
  {Shihui}}, \bibinfo {author} {\bibfnamefont {Zibo}\ \bibnamefont {Miao}},
  \bibinfo {author} {\bibfnamefont {Yu}~\bibnamefont {Pan}}, \bibinfo {author}
  {\bibfnamefont {Sibo}\ \bibnamefont {Tao}}, and\ \bibinfo {author}
  {\bibfnamefont {Yu}~\bibnamefont {Chen}}} (\bibinfo {year} {2025}),\
  \bibfield  {title} {\enquote {\bibinfo {title} {Meta-learning assisted robust
  control of universal quantum gates with uncertainties},}\ }\href
  {https://www.nature.com/articles/s41534-025-01034-9} {\bibfield  {journal}
  {\bibinfo  {journal} {npj Quantum Information}\ }\textbf {\bibinfo {volume}
  {11}}~(\bibinfo {number} {1}),\ \bibinfo {pages} {1--10}}\BibitemShut
  {NoStop}%
\bibitem [{\citenamefont {Zhang}\ \emph {et~al.}(2018)\citenamefont {Zhang},
  \citenamefont {Cui}, \citenamefont {Wang},\ and\ \citenamefont
  {Yung}}]{zhang2018automatic}%
  \BibitemOpen
  \bibfield  {author} {\bibinfo {author} {\bibnamefont {Zhang}, \bibfnamefont
  {Xiao-Ming}}, \bibinfo {author} {\bibfnamefont {Zi-Wei}\ \bibnamefont {Cui}},
  \bibinfo {author} {\bibfnamefont {Xin}\ \bibnamefont {Wang}}, and\ \bibinfo
  {author} {\bibfnamefont {Man-Hong}\ \bibnamefont {Yung}}} (\bibinfo {year}
  {2018}),\ \bibfield  {title} {\enquote {\bibinfo {title} {Automatic
  spin-chain learning to explore the quantum speed limit},}\ }\href
  {https://doi.org/10.1103/PhysRevA.97.052333} {\bibfield  {journal} {\bibinfo
  {journal} {Phys. Rev. A}\ }\textbf {\bibinfo {volume} {97}},\ \bibinfo
  {pages} {052333}}\BibitemShut {NoStop}%
\bibitem [{\citenamefont {Zhang}\ \emph {et~al.}(2019)\citenamefont {Zhang},
  \citenamefont {Wei}, \citenamefont {Asad}, \citenamefont {Yang},\ and\
  \citenamefont {Wang}}]{zhang2019does}%
  \BibitemOpen
  \bibfield  {author} {\bibinfo {author} {\bibnamefont {Zhang}, \bibfnamefont
  {Xiao-Ming}}, \bibinfo {author} {\bibfnamefont {Zezhu}\ \bibnamefont {Wei}},
  \bibinfo {author} {\bibfnamefont {Raza}\ \bibnamefont {Asad}}, \bibinfo
  {author} {\bibfnamefont {Xu-Chen}\ \bibnamefont {Yang}}, and\ \bibinfo
  {author} {\bibfnamefont {Xin}\ \bibnamefont {Wang}}} (\bibinfo {year}
  {2019}),\ \bibfield  {title} {\enquote {\bibinfo {title} {When does
  reinforcement learning stand out in quantum control? a comparative study on
  state preparation},}\ }\href {https://doi.org/10.1038/s41534-019-0201-8}
  {\bibfield  {journal} {\bibinfo  {journal} {npj Quantum Information}\
  }\textbf {\bibinfo {volume} {5}}~(\bibinfo {number} {1}),\ \bibinfo {pages}
  {85}}\BibitemShut {NoStop}%
\bibitem [{\citenamefont {Zhang}\ \emph {et~al.}(2020)\citenamefont {Zhang},
  \citenamefont {Zheng}, \citenamefont {Zhang},\ and\ \citenamefont
  {Deng}}]{zhang2020topological}%
  \BibitemOpen
  \bibfield  {author} {\bibinfo {author} {\bibnamefont {Zhang}, \bibfnamefont
  {Yuan-Hang}}, \bibinfo {author} {\bibfnamefont {Pei-Lin}\ \bibnamefont
  {Zheng}}, \bibinfo {author} {\bibfnamefont {Yi}~\bibnamefont {Zhang}}, and\
  \bibinfo {author} {\bibfnamefont {Dong-Ling}\ \bibnamefont {Deng}}} (\bibinfo
  {year} {2020}),\ \bibfield  {title} {\enquote {\bibinfo {title} {Topological
  quantum compiling with reinforcement learning},}\ }\href
  {https://doi.org/10.1103/PhysRevLett.125.170501} {\bibfield  {journal}
  {\bibinfo  {journal} {Phys. Rev. Lett.}\ }\textbf {\bibinfo {volume} {125}},\
  \bibinfo {pages} {170501}}\BibitemShut {NoStop}%
\bibitem [{\citenamefont {Zhao}\ \emph {et~al.}(2024)\citenamefont {Zhao},
  \citenamefont {Zhao}, \citenamefont {Li}, \citenamefont {Li}, \citenamefont
  {Liu}, \citenamefont {Guo},\ and\ \citenamefont {Yi}}]{zhao2024prepare}%
  \BibitemOpen
  \bibfield  {author} {\bibinfo {author} {\bibnamefont {Zhao}, \bibfnamefont
  {XL}}, \bibinfo {author} {\bibfnamefont {YM}~\bibnamefont {Zhao}}, \bibinfo
  {author} {\bibfnamefont {M}~\bibnamefont {Li}}, \bibinfo {author}
  {\bibfnamefont {TT}~\bibnamefont {Li}}, \bibinfo {author} {\bibfnamefont
  {Q}~\bibnamefont {Liu}}, \bibinfo {author} {\bibfnamefont {S}~\bibnamefont
  {Guo}}, and\ \bibinfo {author} {\bibfnamefont {XX}~\bibnamefont {Yi}}}
  (\bibinfo {year} {2024}),\ \bibfield  {title} {\enquote {\bibinfo {title} {A
  strategy for preparing quantum squeezed states using reinforcement
  learning},}\ }\href {https://arxiv.org/abs/2401.16320} {\bibinfo  {journal}
  {arXiv preprint arXiv:2401.16320}\ }\BibitemShut {NoStop}%
\bibitem [{\citenamefont {Zhou}\ \emph {et~al.}(2023)\citenamefont {Zhou},
  \citenamefont {Ma}, \citenamefont {Kuang},\ and\ \citenamefont
  {Dong}}]{zhou2023auxiliary}%
  \BibitemOpen
\bibfield  {journal} {  }\bibfield  {author} {\bibinfo {author} {\bibnamefont
  {Zhou}, \bibfnamefont {Shumin}}, \bibinfo {author} {\bibfnamefont {Hailan}\
  \bibnamefont {Ma}}, \bibinfo {author} {\bibfnamefont {Sen}\ \bibnamefont
  {Kuang}}, and\ \bibinfo {author} {\bibfnamefont {Daoyi}\ \bibnamefont
  {Dong}}} (\bibinfo {year} {2023}),\ \bibfield  {title} {\enquote {\bibinfo
  {title} {Auxiliary task-based deep reinforcement learning for quantum
  control},}\ }\href {https://doi.org/10.48550/arXiv.2302.14312} {\bibinfo
  {journal} {arXiv preprint arXiv:2302.14312}\ }\BibitemShut {NoStop}%
\bibitem [{\citenamefont {Zhu}\ \emph {et~al.}(2024)\citenamefont {Zhu},
  \citenamefont {Xiao}, \citenamefont {Zeng}, \citenamefont {Chiribella},\ and\
  \citenamefont {Wu}}]{zhu2024controlling}%
  \BibitemOpen
\bibfield  {journal} {  }\bibfield  {author} {\bibinfo {author} {\bibnamefont
  {Zhu}, \bibfnamefont {Yan}}, \bibinfo {author} {\bibfnamefont {Tailong}\
  \bibnamefont {Xiao}}, \bibinfo {author} {\bibfnamefont {Guihua}\ \bibnamefont
  {Zeng}}, \bibinfo {author} {\bibfnamefont {Giulio}\ \bibnamefont
  {Chiribella}}, and\ \bibinfo {author} {\bibfnamefont {Yadong}\ \bibnamefont
  {Wu}}} (\bibinfo {year} {2024}),\ \bibfield  {title} {\enquote {\bibinfo
  {title} {Controlling unknown quantum states via data-driven state
  representations},}\ }\href {https://arxiv.org/abs/2406.05711} {\bibinfo
  {journal} {arXiv preprint arXiv:2406.05711}\ }\BibitemShut {NoStop}%
\end{thebibliography}%
